\documentclass[12pt]{book}

\usepackage[francais]{babel,minitoc}		
\usepackage[T1]{fontenc}					
\usepackage[latin1]{inputenc}				%

\usepackage[left=2.5cm,right=2.5cm,top=2.5cm,bottom=2.5cm]{geometry} 	

\usepackage{xspace} 				

\usepackage{soul}				
\usepackage{textcomp} 			
\usepackage{setspace} 			

\usepackage{graphicx,color}
\usepackage{epic}				
\usepackage{eepic}				

\usepackage{float}

\usepackage{amsmath}			
\usepackage{amssymb}			
\usepackage{mathrsfs}			
\usepackage{stmaryrd}			

\usepackage[
colorlinks=true
,urlcolor=blue
,anchorcolor=blue
,citecolor=blue
,filecolor=blue
,linkcolor=blue
,menucolor=blue
,pagecolor=blue
,linktocpage=true
,pdfproducer=medialab
]{hyperref}

\usepackage{array}
\usepackage{multirow}

\usepackage{cite}				

\usepackage{fancyhdr}
\usepackage[Lenny]{fncychap}

\newcommand{\BE}{\begin{equation}}
\newcommand{\EE}{\end{equation}}
\newcommand{\BC}{\begin{center}}
\newcommand{\EC}{\end{center}}
\newcommand{\BF}{\begin{figure}}
\newcommand{\EF}{\end{figure}}
\newcommand{\bs}{\bigskip}
\newcommand{\ita}[1]{\textit{#1}}

\newcommand{\lp}{\left(}
\newcommand{\rp}{\right)}
\newcommand{\lb}{\left[}
\newcommand{\rb}{\right]}

\newcommand{\dd}{\mathrm{d}}
\newcommand{\mc}[1]{\mathcal{#1}}
\newcommand{\w}{\wedge}
\newcommand{\p}{\partial}
\newcommand{\n}{\nabla}

\newcommand{\al}{\alpha}
\newcommand{\be}{\beta}
\newcommand{\ga}{\gamma}
\newcommand{\om}{\omega}
\newcommand{\de}{\delta}
\newcommand{\si}{\sigma}
\newcommand{\la}{\lambda}
\newcommand{\ka}{\kappa}
\newcommand{\ep}{\epsilon}
\newcommand{\varep}{\varepsilon}
\newcommand{\Ga}{\Gamma}
\newcommand{\Om}{\Omega}
\newcommand{\La}{\Lambda}
\newcommand{\Th}{\Theta}
\newcommand{\Si}{\Sigma}

\newtheorem{proposition}{Résultat}[chapter]
\newtheorem{dico}{Définition}[chapter]
\newtheorem{principe}{Principe}[chapter]
\newtheorem{axiome}{Axiome}[chapter]
\begin{document}
\frontmatter 
\thispagestyle{empty}
\begin{flushright} \small \tt LPT-ORSAY 12-106\end{flushright}
\BF[H] \BC
\includegraphics[height=2.5cm]{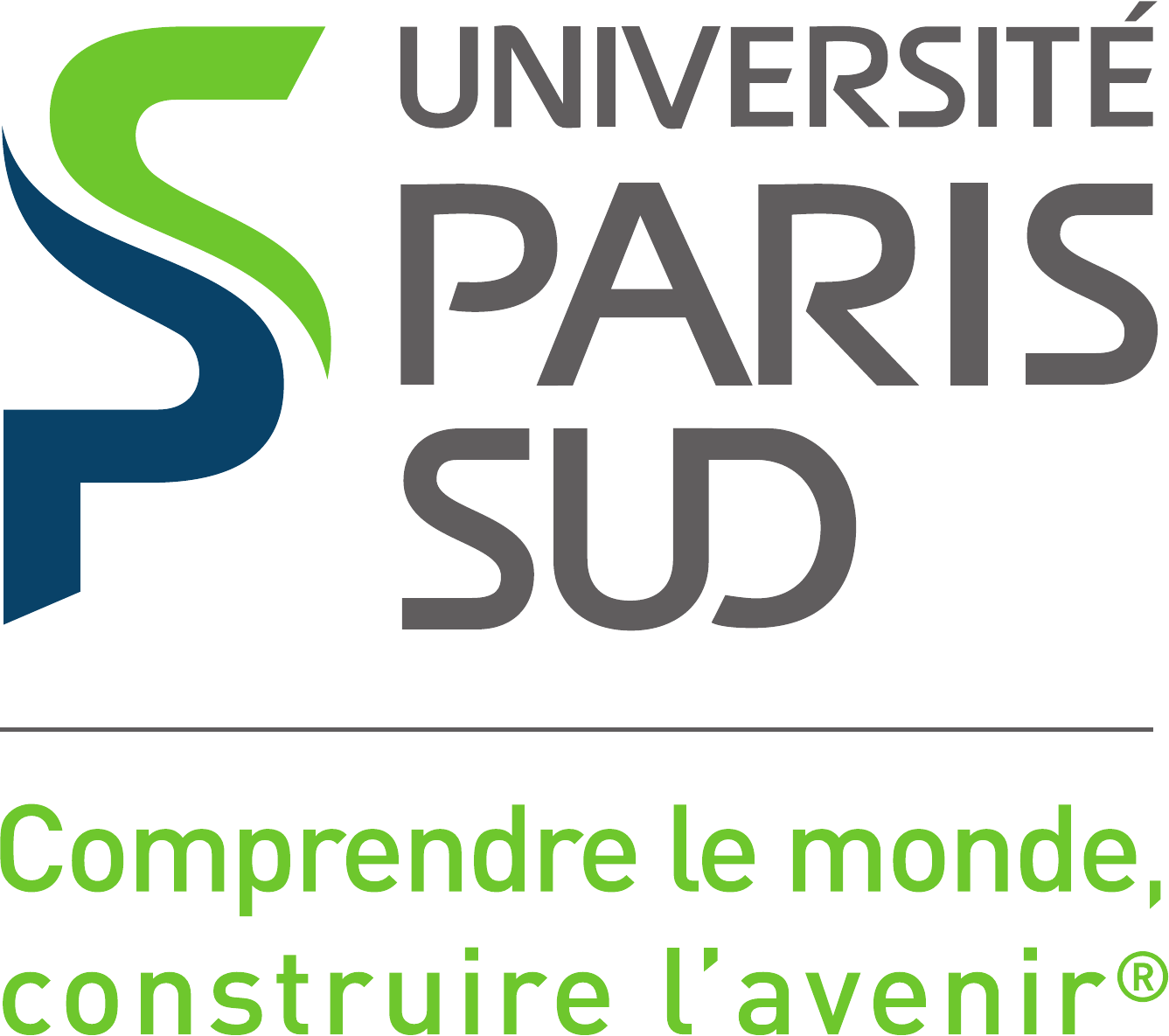} \qquad\qquad\qquad\qquad
\includegraphics[height=2.5cm]{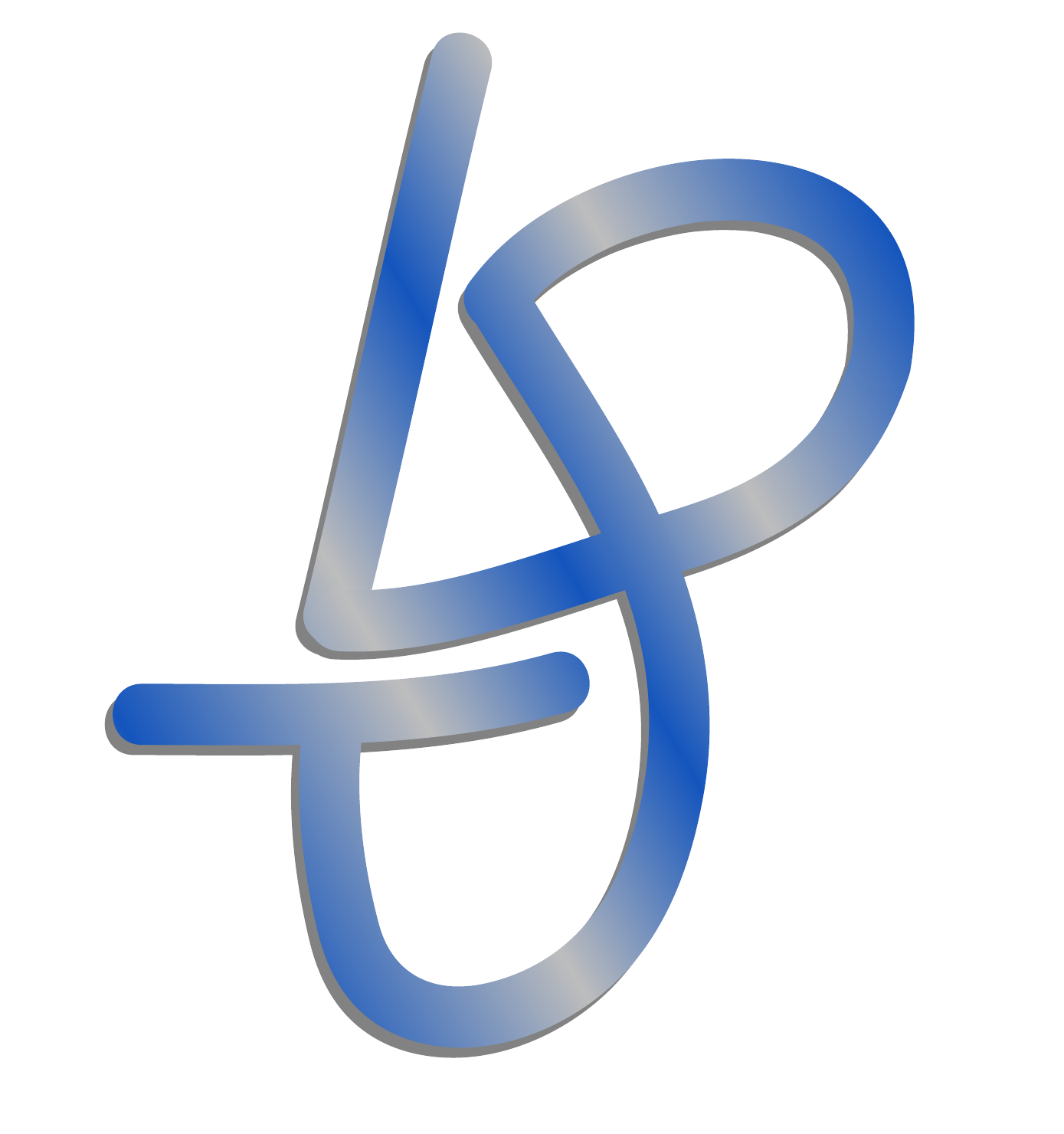} \qquad\qquad\qquad\qquad
\includegraphics[height=2.5cm]{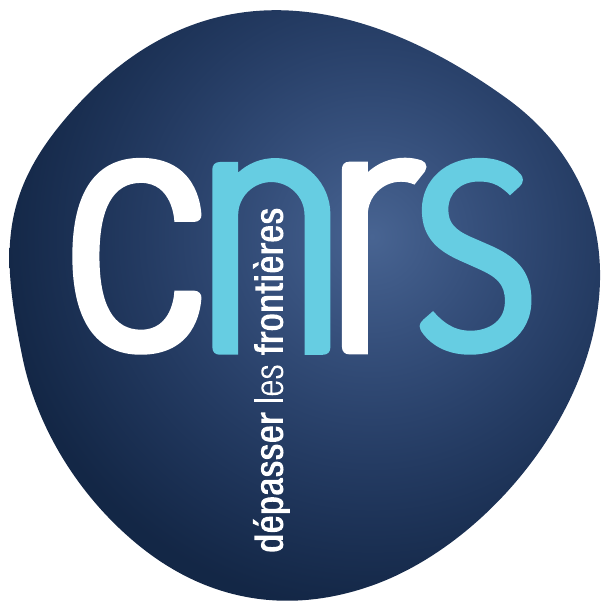} 
\EC \EF
\BC \textbf{TH\`ESE} \EC \medskip
\BC Présentée pour obtenir \EC \medskip
\BC \textbf{LE GRADE DE DOCTEUR EN SCIENCES DE L'UNIVERSIT\'E PARIS-SUD} \EC \medskip
\BC Spécialité: Physique théorique\EC \medskip
\bs\bs\bs\bs
\BC \textbf{\textit{\LARGE{Trous noirs dans des théories modifiées de la gravitation}}} \EC
\bs\bs\bs\bs\bs\bs\bs
\BC par \EC \medskip
\BC \textbf{Yannis Bardoux} \EC
\bs\bs\bs\bs\bs\bs\bs
\BC Soutenue le 24 septembre 2012 devant la commission d'examen : \EC
$$\begin{array}{l l}
\text{Marco Caldarelli} & \text{Invité} \\
\text{Christos Charmousis} & \text{Directeur de thèse} \\
\text{Roberto Emparan} & \text{Examinateur} \\
\text{Karim Noui} & \text{Rapporteur} \\
\text{Renaud Parentani} & \text{Examinateur} \\
\text{Marios Petropoulos} & \text{Examinateur} \\
\text{Simon Ross} & \text{Rapporteur} \\
\end{array}$$

\newpage\thispagestyle{plain}
\hspace{5.5cm}
\begin{minipage}{14cm}
\vspace{21cm}
Thèse préparée dans le cadre de l'Ecole Doctorale 107 au \\
Laboratoire de Physique Théorique d'Orsay (UMR 8627) \\
Bât. 210, Université Paris-Sud 11, 91405 Orsay Cedex
\end{minipage}

\cleardoublepage\thispagestyle{plain}\newpage\thispagestyle{plain}
\BC \it{Remerciements} \EC

{\itshape
Je tiens tout d'abord à remercier très sincèrement Christos Charmousis pour avoir accepté de diriger cette thèse. Il a su me prodiguer des conseils judicieux dans les moments difficiles et être présent durant ces trois ans tout en me laissant une grande liberté dans mon travail de recherche.
\bs

Je souhaite également remercier chaque membre de mon jury en commençant par Karim Noui et Simon Ross qui ont eu la gentillesse d'être les rapporteurs de cette thèse. Un grand merci à Marco Caldarelli avec qui j'ai eu le plaisir de collaborer et qui a su aussi être un pédagogue hors pair. Je voudrais aussi remercier Renaud Parentani qui m'a enseigné en L3 la Relativité Générale avec une passion communicative au sein du magistère de physique d'Orsay. Enfin, la présence de Marios Petropoulos et Roberto Emparan, tous deux spécialistes reconnus de l'interface entre la théorie des cordes et la Relativité Générale, est pour moi un honneur.
\bs

Le bon déroulement de cette thèse au sein du Laboratoire de Physique Théorique d'Orsay a aussi eu lieu grâce à la sympathie des membres de notre groupe Cosmo. J'en profite pour remercier Blaise Goutéraux pour ses conseils au début de ma thèse. J'ai également apprécié les nombreuses discussions avec Robin Zegers, qui m'a fait partager ses connaissances en mathématiques. Le laboratoire ne pourrait pas non plus être ce qu'il est sans la présence précieuse de Patricia Dubois-Violette, qui a toujours su me dénicher les vieux articles, et de toute l'équipe administrative. Je n'oublierai pas non plus les innombrables discussions enrichissantes avec les autres doctorants du laboratoire et nos séminaires Sinje dont celui de Marc Geiller en particulier. 
\bs

Plus généralement, je souhaite remercier l'université Paris-Sud dans laquelle j'ai découvert la physique moderne et où j'ai eu le plaisir d'enseigner dans le cadre de mon monitorat auprès d'Hervé Bergeron et d'Arne Keller. Je dois également beaucoup à tous les enseignants qui, un jour, par une expérience de pensée, ont fait naître en moi une passion pour la physique.
\bs

C'est aussi en ami que je tiens à remercier chaleureusement Antonin Coutant avec qui j'ai eu le plaisir de partager mon bureau. Nos discussions sur la science, l'écologie et la politique m'ont beaucoup apporté. J'ai également apprécié nos réflexions spéculatives autour de la gravité quantique et j'espère qu'il apportera une contribution à ce problème. 
\bs

Je souhaite bien évidemment remercier mes parents pour m'avoir laissé une totale liberté dans mes études et ma grand-mère pour son soutien. 
\bs

Enfin, les mots me manquent pour remercier Caroline qui a eu l'extrême gentillesse de relire ce manuscrit et qui par son amour me permet d'avancer jour après jour.
}

\newpage\thispagestyle{plain}

\BC\begin{minipage}{14cm}
\vspace{12cm}
\BC\includegraphics[scale=0.2]{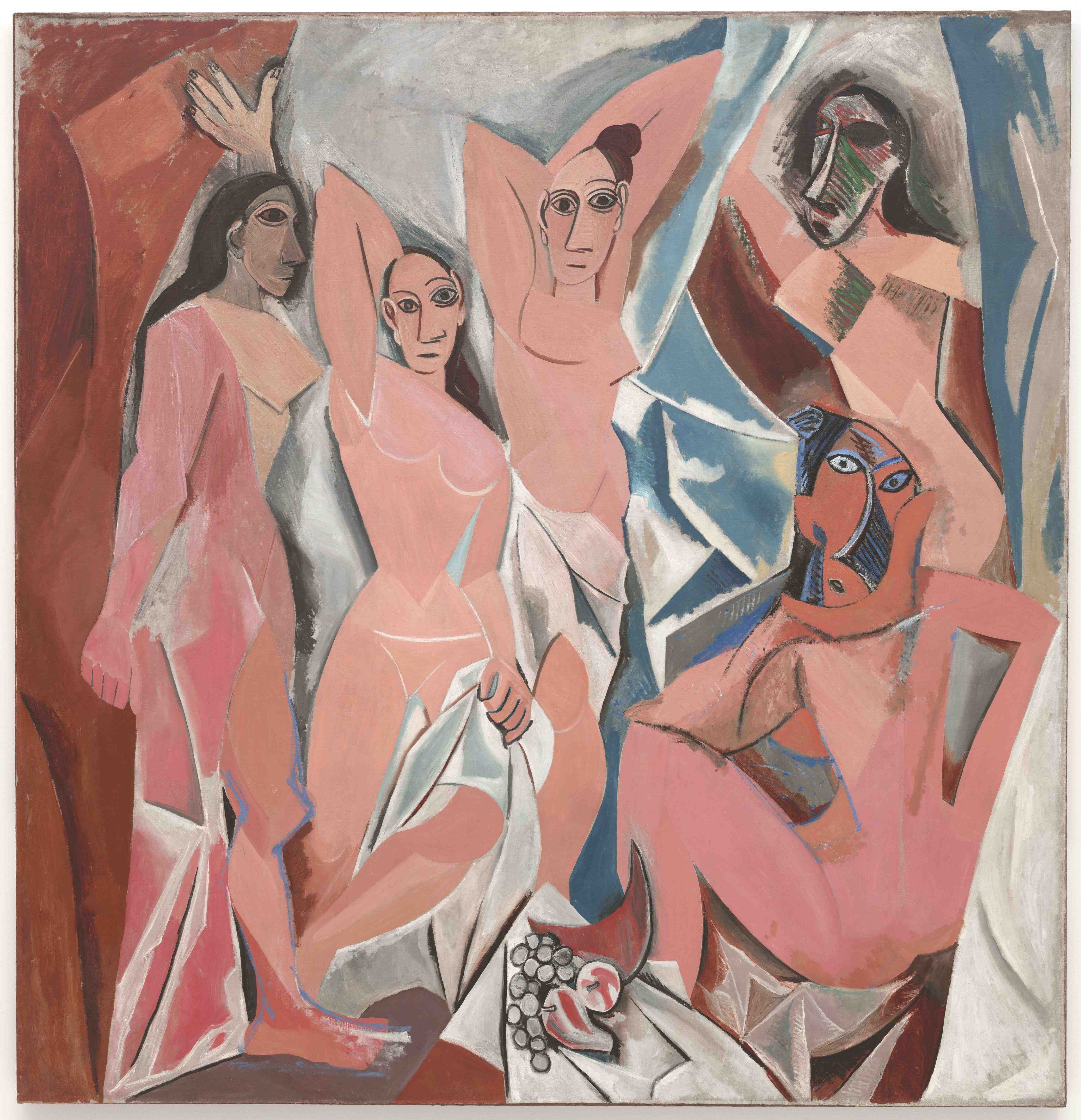} \EC
A l'image des théories physiques qui introduisent des dimensions supplémentaires, le cubisme propose un regard nouveau sur l'espace avec ici \textit{Les demoiselles d'Avignon} de Pablo Picasso (1907). 
\end{minipage} \EC

\dominitoc\tableofcontents
\mainmatter 
\chapter*{Introduction\markboth{\textit{Introduction}}{Introduction} }
\phantomsection\addstarredchapter{Introduction}

Le début du XX$^\text{e}$ siècle a connu deux grandes révolutions scientifiques avec d'une part la \textit{Mécanique Quantique} et d'autre part la \textit{Relativité Générale} qui est au c\oe ur de cette thèse. La formulation finale de cette théorie de la gravitation en Novembre 1915 est due à l'accomplissement intellectuel d'un seul homme : \textit{Albert Einstein}. Après avoir formulé la Relativité Restreinte en 1905, avec d'autres scientifiques contemporains comme Henri Poincaré, dans laquelle les concepts d'espace et de temps sont unifiés en une seule entité appelée l'\textit{espace-temps}, A. Einstein comprit rapidement qu'il n'était pas possible d'inclure les phénomènes gravitationnels dans ce cadre. Citons ce passage d'A. Einstein \cite{einstein} : 

\textit{"Je me décidai à rejeter comme illusoire cet essai que j'exposai plus haut : je ne traiterai plus dès lors le problème de la gravitation dans le cadre de la théorie de la relativité restreinte. Car ce cadre ne correspond absolument pas à la propriété fondamentale de la gravitation. Désormais le principe de l'égalité de la masse inerte et de la masse pesante peut s'expliciter de façon parfaite : dans un champ de gravitation homogène, tous les mouvements s'exécutent comme en l'absence de champ de gravitation, par rapport à un système de coordonnées uniformément accéléré. Si ce principe peut s'appliquer à n'importe quel événement, j'ai une preuve que le principe de relativité pourrait être appliqué à des systèmes de coordonnées qui exécutent un mouvement non uniforme les uns par rapport aux autres. Tout ceci supposait que je veuille aboutir à une théorie naturelle du champ de gravitation. Des réflexions de ce type m'occupèrent de 1908 à 1911 et je m'efforçai d'aboutir à des résultats particuliers dont je ne parlerai pas ici : pour moi j'avais acquis une base solide : j'avais découvert que je ne parviendrais à une théorie rationnelle de la gravitation que par une extension du principe de relativité."}

En effet, il parvint à une telle théorie dans laquelle la gravitation est encodée dans la géométrie de l'espace-temps, qui est désormais une entité dynamique. Cette théorie a véritablement le statut de théorie physique car, en plus de donner une explication au problème de l'époque sur l'avance du périhélie de Mercure, elle fournit un grand nombre de prédictions et de phénomènes nouveaux. De cette théorie naquit par exemple la cosmologie et la physique des trous noirs. Ces derniers objets furent ainsi nommés par John Wheeler et véritablement compris à la fin des années 1950 alors que leurs géométries étaient connues depuis 1916. Alors que les trous noirs sont issus de l'effondrement gravitationnel d'astres très massifs décrits par pléthore de nombres quantiques, ces objets macroscopiques sont simplement décrits par leur masse, leur moment angulaire et leur charge électrique dans le cadre de la Relativité Générale en présence de l'interaction électromagnétique. Ce fait surprenant est résumé dans la fameuse phrase de J. Wheeler : \textit{"A black hole has no hair"}. Ces trous noirs sont en quelque sorte à la Relativité Générale ce que les particules sont à la Mécanique Quantique et ils seront le sujet principal de cette thèse. 

Une solution pour expliquer ce problème de perte d'information jaillit au début des années 1970 : avec d'une part les lois de la mécanique des trous noirs qui sont analogues aux principes de la thermodynamique et avec d'autre part la découverte par Stephen Hawking qu'un trou noir rayonne comme un corps noir dans le cadre de la théorie quantique des champs en espace-temps courbe. De ce constat, une entropie proportionnelle à l'aire de l'horizon d'un trou noir fut alors associée à cet objet pour compenser la perte d'information. Néanmoins, il reste à trouver un cadre théorique dans lequel cette quantité macroscopique est associée aux micro-états d'un système thermodynamique. La théorie des cordes et la théorie de la gravitation à boucles semblent être actuellement des cadres possibles. \bs

L'intérêt majeur des travaux exposés dans cette thèse est d'explorer la "chevelure" des trous noirs dans des cadres plus généraux avec la présence d'une constante cosmologique, de dimensions supplémentaires, de champs de matière plus exotiques ou de termes de courbure de rang plus élevé. Ces extensions de la Relativité Générale peuvent s'inscrire dans le cadre de la théorie des cordes mais cette dernière n'est pas notre motivation première. C'est en étudiant des extensions naturelles de la Relativité Générale que nous pouvons aussi mieux comprendre cette dernière. 
La thèse que nous présentons ici s'appuie sur trois articles \cite{Bardoux:2010sq,Bardoux:2012aw,Bardoux} qui ne sont pas présentés de manière chronologique mais de façon thématique.

Dans un premier temps, nous exposerons la théorie de la Relativité Générale avec notamment les principes sur lesquels elle s'appuie et en donnant tous les éléments mathématiques dont nous avons besoin pour la suite. Puis, une première extension sera présentée au chapitre \ref{BH in higher D} avec l'introduction de dimensions supplémentaires et de champs de $p$-formes qui constituent la généralisation naturelle de l'interaction électromagnétique. Nous construirons dans ce cadre de nouvelles solutions statiques de trous noirs présentées dans \cite{Bardoux:2012aw}. Les $p$-formes vont permettre de modeler la géométrie de l'horizon d'un trou noir. Nous exposerons ensuite au chapitre \ref{EGB} l'extension la plus naturelle de la théorie d'Einstein en dimension quelconque qui génère des équations du second ordre en la métrique : la théorie de Lovelock. Nous déterminerons dans ce contexte une large classe de solutions en dimension 6 pour laquelle la théorie se réduit à celle d'Einstein-Gauss-Bonnet avec toujours la présence de $p$-formes \cite{Bardoux:2010sq}. Enfin, le chapitre \ref{chapter-conforme} sera consacré à l'étude d'une généralisation de la Relativité Générale en dimension 4 dont la modification est induite par un champ scalaire couplé conformément à la gravitation. Nous exhiberons notamment une nouvelle solution de trou noir avec un horizon plat dans cette théorie en présence de champs axioniques \cite{Bardoux}. Pour clore cette thèse, l'aspect thermodynamique des théories présentées aux chapitres \ref{BH in higher D} et \ref{chapter-conforme} sera détaillé dans le dernier chapitre ; ce qui permettra de déterminer la masse et les charges de ces nouvelles solutions et d'étudier des phénomènes de transitions de phase en présence d'un champ scalaire conforme.

\chapter{La Relativité Générale}

Pour une présentation complète de la Relativité Générale, nous renvoyons le lecteur aux ouvrages de référence : \cite{wald1984general,misner1973gravitation,hawking1975large,d1993introducing}. Notre but ici est simplement de présenter les grands principes de cette théorie et les objets mathématiques nécessaires afin d'aborder les chapitres suivants. Les prérequis pour ce chapitre sont la Mécanique Newtonienne et la Relativité Restreinte.

\minitoc
	\section{Les principes physiques}
Cette première section a pour objectif de dresser les principes qui ont mené à la construction de la Relativité Générale formulée par Albert Einstein en 1915. Commençons en rappelant que la seconde loi de Newton s'applique dans un référentiel dit \ita{galiléen} ou \ita{inertiel}; cependant l'introduction de \ita{forces inertielles} dans cette loi permet de traiter le cas des référentiels non-galiléens. Une question fondamentale se pose alors : comment distinguer un référentiel galiléen ? Afin de résoudre cette difficulté, Issac Newton introduisit l'existence d'un \ita{espace absolu} comme référentiel galiléen, qui ne serait pas influencé par la matière présente dans l'univers. A l'inverse, Ernst Mach rejette l'idée que la notion de \ita{mouvement} soit indépendante du contenu matériel de l'univers. Selon lui, seul le \ita{mouvement relatif} par rapport à la distribution des masses présentes dans l'univers a un sens. A. Einstein fut un fervent partisan de cette idée et l'érigea comme un principe afin de construire sa théorie de la gravitation : 
\begin{principe}[de Mach]
La distribution de matière dans l'univers détermine le mouvement des particules.
\end{principe}
Ainsi les forces inertielles que subit un objet seraient induites par la répartition des masses dans l'univers. Soulevons la coïncidence suivante avec la vision newtonienne. Considérons un pendule idéal placé au pôle Nord de la Terre. Le pendule oscille par rapport à l'espace absolu de Newton;  quant à la Terre, elle est en rotation par rapport à cet espace. Un observateur est alors en mesure de déterminer la durée nécessaire pour que le plan dans lequel le pendule oscille effectue un tour. Ce même observateur peut aussi mesurer la période de rotation de la Terre sur elle-même pour effectuer un tour par rapport aux étoiles fixes. Il trouve alors que les deux périodes sont les mêmes. En acceptant le principe de Mach, cette coïncidence est par conséquent effacée dans le cadre newtonien. Précisons un dernier point, la vision de Mach ne rejette pas une \ita{variation de la constante} de gravitation $G$; nous reviendrons par la suite à cette idée avec les théories dites \ita{tenseur-scalaire} à la sous-section \ref{tenseur-scalaire}.
\bs

Passons désormais au principe fondateur de la Relativité Générale : le principe d'équivalence. En Mécanique Newtonienne, \ita{l'universalité de la chute libre} provient de nouveau d'une coïncidence qui correspond à l'égalité entre la masse inertielle $m_I$, qui apparaît dans la seconde loi de Newton, et la masse grave $m_G$ de la force de gravitation : $\vec{F} = - m_G \vec{\n} \phi$ où $\phi$ représente le potentiel gravitationnel. Expérimentalement l'égalité $m_I = m_G$ est vérifiée à $10^{-12}$ près. Actuellement, le projet MICROSCOPE du CNES est en cours pour affiner cette mesure. Le lecteur pourra consulter \cite{Will:2005va} pour plus de détails sur la vérification de cette égalité.  En Relativité Générale, cette coïncidence est élevée de nouveau au rang de principe : le mouvement d'une particule-test dans un champ gravitationnel est indépendant de sa masse et de sa composition. Nous appelons ici \ita{particule-test} une particule qui interagit avec le champ gravitationnel présent mais qui ne l'altère pas. Discutons des conséquences de cette égalité en physique newtonienne en présence du champ de pesanteur terrestre. Considérons une masse $m$ en chute libre dans le référentiel terrestre $\mc R$ considéré galiléen, son accélération est donnée par 
\BE \vec{a}_{/ \mc R} = \vec{g} \EE 
où $\vec{g}$ désigne l'accélération de la pesanteur. Il est aussi loisible de décrire ce mouvement dans un référentiel $\mc R'$ en translation par rapport à $\mc R$ avec une accélération $\vec{a}_{\mc R' / \mc R} = \vec{g}$ d'où : 
\BE \vec{a}_{/ \mc R'} = \vec{a}_{/ \mc R} - \vec{g} \ . \EE
L'introduction de la force d'inertie $-m\vec{a}_{\mc R' / \mc R}$ a donc pour effet d'effacer la force de gravitation du point de vue du référentiel $\mc R'$ en chute libre. Evoquons également une expérience de pensée (Gedankenexperiment en allemand, comme aimait dire A. Einstein) pour illustrer ce résultat. Considérons un astronaute dans une fusée sans fenêtre. Si la fusée est au repos sur Terre, il peut  effectuer l'expérience \ita{locale} qui consiste à faire tomber une masse. Cette situation est en fait indistinguable de celle qui consiste à faire tomber cette même masse pendant que la fusée se déplace par rapport à un observateur inertiel avec une accélération $-\vec{g}$. Ceci illustre de nouveau cette dualité entre force de gravitation et force d'inertie et nous amène à formuler le principe suivant : 
\begin{principe}[d'équivalence]
En chaque point de l'espace-temps et en présence d'un champ de gravitation, il est possible d'adopter un système de coordonnées dit localement inertiel tel que, dans un voisinage suffisamment petit,  toutes les lois de la dynamique prennent la forme qu'elles ont dans l'espace de Minkowski dans un système de coordonnées cartésien.
\end{principe}
Nous commençons à entrevoir par ce principe un lien profond entre gravitation et géométrie. \\

Terminons cette section par un dernier principe. Nous pouvons attacher à chaque observateur un système de coordonnées privilégié ; aussi certaines coordonnées sont plus judicieuses en fonction des symétries du problème. Cependant, les lois de la physique ne doivent pas dépendre du système de coordonnées. Ce qui nous amène au dernier principe : 
\begin{principe}[de covariance]
Les équations de la physique doivent prendre une forme tensorielle.
\end{principe}
Nous voyons qu'une formulation précise est nécessaire. Une traduction mathématique de ces principes par l'introduction d'outils de géométrie différentielle va faire l'objet de la section suivante. Nous établirons également les axiomes de la Relativité Générale durant les deux prochaines sections.
	\section{Eléments de géométrie différentielle}
		\subsection{Variété et champs tensoriels}
L'ensemble des points de l'espace-temps va être encodé dans une nouvelle structure : la variété. Approximativement, une \ita{variété} est un ensemble constitué d'espaces qui ressemblent localement à un ouvert de $\mathbb{R}^D$ et qui peuvent être collés de manière continue. Plus précisément :

\begin{dico}[variété]Une variété différentielle réelle de dimension $D$, notée $\mc M$, est un ensemble muni d'une collection de sous-ensembles $\{\mc O_i \}$ satisfaisant les trois propriétés suivantes : 
\begin{itemize}
	\item Pour chaque élément $P$, appelé point, de $\mc M$, il existe au moins un $\mc O_i$ contenant $P$, ce qui signifie que $\{\mc O_i \}$ couvre $\mc M$.
	\item Pour chaque $i$, il existe une bijection $\psi_i$ de  $\mc O_i$ dans un ouvert $U_i$ de $\mathbb{R}^D$.
	\item Si $\mc O_i \cap \mc O_j \neq \varnothing$ alors $\psi_j \circ \psi_i^{-1}$, amenant les points de $\psi_i \left( \mc O_i \cap O_j \right) \subset U_i$ dans $\psi_j \left( \mc O_i \cap O_j \right)$ $\subset U_j$,  est une application infiniment différentiable de  $\mathbb{R}^D$ dans lui-même. 
\end{itemize}\end{dico}
\BF[H]\BC \includegraphics[scale=0.45,angle=1.5]{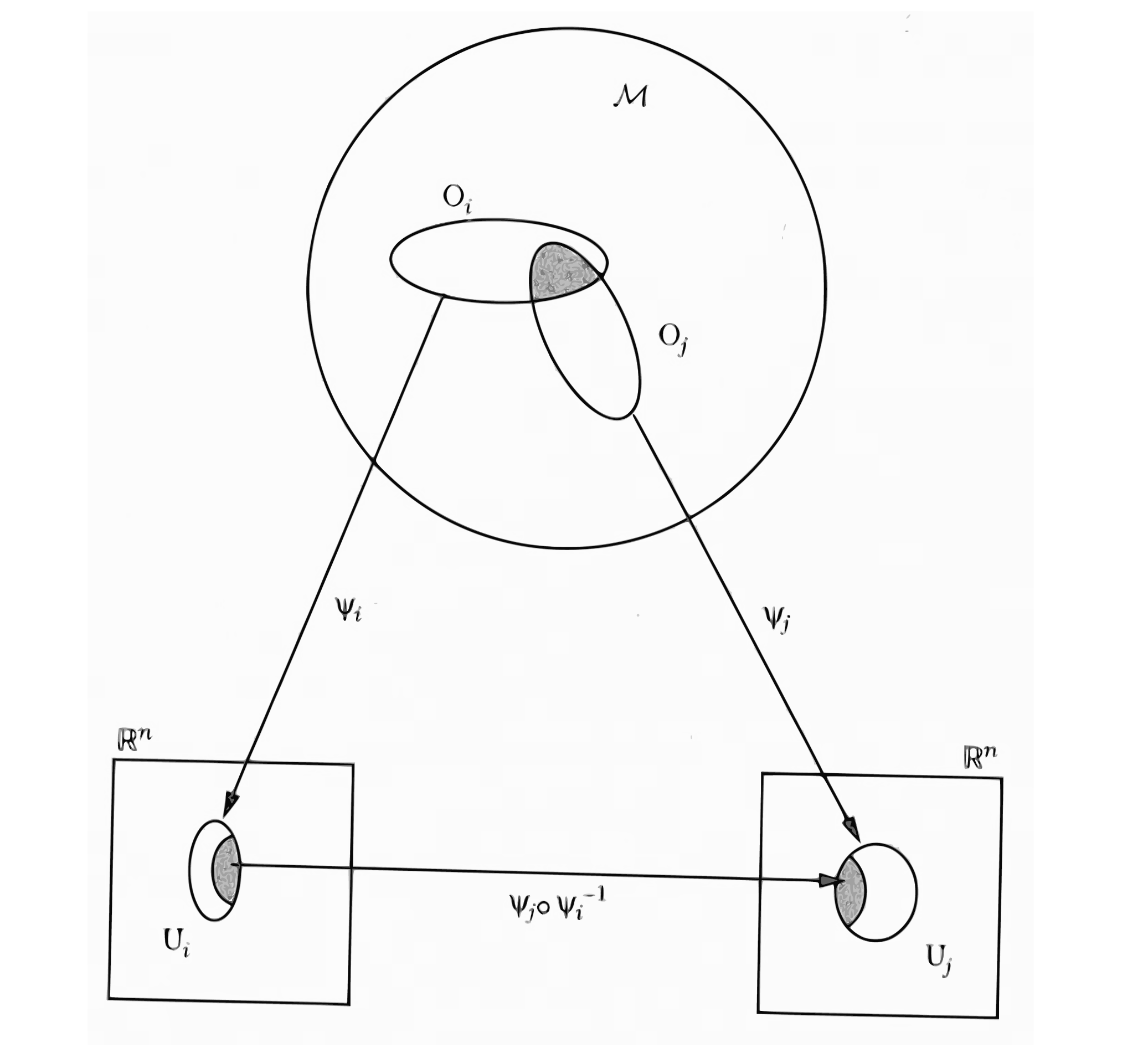} \EC\caption{Relation entre les différentes cartes d'une variété.} \label{manifold} \EF
Communément, la bijection $\psi_i$ est appelée une \ita{carte} ou un \ita{système de coordonnées} et les fonctions $\psi_j \circ \psi_i^{-1}$ sont appelées \ita{fonctions de transitions} et correspondent simplement à un \ita{changement de coordonnées} (voir figure \ref{manifold}). Par la suite, nous rencontrerons également le cas des \ita{variétés complexes}. Pour les définir, il suffit de remplacer $\mathbb{R}^D$ par $\mathbb{C}^D$ dans la définition précédente et d'exiger que les fonctions de transitions soient holomorphes. La sphère $S^2$ est un exemple de variété complexe. Nous reviendrons à la sous-section \ref{cartan} sur ces variétés complexes avec la notion de variété kählérienne.

En présence d'une géométrie courbe, la notion d'espace vectoriel, pour notamment additionner des vecteurs, est perdu \ita{globalement}. Cependant, il est possible de la restaurer en chaque point de la variété en introduisant la notion de \ita{vecteur tangent}. Nous nous inspirons pour cela de la bijection entre vecteurs et dérivées directionnelles dans $\mathbb{R}^D$, en effet chaque vecteur $\vec{v}=\left( v^1, \ldots , v^D \right)$ de $\mathbb{R}^D$ définit une dérivée $v^\mu \frac{\p}{\p x^\mu} $. Nous utilisons dès à présent la \ita{notation d'Einstein} en ce qui concerne les sommes. Introduisons la notation $\mc F = \mc C^\infty\left( \mc M, \mathbb{R} \right)$ et la définition suivante : 
\begin{dico}[vecteur tangent]
Un vecteur tangent $v$ en un point $P$ de $\mc M$ est une fonction de $\mc F$ dans $\mathbb{R}$ linéaire et qui satisfait la règle de Leibniz, c'est-à-dire :
\begin{itemize}
	\item $\forall f,g \in \mc F \ , \forall a,b \in \mathbb{R} \ , v(af+bg)=av(f)+bv(g) \ , $
	\item $\forall f,g \in \mc F \ , v(fg)=f(P)v(g)+g(P)v(f) \ .$
\end{itemize}\end{dico}
Nous parlerons aussi de \ita{vecteur contravariant} pour désigner un vecteur tangent. Nous regagnons ainsi \ita{localement} la notion d'espace vectoriel : 
\begin{proposition}[espace tangent] Soit $\mc M$ une variété différentielle de dimension $D$. L'espace tangent $V_P$ est l'ensemble des vecteurs tangents au point P. $V_P$ est un espace vectoriel de dimension D.
\end{proposition}
En introduisant un système de coordonnées $\psi$, une base $\{ X_\mu \}$ de $V_P$ peut être construite en définissant pour $\mu \in \llbracket 1, D \rrbracket$, les  $X_\mu : \mc F \rightarrow \mathbb{R}$ par $\forall f \in \mc F$,
\BE X_\mu(f) =\left. \frac{\p}{\p x^\mu}\left( f \circ \psi^{-1} \right) \right|_{\psi(P)} \ . \EE
Par conséquent, il est loisible de décomposer un vecteur $v$ de $V_P$ dans cette base : $v=v^\mu X_\mu$. Cette base est appelée \ita{base de coordonnées} et nous la noterons $\{ \p_\mu \}$ dans la suite. Notons que si nous utilisons un autre système de coordonnées, $\psi'$, la nouvelle base associée $\{ \p'_\mu \}$ est reliée à la précédente par 
\BE \p_\mu = \left. \frac{\p x'^\nu}{\p x^\mu}\right|_{\psi(P)} \p'_\nu \EE
où $x'^\mu$ est la $\mu$-ième composante de $\psi' \circ \psi^{-1}$. Nous en déduisons alors la \ita{loi de transformation} suivante pour les composantes de tout vecteur contravariant $v$
\BE v'^\nu = v^\mu \frac{\p x'^\nu}{\p x^\mu} \ . \label{vecteur} \EE Introduisons également la définition suivante : 
\begin{dico}[fibré tangent]
Le fibré tangent $V$ de $\mc M$ est défini par $V = \underset{P\in\mc M}\bigcup V_P$ et un élément de $V$ est appelé champ tangent. 
\end{dico} 
Dans la suite, nous étudierons également \ita{des courbes $\ga$ de $\mc M$}, c'est-à-dire des applications de $\mathbb{R}$ dans $\mc M$. En utilisant un système de coordonnées $\psi$, une courbe $\ga$ paramétrisée par $\la$ est donnée par la paramétrisation $x^\mu(\la)$. En chaque point $P$ de $\mc M$ appartenant à la courbe $\ga$, un vecteur tangent $t \in V_P$ est associé à cette courbe dont les composantes sont données par $t^\mu=\frac{\dd x^\mu}{\dd \la}$  dans le système de coordonnées $\psi$.\bs

Passons à une autre catégorie de vecteur :
\begin{dico}[vecteur dual] Une forme linéaire ou 1-forme $\om : V_P \rightarrow \mathbb{R}$ est appelée vecteur dual ou vecteur covariant de $V_P$. L'ensemble des vecteurs covariants de $V_P$ forme ainsi l'espace vectoriel dual de $V_P$, noté $V_P^*$. 
\end{dico}
$V_P^*$ est de nouveau un espace vectoriel de dimension $D$ et si $\{ \p_\mu \}$ désigne une base de $V_P$ alors la base de $V_P^*$, notée $\{ \dd x^\mu \}$, est définie par
\BE \dd x^\mu \left( \p_\nu \right) = \delta_\nu^\mu \ . \label{basedual}\EE
Par conséquent, un vecteur dual $\om$ de $V_P^*$ est décomposé sous la forme $\om = \om_\mu \dd x^\mu$ où $\om_\mu = \om\left(\p_\mu\right)$. Quant à la loi de transformation pour les composantes de $\om$, elle est donnée par
\BE \om'_\nu = \om_\mu \frac{\p x^\mu}{\p x'^\nu} \EE
que nous déterminons à partir de \eqref{vecteur} et \eqref{basedual}. De manière similaire :
\begin{dico}[fibré cotangent]
Le fibré cotangent $V^*$ de $\mc M$ est défini par $V^* = \underset{P\in\mc M}\bigcup V_P^*$ et un élément de $V^*$ est appelé un champ cotangent. 
\end{dico} 
Nous sommes désormais en mesure de définir la notion de \ita{tenseur} qui encode le principe de covariance.  
\begin{dico}[tenseur] Un tenseur T de type $(k,l)$ sur $V_P$ est une application multilinéaire 
\BC $T : \underbrace{V_P^* \times \cdots \times V_P^*}_{k \text{ fois}} \times \underbrace{V_P \times \cdots  \times V_P}_{l \text{ fois}} \rightarrow \mathbb{R}$ \ .\EC
\end{dico}
L'ensemble des tenseurs $\mc T_P(k,l)$ de type $(k,l)$ forme de nouveau un espace vectoriel et il est loisible de décomposer tout tenseur de type $(k,l)$ sous la forme
\BE T = T^{\mu_1 \ldots \mu_k}_{\ \ \ \ \ \ \ \nu_1 \ldots \nu_l} \p_{\mu_1}\otimes \cdots \otimes \p_{\mu_k} \otimes \dd x^{\nu_1} \otimes \cdots \otimes \dd x^{\nu_l} \EE
où $\otimes$ est le produit tensoriel défini de la manière suivante :  si $T\in \mc T_P(k,l)$, $T'\in \mc T_P(k',l')$, $\om^1, \ldots ,  \om^{k+k'} \in V_P^*$ et $v_1, \ldots , v_{l+l'} \in V_P $ alors $T \otimes T'$ agissant sur ces vecteurs est donné par le produit de $T(\om^1, \ldots , \om^k, v_1, \ldots , v_l)$ par $T'(\om^{k+1}, \ldots , \om^{k+k'}, v_{l+1}, \ldots , v_{l+l'})$. Concernant le vocabulaire, il est courant de dire que $T$ est $k$ fois contravariant et $l$ fois covariant. Quant à la loi de transformation d'un tenseur de type $(k,l)$, elle est déduite de celle des vecteurs tangents et des 1-formes : 
\BE  T^{\mu'_1 \ldots \mu'_k}_{\ \ \ \ \ \ \ \nu'_1 \ldots \nu'_l} =  T^{\mu_1 \ldots \mu_k}_{\ \ \ \ \ \ \ \nu_1 \ldots \nu_l}  \frac{\p x'^{\mu'_1}}{\p x^{\mu_1}} \cdots \frac{\p x'^{\mu'_k}}{\p x^{\mu_k}} \frac{\p x^{\nu_1}}{\p x'^{\nu'_1}} \cdots \frac{\p x^{\nu_l}}{\p x'^{\nu'_l}} \ . \label{tenseur}\EE 
Enfin, nous pouvons introduire la notion de \ita{champ tensoriel} de type $(k,l)$ sur la variété $\mc M $ qui est un élément de $\underset{P\in\mc M}\bigcup \mc T_P (k,l)$ et nous notons par $\mc T(k,l)$ ce dernier ensemble. De plus, un champ tensoriel est dit infiniment différentiable si ses composantes le sont.
\bs

Ajoutons aussi la définition suivante. Etant donné, deux champs tangents $\mc C^\infty$, $v$ et $w$, nous définissons un nouveau champ tangent appelé le \ita{commutateur} de $v$ et $w$, noté $[v,w]$, par
\BE \forall f \in \mc F \ , [v,w](f)= v(w(f)) - w(v(f)) \ . \EE
Indiquons que le commutateur de deux champs tangents d'une base de coordonnées $\{ \p_\mu \}$ est nul. A l'inverse, étant donné $D$ champs tangents non nuls, linéairement indépendants en chaque point et qui commutent entre eux, il est possible de trouver un système de coordonnées dans lequel ces vecteurs forment une base de coordonnées. 
Précisons également que dans un système de coordonnées, les composantes du commutateur $[v,w]$ sont données par 
\BE  [v,w]^\mu = v^\nu \p_\nu w^\mu - w^\nu \p_\nu v^\mu \ . \EE
\bs

Concluons par l'objet central de la Relativité Générale : \ita{la métrique}. 
\begin{dico}[métrique]Une métrique est un tenseur $g$ de type $(0,2)$ sur $V_P$
\begin{itemize}
\item symétrique : $\forall u,v \in V_P \ ,  g(u,v)=g(v,u)$
\item et non dégénéré :  $\forall v \in V_P \ ,  g(u,v)=0 \Rightarrow u=0$ .
\end{itemize}\end{dico}
Quand la signature de la métrique est $(+, +, \ldots , +)$, la métrique est dite \ita{riemannienne} et lorsque la signature est $(-,+, \ldots ,+)$, elle est qualifiée de \ita{lorentzienne} ou \ita{pseudo-riemannien\-ne} ; la signature correspond aux signes des valeurs propres de $g$. La métrique se décompose donc sous la forme suivante sur une base de 1-formes
\BE g = g_{\mu\nu} \dd x^\mu \otimes \dd x^\nu \ . \label{métrique}\EE
Cependant, il est plus courant d'utiliser la notation $ \dd s^2 = g_{\mu\nu} \dd x^\mu \dd x^\nu $ à la place de \eqref{métrique} afin de souligner que cet objet fournit la notion de distance infinitésimale. En Relativité Générale, la métrique devient un objet dynamique qui généralise le potentiel de gravitation de la Mécanique Newtonienne. Nous sommes désormais en mesure de formuler un premier axiome pour la Relativité Générale :  
\begin{axiome} L'espace-temps est une variété différentielle réelle de dimension 4 munie d'une métrique lorentzienne. 
\end{axiome} 		
		\subsection{Connexions, géodésiques et courbures}
Nous avons vu qu'un tenseur en un point $P$ de $\mc  M$ et un autre en un point $Q$ ne peuvent être combinés de manière tensorielle. Par conséquent, l'opération de dérivation sur les tenseurs n'est pas simple à définir. Voici sa définition : 
\begin{dico}[dérivée covariante] L'opération de dérivation sur un champ tensoriel $\mc C^\infty$ de type $(k,l)$, notée $\n$, fournit un champ tensoriel de type $(k,l+1)$. $\n$ est aussi qualifiée de dérivée covariante et est usuellement notée $\n_\mu$ par abus. Elle vérifie les propriétés suivantes : 
\begin{itemize}
\item la linéarité, 
\item la règle de Leibniz, 
\item la commutativité avec l'opération de contraction, 
\item $\forall t \in V_P \ , \forall f \in \mc F \ ,  t(f)= t^\mu \n_\mu f $
\item et $\forall f \in \mc F \ , \n_\mu \n_\nu f =   \n_\nu \n_\mu f $.
\end{itemize}\end{dico}
Cette dernière propriété correspond à l'absence de \ita{torsion} dans la définition de la dérivée covariante. En effet, il existe des théories modifiées de la gravitation en présence de torsion comme celle dite d'\ita{Einstein-Cartan}, pour une revue sur le sujet le lecteur pourra consulter \cite{RevModPhys.48.393}. En utilisant cette définition formelle, il est alors possible de montrer qu'il existe un champ $\Ga_{\mu\nu}^\rho$, appelé la \ita{connexion} ou le \ita{symbole de Christoffel}, qui donne explicitement l'action de la dérivée covariante sur un champ tensoriel. Par exemple, la dérivée covariante d'un tenseur $T$ de type $(1,1)$ est donnée par
\BE \n_\ga T^\al_{\ \be} =  \p_\ga T^\al_{\ \be} + \Ga_{\mu\ga}^\al T^\mu_{\ \be} - \Ga_{\be\ga}^\mu T^\al_{\ \mu}  \ .\EE
Les règles de contraction de cette égalité se généralisent à des tenseurs de rang plus élevé. Par ailleurs, la connexion est symétrique, c'est-à-dire $\Ga_{\mu\nu}^\rho = \Ga_{\nu\mu}^\rho$; cela provient de l'absence de torsion. Précisons également que la connexion ne vérifie pas la loi de transformation des tenseurs \eqref{tenseur}. Introduisons aussi la notion de \ita{transport parallèle} :
\begin{dico}[transport parralèle] Un champ tensoriel de type $(k,l)$ est dit transporté parallèlement le long d'une courbe $\ga$ si
\BE  \frac{D}{\dd \la} T^{\mu_1 \ldots \mu_k}_{\ \ \ \ \ \ \ \nu_1 \ldots \nu_l}  \dot{=}  t^\rho \n_\rho T^{\mu_1 \ldots \mu_k}_{\ \ \ \ \ \ \ \nu_1 \ldots \nu_l} =0\EE
où $t^\mu$ est un vecteur tangent à $\ga$ et $\la$ désigne un paramètre de $\ga$.
\end{dico}
Pour le moment, il existe a priori plusieurs choix possibles pour la connexion. Cependant, étant donné une métrique $g_{\mu\nu}$, il y a un choix naturel. Considérons deux vecteurs $v^\mu$ et $w^\mu$ transportés parallèlement le long d'une courbe $\ga$ et demandons que leur produit scalaire $g_{\mu\nu} v^\mu w^\nu$ reste inchangé le long de $\ga$, alors nécessairement $\n_\rho g_{\mu\nu} = 0$. Il s'ensuit le résultat suivant : 
\begin{proposition} Etant donné une métrique $g_{\mu\nu}$ et une dérivée covariante $\n$ sans torsion vérifiant $\n_\rho g_{\mu\nu} = 0$, il existe une unique dérivée covariante dont la connexion est donnée par
\BE \Ga_{\mu\nu}^\rho = \frac{1}{2} g^{\rho\si} \left( \p_\mu g_{\nu\si} + \p_\nu g_{\mu\si} - \p_\si g_{\mu\nu} \right) \ .\EE
\end{proposition}
Dans toute la suite, nous travaillerons avec une telle connexion dite de \textit{Levi-Civita}. Formulons ainsi un second axiome à la théorie de la Relativité Générale : 
\begin{axiome} A la métrique de l'espace-temps s'ajoute une connexion symétrique $\Ga_{\mu\nu}^\rho$ tel que \BE \n_\rho g_{\mu\nu} = 0 \ .\EE
\end{axiome}
Nous sommes désormais en mesure d'étudier des trajectoires de l'espace-temps. Intéressons nous à une classe privilégiée de courbes de $\mc M$ : les  \ita{géodésiques}. 
\begin{dico}[géodésique] Une courbe de $\mc M$ est une géodésique si elle extrémise la distance entre deux points de $\mc M$.
\end{dico}
Soit $\ga$ une courbe donnée par la paramétrisation $x^\mu (\la)$. La distance entre deux points $P$ et $Q$ le long de cette courbe est donnée par
\BE l = \int_P^Q \sqrt{\pm g_{\mu\nu} \dot{x}^\mu \dot{x}^\nu } \dd \la \EE où $\dot{x}^\mu = \frac{\dd x^\mu}{\dd \la}$. Le signe dans la racine carrée est $+$ pour une courbe de genre-espace et $-$ pour une courbe de genre-temps. Notons que la quantité $l$ est invariante sous la reparamétrisation $\la \rightarrow \la'(\la)$. La courbe pour laquelle $l$ atteint un extremum est déterminée en résolvant l'équation d'Euler-Lagrange
\BE \frac{\dd}{\dd \la} \left( \frac{\p L}{\p \dot{x}^\al} \right) =  \frac{\p L}{\p x^\al} \EE
avec le lagrangien $ L (x^\mu,\dot{x}^\mu)=  \sqrt{\pm g_{\mu\nu} \dot{x}^\mu \dot{x}^\nu } $. Nous déterminons ainsi \ita{l'équation de géodésique}
\BE \ddot{x}^\al + \Ga^\al_{\be\ga} \dot{x}^\be \dot{x}^\ga = \ka(\la) \dot{x}^\al  \EE
avec $\ka(\la)=\frac{\dd \ln L}{\dd \la}$. Il est possible de l'écrire sous la forme $u^\be \n_\be u^\al =  \ka u^\al $ avec $u^\al = \dot{x}^\al$. Afin de simplifier l'équation de géodésique, il est loisible de choisir le temps propre $\tau$, défini par $\dd \tau^2 = - g_{\mu\nu} \dd x^\mu \dd x^\nu $, comme paramètre $\la$ pour une géodésique de genre-temps ; et de choisir la distance propre $s$, définie par $\dd s^2 =  g_{\mu\nu} \dd x^\mu \dd x^\nu $, comme paramètre $\la$ pour une géodésique de genre-espace. Dans ce cas, l'équation de géodésique s'écrit 
\BE u^\be \n_\be u^\al = 0 \ ,\label{géodésique} \EE
c'est-à-dire que le vecteur tangent $u^\al$ est transporté parallèlement le long de la géodésique. Quelque soit le paramètre, les équations de géodésique sont invariantes sous la reparamétrisation $\la \rightarrow  a \la + b$. Les paramètres reliés par une telle transformation aux paramètres $\tau$ et $s$ sont appelés paramètres affines. Notons aussi que l'équation \eqref{géodésique} est restaurée en considérant le lagrangien $ L'= \frac{1}{2}  g_{\mu\nu} \dot{x}^\mu \dot{x}^\nu $, ce qui fournit une méthode rapide pour calculer en pratique des symboles de Christoffel. Pour des géodésiques de genre-lumière, l'équation dite de géodésique est donnée par $u^\be \n_\be u^\al =  \ka u^\al $ pour des raisons de continuité. Cependant, puisque $s=\tau=0$ dans ce cas, il faut utiliser le paramètre affine $\la^*$ défini par 
\BE \frac{\dd \la^*}{\dd \la} = \exp{\int^\la \ka} \EE
pour retrouver l'équation \eqref{géodésique}. Précisons également que le long d'une géodésique $\ga$ paramétrisée par un paramètre affine, la norme du vecteur tangent $\ep = u^\al u_\al$ est constante le long de $\ga$, en effet 
\BE \frac{\dd \ep}{\dd \la} = u^\be \n_\be \left( u^\al u_\al \right) = 2 u_\al u^\be \n_\be u^\al =0 \ . \EE 
Nous pouvons désormais ajouter un autre axiome à la théorie de la Relativité Générale : 
\begin{axiome} Il existe des courbes privilégiées de l'espace-temps : 
\begin{itemize}
 	\item les particules libres voyagent le long des géodésiques de genre-temps
	\item  et les rayons lumineux décrivent des géodésiques de genre-lumière.
\end{itemize}\end{axiome}

Nous pouvons maintenant apporter une formulation précise du principe d'équivalence. En un point $P$ de l'espace-temps, il est toujours possible de trouver un système de coordonnées $x'^\al$ tel que $g_{\mu'\nu'}(P) = \eta_{\mu'\nu'}$ et $\Ga^{\rho'}_{\mu'\nu'}(P)=0$ où $\eta_{\mu'\nu'}=\text{diag}(-1,1,1,1)$ est la métrique de Minkowski. Ce système de coordonnées est dit \ita{localement inertiel en P} puisque l'équation de géodésique se réduit à $\ddot{x}^\mu=0$ au voisinage de $P$. Les forces d'inerties induites par la connexion sont ainsi supprimées et la particule ne voit pas la gravité autour du point $P$ dans ce système de coordonnées. Cependant, il n'est pas possible d'annuler les dérivées de la connexion en $P$. Par conséquent, une expérience \ita{non-locale} permet de voir la gravité. Revenons à l'expérience de pensée de l'astronaute dans sa fusée sans fenêtre pour illustrer ce propos. Dans la première situation, la fusée est au repos sur Terre. Dans la seconde, la fusée se déplace par rapport à un observateur inertiel avec une accélération $-\vec{g}$. Tant que l'astronaute se limite à des expériences locales, il ne peut pas distinguer ces deux situations ; cependant s'il lâche deux masses à la même hauteur, il pourra les différencier : dans la première situation, les deux masses se rapprochent l'une de l'autre durant leur chute car elles se dirigent vers le centre de la Terre; alors que dans la seconde situation, elles restent à distance fixe l'une de l'autre. 

Afin de décrire précisément cette expérience, nous avons besoin d'introduire le \ita{tenseur de Riemann} : 
\begin{dico}[tenseur de Riemann]
 Pour toute 1-forme $\om_\mu$, le tenseur de Riemann est défini par 
\BE \left( \n_\mu \n_\nu - \n_\nu \n_\mu \right) \om_\rho =  R_{\mu\nu\rho}^{\ \ \ \ \si} \om_\si \ . \label{Riemann0} \EE 
\end{dico}
En utilisant la règle de Leibniz de $\n$, le lecteur pourra montrer que, pour tout champ tangent $v^\mu$, 
\BE \left( \n_\mu \n_\nu - \n_\nu \n_\mu \right) v^\rho = - R_{\mu\nu\si}^{\ \ \ \ \rho} v^\si  \EE et que le résultat se généralise aisément pour tout champ tensoriel. Soulignons également l'existence des propriétés suivantes
\BE R_{\mu\nu\rho}^{\ \ \ \ \si} = -R_{\nu\mu\rho}^{\ \ \ \ \si} \ , \label{Riemann1}\EE
\BE R_{[\mu\nu\rho]}^{\ \ \ \ \ \si} = 0 \ , \label{Riemann2} \EE
quelque soit le choix de l'opérateur $\n$. De plus, puisque la connexion est choisie telle que $\n_\rho g_{\mu\nu}=0$,
\BE R_{\mu\nu\rho\si} = -R_{\mu\nu\si\rho} \ . \label{Riemann3} \EE
De toutes ces propriétés, il s'ensuit la symétrie suivante pour le tenseur de Riemann
\BE R_{\mu\nu\rho\si} = R_{\rho\si\mu\nu} \ . \EE
Avec toutes ces symétries, le tenseur de Riemann possède $\frac{D^2 (D^2-1)}{12}$ composantes indépendantes. Donnons également un résultat important appelé \ita{identité de Bianchi} 
\BE \n_{[\mu} R_{\nu\rho]\si}^{\ \ \ \ \de} = 0 \ . \label{bianchi1} \EE
Revenons alors sur l'expérience de pensée précédente pour donner une signification géométrique du tenseur de Riemann en dérivant l'équation dite de \ita{déviation des géodésiques}. Considérons deux géodésiques $\ga_0$ et $\ga_1$ paramétrisées par $x^\al (t)$ où $t$ est un paramètre affine. Introduisons une famille de géodésiques $\ga_s$ qui interpole les courbes $\ga_0$ et $\ga_1$ avec $s \in [0,1]$ tel que $\ga_{s=0}=\ga_0$ et $\ga_{s=1}=\ga_1$. Toutes ces géodésiques peuvent alors être décrites par la paramétrisation $x^\al (s,t)$ où $s$ désigne une géodésique et $t$ est un paramètre affine le long de $\ga_s$. Le vecteur $u^\al = \frac{\p x^\al}{\p t}$ est tangent aux géodésiques et satisfait $u^\be \n_\be u^\al = 0$. A $t$ fixé, les courbes $s \rightarrow x^\al(s,t)$ ont pour vecteur tangent $\xi^\al = \frac{\p x^\al}{\p s} $ et la restriction de ce vecteur à la courbe $\ga_0$ est appelé le \ita{vecteur déviation} entre $\ga_0$ et $\ga_1$. Puisque $\frac{\p u^\al}{\p s} = \frac{\p \xi^\al}{\p t}$, $u^\be \n_\be \xi^\al =  \xi^\be \n_\be u^\al $, ce qui permet de montrer que le produit scalaire $\xi^\al u_\al$ reste constant le long de $\ga_0$, c'est-à-dire 
\BE \frac{\dd (\xi^\al u_\al)}{\dd t}=0 \ .  \EE
Par conséquent, le vecteur déviation peut être choisi orthogonal à $u^\al$, $\xi^\al u_\al = 0$, le long de $\ga_0$. Ceci justifie la dénomination du vecteur déviation $\xi^\al$. Nous désirons maintenant dériver l'accélération relative de $\ga_1$ par rapport à $\ga_0$ donnée par
\BE \frac{D^2 \xi^\al}{\dd t^2} = u^\ga \n_\ga \left( u^\be \n_\be \xi^\al \right) \EE
où les quantités sont évaluées sur $\ga_0$. Dans l'espace-temps plat, $\frac{D^2 \xi^\al}{d t^2}=0$ puisque les géodésiques sont des droites ; en revanche pour une variété différentielle quelconque
\BE \frac{D^2 \xi^\al}{\dd t^2} = - R^\al_{\ \be\ga\de} u^\be \xi^\ga u^\de \label{déviation}  \EE 
après calculs. Le tenseur de Riemann caractérise donc l'accélération relative ou la force de marée entre deux géodésiques voisines. Pour calculer le tenseur de Riemann en pratique, il est utile d'introduire un système de coordonnées et ainsi : 
\BE R^\mu_{\ \nu\rho\si} = \Ga^\mu_{\nu\si,\rho} - \Ga^\mu_{\nu\rho,\si} +  \Ga_{\nu\si}^\de \Ga^\mu_{\rho\de}  - \Ga_{\nu\rho}^\de \Ga^\mu_{\si\de} \ . \EE
Introduisons également le \ita{tenseur de Ricci} défini par
\BE R_{\mu\nu} = R^\rho_{\ \mu\rho\nu} \EE
qui donne la contribution de la trace du tenseur de Riemann. Ce tenseur est symétrique et sa trace est appelée la \ita{courbure scalaire}
\BE R = R^\mu_{\ \mu} \EE
qui est donc un invariant. Quant à la partie sans trace du tenseur de Riemann, elle est donnée par le \ita{tenseur de Weyl}, noté $C_{\mu\nu\rho\si}$, qui est défini pour toutes variétés de dimensions $D \geq 3$ par
\BE R_{\mu\nu\rho\si} = C_{\mu\nu\rho\si} + \frac{2}{D-2} \left( g_{\mu[\rho} R_{\si]\nu} - g_{\nu[\rho} R_{\si]\mu} \right)  - \frac{2}{(D-1)(D-2)} R g_{\mu[\rho} g_{\si]\nu} \ . \label{weyl} \EE
Toute contraction de ce tenseur est nulle. Il partage également les symétries \eqref{Riemann1}, \eqref{Riemann2} et \eqref{Riemann3} du tenseur de Riemann. Le tenseur $C_{\mu\nu\rho}^{\ \ \ \ \si}$ possède aussi la propriété remarquable d'être invariant sous une transformation conforme $g_{\mu\nu} \rightarrow \Om^2 g_{\mu\nu}$ où $\Om$ est une fonction $\mc C^\infty(\mc M, \mathbb{R)}$ strictement positive. Indiquons également que le tenseur de Weyl est nul en dimension 3. \bs

Concernant le vocabulaire, une variété de dimension $D$ est qualifiée d'\ita{espace d'Einstein} si son tenseur de Ricci vérifie
\BE R_{\mu\nu} = \frac{1}{D} R g_{\mu\nu} \ .\EE 
Nous pouvons aussi rencontrer une contrainte plus forte : une variété est dite à \ita{courbure constante} si son tenseur de Riemann vérifie
\BE R_{\mu\nu\rho\la} = \frac{1}{D(D-1)} R \lp g_{\mu\rho} g_{\nu\la} - g_{\mu\la} g_{\nu\rho}  \rp \ . \EE
Par conséquent, une variété à courbure constante est un espace d'Einstein.  
Enfin, introduisons un objet important qui est le \ita{tenseur d'Einstein} défini par
\BE G_{\mu\nu} = R_{\mu\nu} - \frac{1}{2} R g_{\mu\nu} \EE
et qui vérifie l'identité de Bianchi
\BE \n_\mu G^{\mu\nu} = 0 \label{bianchi2} \EE 
d'après l'équation \eqref{bianchi1}. Cette identité joue un rôle important dans l'\ita{équation d'Einstein} que nous étudierons dans la prochaine section. Elle permet aussi de montrer qu'un espace à courbure constante admet nécessairement une courbure scalaire constante.
Passons maintenant à un autre formalisme pour déterminer la connexion et la courbure d'une géométrie courbe.
		\subsection{Formes différentielles et formulation de Cartan\label{cartan}}
Dans cette sous-section, nous allons nous intéresser à l'ensemble des tenseurs totalement antisymétriques de type $(0,p)$, qui est une classe de champs tensoriels très utile en pratique et notamment pour \ita{la théorie de Lovelock} que nous présenterons à la section \ref{lovelock}. Ces tenseurs sont aussi appelés des $p$-formes et l'ensemble de ces champs tensoriels est noté $\La^p(\mc M)$.  Ainsi $\La^1(\mc M)$ correspond au fibré cotangent $V^*$. Par la suite, nous étudierons à la section \ref{BHpforms} des trous noirs en présence de $p$-formes. Le lecteur pourra se tourner vers \cite{straumann2004general} et \cite{eguchi1980gravitation} pour une présentation plus complète de ce formalisme. Commençons par introduire le produit extérieur :
\begin{dico}[produit extérieur] Soit $\{ \dd  x^\mu\}$ une base duale de $V^*_P$ en un point $P$ d'une variété $\mc M$ de dimension $D$ dans un système de coordonnées $\psi$. Le produit extérieur est défini par
\BE \dd  x^{\mu_1} \w \ldots \w \dd  x^{\mu_k} = \sum_{\si \in \mathfrak{S}_k} \ep(\si) \dd  x^{\mu_{\si(1)}} \otimes \cdots \otimes \dd  x^{\mu_{\si(k)}} \EE
où $k\in \mathbb{N}^*$ et $\mathfrak{S}_k$ désigne l'ensemble des permutations de $\llbracket 1,k \rrbracket$.
\end{dico} 
De cette manière, il est loisible de décomposer une $p$-forme $\om $ de la façon suivante
\BE \om = \frac{1}{p!} \om_{\mu_1 \ldots \mu_p }  \dd  x^{\mu_1} \w \ldots \w \dd  x^{\mu_p} \ .  \EE
En particulier, le produit extérieur est bilinéaire, associatif et 
\BE \forall \om \in \La^p(\mc M) \ , \forall \eta \in \La^q(\mc M) \ , \om \w \eta = (-1)^{pq} \eta \w \om \ . \EE
 Remarquons également que la dimension de l'espace $\La^p(\mc M)$ est $\mathrm{C}_D^p = \frac{D!}{p! (D-p)!}$ pour tout  $p$ de $\llbracket 0,D\rrbracket$ et $\La^p(\mc M)=0$ si $p>D$ en particulier. Les espaces $\La^p(\mc M)$ et $\La^{D-p}(\mc M)$ ont donc la même dimension. Dans la suite, nous restreignons notre attention à des $p$-formes infiniment différentiables appelées \ita{formes différentielles} et nous continuons à noter cet ensemble $\La^p(\mc M)$; précisons que $\La^0(\mc M)=\mc F$. Enfin, l'ensemble
$\La^\star(\mc M) = \bigoplus_{p\geq 0} \La^p(\mc M)$
est appelé \ita{l'algèbre extérieure de Cartan des formes différentielles}.\bs

Introduisons maintenant un opérateur différentiel qui permet de passer d'une $p$-forme à une $(p+1)$-forme : 
\begin{dico}[dérivée extérieure] Pour toute $p$-forme différentielle $\om$, la dérivée extérieure $\dd : \La^p(\mc M) \rightarrow \La^{p+1}(\mc M)$ est définie par 
\BE \dd  \om = \frac{1}{p!} \p_\nu \om_{\mu_1 \ldots \mu_p } \dd x^\nu \w \dd  x^{\mu_1} \w \ldots \w \dd  x^{\mu_p} \ . \EE
\end{dico}
Il en découle la propriété suivante : 
\BE \forall \al \in \La^p(\mc M) \ , \forall \be \in \La^q(\mc M) \ ,  \dd (\al \w \be) = \dd \al \w \be + (-1)^p \al \w \dd \be \ . \EE 
Remarquons également l'importante propriété $\dd \circ \dd =0$. Concernant le vocabulaire, une $p$-forme $\om$ est dite \ita{fermée} si $\dd \om = 0$ et l'ensemble des $p$-formes fermées est usuellement noté $Z^p(\mc M)=\text{Ker}\left( \dd : \La^p(\mc M) \rightarrow \La^{p+1}(\mc M) \right)$. Quant à l'ensemble $\text{Im}\left(\dd : \La^{p-1}(\mc M) \rightarrow \La^{p}(\mc M) \right)$, qui est noté $B^p(\mc M)$, il est constitué de $p$-formes dite \ita{exactes}. En particulier $B^p(\mc M)  \subset  Z^p(\mc M)$. Donnons le résultat important suivant : 
\begin{proposition}[Poincaré] Le théorème de Poincaré dit que 
\BE B^p\lp\mathbb{R}^D \rp  =  Z^p \lp \mathbb{R}^D \rp \ . \EE
\end{proposition}
Définissons également le \ita{p-ième groupe de cohomologie de De Rham}, noté $H^p (\mc M)$, défini par
\BE H^p (\mc M) = Z^p (\mc M) / B^p (\mc M) = \{ \om + B^p (\mc M)  \}_{\om \in Z^p (\mc M)} . \EE
La dimension de cet espace est appelée le \ita{nombre de Betti}, noté $b_p$. Par exemple, la suite $(b_p)_{1 \leq p \leq n}$ des nombres de Betti pour la sphère $S^n$ est $(1,0,\ldots,0,1)$.

Nous avons précédemment remarqué qu'il existe une \ita{dualité} entre les espaces $\La^p(\mc M)$ et $\La^{D-p}(\mc M)$. Introduisons alors un opérateur qui transforme une $p$-forme en une $(D-p)$-forme : 
\begin{dico}[dualité de Hodge] Soit $\{ \dd  x^\mu\}$ une base duale de $V^*_P$ en un point $P$ d'une variété $\mc M$ dans un système de coordonnées $\psi$. Le produit $\star : \La^p(\mc M) \rightarrow \La^{D-p}(\mc M) $ est défini de la façon suivante
\BE\star \left( \dd  x^{\mu_1} \w \ldots \w \dd  x^{\mu_p} \right) = \frac{1}{(D-p)!} \ep^{\mu_1 \ldots \mu_p}_{\ \ \ \ \ \ \ \mu_{p+1} \ldots \mu_D} \dd  x^{\mu_{p+1}} \w \ldots \w \dd  x^{\mu_D} \EE
où $ \ep_{\mu_1 \ldots \mu_D}$ est le tenseur dualiseur ou tenseur de Levi-Civita défini par $ \ep_{\mu_1 \ldots \mu_D} = \sqrt{|g|} \varep_{\mu_1 \ldots \mu_D} $.
\end{dico}
 Précisons que $ \varep_{\mu_1 \cdots \mu_D}$ qui est totalement antisymétrique dans ses $D$ indices, avec $ \varep_{1 \cdots D}=1$, n'est pas un tenseur. Pour toute forme différentielle $\om \in \La^p(\mc M) \ , \star (\star \om) =  \pm (-1)^{p(D-p)} \om $ où $\pm$ désigne le signe $+$ si $\mc M$ est variété riemannienne et le signe $-$ si $\mc M$ est une variété lorentzienne. Donnons également la propriété utile suivante : 
\BE  \forall \al,\be \in \La^p(\mc M)  \ ,  \al \w \star \be = \be \w \star \al = \frac{1}{p!} \al_{\mu_1 \cdots \mu_p} \be^{\mu_1 \cdots \mu_p} \star 1 \EE
où $ \star 1 =  \frac{1}{D!} \ep_{\mu_{1} \ldots \mu_D} \dd  x^{\mu_{1}} \w \ldots \w \dd  x^{\mu_D} = \sqrt{|g|} \dd x^1 \w \ldots \w \dd x^D \dot{=} \ep $ est appelé la \ita{forme volume} de la variété $\mc M$. 

Avec ces définitions, nous pouvons également construire une application $\delta$ : $\La^p \lp \mc M \rp \rightarrow \La^{p-1} \lp \mc M \rp $ définie par $\delta = \pm (-1)^{D(p+1)} \star \dd \star$ où $\pm$ désigne le signe $+$ pour une variété riemannienne et le signe $-$ pour une variété lorentzienne, $\delta$ est appelée la co-différentielle. Le \ita{laplacien} $\Delta$ sur une variété peut alors être construit de la manière suivante $\Delta=\dd \circ \delta + \delta \circ \dd$ et une $p$-forme différentielle $\om$ est dite \ita{harmonique} si $\Delta \om=0$. Précisons qu'une forme $\om$ est harmonique si et seulement si elle est fermée et co-fermée, c'est-à-dire $\dd \om = 0$ et $\delta\om=0$ respectivement.\bs

\subsubsection{Equations de structure de Cartan}
 Nous sommes désormais en mesure de présenter les élégantes \ita{équations de structure de Cartan} qui nous seront d'une aide précieuse pour présenter la théorie de Lovelock. Considérons une variété différentielle de dimension $D$ munie d'une métrique. Puisque la métrique est symétrique et réelle, le tenseur $g_{\mu\nu}$ est orthogonalement diagonalisable et se décompose sous la forme $ g_{\mu\nu} = e^A_{\ \mu} \eta_{AB}  e^B_{\ \nu} $
où $\eta_{AB}= \text{diag}(-1, 1, \ldots, 1)$ si $\mc M$ est lorentzienne et $\eta_{AB}=\delta_{AB}$ si $\mc M$ est riemannienne. Les $e^A = e^A_{\ \mu} \dd x^\mu$, appelées les \ita{tétrades}, constituent une base orthonormale de $V_P^*$ en chaque point $P$ de $\mc M$; en effet la métrique est donnée par
\BE g = \eta_{AB} e^A \otimes e^B \ . \EE 
Donnons aussi la relation utile $ \eta^{AB}  = g^{\mu\nu}  e^A_{\ \mu} e^B_{\ \nu} $ et  précisons que les 1-formes $e^A$ ne sont pas forcément des formes exactes. Introduisons la base $\{ E_A \}$ de $V_P$ qui est duale de $\{ e^A \}$ et qui décomposée sous la forme $E_A = E_A^{\ \mu} \p_\mu$ vérifie les relations $ E_A^{\ \mu} e^B_{\ \mu} = \delta_A^B$ et $\eta^{AB}  E_A^{\ \mu} E_B^{\ \nu}  = g^{\mu\nu}$. Les indices grecs sont montés ou descendus avec la métrique $g_{\mu\nu}$ et les indices latins avec la métrique $\eta_{AB}$. 
Ajoutons un outil de calcul pratique pour notamment déterminer les densités lagrangiennes dans la théorie de Lovelock par exemple. Introduisons les $(D-k)$-formes
\BE e^*_{A_1 \ldots A_k} = \star \left( e_{A_1} \w \ldots \w e_{A_k} \right) = \frac{1}{(D-k)!} \varep_{A_1 \ldots A_k A_{k+1} \ldots A_D} e^{k+1} \w \ldots \w e^D \EE
où $e_{A_k} = \eta_{A_k B_k} e^{B_k} $ et par convention $ e^* =  e^1 \w \ldots \w e^D $ correspond à la forme volume de la variété. Le lecteur pourra alors montrer la règle de calcul suivante très utile
\BE e^B \w e^*_{A_1 \ldots A_k} = \delta^B_{A_k} e^*_{A_1 \ldots A_{k-1}} -  \delta^B_{A_{k-1}} e^*_{A_1 \ldots A_{k-2} A_k} + \cdots + (-1)^{k-1}  \delta^B_{A_1} e^*_{A_2 \ldots A_k} \ . 
\label{tetrad}
\EE

Donnons maintenant les équations de structure de Cartan :
\begin{dico}[formulation de Cartan] 
La 1-forme $\om^A_{\ B}$, appelée la connexion de spin affine, est définie par
\BE \dd e^A + \om^A_{\ B} \w e^B = T^A \ , \label{cartan1}\EE 
où $T^A $ est appelée la 2-forme de \ita{torsion}. Quant à la 2-forme de \ita{courbure}, notée $\Th^A_{\ B}$, elle est donnée par
\BE \Th^A_{\ B} = \dd \om^A_{\ B} + \om^A_{\ C} \w \om^C_{\ B}  \ . \label{cartan2}\EE
\end{dico}
La 2-forme de courbure permet de remonter au tenseur de Riemann introduit ultérieurement par les relations
\BE  \Th^A_{\ B}  = \frac{1}{2} R^A_{\ BCD} e^C \w e^D  = \frac{1}{2} R^A_{\ B\mu\nu} \dd x^\mu \w \dd x^\nu   \EE
ainsi
\BE  R^\al_{\ \be\mu\nu} =  R^A_{\ B\mu\nu} E_A^{\ \al} e^B_{\ \be} \ . \EE
Dans le langage précédent, nous avons introduit une dérivée covariante $\n$ sans torsion, ce qui a permis d'obtenir des symboles de Christoffel symétrique $\Ga_{\mu\nu}^\rho = \Ga_{\nu\mu}^\rho$. Dans le langage de Cartan, ceci correspond à considérer que la 2-forme de torsion est nulle, $T^A = 0 $, ce que nous choisirons dans la suite. Nous avons aussi considéré une connexion vérifiant $\n_\rho g_{\mu\nu} = 0$, cette propriété est traduite ici par la symétrie suivante pour la connexion de spin 
\BE \om_{AB} = \om_{BA} \ , \EE
cette dernière est alors qualifiée de connexion de spin de \ita{Levi-Civita}.\bs

\subsubsection{Variétés kählériennes}
Pour finir cette sous-section, nous allons revenir brièvement au cas des variétés complexes en présentant notamment les variétés dites \ita{kählériennes} que nous utiliserons à la sous-section \ref{kähler} en particulier. Fournissons d'abord quelques définitions : 
\begin{dico}[structure complexe] Soit $V$ un espace vectoriel réel de dimension paire ou un espace vectoriel complexe. $V$ possède une structure complexe s'il existe un endomorphisme de $V$, noté $J$, tel que $J \circ J = -\text{Id}$.
\end{dico}

\begin{dico}[structure hermitienne] Soit $V$ un espace vectoriel muni d'une structure complexe $J$. Une structure hermitienne est une application $h$ qui pour tout élément $(u,v)$ de $V^2$ associe un nombre complexe $h(u,v)$ vérifiant
\begin{itemize}
\item $\forall a,b \in \mathbb{C}\ , \forall w \in V  \ , h(au+bv,w) = ah(u,w) +bh(v,w) \ , $
\item $\overline{h(u,v)} = h(v,u) $
\item et $h(Ju,v)=ih(u,v)\ .$
\end{itemize}
\end{dico}
Nous pouvons désormais considérer une variété complexe $\mc M$. Cette dernière est qualifiée de \ita{variété hermitienne} s'il existe pour tout point de $\mc M$ une structure hermitienne $h$ définie sur l'espace tangent lui-même muni d'une structure complexe, il est d'usage de parler de structure presque complexe dans ce cas. Si $\lp x_\al, y_\al \rp$ désigne un système de coordonnées réelles alors il est loisible de considérer aussi les coordonnées conjuguées $z_\al = x_\al + i y_\al$ et $\bar{z}_\al = x_\al - i y_\al$. Ainsi, nous associons à cette variété une \ita{métrique hermitienne} via la relation
\BE h = h_{\al\bar{\be}} \dd z^\al \otimes \dd \bar{z}^\be \EE
où $h_{\al\bar{\be}} = h(\p_\al , \bar{\p}_\be)$ avec des notations évidentes. Il est courant de noter $g=\frac12 (h+\bar{h})$ la partie réelle de $h$, par conséquent $g=\frac12 h_{\al\bar{\be}} \lp \dd z^\al \otimes \dd \bar{z}^\be + \dd \bar{z}^\be \otimes\dd z^\al \rp$; et $\om = \frac{i}{2} (h-\bar{h}) = \frac{i}{2} h_{\al\bar{\be}}  \dd z^\al \w\dd \bar{z}^\be $ est l'opposé de la partie imaginaire de $h$, cette 2-forme réelle est appelée la \ita{forme de Kähler}. Par construction, nous avons $h=g-i\om$. En fait, à partir de la métrique réelle $g$ et de la structure presque complexe $J$, il est possible de construire $\om$ par la relation $\om(u,v) = g(Ju,v)$ pour tout $u,v$ du fibré tangent, puis de construire la métrique hermitienne. Il est possible aussi de commencer avec la forme de Kähler $\om$ en remarquant que $g(u,v)=\om(u,Jv)$. Terminons par la définition suivante : 

\begin{dico}[variété kählérienne] La métrique d'une variété hermitienne $\mc M $ est dite de Kähler si la forme de Kähler $\om$ est fermée, c'est-à-dire $\dd \om =0$. Dans ce cas, $\mc M$ est qualifiée de variété kählérienne. 
\end{dico}
En fait, $\om$ est même harmonique dans ce cas. Avec une métrique de Kähler, toutes les formes $\bigwedge_{k=1}^n \om$, pour $n\in\llbracket 1, D/2 \rrbracket$ où $D$ est la dimension réelle de la variété, sont harmoniques. Citons $\mathbb{R}^2$, $S^2$ et les espaces projectifs $\mathbb{C}P^n$ comme exemples de variétés kählériennes. Le lecteur pourra consulter \cite{chern1979complex} pour de plus amples détails sur les variétés complexes.

		\subsection{Difféomorphismes, dérivée de Lie et vecteurs de Killing}
Nous verrons qu'il est bien souvent nécessaire d'imposer certaines symétries pour trouver des solutions à l'équation d'Einstein. Cette sous-section a pour objectif de définir d'une manière relativement précise les quantités qui décrivent ces symétries.

Considérons dans un premier temps deux variétés $\mc M$ et $\mc M'$ munies de leurs cartes $\{ \psi_i \}$ et $\{ \psi'_i \}$ respectivement. Une fonction $\phi:\mc M \rightarrow \mc M'$ est dite $\mc C^\infty$ si pour tout $i$ et $j$, $\psi'_j \circ \phi \circ \psi^{-1}_i$ est $\mc C^\infty$. Ainsi, si $\phi:\mc M \rightarrow \mc M'$ est $\mc C^\infty$, bijective et $\phi^{-1}$ est $\mc C^\infty$ aussi, alors $\phi$ est appelé un \ita{difféomorphisme} et la variété $\mc M$ est dite \ita{difféomorphe} à $\mc M'$. 
Puis, introduisons un \ita{groupe de difféomorphismes à un paramètre} $\phi_t$ qui est défini comme une fonction $\mc C^\infty$ de $\mathbb{R}\times \mc M \rightarrow \mc M$ tel que $\forall t \in \mathbb{R}, \phi_t : \mc M \rightarrow \mc M$ soit un difféomorphisme et $\forall s,t \in \mathbb{R}, \phi_t \circ \phi_s = \phi_{t+s}$. Un champ tangent $v$ peut alors être associé à $\phi_t$ de la manière suivante. Pour chaque point $P$ de $\mc M$, $\phi_t (P) : \mathbb{R} \rightarrow \mc M$ est une courbe appelée \ita{orbite} de $\phi_t$ passant par $P$ à $t=0$. En particulier, il existe un vecteur dans $V_P$ tangent à cette courbe en $t=0$. Ainsi, à chaque groupe de difféomorphismes à un paramètre est associé un champ tangent $v$, qui est aussi qualifié de \ita{générateur infinitésimal} pour cette transformation.

A partir d'un difféomorphisme $\phi$ de $\mc M$ dans $\mc M$, il est possible de construire naturellement une fonction $\phi^*$ sur les champs tensoriels. Nous ne détaillons pas volontairement cette construction ici. Ainsi, pour tout champ tensoriel $T$, nous pouvons comparer $T$ avec $\phi^* T$ ; si ces deux quantités sont égales alors $\phi$ est une \ita{symétrie} pour le tenseur $T$. En particulier, si $T$ est le tenseur métrique $g$ et $\phi^* g = g$, alors la symétrie est qualifiée d'\ita{isométrie}. Passons à un autre type de dérivation qui va permettre de comparer $T$ et $\phi^* T$: la dérivée de Lie.\bs

Soit $\phi_t$ un groupe de difféomorphisme à un paramètre, généré par un champ tangent $v$ et $T\in \mc T(k,l)$, la \ita{dérivée de Lie} selon $v$ est alors définie par
\BE \pounds_v T^{\mu_1 \ldots \mu_k}_{\ \ \ \ \ \ \ \nu_1 \ldots \nu_l} = \lim_{t \rightarrow 0} \frac{\phi_{t}^* T^{\mu_1 \ldots \mu_k}_{\ \ \ \ \ \ \ \nu_1 \ldots \nu_l} -T^{\mu_1 \ldots \mu_k}_{\ \ \ \ \ \ \ \nu_1 \ldots \nu_l}  }{t} \EE
où chaque tenseur est évalué au même point. Avec cette définition, $\pounds_v$ est linéaire sur les tenseurs, vérifie la règle de Leibniz et $\forall f:\mc M \rightarrow \mathbb{R}, \pounds_v f = v(f)$. Nous remarquons aussi que si $\phi_t$ est une symétrie pour un tenseur alors la dérivée de Lie de ce tenseur selon le générateur infinitésimal de cette transformation est nulle. En utilisant la dérivée covariante $\n$, il est possible de montrer que 
\BE \pounds_v S^\mu_{\ \nu} = v^\rho \n_\rho S^\mu_{\ \nu} -  S^\rho_{\ \nu} \n_\rho v^\mu + S^\mu_{\ \rho} \n_\nu v^\rho \EE
pour un tenseur $S$ de type (1,1); les règles de contraction de cette égalité se généralisent à des tenseurs de rang plus élevé.\bs

Concernant le vocabulaire, un tenseur $T\in \mc T(k,l)$ est dit \ita{Lie transporté} le long d'une courbe $\ga$  paramétrisée par $x^\mu(\la)$ si sa dérivée de Lie selon le vecteur tangent $u^\mu = \frac{\dd x^\mu}{\dd \la}$ à cette courbe est nulle. Si nous choisissons un système de coordonnées tel que $x^1 = \la$ sur $\ga$ et avec les autres coordonnées constantes sur $\ga$ alors $u^\mu\stackrel{*}{=}\delta_1^\mu$; nous indiquons par $*$ que nous utilisons un système de coordonnées particulier pour écrire cette égalité. Ainsi, nous avons
\BE \pounds_u T^{\mu_1 \ldots \mu_k}_{\ \ \ \ \ \ \ \nu_1 \ldots \nu_l} \stackrel{*}{=}  \frac{\p}{\p x^1} T^{\mu_1 \ldots \mu_k}_{\ \ \ \ \ \ \ \nu_1 \ldots \nu_l} \ .\EE
Par conséquent, si ce tenseur est Lie transporté, il ne dépend alors pas de la coordonnée $x^1$. Inversement, si dans un système de coordonnées un tenseur ne dépend pas d'une coordonnée $x^1$ alors la dérivée de Lie selon $u^\mu \stackrel{*}{=}  \delta^\mu_1$ s'annule.
 
En particulier, si dans un système de coordonnées, les composantes de la métrique ne dépendent pas de $x^1$, alors $\pounds_\xi g_{\mu\nu}= 0$ avec $\xi=\p_1$. Dans ce cas, le vecteur $\xi$ est appelé un \ita{vecteur de Killing}. Un vecteur de Killing $\xi$ doit donc vérifier l'équation dite de Killing  
\BE \n_{(\mu} \xi_{\nu)} = 0 \ . \label{killing}\EE
 Une propriété remarquable des vecteurs de Killing est qu'ils permettent de déterminer des constantes du mouvement associées à une géodésique : 
\begin{proposition}Soit $\xi$ un vecteur de Killing et $\ga$ une géodésique paramétrisée par $x^\mu(\tau)$ où $\tau$ est un paramètre affine et $u$ un vecteur tangent. Le produit scalaire $\xi_\mu u^\mu$ est constant le long de $\ga$.   
\end{proposition}
En effet, $u^\nu \n_\nu \left( u^\mu \xi_\mu \right)  = \xi_\mu u^\nu \n_\nu u^\mu +  u^\mu u^\nu \n_\nu \xi_\mu$, le premier terme est nul d'après l'équation de géodésique \eqref{géodésique} et le second terme aussi étant donné l'antisymétrie de $\n_\mu \xi_\nu$ d'après \eqref{killing}. Le lecteur pourra dériver également l'égalité suivante
\BE \n_\mu \n_\nu \xi_\rho = -R^{\ \ \ \ \si}_{\nu\rho\mu} \xi_\si \label{killing-riemann}\EE
qui permet notamment de montrer qu'il y a au plus $D+\frac{D(D-1)}{2} = \frac{D(D+1)}{2}$ vecteurs de Killing linéairement indépendants dans une variété de dimension $D$. Enfin, un espace-temps est dit \ita{maximalement symétrique} s'il possède $\frac{D(D+1)}{2}$ vecteurs de Killing.

		\subsection{Hypersurfaces\label{hypersurface}}
Présentons ici un autre thème. Quand nous aborderons le formalisme lagrangien et hamiltonien et les conditions de jonction, nous serons notamment en présence d'hypersurfaces plongées dans une variété $\mc M$ de dimension $D$. Dans cette sous-section, nous allons donc définir des quantités \ita{intrinsèques} à l'hypersurface et des quantités \ita{extrinsèques}; ces dernières indiquent comment l'hypersurface est plongée dans $\mc M$. Nous restreignons notre attention à des hypersurfaces de genre-espace ou de genre-temps. Le lecteur pourra se tourner vers \cite{poisson2004relativist} pour étudier des hypersurfaces de genre-lumière. 

Une \ita{hypersurface}, notée $\Si$, est une sous-variété de dimension $D-1$, qui peut être décrite par une contrainte
\BE\Phi(x^\mu)=0 \EE
où $x^\mu$ représente un système de coordonnées de $\mc M$, ou par des équations paramétriques sous la forme
\BE x^\mu = x^\mu(y^a) \EE
en introduisant un \ita{système de coordonnées $y^a$ intrinsèques} à l'hypersurface où $a\in\llbracket 1,D-1\rrbracket$. 

Introduisons un vecteur normal à $\Si$, noté $n^\mu$ et imposons le unitaire, c'est-à-dire $n^\mu n_\mu = \varep$ avec $\varep=-1$ pour une hypersurface de genre-espace et $\varep=1$ pour une hypersurface de genre-temps. Remarquons d'abord que $\p_\mu \Phi$ est normal à $\Si$, ainsi en imposant que le vecteur $n^\mu$ pointe dans la direction qui augmente la valeur de $\Phi$, c'est-à-dire $n^\mu \p_\mu \Phi  > 0$, le vecteur normal unitaire est donné par 
\BE  n_\mu = \frac{\varep \p_\mu\Phi}{\sqrt{|\p_\nu\Phi \p^\nu\Phi|}} \ . \EE

\subsubsection{Première forme fondamentale}

Nous allons maintenant déterminer la métrique intrinsèque à l'hypersurface. Pour cela, introduisons dans un premier temps les $D-1$ vecteurs
\BE e_a^\al = \frac{\p x^\al}{\p y^a} \EE
qui correspondent aux vecteurs tangents aux courbes $x^\mu(y^a) $ contenues dans $\Si$. Ainsi, par cette construction, ces vecteurs sont normaux à $n^\al$, c'est-à-dire : $ e_a^\al n_\al =0$. Puis, nous pouvons alors déterminer la restriction à $\Si$ de la métrique de $\mc M$ par
\BE \dd s^2 _{|\Si}= g_{\al\be} \dd x^\al \dd x^\be \stackrel{x^\mu(y^a) }{=}   g_{\al\be} \frac{\p x^\al}{\p y^a} \frac{\p x^\be}{\p y^b}\dd y^a \dd y^b\ .  \EE
Par conséquent, la \ita{métrique induite} ou aussi appelée \ita{première forme fondamentale} est donnée par
\BE h_{ab}=g_{\al\be} e_a^\al e_b^\be \ . \EE
Précisons que la métrique induite est un scalaire sous un changement de coordonnées d'espace-temps $x^\al \rightarrow x'^\al $;  en revanche, du point de vue de la sous-variété $\Si$, elle suit la loi de transformation d'un tenseur $(0,2)$ sous le changement de coordonnées $y^a \rightarrow y'^a $. Pour être complet, donnons également la relation
\BE g^{\al\be} = \varep n^\al n^\be + h^{ab} e_a^\al e_b^\be  \EE
où $h^{ab}$ est l'inverse de la métrique induite, nous introduisons aussi la notation $h^{\al \be} =  h^{ab} e_a^\al e_b^\be$. $h^\al_{\ \be}$ constitue ainsi une projection de $T_P(\mc M)$ sur $T_P(\Si)$ où $P$ est un point de la variété $\mc M$. Partant d'un tenseur $T\in \mc T_P(k,l)$ de la variété $\mc M$, le tenseur 
\BE T_{\parallel\ \ \ \ \ \be_1 \ldots \be_l}^{ \al_1 \ldots \al_k} = h^{\al_1}_{\ \mu_1} \ldots h^{\al_k}_{\ \mu_k} h_{\be_1}^{\ \nu_1} \ldots h_{\be_l}^{\ \nu_l}  T_{\ \ \ \ \ \ \nu_1 \ldots \nu_l}^{ \mu_1 \ldots \mu_k} \EE
est dit \ita{tangent à l'hypersurface} $\Si$, étant donné que toute contraction entre $T_{\parallel\ \ \ \ \ \be_1 \ldots \be_l}^{ \al_1 \ldots \al_k}$ et le vecteur normal $n^\al$ est nulle. Un tenseur $S^{\al\be\ldots}$ tangent à l'hypersurface peut alors être décomposé sur la base des vecteurs tangents de la manière suivante: $S^{\al\be\ldots} = S^{ab\ldots} e_a^\al e_b^\be \cdots$; de cette manière $S^{ab\ldots}$ est un scalaire vis-à-vis de la variété $\mc M$ et c'est un tenseur du point de vue de $\Si$.

\subsubsection{Seconde forme fondamentale}
Etant donné la métrique induite, toute la géométrie intrinsèque de $\Si$ peut alors être construite en déterminant la connexion et les tenseurs de courbures intrinsèques à $\Si$ que nous avons rencontrés dans les sous-sections précédentes. Nous devons désormais préciser comment $\Si$ est plongée dans $\mc M$. Pour cela, nous allons introduire la courbure dite extrinsèque qui décrit ce plongement. Considérons un vecteur tangent à l'hypersurface qui vérifie alors $A^\al = A^a e_a^\al$, $A^\al n_\al = 0$ et $A_a = A_\al e^\al_a$. En notant $\n_\parallel$ et $\Ga_{\parallel ab}^c$ la dérivée covariante et les symboles de Christoffel associés à $\Si$, $\n_{\parallel a} A_b = \p_a A_b - \Ga_{\parallel ab}^c A_c$ par définition, puis après calculs nous trouvons $\n_{\parallel a} A_b = \left(\n_\al A_\be \right) e^\al_a e^\be_b $. $\n_{\parallel a} A_b$ correspond donc à la partie tangentielle de $e^\al_a \n_\al A_\be$. Mais, quelle est sa partie normale ? Le lecteur pourra montrer que $e_b^\be \n_\be A^\al = \left( \n_{\parallel b} A^a \right) e_a^\al - K_{ab}A^a n^\al$ avec 
\BE K_{ab}= \left( \n_\al n_\be \right) e^\al_a e^\be_b \EE
appelée \ita{courbure extrinsèque} ou \ita{seconde forme fondamentale}. En remarquant que les $D-1$ vecteurs $e_a=e_a^\al \p_\al$ sont Lie transportés le long des courbes $x^\mu(y^a)$, c'est-à-dire $\pounds_{e_a} e_b^\al = 0$, et que $e_a^\al n_\al = 0$, nous établissons la symétrie suivante $K_{ab}=K_{ba}$, ainsi
\BE  K_{ab} = e^\al_a e^\be_b \n_{(\al} n_{\be)} = \frac{1}{2} e^\al_a e^\be_b \pounds_{n}g_{\al\be} \ . \EE
Illustrons le caractère extrinsèque de ce tenseur avec les deux exemples suivants. Considérons d'abord le cas de $\Si=S^1$ une sphère de rayon $R$ plongée dans $\mathbb{R}^2$ munie de la métrique $\dd s^2 = \dd r^2 + r^2 \dd\theta^2$ écrite en coordonnées polaires. $\Si$ est paramétrisée par $r=R$, la normale unitaire est $n=n_\al \dd x^\al = \dd r$ et l'unique vecteur tangent est $e_\theta = \p_\theta$. La trace de la courbure extrinsèque, qui est un invariant, vaut alors $K=K_a^a=1/R$, qui est simplement la courbure de la sphère. En revanche, si nous considèrons $\Si=S^1$ plongée dans $S^2$ munie de la métrique $\dd s^2 = \dd \theta^2 + \sin^2\theta \dd\phi^2$ écrite en coordonnées sphériques avec $\Si$ paramétrisée par $\phi=\phi_0$ où $\phi_0$ est une constante, alors un calcul similaire donne le résultat intuitif $K=0$.\bs

Terminons cette sous-section par le \ita{théorème de Stokes} : 
\begin{proposition}[Stokes] \label{th-Stokes}
Considérons un compact $U$ de $\mc M$ avec une frontière $\p U$. Pour tout champ tangent $A^\mu$,
\BE \int_U \n_\mu A^\mu \sqrt{-g} \dd^D x = \oint_{\p U} A^\mu \dd \Si_\mu \label{stokes} \EE
où l'élément de surface orienté est donné par $\dd \Si_\mu  = \varep n_\mu \dd \Si$ avec $\dd \Si = \sqrt{|h|}\dd^{D-1}y$ la forme volume de $\p U$, nous ne discutons pas volontairement du cas où la frontière est de genre-lumière. 
\end{proposition}
Ce résultat se généralise avec des formes différentielles. Si $\mc M$ est une variété de dimension $p$ avec une frontière $\p \mc M$ alors pour toute $(p-1)$-forme $\om$, nous avons 
\BE \int_{\mc M} \dd\om = \oint_{\p \mc M} \om \ . \EE

	\section{Les équations du mouvement}
	Dans la section précédente, nous nous sommes intéressés aux quantités cinématiques. Nous devons maintenant donner la dynamique qui régit la théorie de la gravitation. Celle-ci va être donnée par la fameuse \ita{équation d'Einstein}.
		\subsection{L'équation d'Einstein}
Avant de donner l'équation d'Einstein explicitement, esquissons quelques motivations qui ont mené à l'établir. A travers l'équation \eqref{déviation}, nous avons déterminé la force de marée entre deux particules décrivant des géodésiques voisines, en reliant cette force au tenseur de Riemann. Par ailleurs, pour de faible champ gravitationnel et pour des particules se déplaçant à des vitesses faibles devant celle de la lumière,  la théorie de la gravitation d'Einstein doit restaurer la théorie Newtonienne. Considérons alors deux particules-test proches dans le cadre de la Mécanique Newtonienne, la première paramétrisée par $x^i(t)$ et la seconde par $x^i(t)+\xi^i(t)$. En introduisant le potentiel gravitationnel $\phi$, leur trajectoire est donnée par
\BE \ddot{x}^i = - \left. \frac{\p \phi}{\p x^i} \right|_{x^j} \text{ et } \ddot{x}^i + \ddot{\xi}^i = - \left. \frac{\p \phi}{\p x^i} \right|_{x^j+\xi^j} \EE
respectivement, les indices latins $i$ et $j$ sont ici purement spatiaux. Ainsi, en ne retenant que le premier terme linéaire en $\xi^j$, l'accélération relative entre les particules est donnée par
\BE \ddot{\xi}^i = - \frac{\p^2 \phi}{\p x^i \p x^j} \xi^j \ .\EE
Pour un tel mouvement, nous négligeons dans \eqref{déviation} les composantes $\frac{\dd x^i}{\dd t}$ du vecteur tangent $u^\al$ à la géodésique devant la composante $\frac{\dd x^0}{\dd t}$, ainsi nous établissons la correspondance suivante
\BE  R^i_{\  0 j 0} \leftrightarrow \p^i \p_j \phi \ . \EE
En particulier, $ R^i_{\  0 i 0} $ est relié au laplacien du potentiel gravitationnel $\Delta \phi$ qui est donné par l'équation de Poisson $\Delta \phi = 4 \pi G \rho$ avec $\rho$ la densité volumique de masse. Puis, en remarquant que $ R^i_{\  0 i 0} = R^\mu_{\  0 \mu 0} = R_{00} = R_{\mu\nu} u^\mu u^\nu$ ici et en supposant que la densité $\rho$ soit donnée par $T_{\mu\nu} u^\mu u^\nu$ où $T_{\mu\nu}$ représente le \ita{tenseur énergie-impulsion} qui caractérise les propriétés énergétiques de la matière, nous sommes alors amenés à identifier le tenseur de Ricci au tenseur énergie-impulsion (nous donnerons dans la prochaine sous-section une définition précise du tenseur énergie-impulsion à partir du lagrangien qui caractérise la matière). Cependant, A. Einstein rejeta cette identification qui ne permet pas d'obtenir la conservation du tenseur énergie-impulsion $\n_\mu T^{\mu\nu} =0$, en effet nous verrons en \ref{matière} que ce résultat doit être vérifié pour que l'action représentant la matière soit invariante sous les difféomorphismes. Ainsi, connaissant l'identité de Bianchi \eqref{bianchi2}, A. Einstein proposa le 25 novembre 1915 \cite{miller1997collected} la forme finale suivante pour l'équation qui porte son nom et qui constitue notre dernier axiome concernant sa théorie de la gravitation : 
\begin{axiome}[équation d'Einstein] La métrique de l'espace-temps vérifie l'équation
\BE G_{\mu\nu} = \frac{8\pi G}{c^4} T_{\mu\nu} \label{Einstein}\EE
en rétablissant toutes les unités, où $c$ désigne la vitesse de la lumière dans le vide.
\end{axiome}
Cette équation assure notamment la conservation du tenseur énergie-impulsion. Elle peut aussi être écrite sous la forme 
\BE R_{\mu\nu} = \frac{8\pi G}{c^4} \left( T_{\mu\nu} - \frac{1}{2}g_{\mu\nu}T \right) \label{Einstein2}\EE
où $T=T_\mu^{\ \mu}$. Le résultat newtonien est en fait restauré puisque dans cette limite $T=-\rho=-T_{\mu\nu} u^\mu u^\nu$ et par conséquent $R_{\mu\nu} u^\mu u^\nu = 4\pi G T_{\mu\nu} u^\mu u^\nu$. La première remarque est, qu'à travers l'équation \eqref{Einstein}, nous retrouvons en partie l'idée du principe de Mach. Une seconde remarque est qu'elle est non-linéaire et dépend des dérivées secondes de la métrique, ce qui explique la difficulté de trouver une solution. Enfin, précisons qu'en l'absence de matière, la contribution de la trace du tenseur de Riemann est nulle, c'est-à-dire $R_{\mu\nu}=0$, ainsi les degrés de libertés gravitationnels sont encodés dans le tenseur de Weyl \eqref{weyl}.

Soulevons un dernier point. Après avoir déterminé une solution de l'équation d'Einstein, il est intéressant de regarder le comportement d'une particule-test dans cette géométrie à travers l'équation de géodésique \eqref{géodésique}. De nouveau, nous pouvons préciser ce que devient cette équation dans la limite newtonienne en considérant une particule-test se déplaçant à une faible vitesse devant celle de la lumière dans un champ gravitationnel quasi-stationnaire. Il existe alors un système de coordonnées dans lequel $g_{\mu\nu} = \eta_{\mu\nu} + h_{\mu\nu} $ où $h_{\mu\nu}$ représente une perturbation de la métrique. Ainsi \eqref{géodésique} devient $\frac{\dd^2 x^i}{\dd t^2} = -\frac{1}{2}\p_i h_{00} $ après calculs, or $\frac{\dd^2 x^i}{\dd t^2} = - \p_i \phi$ en Mécanique Newtonienne, il en résulte alors la relation
\BE g_{00} = -1 - 2\phi \ ,\EE
puisque $g_{00}$ et $\phi$ tendent vers $1$ et $0$ à l'infini respectivement. 
		\subsection{Le formalisme lagrangien}
Nous allons présenter dans cette sous-section la formulation lagrangienne de la Relativité Générale valable en dimension $D \geq 4$. En plus de son élégance, ce formalisme est un point de départ possible pour formuler une théorie quantique de la gravitation en utilisant la méthode de l'intégrale de chemin. Nous prendrons notamment le soin de dériver les termes de bord apparaissant dans cette formulation, qui sont d'une importance capitale pour le chapitre \ref{thermo} abordant l'aspect thermodynamique des théories gravitationnelles.

Considérons une théorie décrite par une collection de champs tensoriels définis sur une variété différentielle $\mc M$. Notons par $\psi$ un champ tensoriel de $\mc T(k,l)$ et posons $S:\psi \rightarrow S[ \psi ] \in \mathbb{R}$ une fonctionnelle de $\psi$. Soit $\{ \psi_\la \}$ une famille, $\mc C^\infty$, à un paramètre, de champs tensoriels définis sur un compact $U$ de $\mc M$ tel que $\forall \la \ , \psi_\la = \psi_0 $ sur $\p U$. Nous introduisons également la notation $\de\psi = \frac{\dd \psi_\la}{\dd \la} $. De plus, supposons que $ \frac{ \dd S[\psi_\la]}{\dd \la} $ existe et qu'il existe $\chi \in \mc T(l,k)$ tel que $\frac{\dd S}{\dd \la}  = \int_{\mc M} \chi \de\psi \dd^D x$, dans ce cas la fonctionnelle $S$ est dite \ita{fonctionnellement différentiable} et $\chi$ est appelée la \ita{dérivée fonctionnelle} de $S$, notée  $\frac{\de S}{\de\psi}$. Puis, nous considérons une fonctionnelle $S$ sous la forme
\BE S[\psi] = \int_{\mc M} \mc L[\psi] \ep \EE
où $\mc L$ est une fonction qui en un point $P$ de $\mc M$ ne dépend que de $\psi_{|P}$ et d'un nombre fini de dérivées de $\psi$ évaluées en $P$. Le \ita{principe variationnel} consiste à supposer que $S$ est fonctionnellement différentiable et que les configurations du champ $\psi$ qui extrémisent $S$, c'est-à-dire $\left. \frac{\de S}{\de\psi}\right|_{\psi} = 0$, sont celles qui sont solutions de l'équation du champ $\psi$ en tenant compte de la condition de bord précisée au-dessus. Dans ce cas, $S$ est appelé \ita{l'action} et $\mc L$ la \ita{densité lagrangienne}. 

En ce qui concerne la théorie de la gravitation, elle est décrite par la métrique; plus précisément nous choisissons $ \psi_\la = (g^{\mu\nu})_\la $ et l'action dite d'\ita{Einstein-Hilbert} est la suivante
\BE S_H[g^{\mu\nu}] = \int_U R \ep =\int_U R \sqrt{-g} \dd^D x  \ .\EE
La variation de cette action est donc
\BE \frac{\dd S_H}{\dd \la} = \int_U \left[ R \de(\sqrt{-g}) + \sqrt{-g}g^{\mu\nu}\de{R_{\mu\nu}} +  \sqrt{-g}R_{\mu\nu}\de{g^{\mu\nu}}  \right] \dd^D x \ . \EE
Puis, le lecteur pourra montrer que $ \de(\sqrt{-g}) = -\frac{1}{2}\sqrt{-g}g_{\mu\nu} \de g^{\mu\nu} $ et que la variation du tenseur de Ricci est donnée par
\BE g^{\mu\nu}\de{R_{\mu\nu}} = \n^\mu  \underbrace{  \left( \n^\nu \de g_{\mu\nu} - g^{\rho\si} \n_\mu \de g_{\rho\si}  \right)  }_{\dot{=}v_\mu} \ .\EE
Ainsi, en utilisant le théorème de Stokes \eqref{stokes}, nous trouvons
\BE \frac{\dd S_H}{\dd \la} = \int_U G_{\mu\nu} \de g^{\mu\nu} \ep + \varep \oint_{\p U} v^\mu n_\mu \dd \Si \EE
où $n^\mu$ est le vecteur normal à $\p U$ défini dans la sous-section \ref{hypersurface}, $\varep=-1$ si $\p U$ est de genre-espace et $\varep=1$ si $\p U$ est de genre-temps, nous ne précisons pas volontairement le cas où $\p U$ est de genre-lumière. Après cela, en utilisant l'expression du vecteur $v^\mu$ évalué sur $\p U$ et le fait que $\de g_{\mu\nu}=0$ sur $\p U$, le terme de bord devient $v^\mu n_\mu= -2  \de K $ sur $\p U$ où $K$ est la trace de la courbure extrinsèque de $\p U$. Par conséquent, l'action complète de la théorie de la gravitation couplée à la matière en tenant compte des conditions de bord est
\BE S= \int_U R + 2\varepsilon\int_{\p U} K + \al_m \underbrace{\int_U \mc L_m[g_{\mu\nu},\phi]}_{S_m [g_{\mu\nu},\phi]} \ ,\label{GHY}\EE
où $\al_m$ est une constante de couplage; précisons aussi que nous n'avons pas explicité la forme volume dans cette expression. Le terme de bord est appelé \ita{terme de Gibbons-Hawking-York}. Concernant l'action de la matière $S_m$, il se peut que certains termes de bord apparaissent également en fonction des conditions de bord sur le champ $\phi$, qui est ici un champ tensoriel décrivant la matière. La variation de l'action $S$ par rapport à la métrique reproduit alors l'équation d'Einstein \eqref{Einstein} en définissant le tenseur énergie-impulsion par
\BE T_{\mu\nu} = - \frac{\al_m}{8\pi} \frac{1}{\sqrt{-g}} \frac{\de S_m}{\de g^{\mu\nu}} \ . \label{Tmunu}\EE
Quant à la variation de $S$ par rapport à $\phi$, elle produit l'équation du mouvement du champ tensoriel $\phi$. En partant de ce formalisme lagrangien, nous dériverons, dans le dernier chapitre, la version hamiltonienne en prenant toujours le soin de garder les termes de bord.\bs

Enfin, nous pouvons nous poser la question de l'unicité de la théorie de la gravitation d'Einstein. Quelle est la théorie la plus générale construite à partir de la métrique $g$ qui donne une équation du mouvement sous la forme $D_{\mu\nu}[g]=T_{\mu\nu}$ où $D_{\mu\nu}$ est un champ tensoriel qui ne dépend que de la métrique, de ses dérivées et de ses dérivées secondes au point considéré et tel que $\n_\nu D^{\mu\nu} = 0 $ ? David Lovelock donna une réponse à cette question en 1972 \cite{Lovelock:1972vz} : 
\begin{proposition}[Lovelock] La théorie la plus générale en dimension 4 qui respecte les conditions précédentes est donnée par la densité lagrangienne
$\mc L = \al + \be R $ où $\al$ et $\be$ sont deux constantes. \label{thlovelock}
\end{proposition}
Le premier terme correspondant à la \ita{constante cosmologique} sera présenté à la sous-section suivante. Nous verrons aussi à la section \ref{lovelock} que ce théorème se généralise en dimensions supplémentaires.
		\subsection{La matière\label{matière}}
Nous allons dans cette sous-section diriger notre attention sur le membre de droite de l'équation d'Einstein \eqref{Einstein} qui représente la matière. Observons d'abord que l'action $S_m$ représentant la matière doit être invariante sous les difféomorphismes, c'est-à-dire $S_m [g_{\mu\nu},\phi] = S_m [f^*_\la g_{\mu\nu}, f^*_\la \phi]$ où $f_\la$ est un groupe de difféomorphismes à un paramètre de $\mc M$ dans $\mc M$. Pour de telles variations, 
\BE \frac{\dd S_m}{\dd \la} = 0 = \int_U \frac{\de S_m}{\de g^{\mu\nu}}\de g^{\mu\nu} \dd^D x + \int_U \frac{\de S_m}{\de \phi}\de \phi \dd^D x \EE
où ici $\de g^{\mu\nu} = \pounds_v g^{\mu\nu} = 2 \n^{(\mu} v^{\nu)} $ avec $v$ un vecteur tangent arbitraire. Par conséquent, si $\phi$ vérifie l'équation du mouvement alors $\left. \frac{\de S_m}{\de \phi} \right|_\phi = 0 $ et ainsi $\int_U \sqrt{-g} T_{\mu\nu} \n^\mu v^\nu \dd^D x = 0 $ puisque le tenseur énergie-impulsion est symétrique. Ce qui donne $ \n_\mu T^{\mu\nu} =0 $ après intégration par partie en ignorant le terme de bord. En somme, si une action de matière est invariante sous les difféomorphismes alors le tenseur énergie-impulsion est conservé. 
Dans la suite, nous allons présenter des exemples de tenseur énergie-impulsion. 

\subsubsection{Un fluide parfait}
Commençons par le cas du fluide parfait. En notant par $u^\mu$ la 4-vitesse d'une particule fluide, le tenseur énergie-impulsion est donné par
\BE T_{\mu\nu} = \rho u_\mu u_\nu + P \left( g_{\mu\nu} + u_\mu u_\nu \right) \label{fluide-parfait} \EE
où $\rho$ désigne la masse volumique propre du fluide et $P$ sa pression. Puis, nous obtenons l'équation du mouvement avec $ \n_\mu T^{\mu\nu} =0 $ qui fournit la généralisation de l'équation de continuité et de celle de Navier-Stokes.

\subsubsection{Un champ scalaire} 
Un guide possible pour obtenir l'équation du mouvement pour la matière en Relativité Générale est de remplacer, dans les lois physiques appliquées à la Relativité Restreinte, la métrique de Minkowski $\eta_{\mu\nu}$ par la métrique $g_{\mu\nu}$ et $\p_\mu$ par $\n_\mu$. Cette procédure est parfois appelée \ita{couplage minimal} en référence à la procédure qui s'applique en électromagnétisme. Ainsi, l'équation de Klein-Gordon pour un champ scalaire $\phi$ massif devient $\n_\mu \n^\mu \phi = m^2 \phi $ qui peut être déduite du tenseur énergie-impulsion 
\BE T_{\mu\nu} = \n_\mu \phi \n_\nu \phi - \frac{1}{2} g_{\mu\nu} \left( \n_\rho \phi \n^\rho \phi + m^2\phi^2 \right) \ ,  \EE
lui-même généré via \eqref{Tmunu} par l'action $ S=\int \left( \p_\mu \phi \p^\mu \phi + m^2\phi^2 \right) \ep $ avec $\al_m = - 8 \pi$. Nous verrons à la section \ref{conforme} qu'il est aussi possible de généraliser l'équation de Klein-Gordon sans masse sous la forme suivante $\n_\mu \n^\mu \phi = \frac{1}{6}R\phi $ en tenant compte de l'invariance conforme de cette équation. Dans les deux cas, l'équation de Klein-Gordon sans masse est restaurée si la géométrie est celle de l'espace de Minkowski. 

\bs
Dans le chapitre suivant, nous rencontrerons d'autres exemples de tenseur énergie-impul\-sion : le cas du champ électromagnétique représenté par une 2-forme $F$ et aussi le cas de $p$-formes de rang supérieur. 

\subsubsection{La constante cosmologique}
Terminons cette sous-section par le cas de la constante cosmologique notée $\La$. La présence d'une constante cosmologique apparaît sous la forme suivante dans l'action
\BE S = \int_{\mc M} \sqrt{-g} \lp R-2\La \rp \dd^D x \ .\EE 
Nous avons déjà rencontré cette constante dans le cadre du théorème de Lovelock \ref{thlovelock} en dimension 4 : si nous recherchons une théorie gravitationnelle du second ordre de la métrique qui préserve la conservation du tenseur-énergie impulsion, il n'y a alors aucune raison pour que cette constante soit nulle. L'équation d'Einstein prend dans ce cas la forme suivante
\BE G_{\mu\nu} + \La g_{\mu\nu} = 0 \ .  \EE
La courbure scalaire vaut donc $R=\frac{2D}{D-2}\La$, par conséquent l'équation d'Einstein se simplifie en $ R_{\mu\nu} = \frac{2\La}{D-2}g_{\mu\nu} $ et l'espace-temps est un espace d'Einstein par définition. \bs

Présentons les solutions à courbure constante de cette équation, c'est-à-dire les solutions qui vérifient
\BE  R_{\mu\nu\rho\la} = \frac{2\La}{(D-1)(D-2)} \lp g_{\mu\rho}g_{\nu\la} - g_{\mu\la}g_{\nu\rho} \rp \ .\EE
Tout d'abord, l'espace à courbure constante pour lequel $\La=0$ est simplement l'espace de Minkowski.
En revanche, si la constante cosmologique est strictement positive $(\La>0)$, l'espace est qualifié d'espace-temps de \ita{de Sitter}, noté dS. En dimension 4, cette espace possède la topologie de $\mathbb{R} \times  S^3$ et peut être vu comme l'hyperboloïde $-v^2+w^2+x^2+y^2+z^2=\al^2$ plongé dans l'espace-temps de Minkowski de dimension 5 muni de la métrique $\dd s^2 = -\dd v^2+ \dd w^2 +\dd x^2 + \dd y^2 + \dd z^2 $ où $\al$ est une constante. Les coordonnées  $\lp t,\chi,\theta,\phi \rp$ définies par les relations
\begin{align}
v &= \al \sinh(t/\al) \ , w = \al \cosh(t/\al)\cos\chi \ , x = \al \cosh(t/\al)\sin\chi\cos\theta \ ,\\
y &= \al \cosh(t/\al)\sin\chi\sin\theta\cos\phi \ , z = \al \cosh(t/\al)\sin\chi\sin\theta\sin\phi \ ,
\end{align}
permettent de couvrir entièrement dS avec la métrique
\BE \dd s^2 = - \dd t^2 + \al^2 \cosh^2 \lp t/\al \rp \lb \dd\chi^2 + \sin^2\chi \lp \dd \theta^2 + \sin^2\theta \dd\phi^2 \rp \rb \ .\EE
La généralisation en dimension supérieure est similaire. A l'inverse, si la constante cosmologique est strictement négative $(\La<0)$, l'espace est qualifié d'espace-temps d'\ita{anti-de Sitter}, noté AdS. 
En dimension 4, cet espace possède la topologie de $S^1 \times \mathbb{R}^3$ et peut être vu comme l'hyperboloïde $-v^2- w^2+x^2+y^2+z^2=1$ plongé dans l'espace plat de dimension 5 muni de la métrique suivante $\dd s^2 = -\dd v^2 - \dd w^2 +\dd x^2 + \dd y^2 + \dd z^2 $. Cet espace n'est pas simplement connexe et c'est son recouvrement universel qui est en fait appelé l'espace d'anti-de Sitter et qui possède la topologie de $\mathbb{R}^4$. Cet espace peut être entièrement couvert par un système de coordonnées dans lequel la métrique prend la forme
\BE \dd s^2 = - \cosh^2 r \dd t^2 + \dd r^2 + \sinh^2 r \lp \dd\theta^2 + \sin^2\theta\dd\phi^2 \rp \ . \EE
De plus, cet espace possède une frontière qui a la topologie de $\mathbb{R}\times S^2$. De même, l'espace AdS peut être généralisé en dimensions supérieures.
Précisons que ces trois espaces-temps, celui de Minkowski, dS et AdS, sont maximalement symétriques. Le lecteur pourra aussi consulter \cite{hawking1975large} pour étudier la riche structure causale de ces espaces.
\bs

L'ajout de la constante cosmologique dans les équations du mouvement est en fait dû à A. Einstein en étudiant la cosmologie. Considérons un espace-temps de dimension 4 \ita{spatialement homogène} et \ita{spatialement isotrope}, ce qui est une bonne approximation pour décrire la dynamique de l'univers. L'homogénéité spatiale signifie qu'il existe une foliation de l'espace-temps par une famille d'hypersurfaces $\Si_t$ de genre-espace telle que pour tout $t$ et tout couple de point $(P,Q)$ de $\Si_t$, il existe une isométrie de l'espace-temps amenant $P$ en $Q$. Quant à l'isotropie spatiale, elle suppose  l'existence en chaque point $P$ de l'espace-temps d'une géodésique de genre-temps passant en $P$ et de vecteur tangent $u^\mu$ tel qu'il existe une isométrie laissant $P$ et $u^\mu$ invariants. Le champ de vecteur $u^\mu$ doit alors être nécessairement orthogonal à $\Si_t$. Par conséquent, nous pouvons montrer que chaque espace $\Si_t$ est à courbure constante. Il s'ensuit que la géométrie de $\Si_t$ est soit une sphère $S^3$ de dimension 3 si la courbure scalaire de $\Si_t$ est positive,  soit l'espace euclidien de dimension 3 si la courbure scalaire est nulle ou soit un hyperboloïde $\mathbb{H}^3$ de dimension 3 si la courbure scalaire est négative. En choisissant $t=\tau$, où $\tau$ est le temps propre mesuré par des observateurs isotropiques qui suivent une géodésique avec le vecteur vitesse $u^\mu$, la métrique prend la forme
\BE \dd s^2  = - \dd \tau^2 + a(\tau) \lb \dd \chi^2 + f^2_K(\chi) \lp \dd \theta^2 + \sin^2\theta \dd\phi^2 \rp \rb\EE
où $a(\tau)$ est appelé le \ita{facteur d'échelle}, supposé positif, et 
\[ f_K(\chi)  =  \left\{
\begin{aligned}
&\frac1{\sqrt{K}}\sin\lp\sqrt{K}\chi\rp \quad\quad\quad \text{si} \ \ K>0 \\
&\chi \quad\quad\quad\quad\quad\quad\quad\quad\quad\ \ \text{si} \ \ K=0 \\
&\frac1{\sqrt{-K}}\sinh\lp\sqrt{-K}\chi\rp \quad\text{si} \ \ K<0 
\end{aligned} \right. \] 
$K$ étant une constante proportionnelle à la courbure scalaire de chaque $\Si_t$. Ces métriques sont connues sous le nom de \ita{métriques de Friedmann-Lemaître-Robertson-Walker} (FLRW). Il reste donc à déterminer la dynamique du facteur d'échelle. Pour cela, il suffit de résoudre l'équation d'Einstein
\BE G_{\mu\nu} + \La g_{\mu\nu} = 8\pi G T_{\mu\nu} \ . \EE
Or, le calcul du tenseur d'Einstein dans cette géométrie montre que le tenseur énergie-impulsion se met nécessairement sous la forme
\BE T_{\mu\nu} = \rho u_\mu u_\nu + P h_{\mu\nu}  \EE
où $h_{\mu\nu}=g_{\mu\nu}+ u_\mu u_\nu$ est la métrique spatiale de $\Si_t$. Ce tenseur prend donc la forme de celui d'un fluide parfait avec la densité d'énergie $\rho$ et la pression $P$, ces deux quantités ne dépendent que de $\tau$. L'équation d'Einstein fournit alors deux équations indépendantes, appelées \ita{équations de Friedmann},
\begin{align}
H^2 &= \frac{8\pi G}{3}\rho - \frac{K}{a^2} + \frac{\La}{3} \label{friedmann1}\\
\frac{\ddot{a}}{a} &= -\frac{4\pi G}{3} \lp \rho + 3P \rp + \frac{\La}{3} \label{friedmann2}
\end{align}
où nous avons introduit le \ita{paramètre d'Hubble} $H=\frac{\dot{a}}{a}$ qui caractérise l'expansion de l'univers, le point indique ici la dérivation par rapport à la coordonnée $\tau$. Signalons que nous pouvons supprimer $\ddot{a}$ dans \eqref{friedmann2} en dérivant \eqref{friedmann1} par rapport à $\tau$ pour obtenir l'équation de conservation du tenseur énergie-impulsion
\BE \dot{\rho} + 3H(\rho+P)=0 \ . \EE
Précisons que nous avons seulement deux équations indépendantes pour trois inconnues $(a,\rho,P)$, il faut donc ajouter une équation d'état pour décrire la matière et résoudre le système. Usuellement l'équation de Friedmann \eqref{friedmann1} est écrite sous une forme adimensionnée. En introduisant les quantités
\BE \Om = \frac{8\pi G\rho}{3H^2} \ , \Om_\La =\frac{\La}{3H^2} \ , \Om_K =  - \frac{K}{a^2H^2} \EE
et en décomposant $\Om$ sous la forme $\Om=\sum_X \Om_X$ afin de tenir compte des différents types de matière, l'équation \eqref{friedmann1} s'écrit sous la forme
\BE \sum_X \Om_X + \Om_\La + \Om_K = 1 \ . \EE
Pour une revue complète des modèles cosmologiques, nous conseillons l'ouvrage de référence \cite{peter2005cosmologie}.

Revenons plus en détails sur la constante cosmologique afin d'expliquer son introduction par A. Einstein. Comme Steven Weinberg l'explique dans \cite{Weinberg:1988cp}, A. Einstein tenta en 1917 d'appliquer sa théorie de la gravitation à l'univers. Cependant, il rechercha absolument une solution statique. Pour cela, il dut ajouter ce terme de constante cosmologique $\La$ qui se comporte comme un champ de matière vérifiant $\rho=-P=\frac{\La}{8\pi G}$. En effet, les équations de Friedmann nous montre qu'il existe une telle solution statique avec de la matière sans pression $(P=0)$ pour une valeur spécifique de la constante cosmologique $\La_c = 4\pi G \rho = \frac{K}{a^2}$ de telle façon que la force de gravitation attractive soit compensée par la force répulsive due à la constante cosmologique. En tant que fervent partisan du principe de Mach, A. Einstein fut satisfait par cette solution puisque la quantité géométrique $\La$ est déterminée par la densité de matière. Mais, ce fut alors une surprise pour le père de la Relativité quand son ami de Sitter découvrit une solution cosmologique en présence de $\La$ sans matière : $\rho=P=0$. De plus, les données astronomiques de l'époque commencèrent à indiquer l'expansion de l'univers à travers le décalage vers le rouge du spectre des galaxies. A. Einstein ne vit alors plus aucune raison de garder la constante cosmologique puisqu'il est possible d'obtenir des solutions décrivant un univers en expansion avec $\La=0$. Signalons également que l'univers statique d'Einstein est instable si nous effectuons une perturbation du facteur d'échelle. A. Einstein considéra l'introduction de la constante cosmologique dans l'équation qui porte son nom comme sa plus grosse erreur. Cependant, cette quantité est nécessaire pour décrire notre univers. En effet, les observations des supernovae de type Ia, des grandes structures et du fond diffus cosmologique indiquent que la constante cosmologique représente actuellement 70\%  du contenu de l'univers \cite{Hinshaw:2008kr} : $ \Om_\La^{(0)} \sim 70\% $; cette quantité est appelée couramment l'\ita{énergie noire}. La densité associée à cette énergie est de l'ordre de $\rho_\La \sim 10^{-47}\text{GeV}^4$. Ce qui nous amène à  discuter brièvement de ce que beaucoup de personnes appellent le \ita{problème de la constante cosmologique}. Nous pouvons être tentés d'expliquer l'existence de cette constante en l'assimilant à l'énergie du vide d'une théorie quantique des champs. Une estimation rapide de l'énergie associée aux fluctuations du vide pour un champ de masse $m$ donne 
\BE \langle\rho\rangle_\text{vide}= \int_0^{k_\text{max}} \frac{4\pi k^2 \dd k}{(2\pi)^3} \frac12 \sqrt{m^2+k^2} \sim \frac{k^4_\text{max}}{16\pi^2} \sim 10^{74}\text{GeV}^4 \EE
si nous fixons la fréquence de coupure à l'échelle de Planck, soit un désaccord de 120 ordres de grandeur. La prise en compte de la supersymétrie diminue cette énergie du vide mais avec un désaccord qui reste important : de 60 ordres de grandeur par rapport à $\rho_\La$. Pour de plus amples détails sur ce problème, nous conseillons les références suivantes \cite{Weinberg:1988cp,Carroll:2000fy,Peebles:2002gy,Copeland:2006wr}. Pour finir, notons que cette identification entre la densité d'énergie du vide et la constante cosmologique est loin d'être évidente, comme le souligne \cite{Bianchi:2010uw}, étant donné que nous n'avons toujours pas une théorie quantique des champs en interaction avec la gravité.

\subsubsection{Les conditions d'énergie} 
Terminons cette sous-section en passant en revue les diverses \ita{conditions d'énergie} portant sur le tenseur énergie-impulsion qui fournissent des hypothèses pour certains théorèmes qui seront présentés dans la suite. Ces conditions d'énergie sont habituellement vérifiées pour la matière classique et expriment des hypothèses comme la positivité de la densité d'énergie. Afin d'exprimer ces conditions précisément décomposons le tenseur énergie-impulsion sous la forme
\BE T^{\al\be} = \rho e_0^\al e_0^\be + \sum_{i=1}^{D-1} P_i  e_i^\al e_i^\be \EE
où les vecteurs $\lp  e_0^\al,  \ldots,  e_{D-1}^\al \rp$ forment une base orthonormale, c'est-à-dire qu'ils vérifient $g_{\al\be}e_\mu^\al e_\nu^\be = \eta_{\mu\nu}$ et aussi $g^{\al\be} = e_\mu^\al e_\nu^\be  \eta^{\mu\nu}$.  $\rho$ est appelée la \ita{densité d'énergie} et $P_i$ la \ita{pression} associée à chaque direction. Notons que dans le cas d'un fluide parfait nous avons $P_i=P$ pour tout $i$ et ainsi
\BE T^{\al\be} = \lp \rho + P \rp e_0^\al e_0^\be + P g^{\al\be} \ . \EE
Nous reconnaissons donc le tenseur énergie-impulsion d'un fluide parfait en identifiant le vecteur $e_0^\al$ à la vitesse d'une particule fluide. Donnons les différentes conditions d'énergie.
\begin{itemize}

\item \textit{Condition d'énergie faible} : Cette condition stipule que la densité de matière mesurée par un observateur doit être positive. En introduisant un vecteur vitesse $v^\al$ de genre-temps dirigé vers le futur, cette condition se traduit par
\BE T_{\al\be} v^\al v^\be \geq 0 \Rightarrow \lp \rho \geq 0 \ , \rho+P_i >0 \rp \ . \EE

\item \textit{Condition d'énergie de genre-lumière} : Cette condition d'énergie est similaire à la précédente en remplaçant le vecteur $v^\al$ par un vecteur de genre-lumière dirigé vers le futur, il s'ensuit 
\BE T_{\al\be} k^\al k^\be \geq 0 \Rightarrow  \rho+P_i  \geq 0 \ . \EE
La condition d'énergie faible implique donc la condition d'énergie de genre-lumière.

\item \textit{Condition d'énergie forte} : Etant donné un vecteur $v^\al$ de genre-temps dirigé vers le futur et unitaire, cette condition est 
\BE \lp T_{\al\be} -\frac12 T g_{\al\be} \rp v^\al v^\be \geq 0 \ . \EE 
Au regard de l'équation d'Einstein écrite sous la forme \eqref{Einstein2}, c'est véritablement une condition sur le tenseur de Ricci. La conséquence de cette condition est
\BE  \rho+\sum_i P_i \geq 0 \text{ et } \rho+P_i \geq 0 \ . \EE

\item \textit{Condition d'énergie dominante} : Cette condition encode le fait que la matière doit se propager selon une ligne d'univers de genre-temps ou de genre-lumière. Plus précisément, si nous introduisons un vecteur $v^\al$ de genre-temps dirigé vers le futur représentant la vitesse d'un observateur, la quantité $-T^\al_\be v^\be$ représente alors la densité de moment de la matière mesurée par cet observateur. Cette condition d'énergie demande que ce vecteur $-T^\al_\be v^\be$ soit de genre-temps ou de genre-lumière et dirigé vers le futur. Ce qui se traduit par
\BE \rho\geq |P_i| \ . \EE
\end{itemize}

Le lecteur pourra également consulter la classification de S. W. Hawking et G. F. R. Ellis dans \cite{hawking1975large}.

	\section{Les trous noirs}
Dans la section précédente, nous avons rencontré des solutions du vide à courbure constante de la Relativité Générale. Nous allons ici décrire d'autres solutions : les \ita{trous noirs}. Ce sont les objets fondamentaux de la théorie de la gravitation d'Einstein et ils sont caractérisés par peu de paramètres. 
		\subsection{Le trou noir de Schwarzschild et le théorème de Birkhoff}
Pour déterminer des solutions de l'équation d'Einstein, il est bien souvent utile d'imposer des symétries. Dans cette sous-section, nous allons considérer la symétrie sphérique et étudier ses conséquences. Plus précisément, un espace-temps est à \ita{symétrie sphérique} si son groupe d'isométrie contient un sous-groupe isomorphe au groupe $SO(3)$ de telle façon que les orbites générées par ce sous-groupe forment des sphères bidimensionelles. Nous pouvons alors montrer qu'il est possible d'introduire un système de coordonnées $(t',r',\theta,\phi)$ dans lequel la métrique prend la forme
\BE \dd s^2 = - e^{a(t',r')} \dd t'^2 + e^{b(t',r')} \dd r'^2 + R^2(r',t') \lp \dd \theta^2 + \sin^2\theta \dd\phi^2 \rp \ . \label{sphérique} \EE
Puis, en toute rigueur, nous devrions discuter des différents cas possibles en fonction de la nature des surfaces de niveaux de $R$. Contentons nous de réécrire la métrique lorsque la surface de niveau de $R$ est de genre-temps : 
\BE \dd s^2 = - e^{2\psi(t,r)} f(t,r) \dd t^2 + \frac{\dd r^2}{f(t,r)} + r^2  \lp \dd \theta^2 + \sin^2\theta \dd\phi^2 \rp \ . \EE
Il est alors judicieux d'introduire la \ita{fonction de masse} $m(t,r)$ définie par $f=1-2m/r$. Ainsi l'équation d'Einstein \eqref{Einstein} fournit les équations suivantes
\BE \p_r m = -4\pi r^2 T_t^t \ ,\quad \p_t m = 4\pi r^2 T_t^r \ ,\quad \p_r \psi = \frac{4\pi r}{f}\lp T_r^r - T_t^t \rp \ .\EE

\subsubsection{La solution de Schwarzschild}
Par conséquent, la fonction de masse est une constante, notée $M$, dans le vide et $\psi$ est une fonction de la coordonnée $t$. En redéfinissant le temps, nous pouvons finalement écrire la métrique sous la forme bien connue
\BE \dd s^2 = - \lp 1- \frac{2M}{r} \rp \dd t^2 + \frac{\dd r^2}{1- \frac{2M}{r}} + r^2 \lp \dd \theta^2 + \sin^2\theta \dd\phi^2 \rp \ . \label{schw}\EE
Cette solution porte le nom de \ita{métrique de Schwarzschild} ; elle a été découverte par Karl Schwarzschild seulement deux mois après la publication de l'équation d'Einstein \cite{1999physics...5030S}. Le fait que la symétrie sphérique impose le caractère \ita{statique} de la solution est remarquable. Tout d'abord, la solution est évidemment stationnaire puisqu'elle possède le vecteur de Killing $\xi = \p_t$ mais elle est même statique selon la définition suivante : 
\begin{dico}[espace-temps statique] Un espace-temps est statique si en plus d'être stationnaire, c'est-à-dire qu'il existe un vecteur de Killing $\xi$ de genre-temps,  il existe une hypersurface qui est orthogonale aux orbites générées par le champ tangent $\xi$.
\end{dico}
Il est alors facile de montrer que si $\xi^\mu$ est un champ tangent de Killing vérifiant le critère dit de \ita{Frobenius} $\xi_{[\mu} \n_\nu \xi_{\rho]} = 0 $ alors $\xi^\mu$ est orthogonal à une famille d'hypersurface. La condition de Frobenius s'écrit plus simplement sous la forme
\BE \xi^\flat \w \dd \xi^\flat = 0 \label{frobenius}\EE
en introduisant le champ cotangent $\xi^\flat = \xi_\mu \dd x^\mu$ associé au champ tangent $\xi$. La condition \eqref{frobenius} est donc bien vérifiée pour la solution de Schwarzschild où $\xi^\flat = - \lp 1-\frac{2M}{r}\rp \dd t $. 

En plus du caractère statique, la solution de Schwarzschild est \ita{asymptotiquement plate} \cite{Ashtekar:1978zz}. Cependant, nous ne donnons pas volontairement de définition de cette notion, le lecteur intéressé par ce sujet pourra se tourner vers \cite{wald1984general}. Une autre notion, que nous aborderons dans le chapitre \ref{thermo}, est la \ita{masse} d'une solution. Nous serons capables de montrer dans ce chapitre \ref{thermo} que le paramètre $M$ dans \eqref{schw} est exactement la masse de la solution de Schwarzschild. En somme, retenons que la métrique \eqref{schw} représente la solution à l'extérieur d'un corps à symétrie sphérique de rayon supérieur à $2M$. 
\bs

En effet, le système de coordonnées $(t,r,\theta,\phi)$ ne couvre pas tout l'espace-temps : la métrique diverge en $r=0$ et en $r=2M$. Quant aux singularités en $\theta=0$ et $\theta=\pi$ de la métrique inverse, elles proviennent simplement du choix du système de coordonnées pour la sphère $S^2$. La singularité en $r=0$ est une véritable singularité puisque l'invariant de \ita{Kretschmann}
\BE R_{\mu\nu\rho\si} R^{\mu\nu\rho\si} = \frac{48M^2}{r^6} \EE
y diverge en ce point; nous parlons dans ce cas de \ita{singularité intrinsèque} ou de \ita{singularité de courbure} ou encore de \ita{singularité physique}. En $r=2M$, la signature passe de $(-,+,+,+)$ pour $r>2M$ à $(+,-,+,+)$ pour $r<2M$. Nous allons montrer que cette singularité en $r=2M$ provient du choix du système de coordonnées, nous parlons alors de \ita{singularité de coordonnées} dans ce cas. Pour cela, considérons des particules de masse nulle se propageant radialement sur les géodésiques de la géométrie \eqref{schw}. Il est alors facile de montrer que les particules sortantes et entrantes se déplacent le long des courbes $u=\text{constante}$ et $v=\text{constante}$ respectivement où
\BE u = t - r - 2M\ln \left|\frac{r}{2M}-1\right| \text{ et }   v = t + r + 2M\ln \left|\frac{r}{2M}-1\right| \ . \EE
Puis, il est loisible d'écrire la métrique dans les coordonnées $(u,v,\theta,\phi)$ \cite{Kruskal:1959vx,Szekeres:1960gm}
\BE \dd s^2 = - \lp 1-\frac{2M}{r} \rp \dd u \dd v + r^2 \lp \dd \theta^2 + \sin^2\theta \dd\phi^2 \rp \ , \EE
$r$ étant bien évidemment une fonction de $u$ et $v$. Nous remarquons ainsi que la singularité \ita{apparente} en $r=2M$ n'existe plus dans ce système de coordonnées. Ensuite, en introduisant les coordonnées $U = \mp e^{-\frac{u}{4M}}$ et $V =e^{\frac{v}{4M}}$ où le signe $-$ correspond à la région $r>2M$ et le signe $+$ à la région $r<2M$, nous pouvons écrire la métrique sous la forme 
\BE \dd s^2 =  - \frac{32 M^3}{r} e^{-\frac{r}{2M}}  \dd U \dd V + r^2 \lp \dd \theta^2 + \sin^2\theta \dd\phi^2 \rp \label{Kruscal-metric}\EE
dite de \ita{Kruskal} où $ \lp \frac{r}{2M} -1 \rp e^{\frac{r}{2M}} = - U V $. Enfin, nous pouvons considérer $(U,V)$ dans $\mathbb{R}^2$ et obtenir l'\ita{extension maximale} de la solution de Schwarzschild. Les courbes de niveaux de $U$ et $V$ représentent la trajectoire des photons. Il est alors judicieux de dessiner un diagramme dans le plan $U-V$ pour comprendre la structure causale de la solution comme la figure \ref{kruskal} le montre.
\BF[H]\BC \includegraphics[scale=0.5]{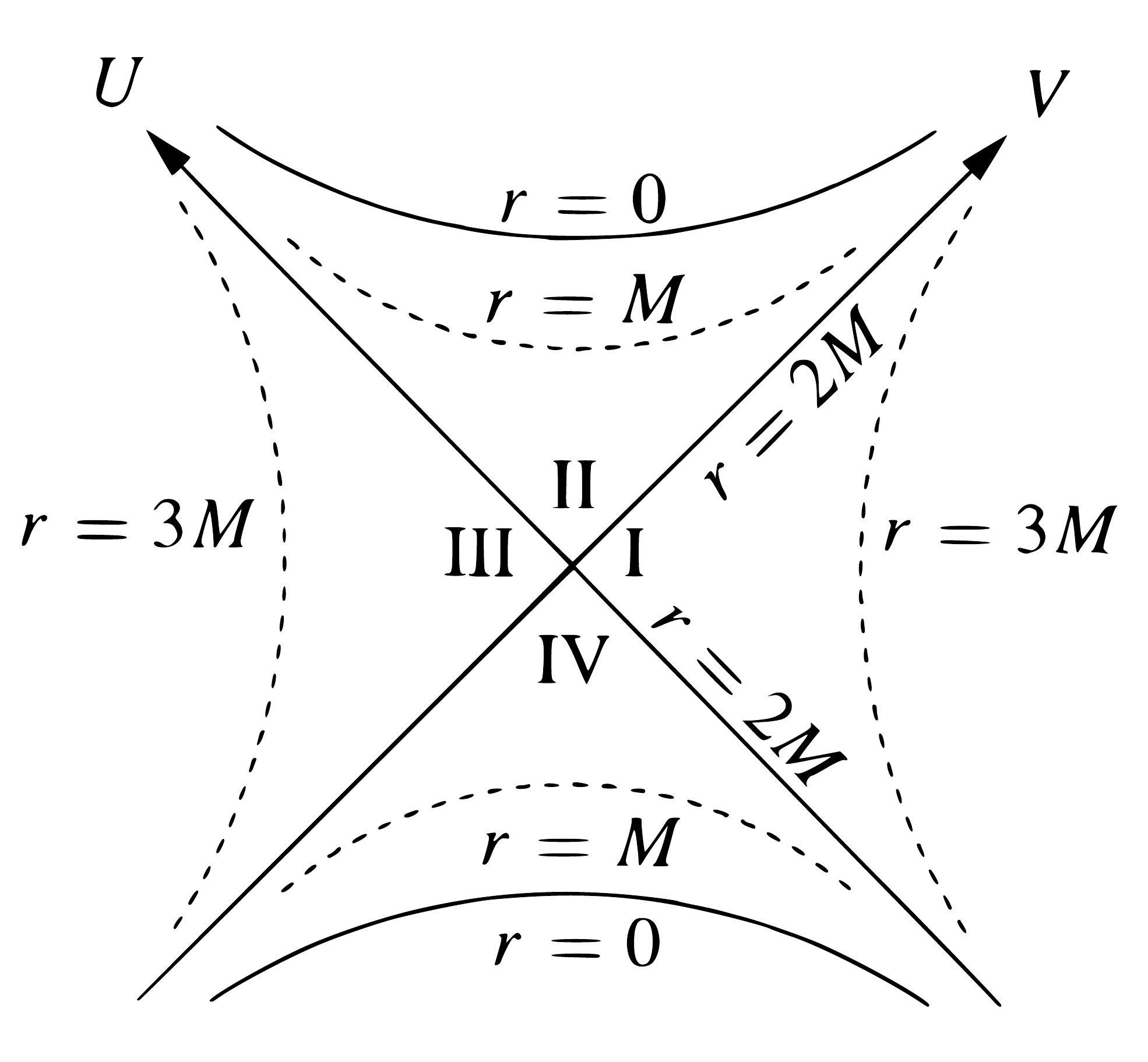} \EC\caption{Diagramme dans les coordonnées de Kruskal-Szekeres $(U,V)$.} \label{kruskal} \EF
La trajectoire d'un observateur positionné à la coordonnée $r$, à l'aide d'une fusée par exemple, est représentée par des hyperboloïdes dans le plan $U-V$. Dans ce système de coordonnées, la métrique \eqref{schw} correspond à la région I pour $r>2M$. Désormais, en plus de pouvoir traverser la surface $r=2M$ en explorant la région II, cette solution admet une continuation dans la région III et IV. Cependant, si le trou noir résulte d'un effondrement gravitationnel, les régions III et IV disparaissent car le diagramme doit être coupé par une hypersurface de genre-temps représentant la surface de l'astre. Puisque les cônes de lumière sur la figure \ref{kruskal} sont orientés de 45 degrés, l'interprétation de la surface $r=2M$ entre la région I et II est claire. Un observateur provenant de la région I et qui traverse la surface $r=2M$ tombe nécessairement sur la singularité $r=0$. De plus, en étant dans la région II, il ne peut même plus envoyer un signal lumineux à un observateur dans la région I. Tout observateur dans la région I ne peut donc pas mesurer une quantité physique dans la région II sans s'y aventurer. Cette surface $r=2M$ entre les régions I et II est alors qualifiée d'\ita{horizon des événements}, la région II est appelée le \ita{trou noir} et la distance $2M$ le \ita{rayon de Schwarzschild}. Nous distinguerons l'\ita{horizon futur} qui est la surface $r=2M$ entre les régions I et II et l'\ita{horizon passé} qui est la surface $r=2M$ entre les régions I et IV; quant à la région IV, elle est qualifiée de \ita{trou blanc}. Précisons que c'est cet horizon futur qui empêche l'apparition d'une \ita{singularité nue}. 

Il est courant de représenter la solution de Kruskal par un diagramme compact. Une manière possible pour cela est d'introduire les coordonnées $\tilde{U}=\arctan U$ et $\tilde{V}=\arctan V$ qui permettent de dessiner la figure \ref{penrose-schw} appelée \ita{diagramme de Penrose-Carter} et de mettre en évidence la structure causale de la solution \cite{Penrose:1962ij,Penrose:1964ge,Carter:1966zz}.
\BF[H]\BC \includegraphics[scale=0.45,angle=-0.5]{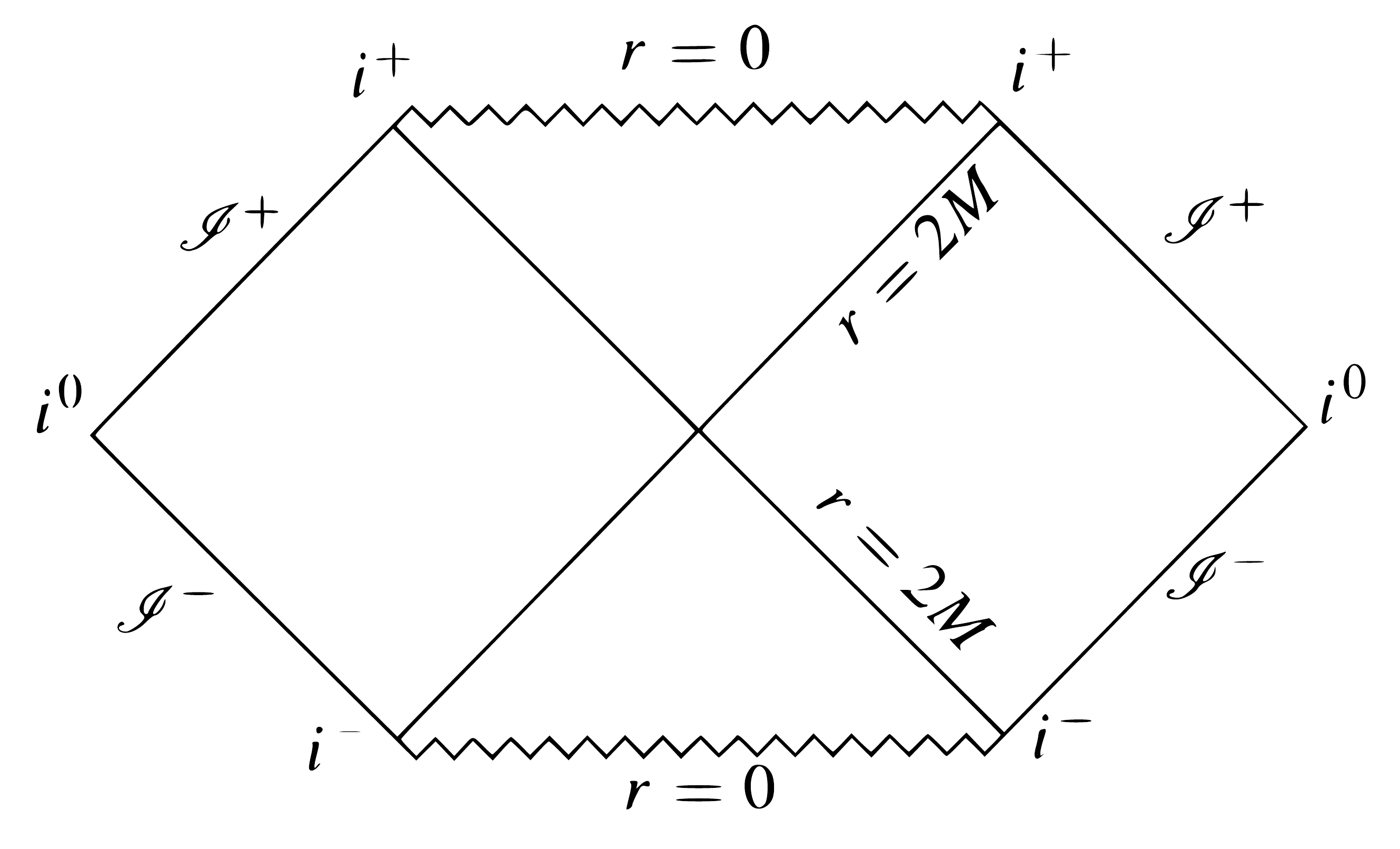} \EC\caption{Diagramme de Penrose-Carter de la solution de Schwarzschild.} \label{penrose-schw} \EF
Concernant le vocabulaire et les notations de ce diagramme :
\begin{itemize}
\item $\mathscr{I}^+$  est appelé l'\ita{infini futur de genre-lumière}, c'est le lieu d'arrivée de toutes les géodésiques de genre-lumière sortantes.
\item $\mathscr{I}^-$  est appelé l'\ita{infini passé de genre-lumière}, c'est le lieu de départ de toutes les géodésiques de genre-lumière entrantes.
\item $i^0$  est appelé l'\ita{infini spatial}, c'est le point d'arrivée de toutes les géodésiques de genre-espace.
\item $i^+$  est appelé l'\ita{infini temporel futur}, c'est le point d'arrivée de toutes courbes de genre-temps qui ne tombent pas en $r=0$.
\item $i^-$ est appelé l'\ita{infini temporel passé}, c'est le point de départ de toutes courbes de genre-temps qui n'émergent pas de la singularité du trou blanc.
\end{itemize}
Signalons que le vecteur de Killing $\xi=\p_t$ est de genre-temps à l'extérieur du trou noir, il est de genre-lumière sur l'horizon des événements et il est de genre-espace à l'intérieur du trou noir. En effet, sa norme est donnée par $\xi_\mu \xi^\mu = -1+\frac{2M}{R}$. La surface $r=2M$ est qualifiée dans ce cas d'\ita{horizon de Killing}.
Ajoutons aussi que la solution de Schwarzschild est stable sous des perturbations gravitationnelles \cite{Regge:1957td}. Enfin, nous souhaitons énoncer un des grands résultats de la Relativité Générale démontré pour la première fois par G. D. Birkhoff en 1923 pour clore sur cette solution : 
\begin{proposition}[théorème de Birkhoff] 
Toute solution $\mc C^2$ des équations d'Einstein dans le vide à symétrie sphérique est localement isométrique à la solution de Kruskal.
\label{Birkhoff}
\end{proposition} 
Ainsi, toute distribution de matière qui préserve la symétrie sphérique ne peut pas rayonner des ondes gravitationnelles ; tout comme en électromagnétisme, la solution de Coulomb est la seule solution à symétrie sphérique des équations de Maxwell dans le vide.

\subsubsection{La solution de Reissner-Nordstrom} 
Mentionnons que ce théorème de Birkhoff se généralise en présence d'un champ de jauge abélien. En effet, considérons l'action d'Einstein-Hilbert couplée minimalement à l'électromagnétisme décrit par la 2-forme de Faraday $F$ tel qu'il existe une 1-forme $A$, appelée le potentiel vecteur, vérifiant $F = \dd A$ : 
\BE S = \int_\mc M R \ep - \frac12 \int_\mc M F \w \star  F  \ . \EE
Les équations du mouvement sont alors données par l'équation d'Einstein \eqref{Einstein} avec le tenseur énergie-impulsion suivant
\BE T_{\mu\nu} = \frac{1}{16\pi G}\lp F_{\mu\rho}F_{\nu}^{\ \rho} - \frac14 g_{\mu\nu} F_{\rho\sigma}F^{\rho\sigma}  \rp \EE
et les équations de Maxwell $\dd F = 0$ et $\dd \star F = 0$. Puis, si le potentiel vecteur possède les mêmes symétries que la métrique, c'est-à-dire $\pounds_\xi A^\mu =0$ pour tout vecteur de Killing $\xi$ de la sphère $S^2$ donné par $\xi_{(1)} = \cos \phi \p_\theta - \cot\theta \sin\phi \p_\phi $, $\xi_{(2)} = \sin \phi \p_\theta + \cot\theta \cos\phi \p_\phi $ et $\xi_{(3)} = \p_\phi$, alors l'unique solution à symétrie sphérique est donnée par la 2-forme de Faraday
\BE F = \frac{Q}{r^2} \dd t \w \dd r \EE 
et la métrique dite de Reissner-Nordstrom \cite{reissner1916eigengravitation,Nordstrom} 
\BE \dd s^2 = - \lp 1- \frac{2M}{r} + \frac{Q}{r^2} \rp \dd t^2 + \frac{\dd r^2}{1- \frac{2M}{r} + \frac{Q}{r^2}} + r^2 \lp \dd \theta^2 + \sin^2\theta \dd\phi^2 \rp \ . \label{RN}\EE
Discutons brièvement de la structure causale de la solution en fonction des valeurs de $M$ et $Q$.
\begin{itemize}

\item Si $M^2 > Q^2$ alors il existe deux horizons de Killing donnés par $r_\pm = M \pm \sqrt{M^2-Q^2}$. De plus, comme le diagramme de Penrose-Carter \ref{penrose-RN} le montre, la singularité de courbure en $r=0$ est désormais de genre-temps et peut être évitée par des observateurs électriquement neutres en chute libre par exemple. Pour dessiner ce diagramme de Penrose, plusieurs systèmes de coordonnées de type $(U,V)$ ont dû être utilisés mettant ainsi en évidence l'existence d'une infinité de trous noirs. Soulignons aussi que seul l'horizon extérieur localisé en $r_+$ est un horizon des événements où nous définissons ce type d'horizon comme le bord du passé causal de $\mathscr{I}^+$.

\BF[H]\BC \includegraphics[scale=0.32,angle=0.2]{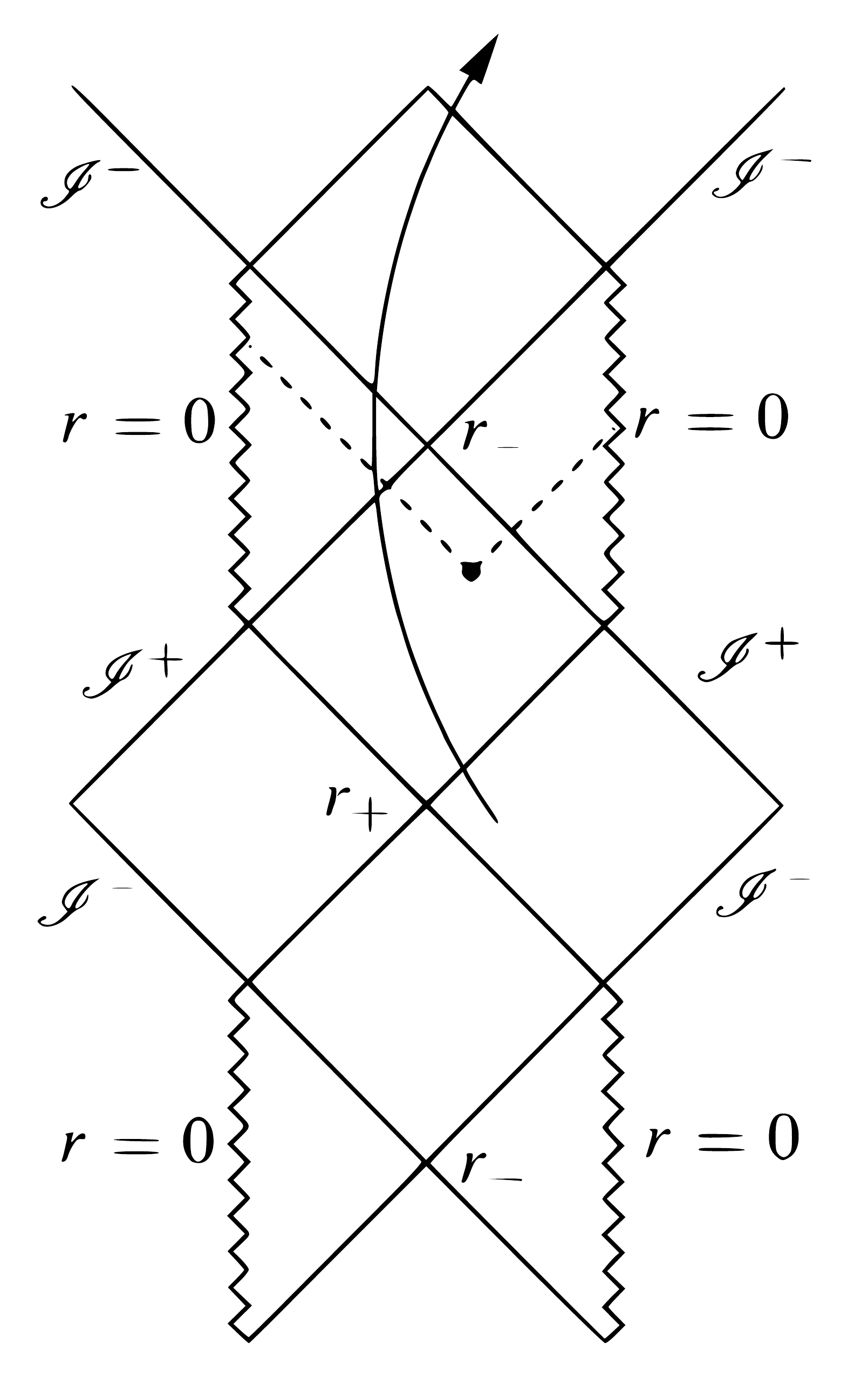} \EC\caption{Diagramme de Penrose-Carter de la solution de Reissner-Nordstrom pour $M^2 > Q^2$ \cite{Graves:1960zz}.} \label{penrose-RN} \EF
\BF[H]\BC \includegraphics[scale=0.45,angle=90]{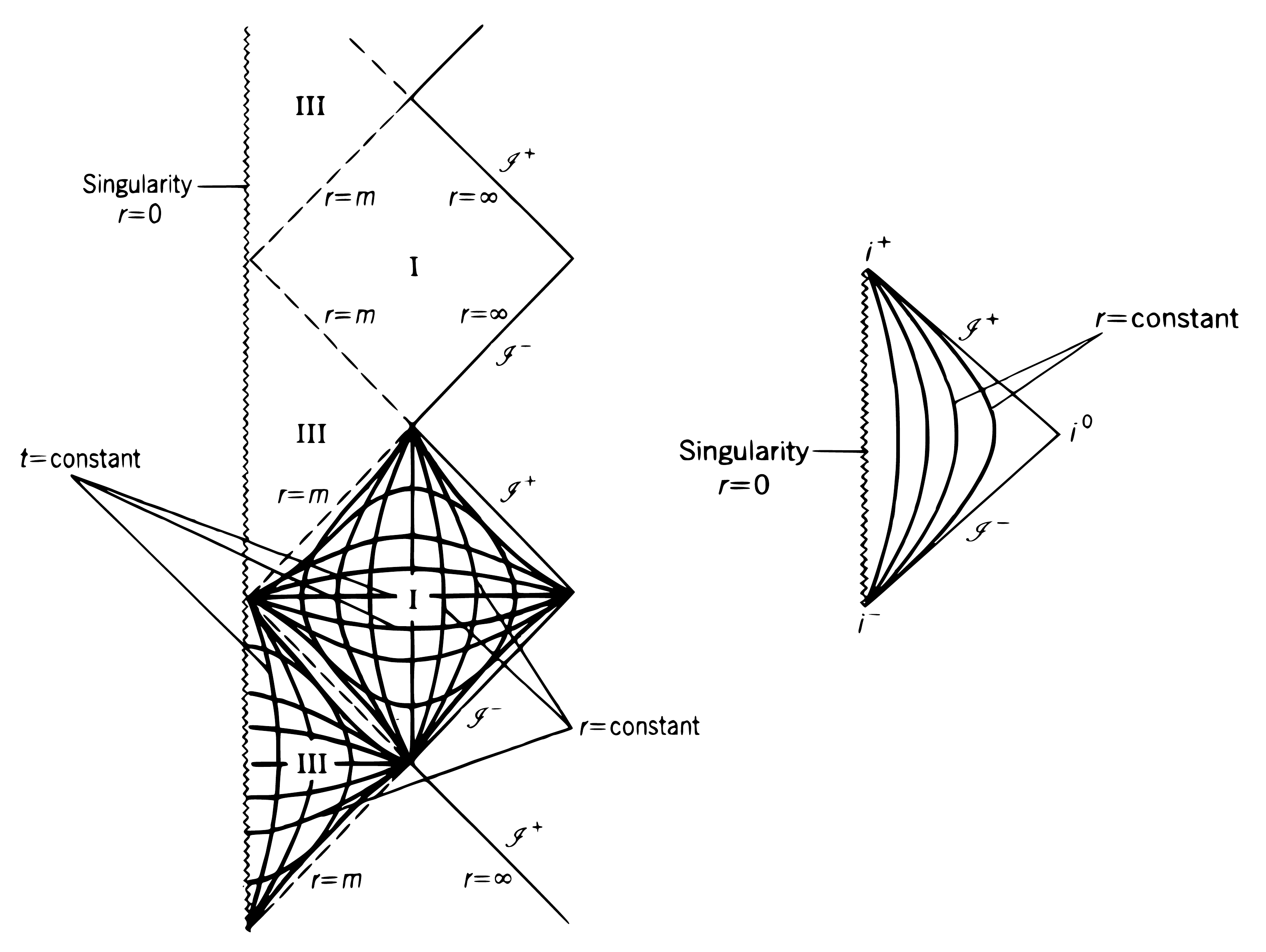} \EC\caption{Diagramme de Penrose-Carter de la solution de Reissner-Nordstrom pour le cas extrémal à gauche \cite{carter1966complete} et pour la singularité nue à droite.} \label{penrose-RN-ext} \EF

\item Si $M^2 = Q^2$ alors le zéro de la norme du vecteur de Killing $\p_t$ est double et vaut $r=M$. Ici les coordonnées $t$ et $r$ n'échangent pas leur rôle à la traversée de l'horizon. Cependant, cette solution garde l'interprétation de trou noir car elle possède toujours un horizon des événements puisque $r=M$ est le bord passé de $\mathscr{I}^+$ comme le montre clairement le diagramme de Penrose de la figure \ref{penrose-RN-ext}. Pour ce cas $M^2=Q^2$, le trou noir est dit \ita{extrémal}.

\item Quant à la situation $M^2 < Q^2$, il n'existe pas d'horizon et nous sommes en présence d'une singularité nue. Le diagramme de Penrose de cette solution est donné à la figure \ref{penrose-RN-ext}.
\end{itemize}
\bs
Revenons sur le potentiel vecteur de la solution de Reissner-Nordstrom donné par
\BE A = \lp \Phi + \frac{Q}{r} \rp \dd t \EE
où $\Phi$ est une constante arbitraire provenant de l'invariance sous le groupe de jauge $U(1)$ et fixant ainsi la valeur du potentiel à l'infini. Contrairement à ce qui se passe dans l'espace de Minkowski, la valeur de cette constante $\Phi$ doit être fixée comme G. W. Gibbons et S. W. Hawking l'ont remarqué dans \cite{Gibbons:1976ue} puisque l'invariant $A_\mu A^\mu$ diverge sur l'horizon extérieur sauf si la constante est fixée à la valeur
\BE \Phi = -\frac{Q}{r_+} \ . \EE
Nous adopterons donc ce choix dans toute la suite et cette valeur prendra une interprétation importante au moment d'étudier la thermodynamique. Pour une discussion concernant la stabilité de la solution de Reissner-Nordstrom, nous renvoyons le lecteur à l'ouvrage \cite{chandrasekhar1983mathematical} écrit par S. Chandrasekhar.

\subsubsection{La solution de Kerr}
Enfin, il est possible de supprimer l'hypothèse de symétrie sphérique. En effet, il existe une solution découverte par Roy Kerr en 1963 \cite{Kerr:1963ud} qui décrit un trou noir en rotation. La version chargée de celle de Kerr, appelée solution de Kerr-Newman, existe également \cite{Newman:1965my}. Donnons simplement la métrique de Kerr dans les coordonnées dites de \ita{Boyer-Lindquist} : 
\BE \dd s^2 = -\frac{\rho^2 \Delta}{\Sigma}\dd t^2 + \frac{\Sigma}{\rho^2}\sin^2\theta \lp \dd\phi-\om\dd t \rp^2 + \frac{\rho^2}{\Delta}\dd r^2 + \rho^2 \dd\theta^2 \EE
où
\BE \rho^2=r^2+a^2\cos^2\theta^2  \ ,  \Delta=r^2-2Mr+a^2 \ , \Sigma = (r^2+a^2)^2-a^2\Delta\sin^2\theta \text{ et }  \om=\frac{2Mar}{\Sigma} \ . \EE
La métrique de Kerr est \ita{stationnaire} et à \ita{symétrie axiale} puisqu'elle admet $\p_t$ et $\p_\phi$ comme vecteurs de Killing. Elle est asymptotiquement plate, sa masse est donnée par $M$ et elle admet un moment angulaire $J$ donné par $J=aM$. 

\subsubsection{Généralités sur les trous noirs} 
Pour clore cette sous-section, donnons brièvement quelques généralités sur les trous noirs. Soit $P$ un point de l'espace-temps $\mc M$. Notons par $J^+(P)$ le \ita{futur causal} de P, constitué par l'ensemble des points de $\mc M$ connectés à $P$ par une courbe de genre-temps ou de genre-lumière dirigée vers le futur. Nous définissons de manière similaire le \ita{passé causal} de $P$ noté $J^-(P)$. Puis, étant donné un sous-ensemble $\mc N$ de $\mc M$, nous introduisons la notation suivante $J^\pm(\mc N)=\bigcup_{P\in\mc N} J^\pm(P)$ pour désigner le futur/passé causal de $\mc N$. 

De manière générale, nous définissons un \ita{trou noir} comme un espace-temps qui possède des géodésiques de genre-lumière qui n'atteignent jamais l'infini futur de genre-lumière noté $\mathscr{I}^+$. La région du trou noir notée $B$ est alors définie par l'ensemble des points de $\mc M$ qui n'appartiennent pas au passé causal de l'infini futur de genre-lumière, c'est-à-dire
\BE B = \mc M \setminus  J^-(\mathscr{I}^+) \ .  \EE
Enfin, nous définissons l'\ita{horizon des événements} $H$ comme la frontière de la région du trou noir : $H=\p B$.
		\subsection{Les théorèmes d'unicité\label{no-hair}}
Dans cette sous-section, nous allons présenter les différents théorèmes d'unicité valables en dimension 4 qui permettent de comprendre la fameuse phrase de John Wheeler : "A black hole has no hair". Commençons en présentant une réciproque du théorème de Birkhoff \ref{Birkhoff} due à Werner Israel en 1967 démontrée dans \cite{Israel:1967wq,bunting1987nonexistence,straumann2004general} :

\begin{proposition}[théorème d'Israel] Toute solution statique et asymptotiquement plate de l'équation d'Einstein dans le vide en dimension 4 doit être à symétrie sphérique et est donc localement isométrique à la métrique de Schwarzschild.
\end{proposition}
Le théorème d'Israel se généralise également en présence d'un champ de jauge abélien \cite{Israel:1967za}. Par conséquent, la solution de Reissner-Nordstrom est la seule solution statique asymptotiquement plate de l'équation d'Einstein en présence de l'interaction électromagnétique en dimension 4. W. Israel, R. Penrose et J. Wheeler conjecturèrent alors que toute solution stationnaire asymptotiquement plate de l'équation d'Einstein en présence d'un champ jauge abélien est nécessairement isométrique à celle de Kerr-Newman. Décrivons les différentes étapes qui ont mené à montrer cette conjecture. Tout d'abord, S. Hawking établit les résultats suivants en 1972\cite{Hawking:1971vc} : 
\begin{proposition}[théorème de rigidité] Si un trou noir est stationnaire alors il doit être soit statique soit à symétrie axiale.
\end{proposition}
\begin{proposition}[topologie de l'horizon]  En supposant la condition d'énergie dominante, l'horizon d'un trou noir stationnaire asymptotiquement plat doit avoir la topologie de la sphère $S^2$.
\end{proposition}
Le théorème de rigidité nous dit que si un trou noir en rotation admet un vecteur de Killing $\xi$ de genre-temps alors il existe un vecteur de Killing $\phi$ de genre-espace dont les orbites sont fermées. En particulier, S. Hawking réussit à montrer que le vecteur de Killing $\xi + \Om_H \phi$ est de norme nulle sur l'horizon où $\Om_H$ représente la vitesse angulaire du trou noir. La situation statique du théorème de rigidité appliqué dans le vide nous ramène alors au théorème d'Israel. L'effondrement gravitationnel en un trou noir d'une distribution de matière quelconque avec un moment angulaire nul conduit nécessairement à la solution de Schwarzschild. Tous les moments multipolaires d'une étoile, qui ne présente pas la symétrie sphérique et qui s'effondre en un trou noir, sont alors expulsés à l'infini ou à l'intérieur du trou noir; ce mécanisme a été décrit par Richard Price en 1972 \cite{Price:1971fb,Price:1972pw}. 
Concernant le cas non statique, nous pouvons être étonnés par la symétrie axiale. En effet, il semble a priori possible de placer une distribution de matière hors d'un trou noir en rotation afin de briser cette symétrie. La compréhension de ce problème a été résolue par S. Hawking and J. Hartle \cite{Hawking:1972hy}. La distribution de matière extérieure va en fait induire une force effective de friction qui aura pour effet de réduire la rotation du trou noir et d'atteindre de manière non stationnaire le régime final statique. Enfin, il resta à étendre le résultat d'Israel au cas non statique : 
\begin{proposition}[unicité de la solution de Kerr] Toute solution stationnaire présentant la symétrie axiale et asymptotiquement plate de l'équation d'Einstein dans le vide en dimension 4 doit être localement isométrique à la solution de Kerr.
\end{proposition}
Ce résultat est dû à Brandon Carter \cite{Carter:1971zc} et D. C. Robinson \cite{Robinson:1975bv}. Il existe également une généralisation tenant compte de la présence d'un champ de jauge abélien par P. Mazur \cite{Mazur:1982db}. Ainsi, la solution de Kerr-Newman est la seule solution stationnaire, à symétrie axiale et asymptotiquement plate de l'équation d'Einstein en présence de l'interaction électromagnétique en dimension 4. 

Finalement, l'effondrement gravitationnel d'une distribution de masse en un trou noir s'accompagne d'une grande perte d'information de l'état initial puisque l'état final supposé stationnaire est simplement décrit par la masse, le moment angulaire et la charge électrique de la solution. Il n'y a pas besoin d'introduire d'autres paramètres pour décrire cet objet macroscopique et c'est ce que laisse transparaître la phrase de J. Wheeler. Ce dernier conjecture que ce résultat se maintient dans un contexte plus général \cite{ruffini1971introducing}, c'est pourquoi nous parlons souvent de \ita{conjecture no-hair}. La sous-section suivante a pour but de décrire les limites de ces résultats d'unicité en se plaçant dans des cadres plus généraux et de tester cette conjecture.

		\subsection{Comment violer les théorèmes d'unicité ?\label{violer}}
Nous avons vu, qu'en Relativité Générale, les trous noirs sont caractérisés par très peu de  paramètres qui correspondent à des charges conservées mesurées à l'infini. Ces objets semblent jouer le rôle analogue des particules dans les théories quantiques. 

Le but ici est de présenter les extensions possibles des théorèmes d'unicité établis au-dessus à une classe de théories plus large et avec des hypothèses plus faibles afin de confirmer ou d'infirmer la conjecture \ita{no-hair}. Ceci constitue ainsi la motivation majeure de cette thèse. Nous allons successivement considérer l'effet de la présence d'un champ de matière, d'une constante cosmologique, de dimensions supplémentaires et enfin de la présence de termes de courbures d'ordre plus élevé dans la théorie. 

\subsubsection{Présence d'un champ de matière}
Nous allons commencer à présenter les limites de la conjecture \ita{no-hair} en incluant un champ scalaire minimalement couplé à la gravitation. Nous utiliserons le terme de cheveu pour désigner un champ de matière couplé au trou noir ; nous pouvons voir le cheveu comme l'analogue d'un état excité pour une particule. Tout d'abord, J. Chase confirma la conjecture \ita{no-hair} pour un champ scalaire de masse nulle \cite{chase1970event}. Le lecteur pourra trouver dans \cite{Bekenstein:1998aw} une revue de J. Bekenstein des théorèmes \ita{no-hair} en présence de matière dans laquelle il fournit en particulier un argument, indépendant des équations d'Einstein, qui permet de rejeter l'existence d'un champ scalaire massif minimalement couplé à la gravité sous la forme
\BE S_\psi = \int \sqrt{-g} \lb \p_\mu\psi \p^\mu\psi + V(\psi^2) \rb \dd^4x \EE
pour tout trou noir stationnaire asymptotiquement plat si $V'>0$ \cite{Bekenstein:1972ny,Bekenstein:1971hc,Bekenstein:1972ky}. Précisons que pour établir ce résultat, le champ est supposé être borné sur l'horizon. Il a été également étendu pour des potentiels positifs $V>0$ \cite{Heusler:1992ss,Bekenstein:1995un,Sudarsky:1995zg} avec des trous noirs asymptotiquement plats statiques, sphériques et neutres; puis au cas chargé et avec un couplage non-minimal dans \cite{Mayo:1996mv}. Cependant, la conjecture de J. Wheeler a été rejeté pour la théorie d'Einstein-Yang-Mills \cite{Bizon:1990sr, Volkov:1989fi, kunzle1990spherically} et pour la théorie d'Einstein-Skyrme \cite{Bizon:1992gb,Droz:1991cx}. Néanmoins, certaines de ces solutions sont instables et nous restaurons la conjecture \ita{no-hair} en ce sens. Pour un regard plus récent sur le sujet, nous renvoyons le lecteur vers \cite{Hertog:2006rr}.

Notons qu'en présence d'un champ scalaire couplé conformément à la gravitation, il existe la solution dite de BBMB découverte par N. Bocharova et al. \cite{Bocharova:1970} et par J. Bekenstein \cite{Bekenstein:1974sf,Bekenstein:1975ts} indépendamment dont la métrique est celle du trou noir de Reissner-Nordstrom extrémal mais avec un champ scalaire qui a la particularité de diverger sur l'horizon. Ce cheveu scalaire est en fait relié à la masse de la solution, il y a ainsi une seule constante qui paramétrise cette solution. Nous parlons dans ce cas de \ita{cheveu secondaire}. Concernant la stabilité, il semble y avoir un débat puisque K. Bronnikov et al. affirment que la solution est instable \cite{Bronnikov:1978mx} alors que P. McFadden et al. soutiennent l'inverse \cite{McFadden:2004ni}. Nous reviendrons plus en détails dans le chapitre \ref{chapter-conforme} sur l'influence d'un champ scalaire conformément couplé à la gravitation. 

Enfin, A. Anabon et al. \cite{Anabalon:2012sn} ont découvert récemment une solution de trou noir dans une théorie de supergravité en dimension 4 en présence de trois champs scalaires dont un paramètre associé à ces champs peut varier en fixant les charges de la solution; nous parlons dans ce cas de \ita{cheveu primaire} et c'est à notre connaissance le premier cheveu scalaire primaire.

\subsubsection{Présence d'une constante cosmologique}
Nous pouvons également nous demander ce qui se passe si nous supprimons l'hypothèse de solutions asymptotiquement plates en introduisant par exemple une constante cosmologique. Il est bien connu que la présence d'une telle constante en dimension 4 permet de construire des solutions asymptotiquement localement AdS avec des horizons sphériques, plats ou hyperboliques \cite{Lemos:1994xp,Mann:1996gj,Vanzo:1997gw}. Nous parlons dans ce cas de \ita{trous noirs topologiques} et nous verrons dans le chapitre suivant que ces solutions se généralisent en dimensions supplémentaires. Présentons les solutions topologiques suivantes de l'équation d'Einstein avec constante cosmologique en dimension 4 données par la métrique
\BE \dd s^2 = - \lp \ka - \frac{2M}{r} - \frac{\La r^2}{3} \rp \dd t^2 + \frac{\dd r^2}{\ka - \frac{2M}{r} - \frac{\La r^2}{3}} + r^2 \lp \dd\theta^2 + \text{si}^2\theta\dd\phi^2 \rp \label{topo}\EE
où $\text{si}\theta=\sin\theta, \theta, \sinh\theta$ si $\kappa=1,0,-1$ respectivement. La géométrie intrinsèque de l'horizon est donc celle d'une 2-sphère $S^2$ lorsque $\ka=1$, d'un 2-tore $T^2$ lorsque $\ka=0$ ou d'un 2-hyperboloïde $H^2$ lorsque $\ka=-1$. Pour $\La=0$, seule la solution avec $\ka=1$ est permise, ce qui donne simplement la solution de Schwarzschild. Si $\La>0$ alors la seule solution est celle de Schwarzschild-de Sitter avec un horizon sphérique $(\ka=1)$, le lecteur trouvera notamment dans \cite{Gibbons:1977mu} la structure causale de cette solution. Quant au cas $\La<0$, toutes les géométries possibles pour l'horizon sont autorisées pour un paramètre de masse $M$ strictement positif alors que seul un horizon hyperbolique $(\ka=-1)$ est permis si le paramètre de masse est négatif.

Tournons nous désormais vers la possibilité d'introduire un champ scalaire en présence d'une constante cosmologique. Tout d'abord, si la constante cosmologique est positive, T. Torii et al. \cite{Torii:1998ir} ont montré qu'il n'existe pas de trou noir régulier statique et à symétrie sphérique en présence d'un champ scalaire de masse nulle ou avec un potentiel $V$ convexe. Cependant, leurs résultats numériques indiquent l'existence de trous noirs munis d'un champ scalaire pour un potentiel positif. Notons aussi l'existence de la solution de C. Martinez et al. en présence d'un champ scalaire couplé conformément à la gravitation \cite{Martinez:2002ru} et de celle d'A. Anabalon et al. \cite{Anabalon:2012tu} plus récemment qui brise cette symétrie conforme.

D'autre part, durant ces dernières années, des solutions de trous noirs en présence d'une constante cosmologique négative ont été découvertes en présence d'un champ scalaire couplé minimalement à la gravitation \cite{Martinez:2004nb,Martinez:2006an} et couplé conformément à la gravitation \cite{Bardoux,Martinez:2005di,Winstanley:2002jt,Winstanley:2005fu}. Nous reviendrons plus en détails sur ces dernières solutions dans le chapitre \ref{chapter-conforme}. Le lecteur trouvera également des résultats numériques en dimensions supplémentaires en présence d'un champ scalaire conformément couplé à la gravitation \cite{Radu:2005bp} avec une constante cosmologique négative.

\subsubsection{Dimensions supplémentaires} Passons maintenant au crible l'influence de dimensions supplémentaires sur les théorèmes d'unicité. Tout d'abord, il faut savoir que le théorème de Birkhoff reste valable en dimension quelconque. C'est-à-dire que toute solution $\mc C^2$ de l'équation d'Einstein dans le vide $R_{\mu\nu}=0$ en dimension $D\geq 4$ est localement isométrique à la solution statique de Tangherlini \cite{Tangherlini:1963bw}, qui est simplement la généralisation en dimension $D$ de la solution de Schwarzschild donnée par la métrique
\BE \dd s^2 = - \lb 1- \lp\frac{r_0}{r}\rp^{D-3} \rb \dd t^2 + \frac{\dd r^2}{1- \lp\frac{r_0}{r}\rp^{D-3} } + r^2 \dd \Om^2_{(D-2)} \EE
où $\dd \Om^2_{(D-2)}$ représente la métrique de la sphère $S^{D-2}$. Nous reviendrons au chapitre suivant à cette solution et à ses versions topologiques.

Concernant la version réciproque, G. Gibbons et al. \cite{Gibbons:2002bh} et S. Hwang \cite{hwang1998rigidity} ont montré indépendamment que le théorème d'Israel présenté dans la sous-section précédente subsiste toujours en dimensions supplémentaires. C'est-à-dire que la seule solution statique et asymptotiquement plate de l'équation d'Einstein dans le vide en dimension quelconque est celle de Tangherlini. Le lecteur pourra se tourner vers \cite{Rogatko:2003kj} pour la généralisation de ce résultat en présence d'un champ de jauge abélien. Précisons que si nous supprimons le caractère asymptotiquement plat de la solution alors nous perdons l'unicité du résultat \cite{Gibbons:2002bh}. En effet, en supprimant cette hypothèse, il est possible d'obtenir des solutions présentant un horizon qui possède la topologie de la sphère $S^{D-2}$ mais avec une métrique induite sur l'horizon qui n'est pas nécessairement à courbure constante mais qui doit être un espace d'Einstein. C. Böhm a construit explicitement une telle classe de métriques de type Einstein qui ne sont pas forcément à courbure constante pour $ 5 \leq D-2 \leq 9$ \cite{bohm}. Nous rencontrerons de nouveau ces métriques dans le chapitre suivant.

En revanche, l'analogue des théorèmes d'unicité dans le cas stationnaire n'existe plus en dimensions supplémentaires. Il est bien connu qu'il n'existe pas une unique solution stationnaire à l'équation d'Einstein $R_{\mu\nu}=0$ en dimension 5 par exemple. En effet, il y a d'une part la solution de Myers-Perry \cite{Myers:1986un} qui est une généralisation de la solution de Kerr en dimension quelconque et d'autre part R. Emparan et H. Reall ont découvert en dimension 5 une solution du vide qui décrit un anneau noir en rotation avec un horizon des événements qui possède la topologie de $S^2 \times S^1$ \cite{Emparan:2001wn}. Pour la même masse et le même moment angulaire, il existe donc deux solutions différentes. La situation est encore plus frappante avec la solution de \ita{Saturne noir} \cite{Elvang:2007rd} qui décrit un trou noir sphérique entouré d'un anneau noir. Ainsi, pour une masse donnée, la solution de Tangherlini en dimension 5 n'est pas unique puisqu'il est possible d'avoir un Saturne noir avec un moment angulaire nul pour lequel la rotation du trou noir compense celle de l'anneau noir qui l'entoure. Dans le chapitre suivant, nous discuterons aussi de l'influence de champ de $p$-formes dans la théorie.

Néanmoins, le théorème de rigidité et celui concernant la topologie de l'horizon ont été étudiés en dimensions supplémentaires. S. Holland et al. \cite{Hollands:2006rj} ont montré qu'un trou noir stationnaire, asymptotiquement plat, non-extrémal dont le vecteur de Killing qui génère la translation dans le temps est de norme non nulle sur l'horizon doit être à symétrie axiale. Au sujet de la topologie de l'horizon, il a été montré dans \cite{Galloway:2005mf,Galloway:2006ws} qu'un trou noir stationnaire, asymptotiquement plat et obéissant à la condition d'énergie dominante admet un horizon dont la courbure scalaire est positive, contraignant ainsi sa topologie.

Pour une discussion exhaustive des théorèmes d'unicité en dimension quelconque, nous renvoyons le lecteur vers la revue de référence de R. Emparan et H. Reall \cite{Emparan:2008eg} traitant des trous noirs solutions de l'équation d'Einstein du vide en dimensions supplémentaires.

\subsubsection{Terme de courbure supplémentaire}
 Pour finir sur ce sujet, discutons brièvement de l'influence de termes de courbure supplémentaire. Nous restreignons notre attention à la théorie de Lovelock que nous présenterons à la section \ref{lovelock}. R. Zegers montra que le théorème de Birkhoff subsiste dans cette théorie \cite{Zegers:2005vx} et que ce résultat se généralise aussi en présence d'un champ de jauge abélien. 
 
Concernant une possible généralisation du théorème d'Israel, il n'en existe pas à notre connaissance. Nous avons vu au-dessus qu'en théorie d'Einstein et en présence de dimensions supplémentaires, il est possible d'obtenir des solutions de trous noirs avec un horizon de type Einstein. En présence de la théorie d'Einstein-Gauss-Bonnet, qui est une généralisation de la théorie d'Einstein mais une restriction de la théorie de Lovelock, il existe des trous noirs statiques dont la géométrie de l'horizon est de type Einstein mais avec une contrainte supplémentaire. Ceci se comprend puisque la théorie d'Einstein-Gauss-Bonnet met en jeu le tenseur de Riemann dans les équations du mouvement. Ainsi, G. Dotti et al. \cite{Dotti:2005rc} ont trouvé des trous noirs statiques avec un horizon de type Einstein et une condition supplémentaire sur le tenseur de Weyl de la géométrie induite de l'horizon, qui autorise des horizons à courbure non constante. Nous reviendrons plus en détails dans le chapitre \ref{EGB} sur les solutions de cette théorie et nous étudierons l'influence de champ de matière dans cette théorie.

\chapter{Trous noirs en dimensions supplémentaires\label{BH in higher D}}

Nous nous intéressons dans ce chapitre à la théorie de la Relativité Générale en dimension $D \geq 4$. Après avoir présenté quelques solutions statiques du vide et en présence d'une constante cosmologique $\La$ négative, nous donnerons des résultats nouveaux, exposés dans \cite{Bardoux:2012aw}, en présence de champs de matière décrits par des $p$-formes libres.

\minitoc
	\section{Solutions statiques du vide en présence d'une constan\-te cosmologique}
		\subsection{La solution de Tangherlini et ses versions topologiques}
	
Les trous noirs topologiques rencontrés à la sous-section \ref{violer} en dimension $D=4$ \eqref{topo} se généralisent facilement en dimension $D \geq 4$ \cite{Vanzo:1997gw,Birmingham:1998nr}. Ce sont des solutions avec des horizons des événements étendus ou compacts dont la géométrie intrinsèque est sphérique, plate ou hyperbolique en dimension 4 et plus généralement la géométrie est celle d'un espace d'Einstein quand $D \geq 4$. Présentons ces trous noirs qui sont des solutions de l'action suivante
\BE S_0 = \frac{1}{16 \pi G} \int_{\mc M} \sqrt{-g}  \left(R-2\La \right)  \dd ^D x \ . \label{action0} \EE
Pour cela, nous considérons une classe de métriques statiques s'écrivant sous la forme
\BE \dd s^2=-V(r) \dd t^2 + \frac{\dd r^2}{V(r)} + r^2\si_{ij}(y^k) \dd y^i \dd y^j \label{ansatz}\EE
solution de \eqref{action0}, le lecteur pourra trouver une discussion de cet ansatz dans \cite{Jacobson:2007tj}. Cette classe de métriques est paramétrisée par la fonction $V(r)$, appelée le \ita{potentiel} du trou noir, tandis que $\si_{ij}(y^k)$ désigne une métrique riemannienne arbitraire associée à la variété transverse, notée $\mc H$, de dimension $D-2$. Nous définissons $n=D-3$ par commodité pour la suite. Les composantes non nulles du tenseur de Riemann associées à cette géométrie sont
\BE R^t_{\ rrt} = \frac{V''}{2V} \ ,\qquad R^t_{\ ij t} = R^r_{\ ij r} = \frac{rV'}{2}\si_{ij}  \ ,\qquad R_{ijkl} = \mc R_{ijkl} - V \lp \si_{ik} \si_{jl} - \si_{il} \si_{jk}  \rp \ , \EE
et celles du tenseur de Ricci sont 
\begin{align}
 R_{tt}  &= -V^2 R_{rr} = \frac{1}{2} V V '' + \frac{n+1}{2r} V V' = \frac{V}{2r^{n+1}} \lp r^{n+1}V' \rp' \ ,\label{Rtt} \\
  R_{ij} &= \mc R_{ij} -  \lp rV'+nV \rp \si_{ij} = \mc R_{ij} - \frac{1}{r^{n-1}} \lp r^n V \rp' \si_{ij} \ , 
\end{align}
avec $\mc R_{ijkl}$ et $\mc R_{ij}$ les tenseurs de Riemann et de Ricci de la variété $\mc H$ construits à partir de la métrique intrinsèque $\si_{ij}$. Etant donné une constante cosmologique négative
\BE \La = -\frac{(D-1)(D-2)}{2 l^2} = - \frac{(n+2)(n+1)}{2 l^2} \label{lambda}\EE
et en l'absence de champ de matière, l'équation d'Einstein \eqref{Einstein} s'écrit alors
\BE R_{\mu\nu}=-\frac{n+2}{l^2}g_{\mu\nu} \ . \EE
D'après \eqref{Rtt}, les composantes $(tt)$ et $(rr)$ de cette équation sont proportionnelles entre elles et sont résolues par le potentiel
\BE V(r) =\ka - \frac{r_0^n}{r^n} + \frac{r^2}{l^2} \label{vacuum} \EE
avec $\ka$ et $r_0$ deux constantes d'intégration. Ainsi, les composantes $(ij)$ de l'équation d'Einstein se réduisent à 
\BE \mc R_{ij}=n\kappa\si_{ij} \label{Einstein manifold} \ ,\EE
c'est-à-dire que $\mc H$ est un espace d'Einstein dont la courbure scalaire vaut $\mc R =  n (n+1) \ka$. Ces solutions possèdent un horizon des événements tant que $r_0$ est suffisamment grand et décrivent la géométrie des trous noirs topologiques présentés dans \cite{Vanzo:1997gw} et \cite{Birmingham:1998nr} en particulier. Remarquons que si $\mc H$ est une variété de dimension 2, c'est-à-dire $D=4$, alors le tenseur de Riemann s'écrit nécessairement sous la forme $R_{ijkl} = R g_{i[k} g_{l]j}$. Par conséquent, l'horizon des événements est à courbure constante si $D=4$ et nous retrouvons ainsi les trous noirs topologiques \eqref{topo} rencontrés à la section \ref{violer}.
\bs

Quand $\ka \leq 0$, la constante cosmologique négative est cruciale pour avoir un horizon des événements qui cache la singularité de courbure située en $r=0$. D'autre part, si l'espace d'Einstein $\mc H$ est à courbure strictement positive $(\ka > 0)$, l'interprétation de la solution en terme de trou noir survit avec une constante cosmologique $\La =0$ ou $\La > 0$. Dans le premier cas, avec $\mc H = S^{n+1}$ et 
\BE V(r)=1-\frac{r_0^n}{r^n} \ ,\EE
nous retrouvons la généralisation de la solution de Schwarzschild en dimension $D \geq 4$ obtenue par F. Tangherlini \cite{Tangherlini:1963bw}. Rappelons que si $\dd \Om^2_{(n+1)}$ désigne la métrique de la sphère $S^{n+1}$, alors elle est construite par récurrence selon la relation suivante
\BE \dd \Om^2_{(n+1)}  = \dd\chi^2 + \sin^2\chi \dd \Om^2_{(n)} \ . \EE
Il est ensuite facile de montrer que $S^{n+1}$ est un espace d'Einstein puisque son tenseur de Ricci est donné par $\mc R_{ij} = n \si_{ij}$, c'est même un espace à courbure constante. Tout autre espace d'Einstein compact avec $\ka > 0$ conduit à une généralisation de la solution de Tangherlini \cite{Gibbons:2002th}. Par exemple, nous pouvons garder $\mc H$ avec la topologie de la sphère $S^{n+1}$ mais avec une métrique de Böhm \cite{bohm} construite pour $5 \leq n+1 \leq 9$. Ces métriques de type Einstein prennent la forme suivante
\BE \dd\theta^2 + a^2(\theta) \dd \Om^2_{(p)} + b^2(\theta) \dd \Om^2_{(q)} \EE
où $p+q=n$, $p>1$ et $q>1$. Ces métriques ne sont pas à courbure constante en général et présentent le groupe d'isométrie $SO(p+1)\times SO(q+1)$. Nous retrouvons cependant la métrique de la sphère $S^{n+1}$ avec $a(\theta)=\sin\theta$ et $b(\theta)=\cos\theta$. De telles solutions sont accompagnées d'instabilités classiques décrites dans \cite{Gibbons:2002th,Gibbons:2002pq}. 
Une autre possibilité, avec une topologie différente, est de considérer $\mc H$ comme un produit de sphères portant chacune une métrique de type Einstein avec une même courbure positive. 
La même construction est possible avec une constante cosmologique strictement positive, donnant des trous noirs de Sitter avec un potentiel
\BE V(r) = \ka-\frac{r_0^n}{r^n} - \frac{r^2}{l^2} \EE
en supprimant le signe moins dans la définition \eqref{lambda} pour ce cas. La présence d'un horizon cosmologique restreint la valeur du paramètre $r_0$ afin d'avoir un trou noir dans de Sitter. Le lecteur doit garder à l'esprit qu'à chaque fois que nous rencontrerons des trous noirs AdS avec $\ka > 0$ dans la section \ref{BHpforms}, il est possible de prolonger ces solutions à $\La \geq 0$ et d'obtenir asymptotiquement un trou noir plat ou localement de Sitter, avec le même $\mc H$ et les mêmes champs de matière. Ceci est vrai en particulier pour les solutions avec $\ka > 0$ exposées dans les sections \ref{single} et \ref{kähler}, bien que nous n'ayons pas souligné cette possibilité à chaque fois par souci de concision.
\bs

La question que nous allons étudier dans la suite de ce chapitre est dans quelle mesure des champs de matière peuvent être ajoutés à ces solutions  pour obtenir des trous noirs décrits par la métrique \eqref{ansatz}. Nous tenterons d'habiller ces trous noirs topologiques avec des $p$-formes qui constituent une généralisation naturelle de l'interaction électromagnétique, ce qui est l'objet de la section suivante.
		\subsection{Une version généralisée du théorème de Birkhoff\label{Birkhoff-généralisé}}
Dans cette sous-section, nous souhaitons revenir brièvement sur le choix de l'ansatz \eqref{ansatz}. Cette classe de métrique peut en fait être vue comme une version faible du théorème de Birkhoff. Au lieu de considérer la classe de métrique \eqref{sphérique} issue de la symétrie sphérique, nous pouvons généraliser à la classe de solutions suivante comme le suggère G. Gibbons et al. dans \cite{Gibbons:2002th} 
\BE \dd s^2 = - 2 e^{2\nu(u,v)} B(u,v)^{-\frac{n}{n+1}} \dd u \dd v + B(u,v)^\frac{2}{n+1} \sigma_{ij}(y^k) \dd y^i \dd y^j   \label{warped} \EE
où les fonctions $B(u,v)$ et $\nu(u,v)$ paramétrisent cette famille de métriques. Dans le vide, cet ansatz nous conduit à une sous-classe de solutions statiques \cite{Gibbons:2002th} qui s'écrivent sous la forme \eqref{ansatz}, ce qui généralise en ce sens le théorème de Birkhoff sans pour autant imposer des symétries. Ce résultat reste notamment vrai en présence de matière décrite par un tenseur énergie-impulsion $T_{\mu\nu}$ non nul tant que $T_{uu}=T_{vv}=0$.

La classification des solutions à partir de l'ansatz \eqref{warped} a notamment été étudiée dans la théorie d'Einstein-Gauss Bonnet dans \cite{Bogdanos:2009pc} en dimension 6 puis dans \cite{Bardoux:2010sq} en présence de champs de matière, ce qui sera au c\oe ur du chapitre suivant.

	\section{Electromagnétisme et $p$-formes}
Avant de modeler l'horizon d'un trou noir avec des $p$-formes, nous allons expliquer succinctement pourquoi ces $p$-formes constituent une généralisation du tenseur de Faraday $F$ qui régit les phénomènes électromagnétiques. Comme nous le savons bien, une particule peut porter une charge électrique s'il existe une interaction entre la particule et le champ de jauge de Maxwell. L'interaction est décrite en remarquant en particulier que la ligne d'univers d'une particule est \ita{unidimensionnelle} et que le champ de jauge $A$ est une forme de \ita{rang 1}. Ainsi, l'interaction d'une particule ponctuelle de charge $q$ est décrite par le terme suivant dans l'action
\BE q \int A_\mu \lp x(\tau) \rp  \frac{\dd x^\mu}{\dd \tau} \dd \tau \ . \label{interactionEM}\EE
Par conséquent, l'action complète, décrivant la particule chargée et le champ de Maxwell, qui redonne l'équation du mouvement en présence de la force de Lorentz et les équations de Maxwell, est 
\BE S = - m\int \dd s - q \int A_\mu (x^\nu) \dd x^\mu - \frac12 \int F \w \star F \EE
où l'intégration des deux premiers termes est faite le long de la trajectoire d'espace-temps. Nous venons donc de voir que les particules ponctuelles sont des sources du champ électromagnétique. 
\bs

Dans le contexte de la théorie des cordes, il est loisible de se demander s'il existe une nouvelle sorte de charge associée à une corde. Une corde décrit une feuille d'univers \ita{bidimensionnelle} dans l'espace-temps et ces coordonnées sont données par $D$ fonctions $X^\mu(\tau,\sigma)$, ainsi les $X^\mu$ constituent des fonctions de l'\ita{espace des paramètres} $(\tau,\si)$ dans l'espace-temps. Si nous choisissons $\tau$ et $\si$ comme les coordonnées intrinsèques de la feuille d'univers, alors $\frac{\p X^\mu}{\p \tau}$ et $\frac{\p X^\mu}{\p \si}$ sont les coordonnées de deux vecteurs tangents à la feuille d'univers linéairement indépendants. Puis, afin de décrire une interaction similaire à celle de l'électromagnétisme \eqref{interactionEM}, nous avons besoin d'un tenseur 2 fois covariant. Or, il existe dans la théorie des cordes fermées une 2-forme $B$ dite de \ita{Kalb-Ramond} de masse nulle. Nous pouvons alors décrire l'interaction dite électrique entre la corde et le champ de Kalb-Ramond par l'action
\BE \int \frac{\p X^\mu}{\p \tau} \frac{\p X^\nu}{\p \si} B_{\mu\nu}\lp X^\rho(\tau,\si) \rp \dd \tau \dd \si \EE
en s'inspirant de l'interaction \eqref{interactionEM} valable dans le cadre électromagnétique. Nous dirons alors que la corde porte une \ita{charge électrique de Kalb-Ramond}. Précisons que l'anti-symétrie du champ $B$ est nécessaire pour assurer l'invariance sous les reparamétrisations $(\tau,\si)$ de ce couplage, tout comme l'action de Polyakov $S_\text{P}$ qui décrit la feuille d'univers d'une corde est invariante sous ces reparamétrisations. Finalement, l'action complète décrivant le système constitué d'une corde chargée et d'un champ de Kalb-Ramond est
\BE S = S_\text{P} - \int \frac{\p X^\mu}{\p \tau} \frac{\p X^\nu}{\p \si} B_{\mu\nu}\lp X^\rho(\tau,\si) \rp \dd \tau \dd \si - \frac12 \int H \w \star H \EE
où le champ $H$ constitue le tenseur de force, analogue au tenseur de Faraday $F$, défini par $H=\dd A$. Nous pourrions continuer l'analogie en introduisant le tenseur anti-symétrique suivant
\BE j^{\mu\nu}(x^\rho) = \int \delta^{(D)} \lp x^\rho - X^\rho(\tau,\si) \rp \frac{\p X^{[\mu}}{\p \tau} \frac{\p X^{\nu]}}{\p \si} \dd \tau \dd \si  \EE
analogue au courant qui apparaît dans les équations de Maxwell. Nous renvoyons le lecteur vers \cite{zwiebach2004first} pour un exposé détaillé et très pédagogique.

\bs
Bien évidemment, il n'y a qu'un pas à faire pour généraliser la construction précédente à des sources possédant plus de dimensions. Dans le cadre de la théorie des cordes, il existe, en plus des cordes, des objets étendus appelés des $Dp$-branes dont la feuille d'univers est de dimension $p+1$. Ces $Dp$-branes peuvent être chargées électriquement avec des $(p+1)$-formes $A$ présentent dans le secteur de Ramond-Ramond des théories des cordes de type I et II, analogues aux 2-formes de Kalb-Ramond. Si $X^\rho(\tau,\si^1, \ldots, \si^p)$ désignent les coordonnées de la $Dp$-brane dans l'espace-temps, alors l'interaction est donnée par l'action
\BE \int \frac{\p X^\mu}{\p \tau} \frac{\p X^{\nu_1}}{\p \si^1} \cdots \frac{\p X^{\nu_p}}{\p \si^p}  A_{\mu\nu_1\ldots\nu_p}\lp X^\rho(\tau,\si^1, \ldots, \si^p) \rp \dd \tau \dd \si^1 \ldots \dd \si^p \ . \EE
Passons, désormais à la construction explicite de trous noirs statiques en présence de formes libres en théorie d'Einstein et en dimension quelconque.
	\section{Modeler l'horizon d'un trou noir avec des $p$-formes \label{BHpforms}}	

A travers cette section, nous allons tenter de contourner les théorèmes d'unicité présentés en \ref{no-hair} en habillant des trous noirs topologiques avec des $p$-formes libres au-delà du cas électromagnétique correspondant à $p=2$. Cela inclut le cas de la constante cosmologique, celui des champs scalaires, celui des champs de 3-formes aussi connus sous le nom d'axions, etc. Nous verrons que ces $p$-formes donnent lieu à des quantités conservées à l'infini qui généralisent la notion de charge électrique pour $p=2$. Ainsi, elles ne sont pas associées à des cheveux. Ces $p$-formes vont permettre de modeler la géométrie de l'horizon des trous noirs en introduisant de nouveaux horizons et en changeant parfois complètement les propriétés de la solution. Un exemple typique est celui du trou noir de Reissner-Nordstrom \eqref{RN} qui possède un horizon interne et une singularité de courbure de genre-temps si la charge est plus petite que la masse et une singularité nue si la charge est supérieure à la masse. De plus, nous constaterons que l'inclusion de $p$-formes change les propriétés asymptotiques des trous noirs dans certains cas. Nous verrons aussi que ces champs peuvent agir comme des champs externes, un peu comme dans la solution de Melvin \cite{ernst1976black} qui habille le trou noir de Schwarzschild avec un champ magnétique homogène.

Récemment, R. Emparan et al. \cite{Emparan:2010ni} ont montré qu'un trou noir statique asymptotiquement plat ne peut pas se vêtir d'une $p$-forme pour $p \in \llbracket (D+1)/2, D-1 \rrbracket$. Leur argument, généralisé plus tard pour $p\geq3$ dans \cite{Shiromizu:2011he}, n'exclut pas que de tels champs soient portés par des horizons avec une topologie non-sphérique, ni même l'existence de $p$-formes en présence d'une constante cosmologique ou avec d'autres conditions asymptotiques. En effet, nous verrons que l'hypothèse d'espace-temps asymptotiquement plat n'est pas adaptée aux trous noirs en présence de $p$-formes, qui modifient en général les propriétés asymptotiques de la solution. Rappelons aussi que R. Emparan donna dans \cite{Emparan:2004wy} une solution explicite d'un anneau noir en rotation en présence d'une 3-forme qui agit comme un dipôle, c'est-à-dire que la charge associée est nulle. Ce champ fournit donc un cheveu primaire qui paramétrise toute une classe de solution. Intuitivement, cette solution existe car la corde associée à la 3-forme subit une tension qui est compensée par la force centrifuge. 
\bs

Dans cette section, nous étudierons avec une certaine généralité le problème qui consiste à habiller des trous noirs statiques avec des $p$-formes et nous expliciterons une pléthore de solutions. En particulier, nous exposerons un trou noir statique en dimension 4 avec des charges axioniques non triviales issues de 3-formes. Nous montrerons comment de tels axions régularisent les solutions chargées électriquement. Nous verrons aussi comment des champs scalaires permettent de construire des trous noirs AdS avec un horizon plat mais avec un potentiel $V(r)$ associé usuellement à un trou noir possédant un horizon hyperbolique. Finalement, certaines de ces solutions avec constante cosmologique négative pourraient avoir des applications en matière condensée via la conjecture AdS/CFT; cependant nous n'avons pas exploré cette direction dans cette thèse.

Dans la première sous-section \ref{fields}, nous allons introduire les champs de matière en les adaptant à la géométrie \eqref{ansatz}, puis nous appliquerons cela au cas des $p$-formes en \ref{pforms}. Les trous noirs habillés par un seul champ de matière seront présentés en \ref{single}, tandis que la sous-section \ref{kähler} traitera du cas spécial des horizons de type Einstein-Kähler. Quand la courbure de l'horizon est positive, nous rappelons que les trous noirs présentés dans ces sections existent quelque soit la valeur de la constante cosmologique $\La$, autrement ce sont des trous noirs AdS. Après cela, nous discuterons en \ref{multiple} le cas des trous noirs habillés par plusieurs champs de matière. Enfin, nous détaillerons en \ref{axionique} le cas particulier des trous noirs en dimension 4 avec deux champs axioniques.


		\subsection{Ansatz sur les champs de matière et effets sur la géométrie\label{fields}}

Nous nous intéressons aux géométries qui gardent la forme \eqref{ansatz} lorsque des champs de matière sont inclus. En particulier, le tenseur énergie-impulsion doit satisfaire les conditions $T_{tr}=0$ et $T_{tt}+V^2T_{rr}=0$ pour satisfaire l'équation d'Einstein en accord avec la discussion de la sous-section \ref{Birkhoff-généralisé}. Il est bien connu que le tenseur énergie-impulsion d'un champ de Maxwell satisfait ces conditions, contrairement à des champs scalaires qui dépendent du temps ou de la coordonnée radiale. En fait, des champs scalaires avec une telle dépendance radiale conduisent à des solutions singulières, alors que les contraintes que nous imposons, mettant à l'écart ces cas, nous fournissent des géométries de trous noirs comme nous allons le voir. 

De plus, nous ne souhaitons pas mettre en jeu des vecteurs ou des tenseurs privilégiés autres que ceux provenant du feuilletage particulier introduit par la métrique \eqref{ansatz}. Nous nous référerons dans la suite à cette hypothèse en parlant d'\ita{isotropie} du tenseur énergie-impulsion. Enfin, nous exigeons aussi que le tenseur énergie-impulsion ne permette pas de distinguer différents points de $\mc H$, nous parlerons alors d'un tenseur énergie-impulsion \ita{homogène}. Sous ces hypothèses, il s'ensuit que le tenseur énergie-impulsion est complètement déterminé par deux fonctions de $r$ seulement, que nous nommons $\ep(r)$ et $P(r)$, et prend la forme diagonale suivante
\BE T_{\mu\nu}=\frac{1}{16\pi G r^{n+1}} \text{diag} \lp V(r)\ep(r), -\ep(r)/V(r) , r^{2}P(r)\si_{ij}  \rp \ .  \label{st} \EE
Avec une telle source, l'équation d'Einstein en présence d'une constante cosmologique $\La$ négative \eqref{lambda},
\BE G_{\mu\nu}+\Lambda g_{\mu\nu} = 8\pi G T_{\mu\nu} \ , \EE
se réduit au système
\begin{align}
\lp r^{n+1} V' \rp ' &= \frac{2(n+2)}{l^2}r^{n+1} + \frac{n-1}{n+1} \ep(r) + P(r) \ , \label{eq1} \\
\mc R_{ij} &= \frac1{r^{n-1}} \lb {\lp r^nV\rp'} - \frac{n+2}{l^2}{r^{n+1}} + \frac{\ep(r)}{n+1} \rb \si_{ij}\ .
\end{align}
Puisqu'à la fois, la métrique $\si_{ij}$ et le tenseur de Ricci $\mc R_{ij}$ de $\mc H$ dépendent seulement des coordonnées dites transverses $y^i$, le facteur de proportionnalité entre les deux doit être constant et nous le dénommons $n \ka$, ainsi
\BE {\lp r^nV\rp'}-\frac{n+2}{l^2}{r^{n+1}}+\frac{\ep(r)}{n+1}=n\kappa{r^{n-1}}\ ; \label{eqV}\EE
et par conséquent $\mc H$ est un espace d'Einstein qui satisfait \eqref{Einstein manifold}. L'équation \eqref{eqV} peut alors être intégrée pour donner le potentiel
\BE V(r)=\kappa-\frac{r_0^n}{r^n}+\frac{r^2}{l^2}-\frac{1}{(n+1) r^n}\int^r \ep \label{V}\EE
avec $r_0$ une constante d'intégration qui a la dimension d'une longueur. Quant à l'équation restante \eqref{eq1} équivalente à la conservation du tenseur énergie-impulsion, elle devient
\BE \ep'(r) + \frac{n+1}{r} P(r) = 0 \ .\label{conservation}\EE

Tant que cette équation est vérifiée, la métrique \eqref{ansatz} avec le potentiel \eqref{V} et avec $\mc H$ un espace d'Einstein \eqref{Einstein manifold} est solution de l'équation d'Einstein avec \eqref{st} comme source. Dans la suite de cette section, nous allons montrer comment des tenseurs énergie-impulsion de cette forme peuvent être obtenus par diverses combinaisons de champs libres. 

\subsubsection{Notations et conventions}
Dans ce qui suit, nous allons considérer des espace-temps statiques de dimension $D$ munis d'une métrique $g_{\mu\nu}$ et nous définissons $n=D-3$ par commodité. Les indices d'espace-temps sont indiqués par des lettres grecques $\mu,\nu,\ldots$. Les sections données par une coordonnée radiale constante et un temps constant de dimension $n+1$ sont notées $\mc H$ et possèdent une métrique induite $\si_{ij}$. Les indices latins $i,j,\ldots$ sont associés à ces sous-variétés. Les indices latins $a,b,\ldots$ sont utilisés pour coller une étiquette à chaque facteur de $\mc H$ quand ce dernier prend la forme d'un produit direct  $\mc H=\mc H^{(a_1)} \times \cdots \times \mc H^{(a_m)}$ et chaque facteur possède une métrique induite notée $\si^{(a_k)}_{ij}$. Les formes volume de l'espace-temps et des sections $\mc H$ sont notées par $\ep$ et $\hat\ep$ respectivement. Finalement, nous considérerons, pour le secteur de la matière, des $p$-formes $H_{[p]}$ exactes, c'est-à-dire qu'il existe $B_{[p-1]} \in \La^{p-1}\lp \mc M \rp$ tel que $H_{[p]}=\dd B_{[p-1]}$, nous abandonnerons l'indice $[p]$ quand le rang de la forme est évident.
		\subsection{$p$-formes libres\label{pforms}}	
		
Pour coupler minimalement une $p$-forme $H_{[p]}=\dd B_{[p-1]}$ au champ gravitationnel, nous ajoutons à l'action d'Einstein-Hilbert \eqref{action0} un terme représentant la matière sous la forme 
\BE S_M  = -\frac1{16\pi G} \int_{\mc M} \frac1{2} H_{[p]} \w \star H_{[p]}  = -\frac1{16\pi G} \int_{\mc M} \sqrt{-g} \frac{1}{2p!} H_{[p]}^2 \dd^Dx \label{SM} \EE
où $H_{[p]}^2 = H_{\mu_1\ldots\mu_p} H^{\mu_1\ldots\mu_p}$.
Les équations du mouvement pour $H_{[p]}$ sont $\dd H_{[p]} =0$ et $\dd \star H_{[p]} =0$, c'est-à-dire 
\BE \nabla_{\mu}H^{\mu\nu_1\ldots\nu_{p-1}}=0 \ ,  \qquad  \nabla_{[\mu}H_{\nu_1\ldots\nu_p]}=0 \ , \label{eqH}\EE
et son tenseur énergie-impulsion est donné par
\BE T_{\mu\nu}=\frac1{16\pi G(p-1)!}\lp H_{\mu\rho_1\ldots\rho_{p-1}}H_\nu{}^{\rho_1\ldots\rho_{p-1}}-\frac1{2p}H^2g_{\mu\nu}\rp \ . \EE
Les contraintes $T_{tr}=0$ et $T_{tt}+V^2T_{rr}=0$ sur un tel tenseur se réduisent à 
\BE H_{t i_1 \ldots i_{p-1}} = H_{r i_1 \ldots i_{p-1}} = 0 \ . \EE
Notons que pour des champs scalaires libres correspondant au cas $p=1$, cette dernière condition signifie que de tels champs sont indépendants des coordonnées $t$ et $r$. Puis, les équations \eqref{eqH} sont résolues avec
\BE H_{tri_1\ldots i_{p-2}}=\frac{1}{r^{n-2p+5}}{\mc E}_{i_1\ldots i_{p-2}}(y^i) \qquad \text{et}\qquad H_{i_1\ldots i_{p}}={\mc B}_{i_1\ldots i_{p}}(y^i) \ .\label{H}\EE
En utilisant la métrique transverse $\si_{ij}$ pour monter et descendre les indices des tenseurs $\mc E$ et $\mc B$, nous avons
\BE H^{tri_1\ldots i_{p-2}}=-\frac{1}{r^{n+1}}{\mc E}^{i_1\ldots i_{p-2}}(y^i) \qquad \text{et} \qquad H^{i_1\ldots i_{p}}=\frac1{r^{2p}}{\mc B}^{i_1\ldots i_{p}}(y^i) \ . \label{Hup} \EE
Ici, $\mc E_{[p-2]}$ et $\mc B_{[p]}$ sont des formes de $\mc H$ de rang $p-2$ et $p$ respectivement tel que
\BE \p_{i_1}\lp\sqrt\si{\mc E}^{i_1\ldots i_{p-2}}\rp=0 \ ,\qquad \p_{[j}{\mc E}_{i_1\ldots i_{p-2}]}=0 \ ,\label{eqE} \EE
\BE \p_{i_1}\lp\sqrt\si{\mc B}^{i_1\ldots i_{p}}\rp=0    \ , \qquad \p_{[j}{\mc B}_{i_1\ldots i_{p}]}=0    \ .\label{eqB}\EE
Avec le langage des formes différentielles, ceci se traduit simplement par
\BE H_{[p]} = \frac1{r^{n-2p+5}} \dd t \w \dd r \w \mc E_{[p-2]} + \mc B_{[p]}  \EE
avec $\dd \mc E =0$ et $\dd \star_{\mc H} \mc E =0$ et de même pour $\mc B$, où $\star_{\mc H}$ est le produit de Hodge construit sur l'espace transverse $\mc H$.
Ces formes harmoniques sur $\mc H$ définissent ce que nous appelons la \ita{polarisation} du champ $H$ en décrivant sa partie électrique avec $\mc E$ et sa partie magnétique avec $\mc B$. Nous verrons aux sous-sections \ref{p-charges-1} et \ref{p-charges-2} qu'elles correspondent à des charges conservées associées au champ $H$. Précisons qu'il y aura bien évidemment des contraintes topologiques : par exemple, il ne sera pas possible de construire une $p$-forme magnétique $\mc B$ sur la sphère $\mc H = S^{n+1}$ avec $p \in \llbracket 2,n \rrbracket$ puisque dans ce cas les nombres de Betti $b_p$ sont nuls. 
\bs

Les composantes non nulles du tenseur énergie-impulsion sont
\begin{align}
T_{tt} &= -V^2T_{rr}=\frac V{16\pi G}\lp \frac{\mc E^2}{2(p-2)!r^{2n-2p+6}}+\frac{\mc B^2}{2p!r^{2p}} \rp \ ,\\
T_{ij} &=\frac1{16\pi G}\lb-\frac{1}{(p-3)!r^{2n-2p+4}}\lp\mc E_{ik_1\ldots}\mc E_{j}{}^{k_1\ldots}-\frac1{2(p-2)}\mc E^2\si_{ij}\rp \right. \notag\\
&\qquad\qquad\qquad\qquad\qquad  + \left. \frac1{(p-1)!r^{2p-2}}\lp\mc B_{ik_1\ldots}\mc B_{j}{}^{k_1\ldots}-\frac1{2p}\mc B^2\si_{ij}\rp\rb \ . 
\end{align}
Pour obtenir un tenseur énergie-impulsion de la forme \eqref{st}, la composante $T_{tt}$ ne doit pas dépendre des coordonnées transverses $y^i$. Par conséquent, quand $2p\neq n+3$,  $\mc E^2$ et $\mc B^2$ sont des constantes ; mais lorsque  $2p=n+3$, le terme électrique et celui magnétique possèdent la même puissance de $r$ et ainsi seulement $\mc B^2+p(p-1)\mc E^2$ doit être constant. Dans ce dernier cas, les invariants $\mc E^2$ et $\mc B^2$ peuvent donc dépendre a priori des coordonnées $y^i$. Cependant, de telles dépendances violent notre hypothèse d'homogénéité; ou dit autrement, nous pourrions construire deux vecteurs $\hat\nabla_i\mc E^2$ et $\hat\nabla_i\mc B^2$ sur $\mc H$ qui briseraient l'isotropie. Dans la suite, nous allons simplement supposer que $\mc E^2$ et $\mc B^2$ sont des constantes pour tout $p$. Nous pouvons alors lire la densité d'énergie $\ep(r)$ à partir de $T_{tt}$
\BE \ep(r) = \frac{\mc E^2}{2(p-2)!r^{n-2p+5}}+\frac{\mc B^2}{2p!r^{2p-n-1}} \ ,\label{ep}\EE
puis définir la pression par $P(r)=16\pi G T_{ij}\si^{ij} r^{n-1}/(n+1)$, d'où
\BE P(r) =\frac1{n+1}\lp \frac{n-2p+5}{2(p-2)!r^{n-2p+5}}\mc E^2 + \frac{2p-n-1}{2p!r^{2p-n-1}}\mc B^2\rp \ . \EE
Ces deux quantités satisfont automatiquement l'équation de conservation \eqref{conservation}. La dernière contrainte pour obtenir un tenseur énergie-impulsion de la forme \eqref{st} provient de l'isotropie et de l'homogénéité sur $\mc H$ qui imposent, quand $2p\neq n+3$,
\BE
\mc E_{ik\ldots}\mc E_j{}^{k\ldots}=\frac{\mc E^2}{n+1}\si_{ij} \qquad \text{et} \qquad
\mc B_{ik\ldots}\mc B_j{}^{k\ldots}=\frac{\mc B^2}{n+1}\si_{ij} \ .
\label{isotropy}\EE
Encore une fois, quand la dimension de l'espace-temps est paire et $2p=n+3$, alors la partie électrique et celle magnétique de $T_{ij}$ possèdent les mêmes puissances de $r$ et la contrainte due à l'isotropie est alors affaiblie 
\BE \mc B_{ik\ldots}\mc B_j{}^{k\ldots}-\frac{\mc B^2}{n+1}\si_{ij}=(p-1)(p-2)\lp\mc E_{ik\ldots}\mc E_j{}^{k\ldots}-\frac{\mc E^2}{n+1}\si_{ij}\rp \ .
\label{dyonicisotropy}\EE
Dans ce dernier cas, nous verrons que des solutions dites \ita{dioniques} existent, c'est-à-dire avec un couple $(\mc E, \mc B)$ non nul. Notons que nous pouvons définir un nouveau tenseur antisymétrique pour des solutions dioniques comme la contraction des composantes des formes électriques et magnétiques $\mc A_{ij}=\mc B_{ijk_1\ldots k_{p-2}}\mc E^{k_1\ldots k_{p-2}}$.
A priori, $\mc A_{ij}$ peut briser l'isotropie. Cependant, comme nous le verrons, il s'avère que les flux électriques et magnétiques des solutions dioniques sont générés par des espaces orthogonaux, ainsi $\mc A_{ij}$ s'annule ; à moins que $\mc H$ soit le produit direct d'espaces de dimension 2 et dans ce cas $\mc A_{ij}$ peut être proportionnel aux formes volumes de ces espaces de dimension 2, sans introduire de directions privilégiées additionnelles. 

Une fois que $\mc E$ et $\mc B$ vérifient \eqref{eqE}, \eqref{eqB} et \eqref{isotropy} ou \eqref{dyonicisotropy}, nous obtenons de cette manière une solution du problème. Nous ne tenterons pas d'établir une classification complète des solutions, mais nous nous contenterons de construire les solutions les plus simples à partir des tenseurs naturels qui sont disponibles sur $\mc H$. S'il n'y a pas une structure supplémentaire présente, le seul tenseur antisymétrique sur $\mc H$ qui peut être construit est la forme volume $\hat\ep$ de $\mc H$. Néanmoins, si l'espace transverse est Kähler, nous pouvons aussi construire les tenseurs de polarisation à partir de la 2-forme de Kähler. Nous commencerons par présenter des solutions en présence d'une seule $p$-forme et nous étendrons la construction à plusieurs $p$-formes. 	
		\subsection{Trous noirs vêtus d'un seul champ\label{single}}
		
Quand le rang de la forme $\mc E$ ou $\mc B$ est égal à la dimension de $\mc H$, la condition d'isotropie \eqref{isotropy} est facilement réalisée en imposant à la forme considérée d'être proportionnelle à la forme volume de $\mc H$. De plus, la forme volume vérifie les équations \eqref{eqE} ou \eqref{eqB}. En notant la forme volume sur $\mc H$ par $\hat\ep_{[n+1]}$, nous pouvons construire une polarisation électrique $\mc E=q_e\hat\ep_{[n+1]}$ quand $p=n+3$ fournissant ainsi une solution chargée électriquement et une polarisation magnétique $\mc B=q_m\hat\ep_{[n+1]}$ quand $p=n+1$ conduisant alors à une solution chargée magnétiquement.

\subsubsection{Solutions électriques $p=2$ pour tout $n$}
Ce cas correspond à la théorie d'Einstein-Maxwell avec $\La < 0$ où $\mc E$ est une 0-forme qui peut être allumée pour tout $n$ sans briser l'isotropie. En prenant $\mc E=q_e$ où $q_e$ est une constante, nous obtenons la solution bien connue de Reissner-Nordstrom-AdS :
\BE
V(r)=\kappa-\frac{r_0^n}{r^n}+\frac{r^2}{l^2}+\frac{q_e^2}{2n(n+1)r^{2n}}
\qquad\text{et}\qquad
H=\frac{q_e}{r^{n+1}} \dd t \w \dd r \ .
\EE

\subsubsection{Solution magnétique $p=n+1=D-2$}
C'est le cas dual à celui d'une 2-forme électrique, la solution est alors duale à la solution de Reissner-Nordstrom-AdS chargée électriquement dans la théorie d'Einstein-Maxwell : 
\BE
V(r)=\kappa-\frac{r_0^n}{r^n}+\frac{r^2}{l^2}+\frac{q_m^2}{2n(n+1)r^{2n}} \ ,
\label{magnetic p=n+1}\EE
\BE
\mc B = q_m\hat\ep_{[n+1]} \qquad\text{et}\qquad
H = q_m \hat\ep_{[n+1]} \ .
\EE

\subsubsection{Solution dionique pour $p=2$ et $n=1$}
En dimension $D=4$, les flux électriques et magnétiques des deux solutions précédentes peuvent être combinés au sein d'une seule 2-forme. C'est encore une fois la solution familière de Reissner-Nordstrom-AdS en dimension 4, portant une charge à la fois électrique et magnétique :
\BE
V(r)=\kappa-\frac{r_0}{r}+\frac{r^2}{l^2}+\frac{q_e^2+q_m^2}{4r^2}
\qquad\text{et}\qquad
H = \frac{q_e}{r^{2}} \dd t \w \dd r + q_m \hat\ep_{[2]} \ .
\label{dyonicRN}\EE

\subsubsection{Solution électrique $p=n+3=D$} 
Dans ce cas, $H$ agit comme une constante cosmologique et la solution est
\BE
V(r)=\kappa-\frac{r_0^n}{r^n}+\frac{r^2}{l^2}\lp1-\frac{q_e^2 l^2}{2(n+1)(n+2)}\rp \ ,
\label{electric p=n+3}\EE
\BE
\mc E = q_e\hat\ep_{[n+1]}\qquad\text{et}\qquad
H = q_e r^{n+1} \dd t \w \dd r \w \hat\ep_{[n+1]} .
\EE
En ajustant la charge électrique $q_e$, il est alors possible d'annuler le terme de constante cosmologique présent dans le potentiel du trou noir et d'obtenir la solution de Schwarzschild-Tangherlini \cite{Tangherlini:1963bw} quand $\ka = 1$. La solution est par conséquent asymptotiquement plate à l'aide de la D-forme qui, agissant comme une constante cosmologique, annule l'effet de $\La$ sur la géométrie de l'espace-temps.

\subsubsection{Produit d'espaces d'Einstein avec de simples flux}
Supposons désormais que $\mc H$ soit le produit direct de $N$ espaces d'Einstein $\mc H^{(a)}$ avec les métriques induites notées $\si^{(a)}_{ij}$ et les tenseurs de Ricci $\mc R^{(a)}_{ij}$ tels que $\mc R^{(a)}_{ij}=\kappa^{(a)}\si^{(a)}_{ij}$. Supposons que tous les $\kappa^{(a)}$ sont égaux à une valeur que nous notons $n\kappa$ par convention, $n+1$ étant la somme des dimensions des espaces $\mc H^{(a)}$. De cette manière, le produit direct $\mc H$ est aussi un espace d'Einstein vérifiant \eqref{Einstein manifold}. Cela ouvre la possibilité d'avoir des flux de $p$-formes avec un plus large spectre concernant les valeurs possibles de $p$.

En effet, supposons que la théorie contienne une $p$-forme $H_{[p]}$. Comme nous avons vu, elle est définie par les deux formes de polarisation $\mc E$ et $\mc B$, de rang $p-2$ et $p$ respectivement. Si tous les espaces $\mc H^{(a)}$ ont la même dimension $d$, ainsi $Nd=n+1$, nous pouvons alors allumer un flux magnétique ou électrique $H_{[p]}$ sur chaque espace d'Einstein de $\mc H$ quand $p=d$ ou $p=d+2$ respectivement de la manière suivante.

Considérons d'abord le cas magnétique. Supposons que $\mc H$ soit le produit direct de $N$ variétés d'Einstein de dimension $p$, ainsi $n+1=Np$. Chacun de ces espaces $\mc H^{(a)}$ possède une forme volume $\hat\ep^{(a)}_{[p]}$. Le champ
\BE \mc B_{[p]} = q_m\sum_{a=1}^N\pm\hat\ep^{(a)} \label{Bprodmag}\EE
résout\footnote{Les équations contraignent la magnitude des flux à être égale sur chaque facteur du produit d'espace d'Einstein, cependant l'orientation relative est libre, d'où les signes arbitraires dans \eqref{Bprodmag}.} alors les équations \eqref{eqB} et \eqref{isotropy} tant que $p\geq2$ (nous avons besoin d'au moins deux indices dans les epsilons pour résoudre \eqref{isotropy}, autrement de simples termes en $\mc B_{[p]}$ brisent l'isotropie) et conduit à une solution en dimension $D=Np+2$ avec
\BE \ep=\frac{Nq_m^2}{2r^{(2-N)p}} \ ,\qquad P = \frac{2-N}2\frac{q_m^2}{r^{(2-N)p}} \ , \EE
et par conséquent
\BE V(r) = \kappa-\frac{r_0^n}{r^n}+\frac{r^2}{l^2}-\frac{q_m^2}{2p(Np-2p+1)}\frac{1}{r^{2(p-1)}} \ . \label{Vprodmag} \EE
Pour $N=1$, nous restaurons la solution magnétique précédente \eqref{magnetic p=n+1} où $p=n+1$. Notons que, quand $N\geq2$, la contribution de la $p$-forme au potentiel du trou noir change de signe et est toujours négative, ce qui signifie qu'il y a des solutions régulières de trous noirs même avec $r_0=0$. Cependant, dans ce cas, la décroissance de ce terme en fonction de $r$ est plus faible que celle du terme de masse et modifie alors la structure asymptotique de l'espace-temps. Ce terme se comporte comme un terme de masse effectif pour une dimension inférieure. 
\bs

Comme simple illustration de cette construction, considérons la théorie d'Einstein-Maxwell en dimension $D=6$. Nous avons par exemple des trous noirs avec un horizon $\mc S^2\times\mc S^2$ qui portent des flux magnétiques sur chaque sphère : 
\BE 
\dd s^2=-V(r) \dd t^2+\frac{\dd r^2}{V(r)}
+r^2\lp \dd\theta_1^2+\sin^2\theta_1\,\dd\phi_1^2\rp
+r^2\lp \dd\theta_2^2+\sin^2\theta_2\,\dd\phi_2^2\rp \ ,
\label{S2xS2m}\EE
\BE
H=q_m\lp\sin\theta_1\,\dd\theta_1\w\dd\phi_1\pm\sin\theta_2\,\dd\theta_2\w\dd\phi_2\rp \qquad\text{et}\qquad
V(r)=\frac13-\frac{r_0^3}{r^3}+\frac{r^2}{l^2}-\frac{q_m^2}{4r^2} \ .
\EE
Cette solution chargée magnétiquement est apparue pour la première fois dans \cite{Maeda:2010qz} en théorie d'Einstein-Gauss-Bonnet et en présence de corrections d'ordres supérieures pour le secteur de la matière.

Une solution légèrement plus compliquée avec une 3-forme libre dans un espace-temps de dimension 8 est donnée par une solution de trou noir avec un horizon $\mc S^3\times\mc S^3$ sous la forme : 
\BE
\dd s^2=-V \dd t^2+\frac{\dd r^2}{V}
+r^2\lp \dd\theta_1^2+\sin^2\theta_1\,\dd\phi_1^2+\cos^2\theta_1\,\dd\psi_1^2\rp
+r^2\lp \dd\theta_2^2+\sin^2\theta_2\,\dd\phi_2^2+\cos^2\theta_2\,\dd\psi_2^2\rp
\label{S3xS3m}\EE
avec
\BE
H=\frac{q_m}{2}\lp\sin2\theta_1\,\dd\theta_1\w\dd\phi_1\w\dd\psi_1
\pm\sin2\theta_2\,\dd\theta_2\w\dd\phi_2\w\dd\psi_2\rp
\quad\text{et}\quad V(r)=\frac25-\frac{r_0^5}{r^5}+\frac{r^2}{l^2}-\frac{q_m^2}{6r^4} \ .
\EE
L'espace $\mc H$ peut également être construit comme le produit d'espaces plats $\mathbb R^p$ ou hyperboliques $\mathbb H^p$ avec une modification évidente pour $H$ et $V$. Des exemples en dimensions supplémentaires peuvent facilement être construits à partir de la forme générale de la solution \eqref{Bprodmag} et \eqref{Vprodmag}.
\bs

Le cas électrique fonctionne de la même manière. Prenons  $\mc H$ comme le produit direct de $N$ espaces d'Einstein de dimension $(p-2)$ avec la même courbure et mettons un flux électrique $H_{[p]}$ de même magnitude sur chaque $\mc H^{(a)}$. Encore une fois, la condition d'isotropie \eqref{isotropy} impose $p\geq4$ et nous obtenons alors
\BE \mc E = q_e\sum_{a=1}^N\pm\hat\ep^{(a)}_{[p-2]}\ ,\qquad \ep=\frac{Nq_e^2}{2r^{(N-2)p-2N+4}} \EE
et
\BE V(r) = \kappa-\frac{r_0^n}{r^n}+\frac{r^2}{l^2}+\frac{q_e^2}{2(p-2)((N-2)(p-2)-1)}\frac1{r^{2(N-1)(p-2)-2}} \ .\EE
La contribution de la charge électrique dans le potentiel du trou noir $V$ est négative pour $N<3$ et positive sinon. Quand $N=1$, nous restaurons le cas $p=n+3$ donné par l'équation \eqref{electric p=n+3}. Pour $N=2$, ces solutions électriques sont duales aux solutions magnétiques précédentes \eqref{Vprodmag}, ces dernières ont un rang $p'=p-2$ pour la forme correspondante donnée par \eqref{Bprodmag}. En revanche, la version duale des solutions électriques avec $N\geq3$ va engendrer des flux magnétiques qui ne sont pas portés par un seul facteur du produit d'espaces d'Einstein, mais par un sous-ensemble de ces espaces. Ce qui nous amène à la classe de solutions suivante.

\subsubsection{Produit d'espaces d'Einstein avec des flux composites}
Si le flux de $\mc B$ ou $\mc E$ n'est pas porté par chaque $\mc H^{(a)}$ uniquement, mais par plusieurs facteurs, la construction se déroule de la même manière, même si la dimension de $\mc H$ n'est pas un multiple du rang de la forme. Soit $d\geq2$ la dimension d'un espace élémentaire $\mc H^{(a)}$ et $p$ le rang de $\mc B$ multiple de $d$ : $p=md$. Pour chaque choix de $m$ espaces élémentaires $\lp \mc H^{(a_1)}, \ldots ,\mc H^{(a_m)} \rp $, nous avons une $p$-forme volume donnée par $\hat\ep^{\{a\}}=\hat\ep^{(a_1)}\w\ldots\w\hat\ep^{(a_m)}$ et nous pouvons définir le flux magnétique par
\BE
\mc B^{\{a\}}_{i_1\ldots i_p}=q_m\frac{p!}{(d!)^m}\hat\ep^{(a_1)}_{[i_1\ldots i_d}
\hat\ep^{(a_{2})}_{i_{d+1}\ldots i_{2d}}\cdots
\hat\ep^{(a_m)}_{i_{p-d+1}\ldots i_{p}]} 
\EE
ou plus simplement par
\BE \mc B^{\{a\}} = q_m \hat\ep^{(a_1)}\w\ldots\w\hat\ep^{(a_m)} \ . \EE
Nous notons ici par $\{a\}=\{a_1,\ldots,a_m\}$ l'ensemble ordonné de $m$ entiers $1\leq a_1<\ldots< a_m\leq N$ qui consiste à choisir $m$ espaces élémentaires $\mc H^{(a)}$ parmi les $N$ disponibles. Un tel flux brise l'isotropie puisque $m$ espaces élémentaires sont privilégiés; néanmoins l'isotropie est aisément restaurée en sommant sur tous les choix possibles de $m$ espaces élémentaires parmi les $N$ qui sont disponibles avec une même magnitude, 
\BE
\mc B = \sum_{\{a\}} \mc B^{\{a\}}
= q_m \sum_{\{a\}}\pm\hat\ep^{(a_1)}\w\ldots\w\hat\ep^{(a_m)} \ .
\label{bbb}\EE
Ceci résout \eqref{isotropy} et nous avons ainsi trouvé une solution chargée magnétiquement. L'orientation de chaque flux dans la somme reste arbitraire, mais la magnitude doit être la même pour chaque flux. Un simple calcul de combinatoire montre que pour un tel flux magnétique, nous avons
\BE \mc B^2=\frac{p!N!}{m!(N-m)!}q_m^2 \ . \EE
Quant à la géométrie de ce trou noir chargé magnétiquement, elle est donnée par \eqref{ansatz} avec le potentiel (un terme logarithmique apparaît pour $2p=n+2$)
\BE V(r) =\kappa-\frac{r_0^n}{r^n}+\frac{r^2}{l^2}+\frac{N!}{2(n+1)(2p-n-2)m!(N-m)!}\frac{q_m^2}{r^{2p-2}} \ , \label{Vbbb}\EE
$\ka$ étant déterminé par la relation \eqref{Einstein manifold}.
Donnons un exemple simple de cette construction en dimension $D=8$ avec $\mc H=\mc S^2\times\mc S^2\times\mc S^2$ et une forme de rang $p=4$. D'après l'équation \eqref{bbb}, nous avons le champ suivant
\BE
H=q_m\lp\hat\ep^{(1)}\w\hat\ep^{(2)}
\pm\hat\ep^{(2)}\w\hat\ep^{(3)}
\pm\hat\ep^{(1)}\w\hat\ep^{(3)}
\rp \ . \EE
Puis, en choisissant un système de coordonnées tel que
\BE
\dd s^2=-V(r) \dd t^2+\frac{\dd r^2}{V(r)}+r^2\lp
\dd\theta_1^2+\sin^2\theta_1\,\dd\phi_1^2
+\dd\theta_2^2+\sin^2\theta_2\,\dd\phi_2^2
+\dd\theta_3^2+\sin^2\theta_3\,\dd\phi_3^2
\rp \ , \label{S3m}\EE
nous avons une solution avec
\begin{align}
&H=q_m\lp\sin\theta_1\sin\theta_2\,\dd\theta_1\w\dd\phi_1\w\dd\theta_2\w\dd\phi_2
\right.
&\quad\text{et}\qquad V(r)=\frac{1}{5}-\frac{r_0^5}{r^5}+\frac{r^2}{l^2}+\frac{q_m^2}{4r^6}\ .\qquad\qquad\nonumber\\
&\qquad\qquad
\pm\sin\theta_2\sin\theta_3\,\dd\theta_2\w\dd\phi_2\w\dd\theta_3\w\dd\phi_3\nonumber\\
&\qquad\qquad\qquad\left.
\pm\sin\theta_3\sin\theta_1\,\dd\theta_3\w\dd\phi_3\w\dd\theta_1\w\dd\phi_1
\rp \ ,
\label{S3H}\end{align}

\bs
L'extension de cette construction à des flux électriques est simple. Soit $H_{[p]}$ une $p$-forme purement électrique, elle est alors définie par la polarisation $\mc E$, qui est une forme de rang $(p-2)$ définie sur $\mc H$. Désormais, la dimension $d\geq2$ des espaces élémentaires $\mc H^{(a)}$ doit être un diviseur de $p-2$ : $p=md+2$. Encore une fois, pour chaque choix de $m$ espaces élémentaires $\lp \mc H^{(a_1)},\ldots,\mc H^{(a_m)} \rp$, nous pouvons définir un flux électrique 
\BE \mc E^{\{a\}} =q_e \hat\ep^{(a_1)} \w \ldots \w \hat\ep^{(a_m)} \EE
avec $\{a\}=\{a_1,\ldots,a_m\}$ définissant la sélection des espaces élémentaires $\mc H^{(a)}$ comme précédemment. Afin de restaurer l'isotropie, nous sommons sur tous les choix possibles de $m$ espaces élémentaires parmi les $N $ qui sont disponibles avec une même magnitude 
\BE
\mc E =\sum_{\{a\}} \mc E^{\{a\}}
=q_e\sum_{\{a\}}\pm \hat\ep^{(a_1)} \w \ldots \w \hat\ep^{(a_m)}  \ .
\label{eee}\EE
Ceci résout l'équation \eqref{isotropy} avec
\BE \mc E^2=\frac{(p-2)!N!}{m!(N-m)!}q_e^2 \EE
et produit une solution chargée électriquement dont la géométrie est donnée par la métrique \eqref{ansatz} avec le potentiel (un terme logarithmique apparaît pour $2p=n+4$)
\BE V(r) =\kappa-\frac{r_0^n}{r^n}+\frac{r^2}{l^2}+\frac{N!}{2(n+1)(n-2p+4)m!(N-m)!}\frac{q_e^2}{r^{2n-2p+4}} \ , \EE
$\ka$ étant déterminé par la relation \eqref{Einstein manifold} comme d'habitude. La solution duale, à celle chargée électriquement par une forme $H_{[p]}$ de rang $p=md+2$, est un trou noir chargé magnétiquement par une forme de rang $p'=D-p=m'd$ avec $m'=N-m$ et une charge $q_m=q_e$. C'est précisément la solution donnée par les équations \eqref{bbb} et \eqref{Vbbb}. Dans le cas particulier $m=N-1$, la solution duale se réduit à celle avec de simples flux donnée par \eqref{Bprodmag} et \eqref{Vprodmag}.

L'exemple précédent en dimension $D=8$ avec un espace d'Einstein de la forme $\mc H=\mc S^2\times\mc S^2\times\mc S^2$ peut alors être étendu au cas électrique avec une même 4-forme. Le trou noir qui en résulte en $D=8$ a pour métrique \eqref{S3m} avec
\BE V(r) =\frac15-\frac{r_0^5}{r^5}+\frac{r^2}{l^2}+\frac{q_e^2}{4r^6}  \EE
et un champ de 4-forme
\BE
H=\frac{q_e}{r^2}\,\dd t\w\dd r\w\lp\sin\theta_1\,\dd\theta_1\w\dd\phi_1
+\sin\theta_2\,\dd\theta_2\w\dd\phi_2
+\sin\theta_3\,\dd\theta_3\w\dd\phi_3\rp \ .
\EE
Il se trouve que dans ce cas, nous avons construit une solution à la fois électrique et magnétique à partir d'une même 4-forme, ceci permet d'entrevoir la possibilité d'obtenir des solutions dioniques que nous allons présenter dans le prochain paragraphe.
\bs

Une autre remarque importante est que la construction de solutions avec des flux composites sur des espaces de dimension $d=1$ ne fonctionne pas ici. A la section \ref{multiple}, nous verrons comment modifier cette construction en introduisant plus de champs indépendants, de telle sorte que toutes les équations du mouvement soient satisfaites quand $\mc H$ est le produit direct d'espaces unidimensionnels.

\subsubsection{Trous noirs dioniques sur des produits d'espaces d'Einstein bidimensionnels}
Dans la construction précédente, nous avons vu comment adapter des flux électriques et magnétiques sur le produit de $N$ espaces d'Einstein de dimension $d$. En particulier, les trous noirs qui en résultent sont chargés électriquement si $p-2$ est un multiple de $d$ et chargés magnétiquement si $p$ est un multiple de $d$. Par conséquent, quand $d=2$, nous pouvons allumer simultanément un flux magnétique et électrique issu d'une forme de rang $p$ pair si $N>p/2$ et le champ total est simplement donné par la somme des parties électriques et magnétiques\footnote{L'autre possibilité $d=1$ nécessite un horizon $\mc H$ plat et sera analysée dans \ref{multiple}.}. La solution de trou noir dionique est donc obtenue facilement en superposant les solutions du paragraphe précédent et, en définissant $m=p/2$, elle prend la forme \eqref{ansatz} avec
\BE
V(r)=\kappa-\frac{r_0^n}{r^n}+\frac{r^2}{l^2}+
\frac{N!}{2(n+1)m!(N-m)!}\lb
\frac{1}{(2p-n-2)}\frac{q_m^2}{r^{2p-2}}+\frac{1}{(n-2p+4)}\frac{q_e^2}{r^{2n-2p+4}}
\rb \ , \EE
et
\BE
H=q_m\sum_{\{a\}}\pm\hat\ep^{(a_1)}\w\ldots\w\hat\ep^{(a_m)}
+\frac{q_e}{r^{n-2p+5}}\sum_{\{a\}}\pm\dd t\w\dd r\w\hat\ep^{(a_1)}\w\ldots\w\hat\ep^{(a_{m-1})} \ .
\EE

Nous pouvons nous demander si ces solutions dioniques possèdent des propriétés d'auto-dualité ou d'anti-dualité quand $D=2p$. Dans ce cas, $N=p-1$ et la dimension de l'espace-temps doit être un multiple de 4 : $D=4m$. La solution précédente se simplifie alors pour donner
\BE V(r) = \kappa-\frac{r_0^n}{r^n}+\frac{r^2}{l^2}+\frac{(p-2)!}{m!(m-1)!}\frac{q_e^2+q_m^2}{4r^{n+1}} \ , \EE
\BE
H=q_m\sum_{\{a\}}\pm\hat\ep^{(a_1)}\w\ldots\w\hat\ep^{(a_m)}
+\frac{q_e}{r^{2}}\sum_{\{a\}}\pm\dd t\w\dd r\w\hat\ep^{(a_1)}\w\ldots\w\hat\ep^{(a_{m-1})} \ .
\EE
Nous pouvons alors facilement montrer qu'aucun choix de signes relatifs peut fournir des propriétés de dualité avec une signature lorentzienne. D'un autre côté, l'\ita{instanton} euclidien associé, obtenu par une double rotation de Wick du temps et de la charge électrique, est auto-dual à condition que les charges soient égales en valeur absolue (c'est-à-dire en prenant $q_e=-iq_m$ avec $q_m$ réel)
et que tous les signes soient égaux. Cette condition impose $q_e^2+q_m^2=0$ de telle façon que la fonction potentiel $V(r)$ coïncide avec celle du vide \eqref{vacuum}. Ces instantons possèdent la même géométrie que les solutions euclidiennes AdS du vide mais avec un champ $H$ non trivial. Ce dernier a donc un tenseur énergie-impulsion nul qui n'influence pas la métrique.
\bs

Explicitons avec un exemple sur $\mc H=\mc S^2\times\mc S^2\times\mc S^2$ et une 4-forme en superposant les deux exemples du paragraphe précédent en dimension 8. La métrique est donnée par \eqref{S3m} avec comme potentiel pour le trou noir
\BE V(r) = \frac15-\frac{r_0^5}{r^5}+\frac{r^2}{l^2}+\frac{q_e^2+q_m^2}{4r^6} \EE
et le champ total est donné par
\begin{align}
&\displaystyle H=\frac{q_e}{r^2}\sin\theta_1\,\dd t\w\dd r\w\dd\theta_1\w\dd\phi_1+q_m\sin\theta_2\sin\theta_3\,\dd\theta_2\w\dd\phi_2\w\dd\theta_3\w\dd\phi_3
\nonumber\\
&\displaystyle\qquad\qquad\quad
+\frac{q_e}{r^2}\sin\theta_2\,\dd t\w\dd r\w\dd\theta_2\w\dd\phi_2+q_m\sin\theta_3\sin\theta_1\,\dd\theta_3\w\dd\phi_3\w\dd\theta_1\w\dd\phi_1\nonumber\\
&\displaystyle\qquad\qquad\qquad\qquad\quad
+\frac{q_e}{r^2}\sin\theta_3\,\dd t\w\dd r\w\dd\theta_3\w\dd\phi_3+q_m\sin\theta_1\sin\theta_2\,\dd\theta_1\w\dd\phi_1\w\dd\theta_2\w\dd\phi_2 \ .
\end{align}
Après cela, nous pouvons prolonger analytiquement le temps et la charge électrique de la solution à des valeurs imaginaires pures et, en imposant $q_e=-iq_m$, nous obtenons ainsi une solution euclidienne avec une 4-forme réelle : 
\begin{align}
\dd s^2 &= V(r) \dd\tau^2 + \frac{\dd r^2}{V(r)}+r^2\lp
\dd\theta_1^2+\sin^2\theta_1\,\dd\phi_1^2
+\dd\theta_2^2+\sin^2\theta_2\,\dd\phi_2^2
+\dd\theta_3^2+\sin^2\theta_3\,\dd\phi_3^2
\rp,\nonumber\\
V(r) &= \frac15 - \frac{r_0^5}{r^5} + \frac{r^2}{l^2}
\label{Vsol}
\end{align}
avec la 4-forme suivante
\begin{align}
&\displaystyle H=\frac{q_m}{r^2}\sin\theta_1\,\dd \tau\w\dd r\w\dd\theta_1\w\dd\phi_1+q_m\sin\theta_2\sin\theta_3\,\dd\theta_2\w\dd\phi_2\w\dd\theta_3\w\dd\phi_3
\nonumber\\
&\displaystyle\qquad\qquad\quad
+\frac{q_m}{r^2}\sin\theta_2\,\dd\tau\w\dd r\w\dd\theta_2\w\dd\phi_2+q_m\sin\theta_3\sin\theta_1\,\dd\theta_3\w\dd\phi_3\w\dd\theta_1\w\dd\phi_1\nonumber\\
&\displaystyle\qquad\qquad\qquad\qquad\quad
+\frac{q_m}{r^2}\sin\theta_3\,\dd\tau\w\dd r\w\dd\theta_3\w\dd\phi_3+q_m\sin\theta_1\sin\theta_2\,\dd\theta_1\w\dd\phi_1\w\dd\theta_2\w\dd\phi_2 \ .
\end{align}

La topologie des sections euclidiennes est $\mathbb R^2\times\mc S^2\times\mc S^2\times\mc S^2$ avec ce qui s'appelle usuellement un "bolt" en anglais qui correspond à la plus grande racine de \eqref{Vsol}. La géométrie de la solution est la même que celle du vide, mais il y a en plus un champ réel $H$ vérifiant $\star H = H$. Dans le prochain paragraphe, nous allons poser notre regard avec plus d'attention sur de tels instantons euclidiens.

\subsubsection{Instantons auto-duaux pour $2p=n+3$}
La raison pour laquelle \eqref{dyonicisotropy} n'est pas résolue facilement en signature lorentzienne quand les parties électriques et magnétiques du champ sont allumées est due à la présence du signe moins devant le terme quadratique en $\mc E$. Il provient de la contraction dans la direction temporelle et est usuellement une obstruction pour obtenir des solutions dioniques lorentziennes avec un seul champ $H$. Cependant, si nous décidons de travailler avec une métrique de signature euclidienne
\BE \dd s^2 = V(r) \dd\tau^2+\frac{\dd r^2}{V(r)}+r^2\si_{ij}(y) \dd y^i \dd y^j \label{emetric}\EE
et avec un champ $H_{[p]}$ donné par \eqref{H}, alors le signe moins disparaît dans la première équation $\eqref{Hup}$ et la condition d'isotropie \eqref{dyonicisotropy} devient
\BE 
\mc B_{ik\ldots}\mc B_j{}^{k\ldots}-\frac{\mc B^2}{n+1}\si_{ij}=
-(p-1)(p-2)\lp\mc E_{ik\ldots}\mc E_j{}^{k\ldots}-\frac{\mc E^2}{n+1}\si_{ij}\rp \ .
\label{euciso}\EE
Nous avons déjà trouvé de tels instantons dans le paragraphe précédent lorsque $\mc H$ est le produit direct d'espaces bidimensionnels. Ici, nous obtenons de nouveaux instantons avec une construction différente.

Soit $\mc H$ le produit direct de deux espaces d'Einstein $\mc H^{(1)}$ et $\mc H^{(2)}$ de dimensions $p-2$ et $p$ respectivement tel que $\mc H$ soit un espace d'Einstein satisfaisant \eqref{Einstein manifold}. Ceci est possible si les métriques $\si^{(1)}_{ij}$ et $\si^{(2)}_{ij}$ et les tenseurs de Ricci $\mc R^{(1)}_{ij}$ et $\mc R^{(2)}_{ij}$ de $\mc H^{(1)}$ et $\mc H^{(2)}$
respectivement sont reliés par
\BE
\mc R^{(1)}_{ij}=n\kappa\si^{(1)}_{ij}\qquad\text{et}\qquad
\mc R^{(2)}_{ij}=n\kappa\si^{(2)}_{ij} \ .
\EE
Puis, en utilisant les formes volumes $\hat\ep^{(1)}$ et $\hat\ep^{(2)}$ sur $\mc H^{(1)}$ et $\mc H^{(2)}$, nous pouvons générer un flux électrique et magnétique en posant
\BE
\mc E=q_e\hat\ep^{(1)}\qquad\text{et}\qquad
\mc B=q_m\hat\ep^{(2)} \ .
\EE
Par construction, ces flux satisfont \eqref{eqE} et \eqref{eqB} et il est facile de vérifier que si $q_e=q_m=q$ alors \eqref{euciso} a lieu aussi. Un simple calcul montre également que le tenseur énergie-impulsion d'un tel champ est nul : $T_{\mu\nu}=0$. 

Le champ $H_{[p]}$ n'influence donc pas la métrique et la géométrie de ces instantons coïncident avec les sections euclidiennes des solutions AdS du vide avec des topologies non triviales : la métrique est donnée par \eqref{ansatz} avec le potentiel \eqref{vacuum}. Cependant, notons que lorsque la solution subit une rotation de Wick pour être exprimée dans la signature lorentzienne, la charge électrique devient purement imaginaire.
Finalement, il est facile de vérifier que si $p$ est pair et $q_e =q_m$ alors la solution est auto-duale, c'est-à-dire $\star H = H$.
\bs

Illustrons cette construction par un exemple en dimension 8 avec une $4$-forme et un espace transverse $\mc H=\mc S^2\times\mc S^4$ avec des rayons relatifs pour les sphères choisis de telle façon que $\mc H$ soit un espace d'Einstein. La solution est auto-duale et s'écrit
\BE  \dd s^2 = V(r)\dd\tau^2+\frac{\dd r^2}{V(r)}+r^2\dd\hat\Om^2_{(2)}+3r^2\dd\hat\Om^2_{(4)} \EE
avec
\BE V(r) = \frac15-\frac{r_0^5}{r^5}+\frac{r^2}{l^2}\qquad\text{et}\qquad H=\frac q{r^2}\dd\tau\w\dd r\w\hat\ep_{[2]}+9q\,\hat\ep_{[4]} \ , \EE
où $d\hat\Omega^2_{(2)}$ et $d\hat\Omega^2_{(4)}$ sont les métriques de la sphère unité $\mc S^2$ et de la sphère unité $\mc S^4$ respectivement avec $\hat\ep_{[2]}$ et $\hat\ep_{[4]}$ les formes volumes correspondantes. 
		\subsection{Trous noirs avec des horizons de type Einstein-Kähler\label{kähler}}
Si la dimension de l'espace-temps est paire, $n+1=2k$, et $\mc H$ est un espace de Kähler, il y a alors une 2-forme harmonique : la 2-forme de Kähler. Le lecteur pourra à cette occasion consulter la sous-section \ref{cartan} qui propose des rappels sur les variétés kählériennes. Pour tout $m\in\llbracket 1,k \rrbracket$, la $2m$-forme $\om^{(m)}$, obtenue en prenant le produit extérieur de $\om$ avec lui-même est harmonique aussi. A l'aide de la structure presque complexe définie sur $\mc H$, la 2-forme $\om$ vérifie la condition d'isotropie \eqref{isotropy}. Par conséquent, tous les $\om^{(m)}$ la vérifient également et nous avons ainsi $k$ formes isotropiques et harmoniques qui peuvent être utilisées pour construire de nouvelles solutions de la même façon que ce qui a été réalisé avec les formes volumes dans les paragraphes précédents. Notons que la forme de rang maximun $\om^{(k)}$ est proportionnelle à la forme volume.

\subsubsection{Trou noir magnétique de type Einstein-Kähler}
Considérons une forme de rang $p$ pair avec $2\leq p\leq 2k$. Ainsi une polarisation magnétique de la forme
\BE \mc B=q_m\om^{(p/2)} \label{KB} \EE
résout les équations \eqref{eqB} et \eqref{isotropy}. Par conséquent, la géométrie donnée par la métrique \eqref{ansatz}, avec le potentiel $V(r)$ obtenu à partir de \eqref{V} et \eqref{ep} où le champ $H$ \eqref{H} est celui qui correspond à la polarisation $\mc B$ mentionnée au-dessus, résout les équations du mouvement.

\subsubsection{Trou noir électrique de type Einstein-Kähler}
La même construction peut être menée dans le cas électrique. Considérons encore une fois une théorie avec une forme de rang $p$ pair mais avec  $4\leq p\leq 2k+2$. Nous pouvons alors choisir la polarisation électrique suivante
\BE  \mc E = q_e \om^{(p/2-1)}\label{KE}\EE
qui résout les équations  \eqref{eqB} et \eqref{isotropy}. Nous obtenons alors une solution avec la métrique donnée par \eqref{ansatz}, \eqref{V} et \eqref{ep}, et le champ $H$ \eqref{H} correspondant à la polarisation $\mc E$.

\subsubsection{Trou noir dionique de type Einstein-Kähler}
En superposant les solutions électriques et magnétiques précédentes, nous pouvons facilement construire des solutions dioniques chargées électriquement et magnétiquement sous une $p$-forme vérifiant $4\leq p\leq n+1$. 

\subsubsection{Produit direct d'espaces de type Einstein-Kähler} Dans les sous-sections précédentes, nous avons expliqué comment construire des solutions lorsque $\mc H$ est le produit d'espaces d'Einstein. La même procédure peut être menée avec le produit d'espaces de type Einstein-Kähler, en utilisant les nombreuses formes harmoniques qui vivent sur ce produit. Observons que le produit direct d'espace de Kähler est lui-même un espace de Kähler, dont la forme de Kähler est donnée par la somme des formes de Kähler de chaque espace. Par conséquent, pour un tel espace $\mc H$, des solutions existent avec des polarisations magnétiques et électriques données par \eqref{KB} et \eqref{KE} respectivement. Toutefois, si le groupe de cohomologie de $\mc H$ l'autorise, des flux plus généraux peuvent être construits. 

En effet, les formes isotropiques et harmoniques de chaque facteur de $\mc H$ permettent de généraliser la procédure que nous avons utilisé pour construire des solutions avec de simples flux sur des produits d'espaces d'Einstein. Tout ce que nous avons à faire est de construire les champs en utilisant les formes $\om^{(k)}$ sur chaque facteur au lieu des formes volumes. Très brièvement, voici comment cela se passe.

Soit $\mc H=\mc K^{(1)}\times\cdots\times\mc K^{(N)}$ un espace d'Einstein formé par le produit direct de $N$ espaces de Kähler $\mc K^{(a)}$, de dimension $d$ chacun, auquel est associé une 2-forme de Kähler $\hat\om^{(a)}$ et soit $H$ une forme de rang $p$ pair. Comme précédemment, nous construisons des formes harmoniques sur $\mc K^{(a)}$  de rang $2m$ notées $\hat\om^{(a,m)}$ définies comme le produit extérieur de $\hat\om^{(a)}$ avec lui-même $m$ fois. 

Si $2\leq p\leq d$, nous pouvons allumer sur $\mc H$ un flux magnétique de la forme
\BE  \mc B_{[p]}=q_m\sum_{a=1}^N\pm\hat\om^{(a,p/2)} \ . \label{KBprodmag}\EE
D'autre part, si $4\leq p\leq d+2$, nous obtenons la version électrique en prenant 
\BE \mc E_{[p-2]} = q_e\sum_{a=1}^N\pm\hat\om^{(a,p/2-1)} \ . \label{KEprodmag}\EE
Il est facile de montrer que ces polarisations sont des formes isotropiques sur $\mc H$ et que la métrique \eqref{ansatz} avec \eqref{V} est obtenue comme d'habitude.
\bs

Des solutions plus générales peuvent être obtenues en utilisant des flux composites, comme nous l'avons vu sur des produits d'espaces d'Einstein, en décomposant les formes de polarisations comme des produits extérieurs de formes harmoniques et en restaurant l'isotropie en sommant sur toutes les permutations possibles. Plutôt que de donner de lourdes formules, illustrons notre propos avec un exemple en dimension $D=10$ qui peut être facilement reproduit à d'autres dimensions. Supposons que $\mc H=\mc K^{(1)}\times\mc K^{(2)}$ soit le produit direct de deux espaces de type Einstein-Kähler de dimension 4 avec des 2-formes de Kähler $\hat\om^{(1)}$ et $\hat\om^{(2)}$ respectivement. Nous avons alors les possibilités suivantes pour les flux
\BE
\begin{array}{l@{\qquad}l}
p=2: &
\mc B =\hat\om^{(1)}\pm\hat\om^{(2)}\\
p=4: &
\mc B =\hat\om^{(1)}\w\,\hat\om^{(2)}\\
 &
\mc B =\hat\om^{(1)}\w\,\hat\om^{(1)}\pm\hat\om^{(2)}\w\,\hat\om^{(2)}\\
p=6: &
\mc B =\hat\om^{(1)}\w\,\hat\om^{(1)}\w\,\hat\om^{(2)}
\pm\hat\om^{(1)}\w\,\hat\om^{(2)}\w\,\hat\om^{(2)}\\
p=8: &
\mc B =\hat\om^{(1)}\w\,\hat\om^{(1)}\w\,\hat\om^{(2)}\w\,\hat\om^{(2)} \ .
\end{array}
\EE
Des combinaisons linéaires sont possibles dans le cas $p=4$, par exemple la solution \eqref{KB} construite à partir de la forme de Kähler de $\mc H$ en étant une
\BE \mc B=\lp\hat\om^{(1)}+\hat\om^{(2)}\rp\w\lp\hat\om^{(1)}+\hat\om^{(2)}\rp \ . \EE
De manière similaire, il est possible de construire des flux électriques.

Un second exemple en dimension $D=14$ avec $\mc H=\mc K^{(1)}\times\mc K^{(2)}\times\mc K^{(3)}$, où les trois facteurs sont des espaces d'Einstein-Kähler de dimension 4, présente les possibilités suivantes
\BE
\begin{array}{l@{\ }l}
p=2: &
\mc B =\hat\om^{(1)}+\hat\om^{(2)}+\hat\om^{(3)}\\
p=4: &
\mc B =\hat\om^{(1)}\w\,\hat\om^{(2)}+\hat\om^{(2)}\w\,\hat\om^{(3)}+\hat\om^{(3)}\w\,\hat\om^{(1)}\\
&
\mc B =\hat\om^{(1)}\w\,\hat\om^{(1)}+\hat\om^{(2)}\w\,\hat\om^{(2)}+\hat\om^{(3)}\w\,\hat\om^{(3)}\\
p=6: &
\mc B =\hat\om^{(1)}\w\,\hat\om^{(2)}\w\,\hat\om^{(3)}\\
&
\mc B =\hat\om^{(1)}\w\,\hat\om^{(1)}\w\,\hat\om^{(2)}
+\hat\om^{(2)}\w\,\hat\om^{(2)}\w\,\hat\om^{(3)}
+\hat\om^{(3)}\w\,\hat\om^{(3)}\w\,\hat\om^{(1)}\\
p=8: &
\mc B =\hat\om^{(1)}\w\,\hat\om^{(1)}\w\,\hat\om^{(2)}\w\,\hat\om^{(3)}
+\hat\om^{(2)}\w\,\hat\om^{(2)}\w\,\hat\om^{(3)}\w\,\hat\om^{(1)}
+\hat\om^{(3)}\w\,\hat\om^{(3)}\w\,\hat\om^{(1)}\w\,\hat\om^{(2)}
\\
&
\mc B =\hat\om^{(1)}\w\,\hat\om^{(1)}\w\,\hat\om^{(2)}\w\,\hat\om^{(2)}
+\hat\om^{(2)}\w\,\hat\om^{(2)}\w\,\hat\om^{(3)}\w\,\hat\om^{(3)}
+\hat\om^{(3)}\w\,\hat\om^{(3)}\w\,\hat\om^{(1)}\w\,\hat\om^{(1)}\\
p=10: &
\mc B =\hat\om^{(1)}\w\,\hat\om^{(1)}\w\,\hat\om^{(2)}\w\,\hat\om^{(2)}\w\,\hat\om^{(3)}
+\hat\om^{(2)}\w\,\hat\om^{(2)}\w\,\hat\om^{(3)}\w\,\hat\om^{(3)}\w\,\hat\om^{(1)}\\
&\qquad\qquad
+\hat\om^{(3)}\w\,\hat\om^{(3)}\w\,\hat\om^{(1)}\w\,\hat\om^{(1)}\w\,\hat\om^{(2)}
\\
p=12:&
\mc B =\hat\om^{(1)}\w\,\hat\om^{(1)}\w\,\hat\om^{(2)}\w\,\hat\om^{(2)}
\w\,\hat\om^{(3)}\w\,\hat\om^{(3)}\ .
\end{array}
\EE
La généralisation à des produits arbitraires d'espaces d'Einstein-Kähler est évidente. Le lecteur intéressé pourra construire des exemples plus explicites sur la métrique de Fubini-Study sur l'espace projectif $\mathbb C{\mathrm P}^n$ ou sur la métrique de Bergman \cite{Zoubos:2002cw,Bogdanos:2009pc}.
		
		\subsection{Trous noirs façonnés par plusieurs champs\label{multiple}}
Considérons une théorie avec plusieurs $p$-formes libres, avec même différents rangs. Le tenseur énergie-impulsion est désormais la somme des tenseurs énergie-impulsion de chaque champ et il peut être sous la forme \eqref{st} même si chaque champ brise l'isotropie. Chaque champ peut être décomposé, selon \eqref{H}, en une partie électrique et une autre magnétique qui résolvent indépendamment les équations \eqref{eqE} et \eqref{eqB}. Puis, une fois la condition d'isotropie vérifiée pour le tenseur énergie-impulsion total, la fonction $\ep(r)$ est simplement la somme des contributions issues de chaque champ et, par conséquent, le potentiel $V(r)$ reçoit simplement des contributions additives. Dans la suite, nous montrerons comment cela peut être achevé en polarisant les champs avec précision.
		
La manière la plus simple pour imposer la condition d'isotropie est que chaque champ vérifie l'équation \eqref{isotropy} ou \eqref{dyonicisotropy}, dans ce cas nous pouvons superposer trivialement les solutions issues de chaque champ.

\subsubsection{Principe de superposition : un exemple}
Considérons une théorie avec une 2-forme $F_{[2]}$ et une $(n+1)$-forme $H_{[n+1]}$. Si nous éteignons la 2-forme, nous pouvons construire une solution magnétique avec $p=n+1$; tandis qu'en l'absence du champ $H$, nous pouvons avoir une solution électrique. En superposant ces champs, nous obtenons une solution chargée électriquement par $F$ et magnétiquement par $H$ : 
\BE V(r) = \kappa-\frac{r_0^n}{r^n}+\frac{r^2}{l^2}+\frac{q_e^2+q_m^2}{2n(n+1)r^{2n}} \ , \EE
\BE
F = \frac{q_e}{r^{n+1}} \dd t \w \dd r \qquad\text{et}\qquad
H = q_m\hat\ep_{[n+1]}\ .
\EE
En particulier, puisque la polarisation électrique de la 2-forme n'a pas d'indice porté par $\mc H$, une telle charge électrique peut être ajoutée à toutes les solutions présentes dans cette section, bien que nous ne l'ayons pas indiquée explicitement.

Plus généralement, quand le rang de la $p$-forme ne correspond pas à la dimension de $\mc H$, cela introduit des directions privilégiées sur $\mc H$ et le tenseur énergie-impulsion n'est alors pas de la forme \eqref{st}. Cependant, il est possible de considérer de multiple copies de ce champ et de les orienter de telle façon que le tenseur énergie-impulsion total prenne la forme souhaitée, ce que nous allons montrer dans la suite. Cette construction devrait rester valide avec des espaces d'Einstein $\mc H$ courbes, à condition de déterminer toutes les formes harmoniques sur $\mc H$. Nous nous limiterons à analyser dans les paragraphes suivants des solutions avec des sections $\mc H$ plates.

\subsubsection{Isotropie issue de plusieurs champs sur $\mathbb R^{n+1}$ : le cas magnétique}
Supposons $1\leq p\leq n+1$ et soit $\{ e^a \}$ une base orthonormale de 1-forme sur $\mathbb R^{n+1}$. Chaque forme admet la décomposition $e^a=e^a{}_i\,\dd y^i$. Par définition, nous avons les relations utiles $\si^{ij}e^a{}_i e^b{}_j=\delta^{ab}$ et $\si_{ij}=\delta_{ab}e^a{}_i e^b{}_j$. Considérons maintenant la partie magnétique d'une $p$-forme $H_{[p]}$. Elle possède $p$ indices sur $\mc H$ et il n'est pas possible d'allumer ce flux sans briser l'isotropie. Cependant, nous pouvons construire un tenseur énergie-impulsion de la forme \eqref{st} à partir de
\BE N_b=\frac{(n+1)!}{p!(n-p+1)!} \EE
$p$-formes libres indépendantes, dont les indices sont distribués de telle sorte que le tenseur énergie-impulsion total, qui est la somme des tenseurs énergie-impulsion de chaque champ, recouvre l'isotropie. Voici comment se fait la construction. A chaque champ est associé un ensemble ordonné $\{a\}=\{a_1,\ldots,a_{p}\}$ d'entiers tels que $1\leq a_1<\ldots<a_{p}\leq n+1$. Ces entiers indiquent dans quelles directions vivent les indices du champ
\BE \mc B^{\{a_1,\ldots,a_{p}\}}_{i_1\ldots i_{p}}=q_mp! e^{a_1}{}_{[i_1}\cdots e^{a_p}{}_{i_p]} \label{ban}\EE
ou plus simplement 
\BE \mc B^{\{a\}}=q_m e^{a_1} \wedge \ldots \wedge e^{a_p} \ . \EE
Cette $p$-forme satisfait trivialement \eqref{eqB} avec une métrique induite plate sur $\mc H$. Puis, la contraction
\BE \mc B^{\{a\}}_{ik_2\ldots k_p}\mc B^{\{a\}}_{j}{}^{k_2\ldots k_p} = q_m^2(p-1)!\sum_{l=1}^p e^{a_l}{}_ie^{a_l}{}_j \EE
devient isotropique une fois que la somme est effectuée sur tous les champs possibles 
\BE \sum_{\{a\}}\mc B^{\{a\}}_{ik_2\ldots k_p}\mc B^{\{a\}}_{j}{}^{k_2\ldots k_p} = q_m^2\frac{n!}{(n-p+1)!}\si_{ij} \ . \EE
Ainsi, le tenseur énergie-impulsion vérifie la forme isotropique \eqref{st} avec
\BE
\ep(r)=\frac{(n+1)!}{2p!(n-p+1)!}\frac{q_m^2}{r^{2p-n-1}}\ ,\qquad
P(r)=\frac{n!(2p-n-1)}{2p!(n-p+1)!}\frac{q_m^2}{r^{2p-n-1}},
\EE
et le potentiel du trou noir devient (un terme logarithmique apparaît quand $2p= n+2$)
\BE V(r) = -\frac{r_0^n}{r^n}+\frac{r^2}{l^2}+\frac{n!}{2p!(n-p+1)!(2p-n-2)}
\frac{q_m^2}{r^{2p-2}} \ ,
\EE
puisque pour un espace $\mc H$ plat nous avons $\ka=0$. Lorsque $p=n+1$, nous avons un seul champ magnétique et nous retrouvons alors le résultat \eqref{magnetic p=n+1} dans le cas particulier $\ka=0$ et $\mc H=\mathbb R^{n+1}$.
\bs

Un cas particulier intéressant a lieu pour $p=1$. La solution décrite correspond alors à $n+1$ champs scalaires $\phi^{(i)}$, un champ pour chaque coordonnée de l'espace transverse plat, et dans ce cas la contribution magnétique dans le potentiel du trou noir ne dépend plus de la coordonnée radiale $r$. La solution est 
\BE
\dd s^2=-V(r) \dd t^2+\frac{\dd r^2}{V(r)}+r^2\sum_{i=1}^{n+1} \lp \dd y^i \rp^2  \ ,\qquad
V(r)=-\frac{q_m^2}{2n}-\frac{r_0^n}{r^n}+\frac{r^2}{l^2}\ ,\qquad
\phi^{(i)}=q_my^i\ ,
\label{magscalar}\EE
décrivant ainsi un trou noir AdS avec un horizon plat, mais avec un potentiel sous la forme usuellement associée à un trou noir possédant un horizon hyperbolique \cite{Birmingham:1998nr}, avec un terme de courbure effectif $\ka_{\rm eff}=-q_m^2/2n<0$. Les champs scalaires sont en fait définis à une constante près puisqu'ils rentrent dans l'action seulement sous forme de dérivées, cela correspond simplement à l'invariance de jauge des $p$-formes. Dans ce cas, nous pouvons compactifier l'horizon en un tore de dimension $n+1$ en utilisant cette invariance. Cela conduit à un trou noir asymptotiquement localement AdS avec un horizon toroïdal habillé par des champs scalaires. En particulier pour $D=4$, ces champs scalaires sont duaux à des axions et nous analyserons ces trous noirs axioniques dans la sous-section suivante. Passons au cas électrique.

\subsubsection{Isotropie issue de plusieurs champs sur $\mathbb R^{n+1}$ : le cas électrique} La même construction peut être développée pour le cas électrique, conduisant à une solution duale à la précédente. Supposons $3\leq p\leq n+3$ et un secteur de matière avec 
\BE N_e=\frac{(n+1)!}{(p-2)!(n-p+3)!} \EE
$p$-formes. A chaque champ est associé un ensemble ordonné $\{a\}=\{a_1,\ldots,a_{p-2}\}$ d'entiers tels que $1\leq a_1<\ldots<a_{p-2}\leq n+1$ comme précédemment. Nous pouvons alors définir les polarisations électriques par
\BE \mc E^{\{a\}}=q_e e^{a_1} \wedge \ldots \wedge e^{a_{p-2}} \ .\label{ean}\EE
Puis, la contraction
\BE
\mc E^{\{a\}}_{ik_2\ldots k_{p-2}}\mc E^{\{a\}}_{j}{}^{k_2\ldots k_{p-2}}=
q_e^2(p-3)!\sum_{l=1}^{p-2} e^{a_l}{}_i e^{a_l}{}_j
\EE
devient isotropique une fois que la somme est effectuée sur tous les champs possibles
\BE
\sum_{\{a\}}\mc E^{\{a\}}_{ik_2\ldots k_p}\mc E^{\{a\}}_{j}{}^{k_2\ldots k_p}=
q_e^2\frac{n!}{(n-p+3)!}\si_{ij} \ .
\EE
Ainsi, le tenseur énergie-impulsion vérifie la forme isotropique \eqref{st} avec
\BE
\ep(r)=\frac{(n+1)!}{2(p-2)!(n-p+3)!}\frac{q_e^2}{r^{n-2p+5}}\ ,\qquad
P(r)=\frac{n!(n-2p+5)}{2(p-2)!(n-p+3)!}\frac{q_e^2}{r^{n-2p+5}}\ ,
\EE
et le potentiel du trou noir devient (un terme logarithmique apparaît quand $2p= n+4$)
\BE
V(r)=-\frac{r_0^n}{r^n}+\frac{r^2}{l^2}+\frac{n!}{2(p-2)!(n-p+3)!(n-2p+4)}
\frac{q_e^2}{r^{2n-2p+4}} \ ,
\EE
puisque pour un espace $\mc H$ plat nous avons $\ka=0$. Lorsque $p=n+3$, nous avons un seul champ électrique et nous restaurons alors le résultat \eqref{electric p=n+3} dans le cas particulier $\ka=0$ et $\mc H=\mathbb R^{n+1}$. 
\bs

Finalement pour $p=n+2$, nous obtenons la solution duale au trou noir muni de $n+1$ champs scalaires magnétiques \eqref{magscalar} présentée au paragraphe précédent avec une métrique qui est déterminée par le potentiel
\BE V(r)=-\frac{q_e^2}{2n}-\frac{r_0^n}{r^n}+\frac{r^2}{l^2} \ .\label{axion1}\EE
Encore une fois, l'espace-temps qui en résulte définit un trou noir AdS avec un horizon plat mais avec un potentiel associé usuellement à un trou noir hyperbolique avec un terme de courbure effectif $\kappa_{\rm eff}=-q_e^2/2n<0$. 
Plus généralement, la dualité électromagnétique relie une solution avec une $p$-forme électrique $H_{[p]}^{\{a\}}$ construite à partir de la polarisation \eqref{ean} à une solution obtenue dans le paragraphe précédent avec une $p_\star$-forme magnétique $H_{[p_\star]}=\star H_{[p]}$ construite à partir de \eqref{ban} avec $q_e=q_m$ où $p_\star=n+3-p$. Il est facile de vérifier que le nombre de champs à utiliser coïncide, c'est-à-dire $N_e(n,p)=N_m(n,p_\star)$.

\subsubsection{Solutions dioniques sur $\mathbb R^{n+1}$ avec $3\leq p\leq n+1$}
Pour ces valeurs de $p$, nous avons à la fois obtenu des solutions magnétiques avec $N_b$ champs et des solutions électriques avec $N_e$ champs. En utilisant le principe de superposition, nous pouvons alors générer des solutions dioniques. Considérons par exemple $N_b+N_e$ $p$-formes, les $N_b$ premiers de la forme \eqref{ban} et les autres sous la forme \eqref{ean}. Nous avons par conséquent la solution
\begin{multline}
V(r)=-\frac{r_0^n}{r^n}+\frac{r^2}{l^2}+\frac{n!}{2p!(n-p+1)!(2p-n-2)}
\frac{q_m^2}{r^{2p-2}} \\
+\frac{n!}{2(p-2)!(n-p+3)!(n-2p+4)}
\frac{q_e^2}{r^{2n-2p+4}}\ .
\end{multline}
Un cas particulier intéressant se produit pour $2p=n+3$. Dans ce cas, le nombre $N$ de flux électriques et magnétiques nécessaires coïncide 
\BE N=N_b=N_e=\frac{(2p-2)!}{p!(p-2)!} \ ,\EE
et compte le nombre de façons de choisir $p-2$ indices parmi $2p-2=n+1$. Nous pouvons alors considérer $N$ $p$-formes avec à la fois une partie électrique et magnétique. Pour chaque choix de $p-2$ indices parmi les $n+1$, nous associons un champ dont les indices choisis déterminent la partie électrique et nous utilisons les $p$ indices restants pour la partie magnétique. La forme explicite de la polarisation, en notant toujours par des ensembles ordonnés $\{ a \}$ de $(p-2)$ entiers, est donnée par
\BE
\mc E^{\{a_1,\ldots,a_{p-2}\}} = q_e e^{a_1} \w \cdots \w e^{a_{p-2}} \quad\text{et}\quad
\mc B^{\{a_1,\ldots,a_{p-2}\}} = q_m\varepsilon^{a_1\ldots a_{p-2}b_1\ldots b_{p}} e^{b_1}\w \cdots \w e^{b_p}
\EE
où $\varepsilon^{a_1\ldots a_{n+1}}$ est totalement antisymétrique dans ces $(n+1)$ indices avec $\varepsilon^{1\ldots n+1} = 1$ et les sommes sur les indices $b_i$ sont sous-entendues.
L'espace-temps qui en résulte possède le potentiel
\BE V(r)=-\frac{r_0^n}{r^n}+\frac{r^2}{l^2}+\frac{N}{2(n+1)}\frac{q_e^2+q_m^2}{r^{n+1}} \EE
avec la $p$-forme suivante
\begin{align}
H^{\{a\}}
=& \frac1{r^2} \dd t \w \dd r \w \mc E^{\{a\}} + \mc B^{\{a\}} \\
=&  \frac{q_e}{r^2} \dd t \w \dd r \w e^{a_1} \w \ldots \w e^{a_{p-2}} + \frac{q_m}{p!}\varepsilon^{a_1\ldots a_{p-2}b_1\ldots b_p} e^{b_1} \w \ldots \w e^{b_p} \ .
\end{align}
Pour $p$ impair, la solution est anti-duale si $q_e=q_m$ et auto-duale si $q_e=- q_m$.
Notons que ces trous noirs généralisent la solution dionique de Reissner-Nordstrom-AdS en dimension 4. Cette dernière, donnée par \eqref{dyonicRN}, est en fait restaurée avec $n=1$ et $p=2$ dans les expressions précédentes.
\bs

Choisissons des coordonnées cartésiennes sur $\mc H$, tel que $\si_{ij}=\delta_{ij}$ ; nous pouvons ainsi choisir les tétrades sous la forme $E^{a}=\dd y^a$. Puis, illustrons notre propos avec le cas de la dimension 6 pour laquelle $n=p=3$. Nous avons besoin dans ce cas de quatre 3-formes. La solution dionique est alors donnée par
\BE \dd s^2=- V(r) \dd t^2 + \frac{\dd r^2}{V(r)} + r^2\sum_{i=1}^{4}(\dd y^i)^2\ ,\qquad V(r) = -\frac{r_0^3}{r^3}+\frac{r^2}{l^2}+\frac{q_e^2+q_m^2}{2r^{4}}\ ,\EE
avec les champs
\begin{align}
H^{\{1\}}=\frac{q_e}{r^2}\,\dd t\w\dd r\w\dd y^1+q_m\,\dd y^2\w\dd y^3\w\dd y^4,
H^{\{2\}}=\frac{q_e}{r^2}\,\dd t\w\dd r\w\dd y^2+q_m\,\dd y^3\w\dd y^4\w\dd y^1,\nonumber\\
H^{\{3\}}=\frac{q_e}{r^2}\,\dd t\w\dd r\w\dd y^3+q_m\,\dd y^4\w\dd y^1\w\dd y^2,
H^{\{4\}}=\frac{q_e}{r^2}\,\dd t\w\dd r\w\dd y^4+q_m\,\dd y^1\w\dd y^2\w\dd y^3,
\end{align}
qui est auto-duale si $q_e=-q_m$ et anti-duale si $q_e=q_m$. Cet espace-temps contient un trou noir avec des charges électriques et magnétiques issues de 3-formes. Cette solution partage une structure similaire au trou noir de dimension 6 de Reissner-Nordstrom, avec ici une décroissance plus lente du terme correspondant aux charges, en $r^{-4}$, alors que celle du terme correspondant à la charge électrique du trou noir de Reissner-Nordstrom est en $r^{-6}$. La solution est un trou noir avec un horizon interne et un horizon externe, localisé à la coordonnée radiale $r_h$, tant que $r_0^3\leq \frac{3(q_e^2+q_m^2)}{5r_h}$. Quant à l'inverse de la température, qui sera défini dans le chapitre \ref{thermo}, il vaut
\BE \beta = \frac{4 \pi}{|V'(r_h)|}=\frac{8 \pi l^2 r_h^5}{10 r_h^6-(q_e^2+q_m^2)\l^2} \ . \EE

\bs

Evidemment, cette procédure peut être généralisée à une superposition de formes avec un rang qui s'étale de $3$ à $n+1$. Ce qui permet de façonner un potentiel de trou noir avec une série de terme pair dans les puissances de $r$ allant de $r^2$ à $1/r^{2n}$ : 
\BE V(r) = -\frac{r_0^n}{r^n}+\frac{r^2}{l^2}+\sum_{m=1}^n\frac{c_m}{r^{2m}} \label{shaping}\EE
où les $c_m$ sont des constantes déterminées par les charges issues des $p$-formes. Des termes logarithmiques supplémentaires apparaissent également lorsque $2p=n+2$ ou $2p=n+4$.		

		\subsection{Trous noirs axioniques et leurs extensions\label{axionique}}

Dans cette section, nous allons revisiter le cas des trous noirs habillés par des 3-formes. Nous nous concentrerons au cas de la dimension $D=4$, qui correspond précisément aux trous noirs axioniques. C'est, à notre connaissance, le premier trou noir statique dans la littérature qui présente des charges axioniques non triviales. Les solutions des équations d'Einstein en présence d'un champ axionique ont été explorées pour la première fois dans \cite{Bowick:1988xh} et \cite{Xanthopoulos:1991mx}, mais dans la première référence la charge est nulle et dans la seconde les solutions sont singulières.
\bs

Considérons le trou noir toroïdal décrit par la métrique
\BE \dd s^2 = -V(r) \dd t^2+\frac{\dd r^2}{V(r)}+r^2(\dd x^2+\dd y^2) \label{tormet}\ ,\EE
avec le potentiel de trou noir
\BE V(r) = -\frac{q_e^2}{2} + \frac{r^2}{l^2}-\frac{r_0}{r} \label{axioniclapse} \ ,\EE
habillé par les deux 3-formes électriques suivantes
\BE H^{(1)}=q_e \dd t \wedge \dd r \wedge \dd x\qquad\text{et}\qquad H^{(2)}=q_e \dd t \wedge \dd r \wedge \dd y \ . \label{Hfields}\EE

Comme noté au-dessus \eqref{axion1}, la charge axionique $q_e$ est associée à un terme de courbure dans le potentiel du trou noir; elle peut également avoir une nature magnétique en étant générée par des champs scalaires \eqref{magscalar} qui sont duaux à des axions. Tout comme les trous noirs avec des horizons hyperboliques, nous devons considérer une constante cosmologique négative afin d'éviter la singularité située en $r=0$. Par conséquent, nous avons un trou noir asymptotiquement localement AdS avec un horizon plat et avec des charges axioniques qui fournissent des propriétés similaires au trou noir non chargé possédant un horizon hyperbolique.

L'inverse de la température est donné par 
\BE  \beta = \frac{4 \pi}{|V'(r_h)|} = \frac{4 \pi l^2 r_h}{3 r_h^2- \frac{1}{2} q_e^2 l^2} \label{conical}\EE
où nous avons remplacé le paramètre de masse $r_0 = r_h\lp\frac{r_h^2}{l^2}-\frac12 q_e^2 \rp $ par la plus grande racine du potentiel du trou noir, notée $r_h$, qui correspond à l'horizon des événements. Rappelons que la solution de Reissner-Nordstrom avec horizon sphérique ou plat conduit à un espace-temps singulier si nous éteignons le paramètre de masse. En fait, au-delà du cas extrémal, la solution de Reissner-Nordstrom est toujours singulière. L'effet des axions est ici très différent. En posant $r_0=0$, la solution axionique reste régulière avec un horizon situé à la coordonnée radiale 
$r_h=\frac{|q_e| l}{\sqrt{2}}$ et avec une singularité de courbure en $r=0$. Encore une fois, tout comme les trous noirs hyperboliques du vide, pour une masse négative, $r_0<0$, la solution présente un horizon interne et le trou noir extrémal est atteint pour le paramètre de masse
\BE r_0^{ext}=-\frac{|q_e|^3 l}{3 \sqrt{6}} \EE 
et avec un horizon des événements situé à la coordonnée radiale $r_{ext}=\frac{|q_e| l}{\sqrt{6}}$. Pour $r_0^{ext}<r_0<0$, nous avons donc un trou noir régulier avec un horizon interne et un horizon des événements. Ainsi, les champs axioniques permettent non seulement d'obtenir des trous noirs avec des faibles masses mais aussi des trous noirs avec un horizon plat et une masse négative, ce qui est usuellement associé à des horizons hyperboliques dans le vide d'AdS. En fait, plus la charge axionique est importante, plus l'espace-temps peut supporter une masse négative. Nous pouvons aussi aller un peu plus loin en ajoutant un champ de Maxwell qui génère une charge électrique $Q$ au trou noir axionique, le potentiel du trou noir est donc
\BE V(r) = -\frac{q_e^2}{2} + \frac{r^2}{l^2}-\frac{r_0}{r}+\frac{Q^2}{r^2}\ . \EE
Nous voyons immédiatement que l'ajout de la charge axionique au trou noir de Reissner-Nordstrom avec un horizon plat conduit à des solutions avec de plus faibles masses. Pour vérifier cela, nous pouvons même éteindre le paramètre de masse $r_0$. Le potentiel du trou noir est
\BE V(r) = -\frac{q_e^2}{2} + \frac{r^2}{l^2}+\frac{Q^2}{r^2} \EE
et cet espace-temps possède deux horizons, un horizon interne et un horizon des événements externe, aux coordonnées 
\BE r_h^2 = \frac14  q_e^2 l^2 \pm \frac14\sqrt{q_e^4 l^4-16Q^2 l^2}  \EE
tant que la charge axionique vérifie $q_e^2\geq \frac{4|Q|}{l}$. Il y a alors une solution extrémale pour $r_h=\frac{|q_e| l}{2}$. L'ajout de la masse ne change pas qualitativement les résultats. Les mêmes propriétés ont lieu en dimension quelconque tant que $p=n+2$, avec par exemple des trous noirs habillés par des champs de 4-formes en dimension 5. Pour résumer, les axions opèrent comme un terme de courbure négatif et ont la tendance de régulariser la géométrie de l'espace-temps.
\bs

Une autre solution intéressante est obtenue en effectuant un prolongement analytique de la solution \eqref{axioniclapse}. Considérons les transformations suivantes $t\rightarrow i \theta$, $x\rightarrow -i \tau$ et $q_e\rightarrow i q_s$. La métrique transformée a de nouveau des composantes réelles et 
\BE \dd s^2 = V(r)  \dd \theta^2 + \frac{\dd r^2}{V(r)}+r^2(-\dd\tau^2 + \dd y^2) \label{straight} \EE
où le potentiel est $V(r)=\frac{q_s^2}{2}+\frac{r^2}{\ell^2}-\frac{r_0}{r}$. Les valeurs de la coordonnée radiale sont limitées à $r \geq r_h$ et la métrique possède désormais le vecteur de Killing axial $\partial_\theta$ avec comme axe $r=r_h$. L'angle $\theta$ possède une périodicité arbitraire fournissant une singularité conique en $r=r_h$. Par une identification de cette période avec $\beta$ \eqref{conical}, nous pouvons supprimer la singularité conique; mais nous souhaitons ici tenir compte de la présence d'une corde cosmique. En effet, cette singularité est générée par la présence d'une corde cosmique dont la feuille d'univers bidimensionnelle vit en $r=r_h$, le lecteur pourra consulter \cite{vilenkin2000cosmic} pour plus de précisions sur ce sujet. La corde est produite par un tenseur énergie-impulsion distributionnel $T_\mu{}^\nu=-\delta^{(2)} T \delta_\mu{}^\nu$ avec une tension $T$ pour la corde donnée par $T=\frac{\Delta}{8\pi G}$ où le défaut conique vaut $\Delta=2\pi-\beta$. Les deux champs scalaires duaux aux champs axioniques peuvent se combiner en un seul champ scalaire complexe $\Phi=q_s(\tau+i y)$. De cette façon, nous avons trouvé une solution de type corde cosmique de l'action
\BE S=\frac{1}{16\pi G} \int  \sqrt{-g} \left( R - 2\Lambda - \frac{1}{2}\partial_\mu\Phi \partial^\mu\Phi^* \right) \dd^4 x \ . \EE
Notons que le terme axionique dans le potentiel du trou noir a changé de signe et indiquons également que, pour le cas $\La = 0$, la tension de la corde est donnée par $T=\frac1{4G} \lp 1 - \frac{8r_0}{q_s^4} \rp$.
\bs

 Finalement, nous souhaitons terminer avec une simple généralisation de ces trous noirs axioniques. En passant dans les coordonnées d'Eddington-Finkelstein, la solution avec les deux champs axioniques est 
\begin{align}
\dd s^2 &= -V(r)\dd u^2 - 2\dd r \dd u+r^2\lp \dd x^2 + \dd y^2\rp\ ,\\
H^{(1)} &= q_e \dd u \wedge \dd r \wedge\dd x\qquad\text{et}\qquad
H^{(2)}=q_e \dd u \wedge \dd r \wedge\dd y,\nonumber
\end{align}
avec le potentiel $V(r)$ toujours donné par \eqref{axioniclapse}. En supprimant l'hypothèse de stationnarité et en ajoutant une source de matière extérieure dont la seule composante non nulle du tenseur énergie-impulsion associé est $T^{\mathrm{ext}}_{uu}=\mu(u,r)/8\pi G$ et en autorisant le paramètre de masse à être une fonction de la coordonnée $u$ retardée (ou avancée), $r_0=r_0(u)$, nous obtenons une solution de type Vaidya \cite{Vaidya:1951zz} représentant un trou noir axionique avec une radiation à condition d'avoir la relation
\BE \mu(u,r) = -\frac{\p_ur_0}{r^2} \ .\EE
En dimension 4, la composante $T^{\mathrm{ext}}_{uu}$ du tenseur énergie-impulsion peut être générée par un champ de Maxwell $F_{[2]}$. 
		\subsection{Conclusion}
		
Dans ce chapitre, nous avons trouvé un grand nombre de trous noirs AdS vêtus de champs scalaires libres ou de champs de $p$-formes libres, dont certains survivent en présence d'une constante cosmologique positive $(\La \geq 0)$. En effet, quand la courbure de l'horizon est positive, ces solutions admettent une continuation à des trous noirs asymptotiquement localement plats ou de Sitter. Nous ne nous sommes pas restreints à un contenu particulier en champs 	pour la théorie dans notre analyse, mais il est important de souligner que ces constructions présentées ici peuvent naturellement être intégrées dans des théories de supergravité admettant un vide AdS, en considérant des compactifications adéquates à la Kaluza-Klein des théories de supergravités $D=10$ ou $D=11$ (voir par exemple \cite{Gauntlett:2007ma} et les références qui s'y trouvent) quand les champs de matière sont libres. De plus, nous prévoyons que ces solutions jouent un rôle dans la correspondance AdS/CFT, où les champs de matière déforment la CFT duale, pour décrire peut-être un système de matière condensée dans lequel les propriétés désirées sont obtenues en façonnant la fonction potentiel $V(r)$ par le contenu en champs \eqref{shaping} comme expliqué dans ce chapitre. Nous ne commentons pas plus cet aspect et nous laissons juste là des pistes d'investigations possibles pour le lecteur intéressé. Indiquons que l'analyse des charges et de la thermodynamique de ces nouvelles solutions sera détaillée dans le chapitre \ref{thermo}.

La géométrie des horizons des événements à courbures positives $(\ka >0)$ des trous noirs décrits dans les sous-sections \ref{single} et \ref{kähler} dévie de la métrique usuelle de la sphère de la solution de Schwarzschild-Tangherlini et de ses généralisations dans les espaces (A)dS. Dans le vide, ces trous noirs avec de tels horizons sont connus pour être classiquement instables \cite{Gibbons:2002th}. Il serait intéressant d'étudier comment la présence des champs de $p$-formes affecte cette instabilité afin de comprendre si ces champs externes peuvent fournir un mécanisme de stabilité. 

Nous avons également trouvé des trous noirs AdS plats \eqref{magscalar} vêtus de $D-2$ champs scalaires où chaque champ est associé à une direction indépendante de l'horizon. En plus de posséder le vecteur de Killing $\p_t$ qui génère les translations temporelles, ces solutions possèdent les symétries du plan euclidien de dimension $D-2$. Sous une isométrie de la métrique générée par un vecteur $\p_x$ par exemple, où $x$ est une coordonnée de l'horizon, la valeur des champs scalaires est déplacée d'une constante ; ce qui provient simplement d'un effet de jauge puisque les quantités physiques, c'est-à-dire les gradients des champs scalaires, restent inchangées au regard de l'action \eqref{SM}. Ainsi, nous pouvons compactifier l'horizon en un tore. Cependant, nous pouvons interpréter les champs scalaires différemment en abandonnant la compactification et en gardant un trou noir avec un horizon plat étendu. Pour cela, nous pouvons choisir les champs scalaires comme des quantités physiques en les couplant à un autre champ scalaire externe par exemple, qui s'annule sur notre classe de solutions. Ainsi, si $\xi$ désigne un vecteur de Killing qui génère les isométries de l'horizon, alors bien que $\pounds_\xi g_{\mu\nu} = 0$ nous avons $\pounds_\xi \phi^{(i)} \neq 0$ en général où $\phi^{(i)}$ désigne un champ scalaire, ce qui ne constitue plus une symétrie pour la solution. En ce sens, la seule symétrie résiduelle est celle de la translation temporelle générée par $\p_t$. En d'autres termes, le trou noir avec un horizon plat étendu vêtu par des champs scalaires \eqref{magscalar} est un exemple très simple de trou noir ne possédant qu'un seul vecteur de Killing en dimension quelconque dans le même esprit que \cite{Dias:2011at,Stotyn:2011ns}, cependant l'horizon est compact dans ces références.

Il serait intéressant de voir à quel point nous pouvons nous écarter de l'ansatz \eqref{ansatz}. En particulier, il est peut-être possible de mettre en rotation les trous noirs axioniques en dimension quatre, $\eqref{tormet}-\eqref{Hfields}$, en habillant les trous noirs topologiques en rotation de \cite{Klemm:1997ea} avec deux champs de 3-formes libres ou de manière équivalente par deux champs scalaires libres. Cet habillement devrait pouvoir s'étendre au cas des trous noirs Taub-NUT-AdS et au cas de la C-métrique dans AdS, ou même plus généralement aux géométries de type D décrites par la métrique de Pleba\'nski-Demia\'nski \cite{Plebanski:1976gy} qui contient toutes ces solutions comme des cas limites.

Il serait aussi d'un intérêt important de généraliser ces solutions à d'autres contenus de matière ou pour des théories différentes. Par exemple, en utilisant les techniques décrites dans ce chapitre, il est possible de construire en dimension quatre des trous noirs axioniques en théorie d'Einstein-Maxwell-AdS en présence d'un champ scalaire conformément couplé à la gravitation et avec deux champs axioniques, conduisant ainsi à une \textit{axionisation} des trous noirs de \cite{Martinez:2005di}. Ces trous noirs possèdent un cheveu secondaire et participent à des transitions de phase avec la solution de Reissner-Nordstrom comme nous le détaillerons au chapitre \ref{chapter-conforme}. Une autre extension de ce travail serait de considérer des potentiels non-triviaux pour les champs de $p$-formes par exemple.

Dans le chapitre suivant, nous allons considérer l'influence de termes de courbure d'ordre plus élevé en présence de champs de $p$-formes libres en dimension 6.

\chapter{Trous noirs en théorie d'Einstein-Gauss-Bonnet\label{EGB}}

	Dans ce chapitre, nous allons nous intéresser à une généralisation de la Relativité Générale en dimension $D\geq 4$ : la \ita{théorie de Lovelock}. Nous verrons que cette théorie préserve les grands principes de la Relativité Générale à travers le théorème de Lovelock. Nous présenterons notamment de nouveaux trous noirs \cite{Bardoux:2010sq} dans cette théorie en présence de $p$-formes libres dans le cas particulier de la dimension $D=6$ pour laquelle la théorie se réduit à celle d'\ita{Einstein-Gauss-Bonnet}.

\minitoc
	\section{La théorie de Lovelock \label{lovelock}}
		Afin de présenter cette théorie en évitant de lourdes formules, le langage adapté est celui des formes différentielles via les équations de structure de Cartan, que nous avons rencontrées à la sous-section \ref{cartan}. Pour une revue bien plus complète sur la théorie de Lovelock, nous conseillons \cite{Deruelle:2003ck,Garraffo:2008hu,Charmousis:2008kc,robin} et les références qui s'y trouvent. Commençons en présentant un résultat de D. Lovelock.
		
		\subsection{Le théorème de Lovelock}		
Nous avons vu qu'en dimension $D=4$, l'action d'Einstein-Hilbert en présence d'une constante cosmologique est la théorie la plus générale construite à partir de la métrique $g$ qui fournit des équations du mouvement sous la forme $D_{\mu\nu}[g]=T_{\mu\nu}$ où $D_{\mu\nu}[g]$ est un champ tensoriel qui ne dépend que de la métrique, de ses dérivées et de ses dérivées secondes au point considéré et tel que $\n_{\nu} D^{\mu\nu}=0$; ce qui constitue le résultat \ref{thlovelock}. D. Lovelock généralisa son théorème en dimension $D$ quelconque pour aboutir au résultat suivant \cite{Lovelock:1971yv,Lovelock:1972vz} :

\begin{proposition}[Lovelock] La théorie la plus générale en dimension D qui respecte les conditions précédentes est donnée par la densité lagrangienne
\BE \mc L = \sum_{k=0}^{\left[ \frac{D-1}{2} \right]} \al_k \mc L_{(k)} \label{lagrangien-lovelock}\EE 
où $\left[ \frac{D-1}{2} \right]$ désigne la partie entière de $\frac{D-1}{2}$, les $\al_k$ sont des constantes réelles et nous définissons les densités lagrangiennes d'ordre $k$ par :
\BE \mc L_{(0)}= e^* \qquad\text{et}\qquad \forall k>0 \ ,\mc L_{(k)} = \lp \bigwedge_{i=1}^k \Theta^{A_i B_i} \rp \w e^*_{A_1 B_1 \ldots A_k B_k} \ . \EE
\end{proposition}
Toutes les notations introduites ici sont celles de la sous-section \ref{cartan}. Tout d'abord, nous remarquons que la théorie d'Einstein en présence de la constante cosmologique correspond à la restriction de la théorie de Lovelock en dimension $D=4$. En effet, $\mc L_{(0)}$, qui est simplement la forme volume de l'espace-temps, fournit le terme de constante cosmologique et le terme $\mc L_{(1)}$ donne celui d'Einstein-Hilbert puisque
\begin{align}
\mc L_{(1)} 
&=  \Theta^{AB} \w e^*_{AB} = \frac12 R^{AB}_{\ \ \ CD} e^C e^D e^*_{AB}  \\
&=  \frac12 R^{AB}_{\ \ \ CD} e^C \lp \delta_B^D e^*_A - \delta_A^D e^*_B \rp \notag \\
&= \frac12 R^{A}_{\ \ C} \delta^C_A e^* + \frac12 R^{B}_{\ \ C} \delta^C_B e^* \notag \\
&= R e^* \label{L1}
\end{align}
où nous avons utilisé la relation \eqref{tetrad}. Puis, en dimension 5 ou 6, il apparaît un terme supplémentaire quadratique en la courbure qui, par un calcul similaire, donne le terme dit de \ita{Gauss-Bonnet} : 
\BE  \mc L_{(2)} = \Theta^{AB} \w \Theta^{CD} \w e^*_{ABCD} = \big( \underbrace{R^2 - 4 R_{AB}R^{AB} + R_{ABCD}R^{ABCD}}_{\dot{=}\hat{G}} \big)e^* \ . \label{GB} \EE
De même, en dimension 7 ou 8, il faut tenir compte d'un terme cubique en la courbure, etc... Précisons également que pour $k>D/2$, la densité lagrangienne d'ordre $k$ est évidemment nulle, $\mc L_{(k)}=0$. Nous voyons ici clairement l'efficacité du formalisme de Cartan pour formuler cette théorie \cite{Zumino:1985dp} ; précisons qu'il existe aussi une méthode diagrammatique pour déterminer ces densités lagrangiennes \cite{Bogdanos:2009yj}. Notons que nous n'avons pas inclus dans le lagrangien \eqref{lagrangien-lovelock} les termes du bord, analogues à celui de Gibbons-Hawking-York \eqref{GHY}, nécessaires pour dériver les équations du mouvement en fixant exclusivement la métrique sur le bord de la variété $\mc M$. Le lecteur pourra trouver ces termes de bord dans \cite{Myers:1987yn,MuellerHoissen:1989yv} qui sont d'une importance capitale pour étudier le formalisme hamiltonien \cite{Padilla:2003qi}, afin de déterminer l'énergie d'une solution, ou pour dériver les conditions de jonction d'Israel, que nous rencontrerons dans le cadre de la gravitation conformément couplée à un champ scalaire dans la sous-section \ref{jonctions}. Comme nous le verrons dans le cas de la gravité d'Einstein dans le chapitre \ref{thermo}, les termes de bords du formalisme hamiltonien permettent d'étudier la thermodynamique ; pour étudier ce domaine dans le cadre de la gravité de Lovelock, nous conseillons les références pionnières \cite{Myers:1988ze,Whitt:1988ax} et la référence \cite{Jacobson:1993xs}.
\bs 
 
Il y a eu un regain d'intérêt pour la théorie de Lovelock dans les années 80 dans le cadre de la théorie des cordes et notamment pour la théorie d'Einstein-Gauss-Bonnet. En effet, en considérant un espace-temps à faible courbure dans lequel les cordes se déplacent, il est possible de déterminer une action effective pour les modes de masse nulle pour chaque ordre en $\al'= \frac1{2\pi T}$, où $T$ est la tension de la corde. Ainsi, pour le secteur de la gravité, en plus du terme d'Einstein-Hilbert à l'ordre $0$, il y a un terme de la forme $R_{\mu\nu\rho\si}R^{\mu\nu\rho\si} + c_1 R_{\mu\nu}R^{\mu\nu} + c_2 R^2 $ qui apparaît à l'ordre 1 en $\al'$ \cite{Gross:1986mw}. Cependant, afin de décrire un graviton de masse nulle en théorie des perturbations autour de l'espace-temps plat, B. Zwiebach remarqua que les coefficients $c_1$ et $c_2$ doivent être fixés aux valeurs $-4$ et $1$ respectivement \cite{Zwiebach:1985uq}, ce qui correspond exactement au terme de Gauss-Bonnet $\hat{G}$.
\bs

Enfin, nous pouvons souligner que nous n'avons pas inclus le terme de Gauss-Bonnet en dimension 4 dans le lagrangien \eqref{lagrangien-lovelock} alors que celui-ci est bien défini. C'est ce que nous allons expliquer dans la suite en donnant notamment une interprétation géométrique de ces densités lagrangiennes.

		\subsection{Interprétation géométrique}
En effet, en dimension $D$ paire, la densité lagrangienne d'ordre $k=D/2$ n'est pas présente dans le lagrangien \eqref{lagrangien-lovelock} alors que $\mc L_{(D/2)}$ est bien défini. Cela provient du fait que ce terme ne modifie pas la dynamique puisque sa variation par rapport aux tétrades $e^A$ est une forme exacte
\BE \delta \mc L_{(D/2)} = \dd \lp \frac{D}{2}  \delta \om^{A_1 B_1} \w \Theta^{A_2 B_2} \w \ldots \w \Theta^{A_{D/2} B_{D/2}} \w e^*_{A_1 B_1 \ldots A_{D/2} B_{D/2}} \rp \ ,\EE
où nous avons utilisé les équations de structure de Cartan \eqref{cartan1} et \eqref{cartan2}. Ainsi, le terme $\int_\mc M \mc L_{(D/2)}$ fournit simplement un terme de bord. Cette intégrale possède en fait une interprétation topologique via le théorème de Chern-Gauss-Bonnet : 
\begin{proposition}[Chern-Gauss-Bonnet] 
Pour tout variété compacte, orientée, de dimension $D$ paire, nous avons le résultat suivant
\BE \chi\lp  \mc M \rp  = \frac1{(4\pi)^{D/2}(D/2)!} \int_\mc M \mc L_{(D/2)} \ ,\EE
où la \ita{caractéristique d'Euler} $\chi\lp  \mc M \rp$ est un invariant topologique définie par
\BE\chi\lp  \mc M \rp = \sum_{p=0}^n (-1)^p b_p \ ,\EE les $b_p$ étant les nombres de Betti rencontrés à la sous-section \ref{cartan}. 
\end{proposition}
Pour une démonstration de ce résultat, le lecteur pourra consulter \cite{robin,chern1,chern2}. La version bidimensionnelle, connue sous le nom de \ita{théorème de Gauss-Bonnet}, donne le résultat suivant
\BE \chi\lp  \mc M \rp  = \frac1{4\pi} \int_\mc M R \ep  = 2(1-g)  \ ,\EE
où $g$ désigne le \ita{genre} de la variété bidimensionnelle $\mc M$ supposée compacte et orientée, $g$ est simplement le nombre de anses de $\mc M$. Par exemple, nous avons $g=0$ pour la 2-sphère $S^2$, $g=1$ pour le 2-tore $T^2$, $g=2$ pour le double tore, etc... Si la variété présente un bord, alors le théorème de Chern-Gauss-Bonnet est modifié par un terme de bord supplémentaire \cite{robin}. Nous comprenons désormais pourquoi les $D$-formes $\mc L_{(k)}$ sont qualifiées de \ita{formes d'Euler continuées dimensionnellement}.

Indiquons qu'il y a ici une connexion avec la théorie des cordes. En effet, la théorie perturbative des cordes est définie comme la somme sur toutes les topologies possibles de la feuille d'univers décrite par la corde durant son mouvement. Pour chaque diagramme de Feynman, c'est-à-dire pour une topologie donnée, la contribution d'un tel terme est donnée par $e^{\la\chi}$, où $e^\la$ désigne la constante de couplage associée à un vertex dans le cas des cordes fermées par exemple.
\bs

Pour clore cette sous-section, retenons que ce théorème de Chern-Gauss-Bonnet nous permet de comprendre géométriquement pourquoi le terme de Gauss-Bonnet $\hat{G}$ ne participe pas à la dynamique gravitationnelle en dimension $D=4$, contrairement aux dimensions $D>4$. De même, le terme suivant, $\mc L_{(3)}$, est topologique en dimension $D=6$ mais il fournit une contribution à la dynamique en dimension $D>6$. Quant au terme d'Einstein-Hilbert qui est topologique en dimension $D=2$, il ne donne pas de dynamique gravitationnelle en dimension $D=3$! En effet, la dimension 3 est particulière puisque toute solution de l'équation d'Einstein sans constante cosmologique dans le vide en dimension 3 est de courbure nulle, $R_{\mu\nu\rho\si}=0$. Pour une présentation complète de ce domaine, \ita{la gravité $2+1$}, le lecteur pourra consulter l'ouvrage de référence \cite{carlip2003quantum} de S. Carlip.
		
		\subsection{Les équations du mouvement\label{EdM-lovelock}}
Il est temps de donner les équations du mouvement de la théorie de Lovelock décrite par l'action $ S=\int_\mc M \mc L $ avec le lagrangien $\mc L$ donné par \eqref{lagrangien-lovelock}. Pour cela, nous allons varier l'action $S$ par rapport aux tétrades $e^A$ en remarquant que
\BE \delta e^* = \delta e^A \w e^*_A \qquad\text{et}\qquad \forall k\in\llbracket1,D \llbracket \ , \delta e^*_{A_1 \ldots A_k} = \delta e^{A_{k+1}} \w e^*_{A_1 \ldots A_k A_{k+1}} \ .  \EE
Ainsi, nous trouvons la variation suivante
\begin{multline} 
\frac{\dd S}{\dd \la} = \al_0 \int_\mc M  \delta e^A \w e^*_A 
+ \sum_{k=1}^{\left[ \frac{D-1}{2} \right]} \al_k \int_\mc M \delta \lp \bigwedge_{i=1}^k \Theta^{A_i B_i} \rp \w e^*_{A_1 B_1 \ldots A_k B_k} \\
+ \sum_{k=1}^{\left[ \frac{D-1}{2} \right]} \al_k \int_\mc M \lp \bigwedge_{i=1}^k \Theta^{A_i B_i} \rp \w \delta e^{A} \w e^*_{A_1 B_1 \ldots A_k B_k A} \ .
\end{multline}
Puis, en utilisant les deux équations de structure de Cartan, \eqref{cartan1} et \eqref{cartan2}, il est facile de montrer que le second terme, à droite de l'égalité ci-dessus, est un terme de bord puisque
\BE \delta \lp \bigwedge_{i=1}^k \Theta^{A_i B_i} \rp \w e^*_{A_1 B_1 \ldots A_k B_k} = \dd \lp k \delta \om^{A_1 B_1} \w \Theta^{A_2 B_2} \w \ldots \w \Theta^{A_k B_k} \w e^*_{A_1 B_1 \ldots A_k B_k}  \rp \ .\EE
Par conséquent, en considérant le lagrangien complet $\mc L_\text{tot}$ de la théorie de Lovelock couplée à la matière, représentée par le lagrangien $\mc L_\text{m}$, avec la constante de couplage $\ka$,
\BE \mc L_\text{tot} = \sum_{k=0}^{\left[ \frac{D-1}{2} \right]} \al_k \mc L_{(k)} + 2 \ka \mc L_\text{m}\EE
et en écrivant les variations du secteur de la matière sous la forme $ \delta \mc L_\text{m} = \delta e^A \w \star T_A$, où nous avons posé $T_A = T_A^{\ B} e_B  $, nous déterminons les équations du mouvement suivantes
\BE \sum_{k=0}^{\left[ \frac{D-1}{2} \right]} \al_k \mc E_{(k)A} = - 2 \ka T_A^{\ B} e^*_B \label{EOM-lovelock}\EE
où les $(D-1)$-formes $\mc E_{(k)A}$ sont données par
\BE \mc E_{(0)A}  = e^*_A \qquad\text{et}\qquad \forall k>0 \ , \mc E_{(k)A} = \lp \bigwedge_{i=1}^k \Theta^{A_i B_i} \rp \w e^*_{A A_1 B_1 \ldots A_k B_k} \ .\EE
Pour revenir dans le langage des composantes, il suffit d'utiliser la relation \eqref{tetrad} et par un calcul similaire à celui de \eqref{L1}, nous trouvons
\BE \mc E_{(1)A} = -2 G_A^{\ B} e^*_B \EE
où $G_{AB}$ est le tenseur d'Einstein, ce qui justifie le choix du facteur 2 dans $\mc L_\text{m}$. De même, en écrivant le terme suivant sous la forme $\mc E_{(2)A} = 2 H_A^{\ B} e^*_B $, nous en déduisons le tenseur de \ita{Lanczos} : 
\BE H_{AB} = \frac12 g_{AB} \hat{G} - 2 R R_{AB} + 4 R_{AC} R^C_{\ B} + 4 R_{CD} R_{\ A\ B}^{C\ D} - 2R_{ACDE}R_B^{\ CDE} \ . \label{lanczos}\EE
En se limitant aux trois premiers termes et en posant $\La=-\al_0/2$, $\al_1=1$ et $\al_2=\al$, nous obtenons les équations de la théorie d'Einstein-Gauss-Bonnet
\BE  G_{AB} + \La g_{AB} - \al H_{AB} = \ka T_{AB} \ . \label{EOM-EGB} \EE

Insistons de nouveau sur le fait que ces équations sont du second ordre en la métrique. En effet, si nous considérons la théorie la plus générale faisant intervenir des termes quadratiques en la courbure, alors l'action prend la forme suivante
\BE S = \int_\mc M \sqrt{-g} \lp R - 2\La + \al R^2 + \be R_{\mu\nu}R^{\mu\nu} + \ga R_{\mu\nu\rho\si}R^{\mu\nu\rho\si} \rp \dd^Dx\EE
avec $\al,\be,\ga$ des constantes. Puis, la variation de cette action par rapport à la métrique fournit les équations du mouvement \cite{Garraffo:2008hu}
\begin{multline}
G_{\mu\nu} 
+ \La g_{\mu\nu} 
+ (\be+4\ga)\Box R_{\mu\nu} 
+ \frac12 (4\al+\be)g_{\mu\nu}\Box R
- (2\al+\be+2\ga) \n_\mu \n_\nu R \\
+ 2\ga R_{\mu\ga\al\be}R_\nu^{\ \ga\al\be} 
+ 2(\be+2\ga) R_{\mu\al\nu\be}R^{\al\be} 
- 4\ga R_{\mu\al} R_\nu^{\ \al} 
+ 2\al R R_{\mu\nu} \\
- \frac12 \lp \al R^2 
+ \be R_{\al\be}R^{\al\be}
+ \ga R_{\al\be\ga\de} R^{\al\be\ga\de} \rp g_{\mu\nu}
=0 \ .
\end{multline}
De manière générale, ces équations différentielles sont d'ordre 4 pour la métrique à cause des termes $\Box R_{\mu\nu} $, $\Box R$ et $\n_\mu \n_\nu R$. Cependant, le choix $\al=\ga=-\be/4$ permet de se limiter à des équations du second ordre et de restaurer ainsi la théorie d'Einstein-Gauss-Bonnet.
\bs

Dans la suite de ce chapitre, nous allons nous concentrer plus particulièrement sur la théorie d'Einstein-Gauss-Bonnet en présentant notamment des solutions de type trou noir dans le vide et en présence de matière. Nous préciserons également l'extension possible de ces résultats dans le cadre de la théorie de Lovelock.

	\section{Solutions du vide}
		
		\subsection{Solutions à courbure constante}
Commençons en présentant les solutions à courbure constante de la théorie d'Einstein-Gauss-Bonnet \eqref{EOM-EGB} dans le vide en dimension $D$. Ces solutions apparaissent pour la première fois dans \cite{Boulware:1985wk} et sont données par le tenseur de Riemann suivant
\BE R_{\mu\nu\rho\la} = \frac{2\La_e}{(D-1)(D-2)} \lp g_{\mu\rho} g_{\nu\la} - g_{\mu\la} g_{\nu\rho}  \rp \ . \label{vacuum1}\EE
La constante $\La_e$ n'est pas égale à la constante cosmologique $\La$ comme dans la théorie d'Einstein, mais elle peut prendre les deux valeurs suivantes 
\BE \La_e^\pm = 2 \La_{CS} \lp 1 \mp\sqrt{1-\frac{\La}{\La_{CS}}} \label{vacuum2} \rp \EE
si $\La \leq \La_{CS} $ où nous avons introduit la constante
\BE \La_{CS} = - \frac{(D-1)(D-2)}{8\al (D-3)(D-4)} \ . \label{vacuum3} \EE
Il est loisible aussi de réécrire ces solutions sous la forme habituelle	
\BE \dd s^2 = - V_\pm (r) \dd t^2 + \frac{\dd r^2}{V_\pm (r)} + r^2 \dd \Om^2_{D-2} \EE
où $\dd \Om^2_{D-2}$ désigne la métrique de la $(D-2)$-sphère unité et avec la fonction potentiel suivante
\BE V_\pm(r) = 1 - \frac{2\La_e^\pm r^2}{(D-1)(D-2)} \ . \EE

Tout d'abord, nous avons deux branches de solutions si $\La < \La_{CS}$, ce qui reflète le fait que l'équation du mouvement \eqref{EOM-EGB} fournit une équation quadratique en $V(r)$. La première branche, donnée par $\La_e^+$, restaure les solutions à courbure constante de la théorie d'Einstein quand le paramètre $\al$ tend vers zéro, puisque $\La^+_e \rightarrow \La$ dans ce cas : la branche $\La_e^+$ est alors qualifiée de \ita{branche d'Einstein}. En revanche, il existe une seconde branche de solutions paramétrisée par $\La_e^-$ et déconnectée des solutions à courbure constante de la théorie d'Einstein puisque, dans ce cas, $\La^-_e \rightarrow \infty$ quand $\al$ tend vers zéro ; nous parlons alors de \ita{branche de Gauss-Bonnet}.
Pour cette classe de solutions, il est intéressant de remarquer qu'en l'absence d'une constante cosmologique $(\La=0)$, les espace-temps (A)dS sont toujours solutions de l'équation \eqref{EOM-EGB} en fonction du signe de la constante $\al$ et avec une constante cosmologique effective $\La_e^-$ donnée par $4\La_{CS}$.
\bs

La stabilité de ces deux branches de solutions a été la source d'un débat raconté et tranché par C. Charmousis et al. dans \cite{Charmousis:2008ce}. Présentons succinctement le résultat de la stabilité linéaire autour des solutions à courbure constante. Pour cela, nous décomposons la métrique sous la forme $g_{\mu\nu}=\bar{g}_{\mu\nu}+h_{\mu\nu}$ où les quantités avec des barres font référence aux solutions à courbure constante \eqref{vacuum1}-\eqref{vacuum3}. Nous allons donner la variation à l'ordre linéaire en $\delta g_{\mu\nu}=h_{\mu\nu}$ de l'équation du mouvement \eqref{EOM-EGB} en utilisant notamment les formules de l'appendice \cite{Deser:2002jk} comme suggéré par B. Goutéraux dans \cite{Gouteraux:2010wh}. Après calculs, nous trouvons
\BE \delta \lp G_{\mu\nu} + \La g_{\mu\nu} - \al H_{\mu\nu} \rp = 
\pm \sqrt{1-\frac{\La}{\La_{CS}}} \delta \lp G_{\mu\nu} +\La_e g_{\mu\nu} \rp = \ka \delta T_{\mu\nu}
\label{stabilité}\EE
où $\delta G_{\mu\nu}$ désigne la variation linéaire du tenseur d'Einstein autour des solutions \eqref{vacuum1}-\eqref{vacuum3}, permettant ainsi de décrire un \ita{graviton} de spin 2\footnote{Plus explicitement, nous avons $\delta  G_{\mu\nu} = R^L_{\mu\nu} - \frac12 g_{\mu\nu} R^L - \frac{D\La_e}{D-2} h_{\mu\nu} $ avec la partie linéaire du tenseur de Ricci donnée par
$ R^L_{\mu\nu} = \frac12 \lp - \bar\Box h_{\mu\nu} - \bar\n_\mu \bar\n_\nu h + \bar\n^\si \bar\n_\nu h_{\si\mu} + \bar\n^\si \bar\n_\mu h_{\si\nu}  \rp $ 
où $h=\bar{g}^{\mu\nu} h_{\mu\nu}$ et $R^L$ désigne la partie linéaire du scalaire de Ricci.}. Par rapport à l'étude de la stabilité linéaire en théorie d'Einstein, la présence du tenseur de Lanczos dans \eqref{EOM-EGB} modifie simplement la valeur de la constante cosmologique et la valeur de la constante de couplage avec la matière, la théorie linéaire est donc toujours décrite par un graviton de spin 2. Pour la branche d'Einstein, cette constante de couplage est positive, cette branche est alors stable en ce sens. En revanche, pour celle de Gauss-Bonnet, la constante de couplage est négative et cette branche est instable. Nous renvoyons vers \cite{Charmousis:2008ce} pour plus de détails sur ce sujet.

\bs
Par ailleurs, si $\La=\La_{CS}$, il y a alors une seule branche de solutions. Pour cette relation particulière entre $\al$ et $\La$, la théorie d'Einstein-Gauss-Bonnet est en relation avec celle de \ita{Chern-Simons} en dimension $D$ impaire et avec celle de \ita{Born-Infeld} en dimension $D$ paire. Pour une revue de ces théories, invariantes sous le groupe d'anti-de Sitter, nous conseillons \cite{Banados:1993ur} et les références qui s'y trouvent. Concernant la théorie de perturbations autour de cette branche, la stabilité linéaire ne permet pas de conclure au regard de l'équation \eqref{stabilité}, nous renvoyons le lecteur vers \cite{Charmousis:2008ce} pour ce problème.

		\subsection{Solutions à symétrie sphérique}
Nous allons présenter brièvement dans cette sous-section des solutions statiques du vide de type trou noir et possédant notamment la symétrie sphérique. Ces solutions ont été découvertes pour la première fois par D. Boulware et S. Deser \cite{Boulware:1985wk} et indépendamment par J. Wheeler \cite{Wheeler:1985nh,Wheeler:1985qd}. 
En l'absence de constante cosmologique $\La$, ces géométries qui sont solutions de l'équation du mouvement \eqref{EOM-EGB} sont données par la métrique
\BE \dd s^2 = - V(r) \dd t^2 + \frac{\dd r^2}{V(r)} + r^2 \dd \Om^2_{D-2} \EE
avec la fonction potentiel suivante
\BE V(r) = 1 + \frac{r^2}{2\al (D-3)(D-4)} \lb 1 \mp \sqrt{1+\frac{8\al(D-3)(D-4)m}{r^{D-1}}} \rb \ . \EE
Le paramètre $m$ est ici relié à la masse de la solution par un facteur numérique \cite{Deser:2002jk}.
Dans la limite où le paramètre $\al$ tend vers $0$, la branche d'Einstein, correspondant au signe $-$ devant la racine carrée, restaure la solution de Tangherlini \cite{Tangherlini:1963bw} puisque
\BE V(r) = 1 - \frac{2m}{r^{D-3}} + \mc O(\al) \ ; \EE
alors que la second branche fournit une solution asymptotiquement (A)dS, en effet
\BE V(r) = \frac{r^2}{\al (D-3)(D-4)} + 1 + \frac{2m}{r^{D-3}} + \mc O (\al)  \ . \EE
Concernant la présence des horizons et des singularités, nous renvoyons le lecteur vers \cite{Garraffo:2008hu} pour le cas de la dimension 5 et vers \cite{Wiltshire:1988uq} pour les dimensions supérieures. Précisons qu'en plus de la singularité de courbure en $r=0$, il existe également une possible singularité dite de branche correspondant à la valeur de la coordonnée radiale pour laquelle la fonction dans la racine carrée s'annule.
\bs

Ces solutions statiques et possédant la symétrie sphérique peuvent être généralisées à la théorie de Lovelock \cite{Wheeler:1985nh,Wheeler:1985qd} et elles admettent également une extension pour la symétrique planaire et la symétrie hyperbolique \cite{Cai:2001dz,Cai:2003kt}. Ces géométries sont solutions de l'équation \eqref{EOM-lovelock} dans le vide et prennent la forme suivante
\BE \dd s^2 = - V(r) \dd t^2 + \frac{\dd r^2}{V(r)} + r^2 \sigma_{ij} \dd x^i \dd x^j \label{lovelock-1}\EE
où $\si_{ij}$ est la métrique de la sphère, du plan euclidien ou de l'espace hyperbolique pour la symétrie sphérique, planaire ou hyperbolique respectivement. Puis, en introduisant le polynôme suivant construit à partir des coefficients du lagrangien \eqref{lagrangien-lovelock}
\BE P[X]=\sum_{k=0}^{\left[ \frac{D-1}{2} \right]} \frac{\al_k}{(D-2k-1)!} X^k \ , \label{lovelock-2}\EE
le potentiel $V(r)$ de la solution doit vérifier
\BE P \lb \frac{k-V(r)}{r^2} \rb = \frac{\mu}{(D-1)! r^{D-1}} \label{lovelock-3}\EE
où $\mu$ est une constante d'intégration de nouveau reliée à la masse de la solution et $k$ correspond au signe du scalaire de Ricci déterminé à partir de la métrique $\si_{ij}$. Enfin, en introduisant $\La\lp \al_0, \ldots , \al_{\left[ (D-1)/2 \right]} \rp$ une racine réelle de $P[X]$, il est alors loisible de réécrire le potentiel sous la forme
\BE V(r)=k-r^2 \La\lp \al_0 - \frac{\mu}{r^{D-1}}, \al_1, \ldots , \al_{\left[ (D-1)/2 \right]} \rp \ . \EE
Pour une étude de la structure des horizons de ces trous noirs de Lovelock, nous conseillons \cite{Myers:1988ze}. Indiquons que la version chargée de ces solutions existent aussi. Les solutions de la théorie d'Einstein-Gauss-Bonnet en présence de l'interaction électromagnétique sont présentées dans \cite{Wiltshire:1985us} et l'extension à la théorie de Lovelock se trouve dans \cite{Myers:1988ze}. L'étude de ces trous noirs en présence de l'interaction de \ita{Born-Infeld}, qui est une théorie non-linéaire de l'électromagnétisme, existe aussi dans le cadre de la théorie de Gauss-Bonnet \cite{Wiltshire:1988uq,Aiello:2004rz}.
\bs

Pour clore cette sous-section, il nous semble important de signaler que le théorème de Birkhoff subsiste toujours en théorie de Lovelock : 
\begin{proposition}[théorème de Birkhoff]
Toute solution de l'équation générale du mouvement \eqref{EOM-lovelock} dans le vide à symétrie sphérique, planaire ou hyperbolique est localement isométrique à la solution \eqref{lovelock-1}-\eqref{lovelock-3} avec $k=1,0,-1$ respectivement.
\end{proposition}
Une preuve de ce résultat est d'abord apparue dans \cite{Wiltshire:1985us,Charmousis:2002rc} dans le cadre de la théorie d'Einstein-Gauss-Bonnet, puis le lecteur trouvera une élégante démonstration dans \cite{Zegers:2005vx} pour la théorie de Lovelock (voir aussi \cite{Deser:2005gr}). Précisons que pour un choix spécifique des coefficients $\al_k$, il est possible de contourner ce théorème \cite{Zegers:2005vx}, c'est pourquoi nous avons parlé d'équation \ita{générale} du mouvement dans l'énoncé de ce théorème. Le théorème de Birkhoff se généralise également en présence d'un champ de jauge abélien \cite{Zegers:2005vx}.

			\subsection{Solutions topologiques}
Passons à une classe de solutions statiques du vide plus générale. Dans le cadre de la théorie d'Einstein en dimension quelconque, nous avons vu, au chapitre précédent, qu'il existe des solutions statiques de type trou noir, s'écrivant sous la forme \eqref{lovelock-1}, dont la géométrie de l'horizon n'est pas nécessairement à courbure constante. En effet, la géométrie de l'horizon peut être plus généralement celle d'un espace d'Einstein. Dans le cadre de la théorie d'Einstein-Gauss-Bonnet, ce résultat n'est plus valable. G. Dotti et R. Gleiser ont montré \cite{Dotti:2005rc} que les métriques solutions de \eqref{EOM-EGB} dans le vide, qui s'écrivent sous la forme \eqref{lovelock-1} où $\si_{ij}$ désigne une métrique riemannienne arbitraire associée une variété transverse $\mc H$ de dimension $D-2$ et Einstein, doivent vérifier une contrainte supplémentaire sur le tenseur Weyl de $\mc H$ : 
\BE C_{iklm}C_j^{\ klm} = \frac{\Theta}{4}\sigma_{ij} \label{weyl-contrainte} \EE
avec le scalaire $\Theta \dot{=} C_{ijkl} C^{ijkl}$ constant. 
Ce résultat n'est pas surprenant puisque le tenseur de Riemann est entièrement mis en jeu dans les équations du mouvement dans la théorie d'Einstein-Gauss-Bonnet, alors que seul le tenseur de Ricci intervient dans la théorie d'Einstein. Précisons que le caractère Einstein de l'espace $\mc H$ implique la relation suivante entre le tenseur de Riemann intrinsèque de $\mc H$ et son tenseur de Weyl : 
\BE R_{ijkl} = C_{ijkl} + \frac{2R}{D(D-1)} \si_{i [ k} \si_{l ] j} . \label{einstein-weyl}\EE
Par conséquent, l'espace transverse $\mc H$ est nécessairement à courbure constante en dimension $D=5$\cite{Dotti:2007az} , puisque le tenseur de Weyl est nul en dimension 3, ou lorsque le paramètre $\Theta$ est nul. Indiquons que la relation \eqref{weyl-contrainte} est toujours satisfaite pour tout espace d'Einstein $\mc H$ de dimension 4 avec un scalaire $\Theta$ qui n'est pas nécessairement constant. Pour montrer cette relation, il suffit de remarquer que le tenseur de Lanczos \eqref{lanczos} est nul en dimension 4.
\bs

La thermodynamique de ces trous noirs topologiques en théorie d'Einstein-Gauss-Bonnet a également été étudiée par H. Maeda \cite{Maeda:2010bu}. Le lecteur trouvera également dans \cite{Bogdanos:2009pc} des exemples explicites d'espaces d'Einstein $\mc H$ en dimension $D=6$ comme le produit de 2-sphères $S^2 \times S^2$ et l'espace de Bergman. Dans la suite, nous allons considérer un ansatz sur la métrique qui généralise cette classe de solutions en tenant compte de la présence de champ de matière.

	\section{Trous noirs en présence de $p$-formes}
Nous détaillerons dans cette section les nouveaux résultats de \cite{Bardoux:2010sq} en suivant une présentation similaire à cette référence. Pour une large classe de géométries présentant une dépendance spatiale et temporelle, nous allons donner la solution générale de l'équation d'Einstein-Gauss-Bonnet en dimension 6 et en présence de champs de matière décrits par des $p$-formes libres. Pour cela, nous supposerons deux conditions sur les champs de matière permettant d'intégrer complètement les équations du mouvement. Puis, ces solutions seront classifiées et nous verrons que certaines limites de ces géométries restaurent des trous noirs connus. En particulier, nous montrerons que la gravitation de Lovelock restreint drastiquement la géométrie des horizons possibles tout en autorisant des sources de matière. En fait, si nous retenons exclusivement les solutions qui sont asymptotiquement plates alors il en existe une seule, parmi notre classe de géométries, qui est celle à symétrie sphérique avec une charge électromagnétique, rencontrée à la section précédente. En ce sens, la situation de la gravitation de Lovelock en dimension 6 est presque identique à celle de la Relativité Générale en dimension 4. 

Nous présenterons notamment des solutions statiques mettant en jeu des champs de 3-formes dans l'espace d'anti-de Sitter qui correspondent à la généralisation des solutions que nous avons rencontrées dans le chapitre précédent dans le cadre de la gravitation d'Einstein. Nous présenterons également des solutions de type corde noire et de type cosmologique.

		\subsection{Introduction}
		
Nous avons expliqué à la sous-section \ref{Birkhoff-généralisé} que le théorème de Birkhoff se généralise en théorie d'Einstein en dimension quelconque en considérant une plus large classe \eqref{warped} de géométries possibles pour l'horizon. En dimension $D=4$, la géométrie bidimensionnelle de l'horizon est nécessairement à courbure constante ; en revanche, en dimension $D \geq 6$, la géométrie intrinsèque de l'horizon est celle d'un espace d'Einstein \cite{Gibbons:2002th} comme nous avons pu le voir au chapitre précédent. Par exemple, en théorie d'Einstein dans le vide en dimension 6, il existe la solution suivante
\BE \dd s^2 = -V(r) \dd t^2 + \frac{\dd r^2}{V(r)} + r^2 \lb f(\rho)\dd\tau^2 + \frac{\dd\rho^2}{f(\rho)} + \rho^2 \lp \dd \theta^2 + \sin^2\theta\dd \phi^2 \rp \rb \label{euclidean} \EE
avec $f(\rho)=1-\frac{\mu}{\rho}$ le potentiel de la version euclidienne de la solution de Schwarzschild. La géométrie intrinsèque de l'horizon peut aussi être plus simplement celle d'un 4-tore avec la solution suivante
\BE  \dd s^2 = -V(r) \dd t^2 + \frac{\dd r^2}{V(r)} + r^2 \lp \dd w^2 + \dd x^2 + \dd y^2 + \dd z^2 \rp \ . \EE
Précisons que ces métriques partagent le même potentiel $V(r)=-\frac{\Lambda}{10} r^2-\frac{m}{r^3}$, ces deux solutions ne sont alors pas influencées par la géométrie intrinsèque de leur horizon en ce sens. D'autres exemples de type corde noire sont aussi obtenus en effectuant une double rotation de Wick sur la solution \eqref{euclidean} par exemple ou en ajoutant simplement $N$ "dimensions plates" à la métrique de Schwarzschild de dimension 4 :
\BE \dd s^2 = \sum_i \dd z_i^2 + \lb -f(\rho) \dd\tau^2 + \frac{\dd\rho^2}{f(\rho)} + \rho^2 \lp \dd \theta^2 + \sin^2\theta\dd \phi^2 \rp \rb \ . \label{euclidean2}\EE
Indiquons que toutes ces solutions sont accompagnées d'instabilités classiques décrites dans \cite{Gibbons:2002th} et \cite{Gregory:1993vy}.
\bs

Nous pouvons désormais nous demander quelle est la situation dans le cadre de la théorie de Lovelock. Est-ce que cette dégénérescence de la géométrie de l'horizon subsiste toujours en tenant compte en particulier du terme de Gauss-Bonnet ? Récemment, deux articles ont répondu à cette question. Il y a d'abord eu celui de G. Dotti et R. Gleiser \cite{Dotti:2005rc}, comme nous l'avons expliqué à la section précédente, puis la référence \cite{Bogdanos:2009pc} dans laquelle une plus large classe de solutions est exhibée. Le résultat important de ces études est que la dégénérescence est en partie levée lorsque nous tenons compte du terme de Gauss-Bonnet. Parmi ces solutions, de nouveaux trous noirs topologiques présentant un horizon de type Einstein et avec une contrainte sur le tenseur de Weyl de l'horizon \eqref{weyl-contrainte} sont solutions avec un potentiel $V(r)$ qui diffère du cas où l'horizon est simplement à courbure constante. 
\bs

Dans cette section, nous souhaitons répondre à la question suivante : comment la présence de matière modifie-t-elle la géométrie de ces solutions topologiques dans un espace-temps de dimension 6 en théorie de Lovelock ? Soulignons le fait que l'horizon doit être au moins de dimension 4 afin que son tenseur de Weyl soit non nul. En se basant sur l'intégrabilité du système trouvé dans \cite{Bogdanos:2009pc} et en imposant ainsi des conditions sur la matière, nous allons présenter la solution générale mettant en jeu un champ scalaire, un champ de jauge abélien et un champ de 3-forme (les formes de rang plus élevé sont quant à elles obtenues par une relation de dualité). Certaines nouvelles solutions constituent une généralisation des trous noirs façonnés par plusieurs champs rencontrés au chapitre précédent à la sous-section \ref{multiple}.
L'action de la matière mettra en jeu une $p$-forme exacte $\mc F$ libre par un terme $\int_\mc{M} \mc F \w \star \mc F$ pour lequel il existe $\mc A \in \La^{p-1} \lp \mc M \rp $ tel que $\mc F = \dd \mc A$. Puisque nous restreignons notre attention à la dimension $D=6$, le rang $p$ de la forme va de 1 à 6. Le cas $p=1$ correspond à un champ scalaire identifié au potentiel $\mc A$, $p=2$ correspond à l'interaction électromagnétique et $p=3$ à un champ axionique. Les formes de rang plus élevé sont reliées aux précédentes par une transformation de dualité, ce qui inclut le cas de la constante cosmologique pour $p=6$ en particulier.
\bs

Dans la prochaine sous-section, nous allons donner nos hypothèses et les équations du mouvement. Puis, nous détaillerons trois classes de solutions en décrivant le potentiel $V(r)$ et les contraintes sur la géométrie intrinsèque de l'horizon. Enfin, nous construirons quelques exemples spécifiques avant de conclure.

		\subsection{Ansatz sur la géométrie et les champs de matière}
Considérons la théorie d'Einstein-Gauss-Bonnet en dimension 6 en présence d'une $p$-forme exacte $\mc F = \frac1{p!} F_{A_1 \ldots A_p} \dd y^{A_1} \w \ldots \w \dd y^{A_p}$ donnée par
\BE S^{(6)} =  \int_\mc M  \sqrt{-g^{(6)}} \lb R - 2\La + \al \hat{G} - \frac{\ka}{p!} F_{A_1 \ldots A_p}F^{A_1 \ldots A_p} \rb \dd^6 x \label{action} \EE
avec $\ka$ le couplage entre la gravitation et la matière, et $\hat{G}$ le terme de Gauss-Bonnet \eqref{GB} : 
\BE \hat{G} = R^2 - 4R_{AB}R^{AB} + R_{ABCD}R^{ABCD} \ . \EE
Les indices en majuscule font ici référence aux coordonnées de l'espace-temps. Comme nous l'avons vu à la sous-section \ref{EdM-lovelock}, il est facile de dériver l'équation du mouvement en variant l'action $S^{(6)} $ par rapport à la métrique pour obtenir
\BE \mc E _{AB} = G_{AB} + \La g_{AB} - \al H_{AB} = \ka T_{AB} \label{EOM1} \EE
avec le tenseur énergie-impulsion
\BE T_{AB} = \frac1{(p-1)!} F_{A C_1 \ldots C_{p-1} } F_B^{\ C_1 \ldots C_{p-1}} - \frac1{2p!} g_{AB} F_{C_1 \ldots C_p} F^{C_1 \ldots C_p} \label{matter}  \EE
et nous redonnons l'expression du tenseur de Lanczos
\BE H_{AB} = \frac12 g_{AB} \hat{G} - 2 R R_{AB} + 4 R_{AC} R^C_{\ B} + 4 R_{CD} R_{\ A\ B}^{C\ D} - 2R_{ACDE}R_B^{\ CDE} \ . \EE
Par ailleurs, la variation de l'action $S^{(6)}$ par rapport au potentiel $\mc A$ donne $\dd \star \mc F=0$, c'est-à-dire
\BE \p_{A_p} \lp \sqrt{-g^{(6)}} F^{A_1 \ldots A_{p}} \rp = 0 \ . \label{EOM2} \EE
\bs

Puis, nous allons poursuivre en donnant l'ansatz pour la métrique et les champs de matière. Nous ferons la distinction entre l'espace bidimensionnel, qui porte la coordonnée temporelle $t$ et la coordonnée radiale $r$, et les sections transverses euclidiennes de dimension 4, notées $\mc H$, qui représenteront la géométrie intrinsèque de l'horizon d'un trou noir dans la suite. Nous supposerons que la variété transverse $\mc H$ est munie d'une métrique arbitraire $h_{\mu\nu}(x^\rho)$. Ainsi, en tentant de donner une version faible du théorème de Birkhoff similaire à la sous-section \ref{Birkhoff-généralisé}, nous choisissons la métrique sous la forme suivante
\BE \dd s^2 = e^{2\nu(t,z)} B(t,z)^{-3/4} \lp -\dd t^2 + \dd z^2 \rp + B(t,z)^{1/2} h_{\mu\nu}(x^\rho) \dd x^\mu \dd x^\nu \ .  \label{metric0} \EE
Les indices grecs font ici référence aux coordonnées intrinsèques de l'espace $\mc H$. Nous pouvons ensuite changer les coordonnées de l'espace bidimensionnel par 

\BE u = \frac{t-z}{\sqrt{2}} \qquad\text{et}\qquad v=\frac{t+z}{\sqrt{2}} \EE
pour réécrire la métrique sous la forme 
\BE \dd s^2 = - 2 e^{2\nu(u,v)} B(u,v)^{-3/4} \dd u\dd v  + B(u,v)^{1/2} h_{\mu\nu}(x^\rho) \dd x^\mu \dd x^\nu  \ .  \label{metric} \EE
Il est loisible aussi d'effectuer une double rotation de Wick pour la coordonnée temporelle $t=i\theta$ et pour une des coordonnées intrinsèques de $\mc H$ :
\BE\dd s^2 = e^{2\nu(\theta,z)} B(\theta,z)^{-3/4} \lp \dd \theta^2 + \dd z^2 \rp + B(\theta,z)^{1/2} h_{\mu\nu}(x) \dd x^\mu \dd x^\nu \label{bstrings} \EE
où $h_{\mu\nu}$ est désormais une métrique lorentzienne. Puis, nous pouvons retrouver une métrique analogue à \eqref{metric} par l'introduction des coordonnées complexes $u=\frac{-\theta+i z}{\sqrt{2}}$ et $v=\frac{\theta+i z}{\sqrt{2}}$. Dans la suite, nous donnerons la résolution à partir de la métrique \eqref{metric0} puisque les équations du mouvement se résolvent de la même façon à partir de la métrique \eqref{bstrings}. Nous donnerons néanmoins des exemples pour ces deux classes de métriques.

\bs
En utilisant la géométrie \eqref{metric}, nous sommes désormais en mesure d'écrire les composantes $(uu)$ et $(vv)$ du membre de gauche de \eqref{EOM1} : 
\begin{align}
\mc E_{uu}&=\frac{2 \nu_{,u} B_{,u} - B_{,uu}}{B} 
\left[ 1+\alpha \left( B^{-1/2} R^{(4)}+\frac{3}{2} e^{-2\nu} B^{-5/4} B_{,u} B_{,v} \right) \right] \ ,\label{uu} \\
\mc E_{vv}&=\frac{2 \nu_{,v} B_{,v} - B_{,vv}}{B} 
\left[ 1+\alpha \left( B^{-1/2} R^{(4)}+\frac{3}{2} e^{-2\nu} B^{-5/4} B_{,u} B_{,v} \right) \right] \label{vv} \ .
\end{align}
Comme nous allons le voir, la factorisation de ces équations est capitale pour l'intégrabilité du problème et l'obtention de solutions. En l'absence de champs de matière, il a été montré \cite{Bogdanos:2009pc} qu'il existe trois classes de solutions en considérant nul l'un des deux facteurs de \eqref{uu} et \eqref{vv} ou en considérant $B$ comme un champ scalaire constant sur $\mc M$. Notre seconde hypothèse de travail consiste à préserver $\mc E_{uu}=0$ et $\mc E_{vv}=0$ même en présence de matière. Pour le cas électromagnétique $(p=2)$, cela est très naturel et une version du théorème de Birkhoff peut être obtenue. Que se passe-t-il cependant pour une $p$-forme de manière générale ? En d'autres termes, si nous imposons $T_{uu}=0$ et $T_{vv}=0$ pour la métrique $\eqref{metric}$, obtenons-nous une hypothèse de travail raisonnable ?

A l'aide de l'expression \eqref{matter}, la contrainte $T_{uu}=0$ devient
\BE h^{\rho_1 \si_1} \cdots h^{\rho_{p-1} \si_{p-1}} F_{u \rho_1 \ldots \rho_{p-1}}  F_{u \si_1 \ldots \si_{p-1}} = 0  \quad (u \leftrightarrow v) \ . \EE
Puis, étant donné que $h_{\rho\sigma}$ est une métrique riemanienne, nous obtenons
\BE F_{u \si_1 \ldots \si_{p-1}} = 0 = F_{v \si_1 \ldots \si_{p-1}} \ . \label{IC} \EE
En particulier, pour un champ scalaire libre ($p=1$), cette notation signifie $F_u = F_v=0$. C'est-à-dire que nous devons considérer un champ scalaire, $\phi=\mc A$, qui dépend seulement des coordonnées intrinsèques $x^\mu$ de $\mc H$. Nous verrons et nous en avons eu un aperçu au chapitre précédent que ce cas fournit néanmoins des possibilités intéressantes. Pour une dépendance selon les coordonnées $(t,z)$ du champ scalaire, nous renvoyons le lecteur vers \cite{Charmousis:2001nq} dans le cadre de la gravitation d'Einstein en présence d'un dilaton avec un potentiel de Liouville.

Maintenant pour un rang $p$ arbitraire, si nous utilisons \eqref{EOM2} et le fait que la $p$-forme $\mc F$ soit fermée, alors  les composantes $F_{\si_1 \ldots \si_p}$ ne dépendent que des coordonnées $x^\mu$ et vérifient
\BE \p_{[\si_1} F_{\si_2 \ldots \si_{p+1}]} = 0 \qquad\text{et}\qquad \n_{\si_p}^{(4)} F^{\si_1 \ldots \si_p} = 0 \ . \label{yannis1} \EE
De plus, si $p \geq 2$, nous pouvons alors définir un nouveau tenseur
\BE J^{\si_1 \ldots \si_{p-2}} \dot{=} e^{-2\nu} B^{7/4} F^{\ \ \si_1 \ldots \si_{p-2}}_{uv} \EE 
qui encore une fois ne dépend que des coordonnées intrinsèques $x^\mu$ de $\mc H$ et vérifie
\BE \p_{[\si_1} J_{\si_2 \ldots \si_{p-1}]} = 0 \qquad\text{et}\qquad \n_{\si_{p-2}}^{(4)} J^{\si_1 \ldots \si_{p-2}} = 0 \ , \label{yannis2}\EE
où $ J_{\si_1 \ldots \sigma_{p-2}} \dot{=} h_{\si_1 \rho_1} \cdots h_{\si_{p-2} \rho_{p-2}} J^{\rho_1 \ldots \rho_{p-2}} $. Dans le langage des formes différentielles, nous avons le résultat suivant : 
\textit{étant donné les conditions d'intégrabilité $\eqref{uu}$ et $\eqref{vv}$ et la métrique \eqref{metric}, il existe $\mc F^{(4)} \in \La^p(\mc H)$ et $\mc J^{(4)} \in \La^{p-2}(\mc H)$ harmoniques où 
\BE \mc F^{(4)} \dot{=} \frac1{p!} F_{\si_1 \ldots \si_p } \dd x^{\si_1} \w \ldots \w \dd x^{\si_p} \qquad\text{et}\qquad
\mc J^{(4)} \dot{=} \frac1{(p-2)!} J_{\si_1 \ldots \si_{p-2}} \dd x^{\si_1} \w \ldots \w \dd x^{\si_{p-2}} \ . \EE
}

Les formes $\mc J^{(4)}$ et $\mc F^{(4)}$ définissent la polarisation électrique et magnétique de $\mc F$ respectivement comme nous avons pu le voir au chapitre précédent. $\mc J^{(4)}$ est une fonction constante dans le cas électromagnétique $(p=2)$ et $\mc J^{(4)} =0$ pour le cas d'un champ scalaire. Tournons nous vers les autres composantes de l'équation du mouvement \eqref{EOM1}.
\bs

La composante $(uv)$ du membre de gauche de \eqref{EOM1} est 
\begin{align}
	\mathcal{E}_{uv}
	= \frac{{B_{,uv} }}{B} - \Lambda e^{2\nu } B^{ - 3/4}  &+ \frac{\alpha }{2}e^{2\nu } B^{ - 7/4} \hat G^{(4)} 
	+ R^{(4)} \left[ {\frac{1}{2}e^{2\nu } B^{ - 5/4}  - \alpha B^{ - 3/2} \left( {\frac{1}{2}\frac{{B_{,u} B_{,v} }}{B} - B_{,uv} } \right)} \right] \notag \\
	&+  \alpha e^{ - 2\nu } B^{ - 5/4} \left[ { - \frac{{15}}{{16}}\left( {\frac{{B_{,u} B_{,v} }}{B}} \right)^2  + \frac{3}{2}\frac{{B_{,u} B_{,v} }}{B}B_{,uv} } \right]
\label{Euv}\end{align}
tandis que
\BE T_{uv} = \frac{e^{2\nu}}{2} \left[ \frac{B^\frac{2p-15}{4}}{(p-2)!}\left(J^{(4)}\right)^2 + \frac{B^\frac{-2p-3}{4}}{p!}\left(F^{(4)}\right)^2  \right]  \EE
où
$\lp J^{(4)}\rp ^2 \dot{=} h^{\si_1 \rho_1} \cdots h^{\si_{p-2} \rho_{p-2}} J_{\si_1 \ldots \si_{p-2}} J_{\rho_1 \ldots \rho_{p-2}}$ et
$\lp F^{(4)}\rp ^2 \dot{=} h^{\si_1 \rho_1} \cdots h^{\si_p \rho_p} F_{\si_1 \ldots \sigma_p} F_{\rho_1 \ldots \rho_p} \ . $ Quant aux composantes $(\mu\nu)$ du membre de gauche de \eqref{EOM1}, nous pouvons les écrire sous la forme
\begin{align}
	\mathcal{E}_{\mu \nu}
	&= G_{\mu \nu }^{(4)}  - e^{ - 2\nu } B^{1/4} \left( {\frac{3}{4}B_{,uv}  + 2B\nu _{,uv} } \right)h_{\mu \nu }  + \Lambda B^{1/2} h_{\mu \nu } \notag \\
	&+ \frac{3}{2}\alpha e^{ - 4\nu } \left( {B_{,uu}  - 2\nu _{,u} B_{,u} } \right)\left( {B_{,vv}  - 2\nu _{,v} B_{,v} } \right)h_{\mu \nu } \notag \\
	&-\alpha e^{ - 4\nu } \left[ {\frac{{45}}{{32}}\left( {\frac{{B_{,u} B_{,v} }}{B}} \right)^2  - \frac{{21}}{8}\frac{{B_{,u} B_{,v} }}{B}B_{,uv}  + \frac{3}{2}B_{,uv} ^2   + 3B_{,u} B_{,v} \nu _{,uv} } \right]h_{\mu \nu } \notag \\
	&+ 2 \alpha e^{ - 2\nu } B^{ - 1/4} \left( {\frac{3}{4}\frac{{B_{,u} B_{,v} }}{B} - \frac{1}{2}B_{,uv}  + 4B\nu _{,uv} } \right) G_{\mu \nu }^{(4)} 
\label{Emunu}\end{align}
tandis que
\BE T_{\mu\nu} = B^\frac{1-p}{2} T_{\mu\nu}\lp \mc F ^{(4)}\rp - B^\frac{p-5}{2} T_{\mu\nu}\lp \mc J^{(4)} \rp \label{TmunuII} \EE
où nous définissons
\begin{align}
 T_{\mu\nu}\left(\mathcal{F}^{(4)}\right) &= \frac{1}{(p-1)!} h^{\sigma_1 \rho_1} \cdots h^{\sigma_{p-1} \rho_{p-1}} F_{\mu\sigma_1 \ldots \sigma_{p-1}} F_{\nu\rho_1 \ldots \rho_{p-1}} - \frac{1}{2 p!}h_{\mu\nu} \left(F^{(4)}\right)^2 \label{TM} \\
 T_{\mu\nu}\left(\mathcal{J}^{(4)}\right) &= \frac{1}{(p-3)!} h^{\sigma_1 \rho_1} \cdots h^{\sigma_{p-3} \rho_{p-3}} J_{\mu\sigma_1 \ldots \sigma_{p-3}} J_{\nu\rho_1 \ldots \rho_{p-3}} - \frac{1}{2 (p-2)!}h_{\mu\nu} \left(J^{(4)}\right)^2 \ . \label{TE}
\end{align}
Notons que $R^{(4)}$, $G_{\mu\nu}^{(4)}$ et $\hat{G}^{(4)}$, qui caractérisent la géométrie de $\mc H$, apparaissent dans les équations du mouvement. Finalement, les composantes $(u\rho)$ et $(v\rho)$ de \eqref{EOM1} ne fournissent aucune information. \bs

Avant de résoudre les équations du mouvement, rappelons la dualité qui permet de nous concentrer exclusivement aux cas des 1, 2 ou 3-formes en dimension $D=6$. Considérons pour cela l'application suivante
\[ \sim
\left\{
	\begin{aligned}
	p   &\rightarrow \tilde{p}=6-p \\
	\mc F^{(4)}\in\La^p\lp \mc H \rp 	&\rightarrow  \star_{(4)}\tilde{\mc J}^{(4)}\in\La^{6-\tilde{p}}\lp\mc H\rp \\
	\mc J^{(4)}\in\La^{p-2}\lp \mc H \rp &\rightarrow  \star_{(4)}\tilde{\mc F }^{(4)}\in\La^{4-\tilde{p}}\lp\mc H\rp
	\end{aligned}
\right. \]
Cette fonction transforme une $p$-forme en une $(6-p)$-forme et $\star_{(4)}$ est le produit Hodge sur l'espace $\mc H$. Quand nous appliquons $\sim$ aux équations du mouvement, nous trouvons les mêmes équations du mouvement pour les quantités munies du $\sim$. Ainsi, toute solution en présence de $p$-forme génère une solution en présence d'une $(6-p)$-forme. Pour esquisser la façon dont cela se produit, commençons avec l'équation du mouvement de la $p$-forme $\mc F^{(4)}$
\begin{align}
\dd \mc F^{(4)} \stackrel{\sim}{\longrightarrow} \dd \star_{(4)}\tilde{\mc J}^{(4)} &= 0 \qquad\text{ie}\qquad \delta \tilde{\mc J}^{(4)} = 0 \\
\delta \mc F^{(4)} \stackrel{\sim}{\longrightarrow} \delta\star_{(4)}\tilde{\mc J}^{(4)} &= 0 \qquad\text{ie}\qquad \dd \tilde{\mc J}^{(4)} = 0 \ . \end{align}
Puis, de la même façon, nous avons ce résultat pour $\tilde{\mc F}^{(4)}$ en appliquant $\sim$ à $\mc J^{(4)}$. $\mc J^{(4)}$ et $\mc F^{(4)}$ ont simplement échangé leur rôle. Il est ensuite facile de vérifier que $T_{uv}$ se transforme ainsi
\BE T_{uv} \stackrel{\sim}{\longrightarrow} \frac{e^{2\nu}}{2} \left[ \frac{B^\frac{2\tilde{p}-15}{4}}{(\tilde{p}-2)!}\left(\tilde{J}^{(4)}\right)^2 + \frac{B^\frac{-2\tilde{p}-3}{4}}{\tilde{p}!}\left(\tilde{F}^{(4)}\right)^2  \right]\EE
étant donné que
\begin{align}
\frac{\left(J^{(4)}\right)^2}{(p-2)!} \star_{(4)}1 = \mathcal{J}^{(4)}\wedge\star_{(4)}\mathcal{J}^{(4)} \stackrel{\sim}{\longrightarrow}
\star_{(4)}\tilde{\mathcal{F}}^{(4)}\wedge\star_{(4)}\star_{(4)}\tilde{\mathcal{F}}^{(4)}
=&(-1)^{\tilde{p}(4-\tilde{p})}\star_{(4)}\tilde{\mathcal{F}}^{(4)}\wedge\tilde{\mathcal{F}}^{(4)} \notag \\
=&\tilde{\mathcal{F}}^{(4)}\wedge\star_{(4)}\tilde{\mathcal{F}}^{(4)} \notag \\
=& \frac{1}{\tilde{p}!}\left(\tilde{F}^{(4)}\right)^2 \star_{(4)}1
\end{align}
où nous avons utilisé le fait que $h_{\mu\nu}$ est une métrique riemannienne et $\star_{(4)}1 \dot{=} \sqrt{h}\dd x^1 \w \ldots \w \dd x^4$ est la forme volume sur $\lp \mc H,h \rp$. Finalement, il nous reste à étudier la transformation de $T_{\mu\nu}$ sous $\sim$ en utilisant \eqref{TmunuII}. En particulier, nous avons
\BE  F_{\mu \si_1 \ldots \si_{p-1}} \stackrel{\sim}{\longrightarrow} \left(\star_{(4)}\tilde{\mathcal{J}}^{(4)}\right)_{\mu\si_1 \ldots \si_{5-\tilde{p}}}
= \frac{1}{(\tilde{p}-2)!} \ep_{\mu\si_1 \ldots \si_{5-\tilde{p}}}^{\ \ \ \ \ \ \ \ \ \ \ \ \al_1 \ldots \al_{\tilde{p}-2}} \tilde{J}_{\al_1 \ldots \al_{\tilde{p}-2}} \ .
\EE
Puis, en introduisant le tenseur dualiseur $\ep_{\si_1 \ldots \si_4} = \sqrt{h} \varep_{\si_1 \ldots \si_4}$, l'identité suivante
\BE \ep^{\mu_1 \ldots \mu_m \si_1 \ldots \si_n} \ep_{\nu_1 \ldots \nu_m \si_1 \ldots \si_n} = n!m! \delta_{[\nu_1}^{\mu_1} \ldots \delta_{\nu_m]}^{\mu_m} \qquad\text{où}\qquad m+n=4\quad  \EE
permet de montrer que $T_{\mu\nu}$ se transforme en $\tilde{T}_{\mu\nu}$. De cette manière, l'étude pour les 5 et 4-formes se réduit à celle des 1 et 2-formes respectivement.
\bs

Dans les trois prochaines sous-sections, nous donnons une classification des solutions des équations du mouvement. Les équations $\mc E_{uu}=\mc E_{vv}=0$ fournissent trois classes de solutions : il y a deux classes de solutions (classe I et II) en fonction de l'annulation d'un des deux facteurs de \eqref{uu} et \eqref{vv} et une troisième classe (classe III) émerge lorsque la fonction $B$ de \eqref{metric} est constante. Les solutions de la classe I sont présentes pour un paramètre $\al$ non nul alors que les classes II et III se retrouvent aussi en gravité d'Einstein $(\al=0)$. 
		
		\subsection{Classe I}
Pour cette première classe de solutions, nous supposons que la fonction $B$ n'est pas constante et que le second facteur de \eqref{uu} et \eqref{vv} s'annule, ainsi
\BE  1 + \al {B^{ - 1/2} R^{(4)}  + \frac{3}{2} \al e^{ - 2\nu } B^{ - 5/4} B_{,u} B_{,v} }  = 0\ . \label{classI}\EE
Par conséquent, nous pouvons exprimer la fonction $\nu(u,v)$ en fonction de $B(u,v)$ de la façon suivante :
\BE \nu(u,v) = \frac{1}{2}\ln \lp -\frac{3\al }{2}\frac{{B_{,u}B_{,v}}}{B^{5/4}\lp 1+\al B^{-1/2}R^{(4)}\rp}\rp \ . \label{nu(u,v)}\EE
Notons que cette équation contraint le scalaire de Ricci $R^{(4)}$ de $\mc H$ à être constant. Puis, en substituant l'expression précédente pour $\nu(u,v)$ dans \eqref{Euv}, nous avons la contrainte
\BE B\lp 5+12\al\La\rp + \al^2\lb \lp R^{(4)}\rp^2 - 6 \hat{G}^{(4)} \rb + 6\al\ka \lb \frac{B^{\frac{p-4}{2}}}{(p-2)!} \lp J^{(4)}\rp ^2 + \frac{B^{\frac{2-p}{2}}}{p!} \left(F^{(4)}\right)^2 \rb = 0 \ .  \label{uvI} \EE
Ensuite, en prenant la trace de l'équation \eqref{Emunu} et en effectuant la même substitution, nous aboutissons à l'équation
\BE 5+12\al\La + 3\al\ka \left[ \frac{(p-4) B^{\frac{p-6}{2}}}{(p-2)!} \left(J^{(4)}\right)^2 + \frac{(2-p) B^{-\frac{p}{2}}}{p!} \left(F^{(4)}\right)^2 \right] =0 \ . \label{trace} \EE
Dans le cas $p=1$ et $p=3$, puisque $B$ n'est pas constant et $h_{\mu\nu}$ est une métrique riemannienne, nous montrons que $\mc F^{(4)}=0=\mc J^{(4)}$ à partir de \eqref{trace}. Nous obtenons aussi la limite dite de \ita{Born-Infeld} $5+12\al\La=0$ qui correspond à $\La=\La_{CS}$ où la constante $\La_{CS}$ a été introduite à la section précédente \eqref{vacuum3}. De plus, l'équation \eqref{uvI} fournit la condition géométrique $\hat{G}^{(4)}=\frac{1}{6}\lp R^{(4)}\rp^2$ et nous restaurons ainsi le cas de la gravitation dans le vide de \cite{Bogdanos:2009pc}. Nous en concluons donc que c'est impossible d'ajouter un champ scalaire ou une 3-forme dans la théorie pour cette classe de solutions.
\bs

Par conséquent, nous restreignons notre attention au cas $p=2$ (et $p=4$ par dualité) en analysant les équations $(\mu\nu)$ de \eqref{EOM1}. Pour cela, nous réécrivons l'équation \eqref{Emunu} à l'aide de sa trace
\BE \mc E_{\mu\nu} = \frac{1}{4} B^{1/2} \mc E h_{\mu\nu} + \lp R^{(4)}_{\mu\nu}-\frac{1}{4}R^{(4)}h_{\mu\nu} \rp
\left[ 1 + 2\alpha e^{-2\nu} B^{-1/4} \left(\frac{3}{4}\frac{{B_{,u} B_{,v} }}{B} - \frac{1}{2}B_{,uv} + 4B\nu_{,uv}\right)\right]
\label{munu} \EE
où $\mc E = g^{\mu\nu}\mc E_{\mu\nu}$. En fait, $\mc E =0$ puisque le tenseur énergie-impulsion \eqref{TM} est de trace nulle pour $p=2$. Ainsi, l'équation $(\mu\nu)$ se réduit à la forme factorisée suivante : 
\BE
\left(R^{(4)}_{\mu\nu}-\frac{1}{4}R^{(4)}h_{\mu\nu} \right)
\left[ 1 + 2\alpha e^{-2\nu} B^{-1/4} \left(\frac{3}{4}\frac{{B_{,u} B_{,v} }}{B} - \frac{1}{2}B_{,uv} + 4B\nu_{,uv}\right)\right]
=\kappa B^{-1/2} T_{\mu\nu}\left(\mathcal{F}^{(4)}\right) \ .
\label{munuI} \EE
Etant donné l'équation ci-dessus, nous avons trois branches de solutions dans cette classe I. Pour ces branches, la fonction $\nu(u,v)$ est donnée par \eqref{nu(u,v)}, $R^{(4)}$ est constant, nous devons nous placer à la limite de Born-Infeld et nous avons la condition géométrique suivante : 
\BE \left(R^{(4)}\right)^2 - 6 \hat{G}^{(4)} + \frac{3\kappa}{\alpha}\left(F^{(4)}\right)^2 = 0 \qquad\text{et}\qquad \mathcal{J}^{(4)}=0 \ , \label{scalar1}\EE
aucune charge électrique n'est donc possible pour cette classe. Les trois sous-classes de solutions sont les suivantes : 
\begin{itemize}
	\item[\textbullet] \textbf{Classe Ia} : Nous pouvons avoir $R^{(4)}_{\mu\nu}=\frac{1}{4}R^{(4)}h_{\mu\nu}$, c'est-à-dire que $\mc H$ est un espace d'Einstein et ainsi $T_{\mu\nu}=0$. Cette condition ne signifie pas qu'il n'existe pas de champ magnétique comme nous le verrons à la sous-section \ref{exemples} consacrée aux exemples. Pour cette sous-classe, la fonction $B(u,v)$ est arbitraire. Cette dégénérescence correspond à celle déjà présente dans \cite{Charmousis:2002rc}.

	\item[\textbullet] \textbf{Classe Ib} : Nous pouvons aussi avoir $T_{\mu\nu}=0$ avec cette fois-ci le second facteur du membre de gauche de \eqref{munuI} nul. En utilisant \eqref{nu(u,v)}, nous obtenons une équation différentielle du troisième ordre en $B(u,v)$ qui est
\begin{align}
&\left( {1 + \alpha B^{ - 1/2} R^{\left( 4 \right)} } \right)^2 \left( {B_{,u} ^2 B_{,vv} B_{,uv}  + B_{,v} ^2 B_{,uu} B_{,uv}  - B_{,u} ^2 B_{,v} B_{,uvv}  - B_{,v} ^2 B_{,u} B_{,uuv} } \right) \notag\\
&+ \frac{{B_{,uv} }}{B}B_{,u} ^2 B_{,v} ^2 \left[ {\frac{3}{2} + \frac{5}{2}\alpha B^{ - 1/2} R^{\left( 4 \right)}  + \left( {\alpha B^{ - 1/2} R^{\left( 4 \right)} } \right)^2 } \right] \notag\\
&- \frac{{B_{,u} ^3 B_{,v} ^3 }}{{B^2 }}\left[ {\frac{5}{4} + \frac{{17}}{8}\alpha B^{ - 1/2} R^{\left( 4 \right)}  + \frac{9}{8}\left( {\alpha B^{ - 1/2} R^{\left( 4 \right)} } \right)^2 } \right] = 0 \ .
\label{Bfield0}\end{align}
Cette équation fixe complètement la métrique. Nous pouvons trouver des solutions statiques en imposant par exemple $B_{,u}=-B_{,v}$ après quoi \eqref{Bfield0} devient une équation différentielle du second ordre en $B'$, où le signe $'$ désigne la dérivation par rapport à la variable $u-v$. Puis, en introduisant la coordonnée radiale $r=B^{1/4}$, nous pouvons écrire la métrique sous la forme
\BE \dd s^2=-U(r) \dd t^2 +24 \alpha \frac{\dd r^2}{\alpha R^{(4)}+r^2}+r^2 h_{\mu\nu} \dd  x^\mu \dd x^\nu \label{0} \EE
avec 
\BE U(r)=(r^2+\alpha R^{(4)})\left[C_1+C_2\left(\sqrt{\frac{|\alpha R^{(4)}|}{r^2+\alpha R^{(4)}}}- \text{arctanh} \sqrt{\frac{|\alpha R^{(4)}|}{r^2+\alpha R^{(4)}}} \right)\right]^2 \label{staticsol}\EE
où $C_1$ et $C_2$ sont des constantes d'intégration. Enfin, notons que pour le cas $\alpha R^{(4)}<0$, il est possible de construire des solutions possédant des horizons.

	\item[\textbullet] \textbf{Classe Ic} : Il existe une constante $\la\neq 0$ telle que
\BE \la\left(R^{(4)}_{\mu\nu}-\frac{1}{4}R^{(4)}h_{\mu\nu}\right) = \ka T_{\mu\nu}\left(F^{(4)}\right) \qquad\text{pour}\qquad p=2 \label{cl1} \EE
avec $B$ une solution de l'équation différentielle
\BE 1 + 2\al e^{-2\nu} B^{-1/4} \left(\frac{3}{4}\frac{{B_{,u} B_{,v} }}{B} - \frac{1}{2}B_{,uv} + 4B\nu_{,uv}\right) = \la B^{-1/2} \label{Bfield} \ . \EE
\end{itemize}
L'équation \eqref{Bfield} est simplement l'équation \eqref{Bfield0} avec un terme supplémentaire. Nous pouvons aussi trouver des solutions statiques pour ce cas ; cependant, les expressions sont plus compliquées et vont au delà de l'objectif de ce chapitre.
\bs

Dans la sous-section \ref{exemples}, nous construirons des solutions magnétiques pour cette classe I en considérant $\mc H = S^2 \times S^2$. Retenons que pour cette classe I, le caractère statique n'est pas imposé contrairement à la classe de solutions de la sous-section suivante.

		\subsection{Classe II}
			
			\subsubsection{Solutions localement statiques}
Les solutions de cette classe II sont obtenues par l'annulation du premier facteur de \eqref{uu} et \eqref{vv} : 
\BE  2\nu_{,u}B_{,u}-B_{,uu} = 0 \qquad (u \leftrightarrow v) \ . \label{int.cond.}\EE
Ces conditions d'intégrabilité se retrouvent également dans le cas de la gravitation d'Einstein \cite{Bowcock:2000cq}. Nous supposons aussi que la fonction $B$ n'est pas constante. L'équation \eqref{int.cond.} implique
\BE e^{2\nu} = B_{,u} f(v) = B_{,v} g(u) \EE
avec $f$ et $g$ deux fonctions arbitraires. Ce qui conduit à $B=B(U+V)$ après le changement de coordonnées
\BE (u,v) \longrightarrow \lp U(u)= \int_0^u g(\tilde{u})\dd\tilde{u} \ ,  V(v)=\int_0^v f(\tilde{v})\dd\tilde{v} \rp \ . \EE
La métrique devient alors
\BE \mathrm{d}s^2=- 2 B'(U+V) B^{-3/4} \mathrm{d}U\mathrm{d}V + B^{1/2} h_{\mu\nu} \mathrm{d}x^\mu \mathrm{d}x^\nu \ . \EE
Enfin, en effectuant successivement les changements de coordonnées suivants
\BE (U,V) \longrightarrow (\bar{z}=U+V,  \bar{t}=V-U) \longrightarrow (\bar{t},B(\bar{z})) \longrightarrow \left(\bar{t}=t/2,r=B^{1/4}\right) \label{bhmetric} \EE
et en posant
\BE V(r) \dot{=} -\frac{B'(\bar{z})}{8 r^3} \label{clapton} \ ,\EE
la métrique prend la forme usuelle
\BE \mathrm{d}s^2= -V(r)\mathrm{d}t^2 + \frac{\mathrm{d}r^2}{V(r)} + r^2 h_{\mu\nu} \mathrm{d}x^\mu  \mathrm{d}x^\nu \ . \EE
Par conséquent, les géométries de cette classe II sont localement statiques puisque $\p_t$ est un vecteur de Killing de genre-temps lorsque $V>0$. Ce sont les équations $(uu)$ et $(vv)$ qui ont permis de déterminer le caractère statique et ceci reste vrai tant que $T_{uu}=T_{vv}=0$. Nous allons dans la suite résoudre les équations restantes et caractériser en détail les champs de matière pour chaque cas.			
			
			\subsubsection{Equations pour tout $p$}
Nous pouvons désormais déterminer la fonction $B$ à partir de l'équation $(uv)$ \eqref{Euv} : 
\begin{align}
B'' - \Lambda B^{1/4}B' + \frac{\alpha}{2}B^{-3/4}B'\hat{G}^{(4)} + R^{(4)}\left[\frac{1}{2}B^{-1/4}B'+2\alpha\left(B^{1/2}\right)''\right] \notag \\
+ \frac{3\alpha}{4}\left(B^{-5/4}B'^2\right)' -\frac{\kappa}{2}\left[ \frac{B^{\frac{2p-11}{4}}B'}{(p-2)!} \left(J^{(4)}\right)^2 + \frac{B^{\frac{1-2p}{4}}B'}{p!} \left(F^{(4)}\right)^2 \right] = 0
\end{align}
où le signe $'$ correspond à la dérivation par rapport à la variable $\bar{z}$. Puis, nous intégrons l'équation ci-dessus par rapport à $\bar{z}$ ; il existe ainsi une fonction $h$, qui dépend seulement des coordonnées intrinsèques de $\mc H$, telle que 
\begin{align}
B'- \frac{4\Lambda}{5}B^{5/4} + 2\alpha\hat{G}^{(4)}B^{1/4} + 2R^{(4)}\left[\frac{1}{3}B^{3/4}+\alpha\left(B^{1/2}\right)'\right] + \frac{3\alpha}{4}B^{-5/4}B'^2 \notag\\
+ \frac{2\kappa}{(7-2p)(p-2)!}B^{\frac{2p-7}{4}}\left(J^{(4)}\right)^2 + \frac{2\kappa}{(2p-5)p!}B^{\frac{5-2p}{4}}\left(F^{(4)}\right)^2 = h(x^\mu) \ .
\label{uvII}
\end{align}
Puisque $B=B(\bar{z})$, l'équation \eqref{uvII} impose des contraintes sur la géométrie de l'horizon $\mc H$. De plus, \eqref{uvII} conduit à une équation du second degré en $B'$ qui nous permettra de déterminer le potentiel $V$ à l'aide de \eqref{clapton}. Avant cela, regardons les équations $(\mu\nu)$ restantes. En utilisant \eqref{munu}, nous obtenons 
\begin{align} \label{munuII}
\left(R^{(4)}_{\mu\nu} -  \frac{1}{4}R^{(4)}h_{\mu\nu}\right) 
\left[1+\frac{4\alpha B^{1/4}}{B'}\left(B^{1/2}\frac{U'}{U}\right)'\right] 
=& \kappa B^{\frac{1-p}{2}} \left[ T_{\mu\nu}\left(\mathcal{F}^{(4)}\right) -\frac{1}{4} T^{(4)}\left(\mathcal{F}^{(4)}\right) h_{\mu\nu} \right]  \\ \notag
&- \kappa B^{\frac{p-5}{2}} \left[ T_{\mu\nu}\left(\mathcal{J}^{(4)}\right) -\frac{1}{4} T^{(4)}\left(\mathcal{J}^{(4)}\right) h_{\mu\nu} \right] 
 \end{align}
où $U(z) = -\frac{B'(\bar{z})}{8 B^{3/4}(\bar{z})}$, $T^{(4)}\left(\mathcal{F}^{(4)}\right) = h^{\rho\sigma}T_{\rho\sigma}\left(\mathcal{F}^{(4)}\right)$ et de même pour $\mathcal{J}^{(4)}$. Les tenseurs du membre de droite sont donnés par
\BE T_{\mu\nu}\left(\mc F^{(4)}\right) -\frac{1}{4} T^{(4)}\left(\mc F^{(4)}\right) = \frac{1}{(p-1)!}\left[ h^{\si_1 \rho_1} \cdots h^{\si_{p-1} \rho_{p-1}} F_{\mu\si_1 \ldots \si_{p-1}} F_{\nu\rho_1 \ldots \rho_{p-1}} - \frac{1}{4}h_{\mu\nu} \left(F^{(4)}\right)^2 \right] \EE
et
\BE T_{\mu\nu}\left(\mc J^{(4)}\right) -\frac{1}{4} T^{(4)}\left(\mc J^{(4)}\right) = \frac{1}{(p-3)!}\left[ h^{\si_1 \rho_1} \cdots h^{\si_{p-3} \rho_{p-3}} J_{\mu\si_1 \ldots \si_{p-3}} J_{\nu\rho_1 \ldots \rho_{p-3}} - \frac{1}{4}h_{\mu\nu} \left(J^{(4)}\right)^2 \right] \ .\EE
Ils sont de trace nulle tout comme le premier facteur du membre de gauche. La trace des équations $(\mu\nu)$ de \eqref{EOM1} peut être dérivée à partir de l'équation \eqref{uvII} via l'identité de Bianchi. Finalement, toute solution de \eqref{uvII}, \eqref{munuII} et des équations du secteur de la matière \eqref{yannis1} et \eqref{yannis2} est une solution des équations du mouvement en présence d'une forme de rang $p$. Nous allons désormais détailler les cas $p=1,2$ et 3 puisque les cas $p=4,5$ et 6 sont obtenus par la dualité $\sim$.

			\subsubsection{Le champ scalaire libre}
Pour ce cas $(p=1)$, nous avons $\mc J^{(4)}=0$ et $\mc F^{(4)}=\p_\mu \phi \dd x^\mu$ où $\phi$ est un champ scalaire qui ne dépend que des coordonnées intrinsèques $x^\mu$ d'après \eqref{IC}. L'équation \eqref{uvII} implique que $h(x^\mu)=m$ est une constante et nous avons l'équation du second degré pour $B'$ suivante : 
\BE \frac{3\al}{4}B^{-5/4}B'^2 + \left(1+\al R^{(4)}B^{-1/2}\right)B' -\frac{4\La}{5}B^{5/4} + \frac{2}{3}\left[R^{(4)}-\ka\left(F^{(4)}\right)^2\right] B^{3/4} + 2\al\hat{G}^{(4)} B^{1/4}- m = 0 \ . \EE
Ainsi, nous obtenons le potentiel
\BE \label{v1}
V(r) = \frac{R^{(4)}}{12} + \frac{r^2}{12\alpha}
\left[
1\pm\sqrt{1 + \frac{12\alpha\Lambda}{5} + \frac{2\alpha\kappa\left(F^{(4)}\right)^2}{r^2} + \alpha^2\frac{\left(R^{(4)}\right)^2-6\hat{G}^{(4)}}{r^4} + \frac{3\alpha m}{r^5} }
\right] \ . \EE
Puisque $V=V(r)$, tout coefficient devant une puissance de $r$ est nécessairement constant. A ce titre, le scalaire $\lp F^{(4)}\rp^2=\p_\mu\phi \p^\mu \phi$ est constant. Mais, nous devons aussi tenir compte de l'équation \eqref{munuII} qui se réécrit sous la forme factorisée :
\BE \left(R^{(4)}_{\mu\nu} - \frac{1}{4}R^{(4)}h_{\mu\nu}\right)\left[1+\frac{4\alpha B^{1/4}}{B'}\left(B^{1/2}\frac{U'}{U}\right)'\right]
= \kappa\left[F_\mu F_\nu - \frac{1}{4} h_{\mu\nu}\left(F^{(4)}\right)^2\right] \ . \EE
Notons que le terme supplémentaire généré par $\al$ fournit une contrainte supplémentaire sur le potentiel $V$ que nous avons déjà déterminé \eqref{v1}. La situation est donc plus contraignante que dans le cas de la gravité d'Einstein pour laquelle $\al=0$. Nous avons deux possibilités : 

\begin{itemize}
\item[\textbullet] Si $F_\mu F_\nu - \frac{1}{4} h_{\mu\nu}\lp F^{(4)}\rp^2 \neq 0 $ alors il existe une constante $\la\in\mathbb{R}^*$ telle que
\begin{align}
\lambda\left(R^{(4)}_{\mu\nu} - \frac{1}{4}R^{(4)}h_{\mu\nu}\right) =& \kappa\left[F_\mu F_\nu - \frac{1}{4} h_{\mu\nu}\left(F^{(4)}\right)^2\right] \ ,\label{ein1}  \\
 1+\frac{4\alpha B^{1/4}}{B'}\left(B^{1/2}\frac{U'}{U}\right)' =& \lambda \ . \label{pot}
\end{align}
La constante $\la$ est positive puisque nous imposons que le couplage $\ka/\la$ soit positif dans \eqref{ein1}. Puis, en intégrant l'équation \eqref{pot} et en la comparant à \eqref{v1}, nous trouvons
\BE \label{bar} V(r) = \frac{1-\lambda}{12\alpha}r^2 + \rho \qquad\text{avec}\qquad \rho=\frac{1}{12}\left[R^{(4)} - \frac{\kappa}{\lambda} \left(F^{(4)}\right)^2 \right] \EE
où $\rho$ est une constante et nous avons la contrainte additionnelle sur $\mc H$
\BE \label{dressing} \frac{\kappa}{\lambda} \left(F^{(4)}\right)^2 =\sqrt{\left(R^{(4)}\right)^2 - 6\hat{G}^{(4)}} \EE
où $\la$ est fixé par la relation
\BE 5\left(1-\lambda^2\right) + 12\alpha\Lambda = 0 \ . \EE
Puis, en utilisant \eqref{ein1} et la contrainte mettant en jeu $\rho$, nous déterminons une équation d'Einstein sur $\mc H$ en présence de matière
\BE \label{ein2} G^{(4)}_{\mu\nu} +3 \rho h_{\mu\nu}=\frac{\kappa}{\lambda}T_{\mu\nu}^{(4)}(\mathcal{F}^{(4)}) \EE
où $T_{\mu\nu}^{(4)}(\mathcal{F}^{(4)})$ est le tenseur énergie-impulsion usuel d'un champ scalaire. Il y a deux effets notables ici : le champ scalaire induit une constante cosmologique effective $3\rho$ dans l'équation précédente qui explose lorsque nous approchons de la limite de Born-Infeld ($\la\rightarrow 0$) et, par ailleurs, le champ scalaire permet de s'écarter de cette limite. Notons que nous avons également la seconde contrainte sur la géométrie de $\mc H$ via \eqref{dressing}. En utilisant l'identité suivante
\BE 2 R^{(4)}_{\mu\nu} R^{(4)\mu\nu} - \frac{2}{3}\left(R^{(4)}\right)^2=\left(C^{(4)}\right)^2-\hat{G}^{(4)} \ , \label{geo} \EE
où $\left(C^{(4)}\right)^2$ correspond au "carré" du tenseur de Weyl de $\mc H$, nous pouvons formuler la seconde contrainte sous la forme 
\BE\label{teos1} \frac{4\ka^2}{3\la^2} \left[\left(F^{(4)}\right)^2\right]^2 = \left(C^{(4)}\right)^2 \ . \EE	
La métrique $h_{\mu\nu}$ doit donc vérifier l'équation d'Einstein euclidienne \eqref{ein2} mais avec une contrainte entre son tenseur de Weyl et le champ scalaire donnée par \eqref{teos1}, ce qui est plutôt restrictif. Nous illustrerons à la sous-section \ref{exemples} avec une géométrie possédant un tenseur de Weyl non trivial.

\item[\textbullet] Par ailleurs, nous pouvons avoir $F_\mu F_\nu - \frac{1}{4} h_{\mu\nu}\left(F^{(4)}\right)^2 = 0$. Comme nous avons pu le voir au chapitre précédent cela implique que $\mc F^{(4)}=0$ et il n'est donc pas possible de construire de solution en présence d'\ita{un seul} champ scalaire. En effet, nous avons $F_\mu F^\nu = \frac{1}{4} \left(F^{(4)}\right)^2 \delta_\mu^\nu$ . Par conséquent
\BE  F^{\rho_1} F _{\rho_1} = \cdots = F^{\rho_4} F _{\rho_4} = \frac{1}{4}\left(F^{(4)}\right)^2 \label{F^2} \ . \EE
De plus, tout produit mixte est nul, par exemple $ F_{\rho_1} F^{\rho_2} = 0 $. Ainsi, nous avons deux possibilités : soit $F_{\rho_1}=0$ et \eqref{F^2} conduit à $\mc F^{(4)}=0$ puisque $h_{\mu\nu}$ est une métrique riemannienne ; soit $F^{\rho_2}=0$ et nous en déduisons la même conclusion. Nous verrons à la sous-section \ref{exemples} qu'il est possible de contourner ce résultat en utilisant 4 champs scalaires,  ceci sera simplement une illustration de trous noirs façonnés par plusieurs champs rencontrés à la sous-section \ref{multiple}.
\end{itemize}

			\subsubsection{L'interaction électromagnétique}
Dans ce cas $(p=2)$, $\mc J^{(4)}$ est une fonction constante correspondant à la charge électrique et nous avons le champ électrique coulombien usuel pour un espace-temps de dimension 6 : $F_{rt} = \mathcal{J}^{(4)}/r^4$. L'équation $(uv)$ fournit l'équation du second degré suivante en $B'$ :
\BE \frac{3\alpha}{4}B^{-5/4}B'^2 + \left(1+\alpha R^{(4)}B^{-1/2}\right)B' -\frac{4\Lambda}{5}B^{5/4} + \frac{2}{3}R^{(4)} B^{3/4} + 2 a B^{1/4}- m + \frac{2}{3}\kappa\left(J^{(4)}\right)^2 B^{-3/4}= 0 \ . \EE
Nous obtenons par conséquent le potentiel
\BE V(r) = \frac{R^{(4)}}{12} +\frac{r^2}{12\alpha}
	 \left[
	 1\mp\sqrt{1 + \frac{12\alpha\Lambda}{5} + \frac{ W}{r^4} + \frac{3\alpha m}{r^5} - \frac{2\alpha\kappa\left(J^{(4)}\right)^2}{r^8}}
	 \right] \label{pot1}
\EE
où \BE \label{miles} W=\alpha^2 \left(R^{(4)}\right)^2-6\alpha^2\hat{G}^{(4)} + 3\kappa\alpha\left(F^{(4)}\right)^2 \EE
avec $W$ une constante. Notons que la partie électrique et celle magnétique du champ interviennent avec des puissances de $r$ différentes. Encore une fois, tous les coefficients devant les puissances de $r$ sont constants. Quant aux équations $(\mu\nu)$ \eqref{munuII}, elles donnent la forme factorisée suivante
\BE \left(R^{(4)}_{\mu\nu} - \frac{1}{4}R^{(4)}h_{\mu\nu}\right)\left[1+\frac{4\alpha B^{1/4}}{B'}\left(B^{1/2}\frac{U'}{U}\right)'\right]
=  \kappa B^{-1/2} \left[ h^{\rho\sigma}F_{\mu\rho}F_{\nu\sigma} - \frac{1}{4}h_{\mu\nu}\left(F^{(4)}\right)^2 \right] \ .\EE
Notons qu'il n'y a ici aucune condition sur la partie électrique $\mc J^{(4)}$.
Supposons tout d'abord que  $h^{\rho\sigma}F_{\mu\rho} F_{\nu\sigma} - \frac{1}{4} h_{\mu\nu}\left(F^{(4)}\right)^2 \neq 0 $, il existe alors dans ce cas une constante $\la$ telle que
\begin{align}
\lambda\left(R^{(4)}_{\mu\nu} - \frac{1}{4}R^{(4)}h_{\mu\nu}\right) &= \kappa\left[h^{\rho\sigma}F_{\mu\rho} F_{\nu\sigma} - \frac{1}{4} h_{\mu\nu}\left(F^{(4)}\right)^2\right] \ , \\
 1+\frac{4\alpha B^{1/4}}{B'}\left(B^{1/2}\frac{U'}{U}\right)' &= \lambda B^{-1/2} \ . \label{pot2}
\end{align}
Puis, en intégrant \eqref{pot2}, nous trouvons le potentiel
\BE V(r) = \frac{r^2}{12\alpha} + p + \frac{q}{2r} - \frac{\lambda}{2\alpha}\ln r \EE
où $p$ et $q$ sont des constantes. Mais, par comparaison avec \eqref{pot1}, $\la=0$ ! Ce cas est donc à exclure.
Ce qui nous conduit nécessairement à la contrainte suivante pour la partie magnétique
\BE h^{\rho\sigma}F_{\mu\rho} F_{\nu\sigma} - \frac{1}{4} h_{\mu\nu}\left(F^{(4)}\right)^2 = 0 \label{ref}\EE
avec une charge magnétique $\mc {F}^{(4)}\neq 0$. Nous avons alors 
\BE \left(R^{(4)}_{\mu\nu} - \frac{1}{4}R^{(4)}h_{\mu\nu}\right)\lb 1+\frac{4\al B^{1/4}}{B'}\left(B^{1/2}\frac{U'}{U}\right)'\rb = 0 \ .\label{integrability}\EE
Le premier facteur de l'équation ci-dessus peut être nul et $\mc H$ est alors un espace d'Einstein. Nous sommes dans ce cas en présence d'un trou noir dionique donné par le potentiel \eqref{pot1}. Cette classe de solutions fournit en particulier des trous noirs magnétiques qui ont récemment été étudiés dans \cite{Maeda:2010qz}. Les auteurs de cette référence ont aussi considéré des corrections d'ordre plus élevé pour le champ magnétique et nous renvoyons le lecteur à cet article pour plus de détails. Comme nous avons déjà pu le voir au chapitre précédent, l'équation \eqref{ref} fournit la remarquable propriété pour $\mc H$ d'être une variété kählérienne. Pour expliciter cela, posons $I_\mu^{\ \nu} = 2 F_\mu^{(4)\nu}/\left(F^{(4)}\right)^2$ ; ainsi $I$ est un endomorphisme du fibré tangent $T(\mathcal{H})$ tel que $I^2=-\text{Id}$ puisque $I_\mu^{\ \rho} I_\rho^{\ \nu} = - \delta_\mu^\nu$ d'après \eqref{ref}, $I$ correspond donc à la structure presque complexe de $\mc H$. Nous pouvons alors construire la forme de Kähler $\omega$, puisque pour tout $X,Y \in T(\mathcal{H)}$, nous pouvons montrer que $\omega(X,Y)=h(IX,Y)$ où $h$ est la métrique sur $\mc H$. De cette manière, nous retrouvons la métrique hermitienne $k$ par $k=h-i\omega$. En termes de composantes, nous obtenons $\omega_{\mu\nu} = I_\mu^{\ \rho}h_{\rho\nu}=I_{\mu\nu}=2 F_{\mu\nu}^{(4)}/\left(F^{(4)}\right)^2$. Par conséquent, si $\left(F^{(4)}\right)^2$ est constant sur $\mathcal{H}$, ce qui est le cas en gravité d'Einstein pour laquelle $\al=0$, alors $\om$ est fermée $\dd\omega=0$. Ce qui montre que $\left(\mathcal{H},k\right)$ est une variété kählérienne. Le lecteur pourra consulter les rappels sur ces variétés à la sous-section \ref{cartan}.

Par ailleurs, $\mc H$ n'est pas nécessairement un espace d'Einstein mais dans ce cas $ V(r) = \frac{r^2}{12\alpha} + p + \frac{q}{2r} $ où $p$ et $q$ sont des constantes d'après \eqref{integrability}. En comparant avec \eqref{pot1}, nous avons 
\BE q=0 \ , \mathcal{J}^{(4)}=0 \ ,  p = \frac{1}{12}\left[R^{(4)} \pm \sqrt{\left(R^{(4)}\right)^2 - 6\hat{G}^{(4)} + \frac{3\kappa}{\alpha} \left(F^{(4)}\right)^2 } \right] \quad\text{et}\quad 5+12\alpha\Lambda=0 \ .\EE
La géométrie de $\mc H$ doit donc vérifier une contrainte à la limite de Born-Infeld.

Enfin, la dernière possibilité est de garder seulement une composante électrique, $\mathcal{J}^{(4)}$, avec une charge magnétique nulle $\mathcal{F}^{(4)}=0$. Tout espace d'Einstein est alors une solution possible pour $\mc H$ et le potentiel du trou noir est
\BE V(r) = \frac{R^{(4)}}{12} + \frac{r^2}{12\alpha} \left[ 1\mp\sqrt{1 + \frac{12\alpha\Lambda}{5} + \frac{ \alpha^2 \left(R^{(4)}\right)^2-6\alpha^2\hat{G}^{(4)} }{r^4} + \frac{3\alpha m}{r^5} - \frac{2\alpha\kappa\left(J^{(4)}\right)^2}{r^8}}	 \right] \label{pot21} \EE
avec les scalaires $R^{(4)}$ et $\hat{G}^{(4)}$ constants. Dans ce dernier cas sans charge magnétique, l'espace-temps est asymptotiquement plat, c'est-à-dire $V(r)=1 + \mathcal{O}\left(\frac{1}{r^3}\right)$, si $\left(R^{(4)}\right)^2=6\hat{G}^{(4)}$ où $R^{(4)}=12$. Or, puisque $\mc H$ est un espace d'Einstein, l'identité \eqref{geo} montre que le tenseur de Weyl de $\mc H$ est nécessairement nul et par conséquent $\mc H$ est à courbure constante d'après \eqref{einstein-weyl}. De cette manière, nous restaurons la solution de Boulware et Deser avec une charge électrique où la métrique intrinsèque de l'horizon est celle de la sphère $S^4$.

			\subsubsection{Le champ axionique}
Pour ce cas dans lequel $p=3$, l'équation $(uv)$ donne le potentiel
\BE \label{pot31} V(r) = \frac{R^{(4)}}{12} +\frac{r^2}{12\alpha} \left[1\mp\sqrt{1 + \frac{12\alpha\Lambda}{5} + \alpha^2\frac{\left(R^{(4)}\right)^2-6\hat{G}^{(4)}}{r^4} + \frac{3\alpha m}{r^5} - \frac{6\alpha\kappa\left[\left(J^{(4)}\right)^2 + \frac{\left(F^{(4)}\right)^2}{6} \right]}{r^6} }\right] \EE
Notons que la polarisation électrique, qui est une 1-forme, et celle magnétique, qui est une 3-forme, du champ axionique interviennent avec la même puissance de $r$ dans le potentiel. Par ailleurs, l'équation restante $(\mu\nu)$ est 
\begin{align}
\left(R^{(4)}_{\mu\nu} - \frac{1}{4}R^{(4)}h_{\mu\nu}\right) & \left[1+\frac{4\alpha B^{1/4}}{B'}\left(B^{1/2}\frac{U'}{U}\right)'\right] \notag \\
&= \kappa B^{-1} \left[
\frac{1}{2} h^{\rho\sigma}h^{\alpha\beta}F_{\mu\rho\alpha} F_{\nu\sigma\beta} - \frac{1}{8} h_{\mu\nu}\left(F^{(4)}\right)^2 - J_\mu J_\nu + \frac{1}{4} h_{\mu\nu}\left(J^{(4)}\right)^2 \right] \ .
\end{align}
Si le terme à droite de l'équation ci-dessus est non nul, il existe alors une constante $\la\neq 0$ telle que
\begin{align}
\lambda\left(R^{(4)}_{\mu\nu} - \frac{1}{4}R^{(4)}h_{\mu\nu}\right) &= \kappa\left[ \frac{1}{2} h^{\rho\sigma}h^{\alpha\beta}F_{\mu\rho\alpha} F_{\nu\sigma\beta} - \frac{1}{8} h_{\mu\nu}\left(F^{(4)}\right)^2 - J_\mu J_\nu + \frac{1}{4} h_{\mu\nu}\left(J^{(4)}\right)^2 \right] \ , \\
 1+\frac{4\alpha B^{1/4}}{B'}\left(B^{1/2}\frac{U'}{U}\right)' &= \lambda B^{-1} \ . \label{pot3}
\end{align}
Nous pouvons alors de nouveau intégrer facilement \eqref{pot3} pour trouver le potentiel suivant
\BE V(r) = \frac{r^2}{12\alpha} + p + \frac{q}{2r} - \frac{\lambda}{4\alpha r^2} \label{pot4}\EE
où $p$ et $q$ sont des constantes. Puis, par comparaison avec le potentiel \eqref{pot31}, nous devons imposer $\la=0$!
Le seul cas possible est donc
$\frac{1}{2} h^{\rho\sigma}h^{\alpha\beta}F_{\mu\rho\alpha} F_{\nu\sigma\beta} - \frac{1}{8} h_{\mu\nu}\left(F^{(4)}\right)^2 = J_\mu J_\nu - \frac{1}{4} h_{\mu\nu}\left(J^{(4)}\right)^2$. Cela s'avère très restrictif pour $\mathcal{F}^{(4)}$ et $\mathcal{J}^{(4)}$. Détaillons cela en introduisant la 1-forme $\mathcal{K}^{(4)}=K_\mu \mathrm{d}x^{\mu}$ tel que $\mathcal{F}^{(4)} = \star_{(4)} \mathcal{K}^{(4)}$. Ce qui nous ramène au cas $p=1$ avec deux champs scalaires. Nous avons alors
\BE \frac{1}{2} h^{\rho\sigma}h^{\alpha\beta}F_{\mu\rho\alpha} F_{\nu\sigma\beta} - \frac{1}{8} h_{\mu\nu}\left(F^{(4)}\right)^2 = - \left[ K_\mu K_\nu - \frac{1}{4} h_{\mu\nu}\left(K^{(4)}\right)^2 \right] \ , \EE
ce qui conduit à la relation
\BE J_\mu J_\nu + K_\mu K_\nu = \frac{1}{4} h_{\mu\nu}\left[ \left(J^{(4)}\right)^2 + \left(K^{(4)}\right)^2 \right] \ . \label{J^2,K^2} \EE
Après cela, si $\mu \neq \nu$, il est alors facile de montrer que
\BE \left( K^\mu K_\mu + J^\mu J_\mu \right)\left( K^\mu K_\mu - J^\nu J_\nu \right) =0 \EE
où il n'y a pas de somme sur les indices dans cette dernière égalité. Soit nous avons $K^\mu K_\mu + J^\mu J_\mu = 0 $, ce qui implique $\left(J^{(4)}\right)^2 + \left(K^{(4)}\right)^2=0$ et ainsi $\mathcal{F}^{(4)}=0$ et $\mathcal{J}^{(4)}=0$. Ou pour tout $\mu \neq \nu$ nous avons $K^\mu K_\mu = J^\nu J_\nu$ qui fournit
\BE K^{\rho_1} K_{\rho_1} = \cdots = K^{\rho_4} K_{\rho_4} = J^{\rho_1} J_{\rho_1} = \cdots = J^{\rho_4} J_{\rho_4}\EE
puisque $\text{dim}\left(\mathcal{H}\right) > 2$ ; \eqref{J^2,K^2}  nous ramène alors au même résultat que pour $p=1$, ce qui permet de conclure que $\mathcal{F}^{(4)}=0$ et $\mathcal{J}^{(4)}=0$.
\bs

Nous avons montré qu'il n'est donc pas possible d'ajouter \ita{un seul} champ axionique dans cette classe II. Nous verrons cependant dans la sous-section \eqref{exemples} que l'introduction d'un autre champ axionique permet de contourner ce résultat.

		\subsection{Classe III}
Passons à la dernière classe de solutions pour laquelle $B$ est une fonction constante et nous posons $B \dot{=} \beta^4 \neq 0$. Les équations $(uv)$ et $(\mu\nu)$ se réduisent à
\BE -2\Lambda + \alpha\hat{G}^{(4)}\beta^{-4} + R^{(4)}\beta^{-2} = \frac{\kappa\beta^{2(p-6)}}{(p-2)!}\left(J^{(4)}\right)^2 + \frac{\kappa\beta^{-2p}}{p!}\left(F^{(4)}\right)^2 \label{1}  \EE
et
\begin{align} 
\left(R^{(4)}_{\mu\nu} - \frac{1}{4}R^{(4)}h_{\mu\nu}\right) \left(1+ 8\alpha\beta^3 e^{-2\nu}\nu_{,uv}\right)
=& \kappa \beta^{2(1-p)} \left[ T_{\mu\nu}\left(\mathcal{F}^{(4)}\right) -\frac{1}{4} T^{(4)}\left(\mathcal{F}^{(4)}\right) h_{\mu\nu} \right] \notag \\
&- \kappa \beta^{2(p-5)} \left[ T_{\mu\nu}\left(\mathcal{J}^{(4)}\right) -\frac{1}{4} T^{(4)}\left(\mathcal{J}^{(4)}\right) h_{\mu\nu} \right]
\label{2}\end{align}
où la trace de $\mathcal{E}_{\mu\nu}=\kappa T_{\mu\nu}$ est donnée par
\BE 4\Lambda\beta^2 - R^{(4)} - 8\beta^3 e^{-2\nu} \nu_{,uv} \left(\beta^2+\alpha R^{(4)}\right) = \kappa\beta^{2(1-p)}T^{(4)}\left(F^{(4)}\right) - \kappa\beta^{2(p-5)}T^{(4)}\left(J^{(4)}\right) \ . \label{3}\EE

La forme de \eqref{2} implique qu'il existe une constante $\la\neq 0$ telle que
\begin{align}
\lambda\left( R^{(4)}_{\mu\nu} -\frac{1}{4}R^{(4)}h_{\mu\nu} \right) &= \kappa \beta^{2(1-p)} \left[ T_{\mu\nu}\left(\mathcal{F}^{(4)}\right) -\frac{1}{4} T^{(4)}\left(\mathcal{F}^{(4)}\right) h_{\mu\nu} \right] \notag \\
&- \kappa \beta^{2(p-5)} \left[ T_{\mu\nu}\left(\mathcal{J}^{(4)}\right) -\frac{1}{4} T^{(4)}\left(\mathcal{J}^{(4)}\right) h_{\mu\nu} \right] \ , \\ 
1 + 8\alpha\beta^3 e^{-2\nu} \nu_{,uv} &= \lambda \ . \label{liouville}
\end{align}
A partir de \eqref{liouville}, nous observons que pour $\al\neq 0$ et $\la\neq 1$ la fonction $\nu$ vérifie une équation dite de \ita{Liouville} : $\nu_{uv}=\frac{\lambda-1}{8\alpha\beta^3}e^{2\nu}$. Cette dernière conduit à $e^{2\nu}=\frac{8\alpha\beta^3}{\lambda-1}\frac{U'V'}{(U+V)^2}$ avec $U=U(u)$ et $V=V(v)$ deux fonctions arbitraires. Puis, en effectuant le changement de coordonnées $\left(z=U+V,t=V-U\right)$, nous obtenons
\BE \mathrm{d}s^2 = \frac{4\alpha}{(1-\lambda)z^2} \left( -\mathrm{d}t^2 + \mathrm{d}z^2 \right) + \beta^2 h_{\mu\nu}(x) \mathrm{d}x^\mu \mathrm{d}x^\nu \ . \label{ds1}
\EE
Par ailleurs, si $\lambda=1$ alors $\nu_{uv}=0$ et la fonction $\nu$ se décompose sous la forme $\nu=f(u)+g(v)$ avec $f$ et $g$ deux fonctions arbitraires. Puis, nous effectuons le changement de coordonnées $U=-\int_0^u e^{2f(x)}\mathrm{d}x$ et $V=\int_0^v e^{2g(x)}\mathrm{d}x$. Enfin, le même changement de coordonnées que précédemment $(U,V)\rightarrow(z,t)$ conduit à la métrique suivante
\BE \mathrm{d}s^2 = \frac{1}{2\beta^3} \left( -\mathrm{d}t^2 + \mathrm{d}z^2 \right) + \beta^2 h_{\mu\nu}(x) \mathrm{d}x^\mu  \mathrm{d}x^\nu \ .\label{ds2} \EE
Par conséquent, les solutions de cette classe III sont statiques. La métrique \eqref{ds2} coïncide avec le cas de la gravité d'Einstein. En fait, $\al=0$ conduit directement à $\la=1$. Nous verrons dans la suite que cette classe de solutions présente plus d'intérêt dans sa version \eqref{bstrings} après une double rotation de Wick.

Concernant la métrique $h_{\mu\nu}(x^\rho)$ de $\mc H$, elle doit vérifier l'équation d'Einstein euclidienne suivante
\BE \label{einstein2}
G^{(4)}_{\mu\nu}-\beta^2 \left(\frac{\lambda-1}{4\lambda}-\frac{\Lambda}{\lambda}\right)h_{\mu\nu}=\frac{\kappa}{\lambda} \beta^{2(1-p)} T_{\mu\nu}\left(\mathcal{F}^{(4)}\right)- \frac{\kappa}{\lambda} \beta^{2(p-5)} T_{\mu\nu}\left(\mathcal{J}^{(4)}\right)
\EE
avec la contrainte \eqref{1}. Notons que la constante cosmologique effective de cette équation dépend de la constante cosmologique $\La$ mais aussi de $\la$ et $\be$. Détaillons désormais pour chaque rang $p$.

			\subsubsection{Le champ scalaire libre}
Nous supposons $\la\neq 0$ pour avoir une solution non triviale. Au lieu de considérer simplement un champ scalaire libre, le caractère constant de $B$ nous permet d'ajouter un potentiel arbitraire $V(\phi)$ afin d'obtenir un tenseur énergie-impulsion de la forme
\BE T_{AB} = \partial_A \phi \partial_B \phi - g_{AB}\left[\frac{1}{2}\partial^C \phi \partial_C \phi + V(\phi) \right] \ . \EE
Les conditions d'intégrabilité $\mc E_{uu}=0$ et $\mc E_{vv}=0$ fournissent encore une fois un champ scalaire $\phi=\phi(x^\mu)$ qui ne dépend que des coordonnées intrinsèques de $\mc H$. Nous avons alors toujours l'équation d'Einstein euclidienne \eqref{einstein2} avec le tenseur énergie-impulsion $T_{\mu\nu}\left(\mathcal{F}^{(4)}\right)$ complété par la présence du potentiel $V(\phi)$. La trace de cette équation est donnée par
\BE 4\left[ \Lambda + \kappa V(\phi) \right]\beta^2-R^{(4)}+\kappa\left(\partial \phi \right)^2  = \frac{\lambda-1}{\alpha}\left(\beta^2+\alpha R^{(4)}\right) \label{constraint1}\EE
et l'équation \eqref{1} fournit la contrainte supplémentaire
\BE \alpha \hat{G}^{(4)}=2\beta^4 \left[ \Lambda + \kappa V(\phi) \right] -\beta^2  R^{(4)}+  \kappa\beta^2\left(\partial \phi \right)^2 \label{constraint2} \ . \EE
En effectuant une double de rotation de Wick sur la métrique $g$, nous décrirons un exemple dans la sous-section \eqref{exemples} dans lequel la métrique $h$ est celle de FLRW.
						
			\subsubsection{L'interaction électromagnétique}
Ce cas présente un intérêt particulier. Tout d'abord, l'équation \eqref{einstein2} se réduit à	
\BE \label{ein3} G^{(4)}_{\mu\nu}-\beta^2 \left(\frac{\lambda-1}{4\lambda}-\frac{\Lambda}{\lambda}+\frac{\kappa\left(J^{(4)}\right)^2}{2\lambda \beta^8}\right)h_{\mu\nu}=\frac{\kappa}{\lambda} \beta^{-2} T_{\mu\nu}\left(\mathcal{F}^{(4)}\right) \EE
qui est une équation d'Einstein euclidienne en dimension 4 en présence du champ de matière $\mathcal{F}^{(4)}$ et avec une constante cosmologique effective qui inclut la charge électrique de l'espace-temps de dimension 6. De plus, si $h_{\mu\nu}$ est une métrique lorentzienne après une double rotation de Wick, nous pouvons interpréter $\mathcal{F}^{(4)}$ comme une 2-forme de Faraday effective. L'équation \eqref{1} implique la contrainte additionnelle
\BE \label{teos} \hat{G}^{(4)} - \frac{\kappa}{2\alpha}\left(F^{(4)}\right)^2 =\frac{4\Lambda\beta^4}{\alpha} - \frac{3R^{(4)}\beta^2}{2\alpha} - \frac{\beta^2}{2\alpha^2}(\lambda-1)\left(\beta^2+\alpha R^{(4)}\right) \ . \EE
Ajoutons le commentaire suivant. Si $\mc H$ est une variété lorentzienne, il est loisible de considérer une solution à symétrie sphérique pour $h$ en présence du champ électromagnétique $\mathcal{F}^{(4)}$. La seule solution est celle de Reissner-Nordstrom avec constante cosmologique. Cependant, contrairement au cas de la gravité d'Einstein en dimension 6, ce cas est à exclure puisque la contrainte \eqref{teos} n'est pas compatible avec la solution de Reissner-Nordstrom.
Par ailleurs, contrairement au cas $p=1$, nous pouvons ici imposer $\la=0$ avec une 2-forme de Faraday pour l'espace-temps de dimension 6 non nulle. Ainsi l'équation d'Einstein euclidienne se réduit à sa trace
\BE \label{trace2} 4\alpha \Lambda+1=2\kappa\alpha\beta^{-8}\left(J^{(4)}\right)^2 \EE
où la charge électrique $\left(J^{(4)}\right)^2$ permet d'éviter un réglage fin entre $\al$ et $\La$. Il faut aussi tenir compte de la contrainte \eqref{teos}. Nous présenterons dans la suite l'exemple d'une corde noire avec $\la=0$.

			\subsubsection{Le champ axionique}
Nous considérons de nouveau le cas $\la\neq 0$ pour ne pas avoir un champ axionique trivial. Pour ce cas $(p=3)$, nous avons toujours l'équation d'Einstein euclidienne \eqref{einstein2} dont la trace est donnée par
\BE 4\Lambda\beta^2-\lambda R^{(4)}-\kappa\beta^{-4}\left[\frac{1}{6}\left(F^{(4)}\right)^2+\left(J^{(4)}\right)^2\right] -\frac{(\lambda-1)\beta^2}{\alpha}=0 \ . \EE
Enfin, nous avons la contrainte supplémentaire
\BE -2\Lambda + \alpha\hat{G}^{(4)}\beta^{-4} + R^{(4)}\beta^{-2} =\kappa\beta^{-6}\left[\frac{1}{6}\left(F^{(4)}\right)^2+\left(J^{(4)}\right)^2\right] \ . \label{11}\EE

		\subsection{Exemples de solutions\label{exemples}}
Dans cette sous-section, nous allons construire des exemples pour les trois classes de solutions que nous venons de présenter.

			\subsubsection{Champ magnétique pour classes I et II}
Comme nous avons déjà pu le voir dans le chapitre précédent, des solutions magnétiques pour $p=2$ peuvent être construites en considérant par exemple $\mathcal{H}=S^2 \times S^2$. L'idée est d'associer une composante du champ magnétique pour chaque sphère.
Contentons nous du cas de la classe I même si la description s'applique de manière similaire aux autres classes. Considérons la métrique suivante sur 
$\mathcal{H}=S^2 \times S^2$ : 
\BE ds^2=\rho_1^2(d\theta_1^2+\sin^2 \theta_1 d\phi_1^2)+\rho_2^2(d\theta_2^2+\sin^2 \theta_2 d\phi_2^2) \EE
où $\rho_1$ et  $\rho_2$ désignent les rayons de chaque sphère. Nous pouvons remarquer que
\BE R^{(4)}=2\frac{\rho_1^2+\rho_2^2}{\rho_1^2\rho_2^2} \qquad\text{et}\qquad \hat{G}^{(4)}=\frac{8}{\rho_1^2 \rho_2^2} \ . \EE
Puis, nous introduisons le champ magnétique suivant 
\BE \mathcal{F}^{(4)}=Q_1\sin \theta_1 \mathrm{d}\theta_1 \wedge \mathrm{d}\phi_1+Q_2\sin \theta_2 \mathrm{d}\theta_2 \wedge \mathrm{d}\phi_2 \ . \EE
Si $\rho_1=\rho_2$ et $Q_1=Q_2$ alors $\mc H$ est un espace d'Einstein et $T_{\mu\nu}\left(F^{(4)}\right)=0$, ce qui satisfait \eqref{munuI}. C'est donc une solution pour la classe Ia où $B$ est une fonction arbitraire. Par ailleurs, si $\mc H$ n'est pas un espace d'Einstein, nous pouvons trouver une solution de la classe Ic vérifiant l'équation \eqref{cl1} avec la contrainte géométrique \eqref{scalar1} en imposant 
\begin{align}
2\kappa Q_1^2&=\frac{\rho_1^2}{\rho_2^2}\left[(\rho_2^2-\rho_1^2) \lambda+\frac{2\alpha}{3}\frac{12\rho_1^2\rho_2^2-(\rho_1^2+\rho_2^2)^2}{\rho_1^2\rho_2^2}\right] \ ,\nonumber\\
2\kappa Q_2^2&=\frac{\rho_2^2}{\rho_1^2}\left[(\rho_1^2-\rho_2^2)\lambda+\frac{2\alpha}{3}\frac{12\rho_1^2\rho_2^2-(\rho_1^2+\rho_2^2)^2}{\rho_1^2\rho_2^2}\right] \ .
\end{align}
Rappelons que si $\mc H$ n'est pas un espace d'Einstein, la fonction $B$ n'est plus arbitraire mais doit résoudre \eqref{Bfield}. Alors qu'il y a une dégénérescence de la fonction $B$ lorsque $\rho_1=\rho_2$, ce n'est plus le cas dès que $\rho_1\neq\rho_2$.

			\subsubsection{Solution de la classe II en présence d'un champ scalaire}
Considérons le cas de la classe II en présence d'un seul champ scalaire libre. De plus, nous supposons pour simplifier que $\rho=0$ dans l'équation \eqref{bar}. Par conséquent, le potentiel est $V(r)=\frac{1-\lambda}{12\alpha}r^2$ et l'équation d'Einstein euclidienne est $R^{(4)}_{\mu\nu}=\frac{\kappa}{\lambda}\partial_\mu\phi\partial_\nu\phi$ où $\phi$ est harmonique sur $\mathcal{H}$. Nous souhaitons examiner les solutions statiques et à symétrie sphérique de cette équation qui sont données par
\BE h = \left(1-\frac{2\eta}{R}\right)^{\cos\chi}\mathrm{d}\tau^2 + \frac{\mathrm{d}R^2}{\left(1-\frac{2\eta}{R}\right)^{\cos\chi}} + \left(1-\frac{2\eta}{R}\right)^{1-\cos\chi}R^2\left(\mathrm{d}\theta^2+ \sin^2\theta\mathrm{d}\phi^2\right)\EE
avec le champ scalaire
\BE \phi = \sqrt{\frac{\lambda}{2\kappa}}\sin\chi\ln\left(1-\frac{2\eta}{R}\right) \ . \EE
Le lecteur pourra trouver la version lorentzienne de ces solutions dans \cite{Agnese:1985xj}. En particulier, nous retrouvons la solution de Schwarzschild pour $\chi=0$ alors que la solution pour laquelle $\chi=\pi/3$ est conformément reliée à la solution de BBMB \cite{Bocharova:1970,Bekenstein:1974sf,Bekenstein:1975ts}. Ici nous avons aussi la contrainte supplémentaire $\left(C^{(4)}\right)^2 = \frac{4\kappa^2}{3\lambda^2}\left(\partial_\mu\phi\partial^\mu\phi\right)^2$ à prendre en compte, qui est seulement vérifiée pour $\chi=\pi/2$ : 
\BE h = \mathrm{d}\tau^2 + \mathrm{d}R^2 + \left(1-\frac{2\eta}{R}\right)R^2\left(\mathrm{d}\theta^2+ \sin^2\theta\mathrm{d}\phi^2\right)\EE
avec
\BE \phi = \sqrt{\frac{\lambda}{2\kappa}}\ln\left(1-\frac{2\eta}{R}\right) \ . \EE
Cette solution est en fait singulière puisque le scalaire de Ricci $R^{(4)}=\frac{2\eta^2}{(2\eta-R)^2 R^2}$ diverge quand $R$ tend vers $2\eta^+$ et il n'y a pas d'horizon des événements pour cacher cette singularité.
			
			\subsubsection{Trou noir avec deux champs axioniques avec $\mc H=T^4$}
Nous avons vu qu'il n'est pas possible de construire un trou noir statique en présence d'un seul champ axionique dans la classe II. Contournons ce problème en considérant deux 3-formes dans la théorie : 			
\BE S^{(6)} =\int_\mathcal{M} \mathrm{d}^6 x \sqrt{-g^{(6)}} \left[ R - 2\Lambda + \alpha \hat{G} - \frac{\kappa_1}{6} F_{(1)ABC}F_{(1)}^{ABC} - \frac{\kappa_2}{6} F_{(2)ABC}F_{(2)}^{ABC} \right] \ . \EE
En appliquant la même méthode qu'auparavant, nous trouvons que ces champs de matière doivent vérifier la contrainte suivante
\BE \label{sum1} \sum_{i=1}^2 \kappa_i \left[\frac{1}{2} h^{\rho\sigma}h^{\alpha\beta}F_{(i)\mu\rho\alpha} F_{(i)\nu\sigma\beta} - \frac{1}{8} h_{\mu\nu}\left(F_{(i)}^{(4)}\right)^2 - J_{(i)\mu} J_{(i)\nu} + \frac{1}{4} h_{\mu\nu}\left(J_{(i)}^{(4)}\right)^2 \right]= 0 \ . \EE
En fait, chaque champ axionique fournit deux champs scalaires. Plus explicitement, introduisons les 1-formes $\mathcal{K}_{(i)}^{(4)}=K_{(i)\mu}\mathrm{d}x^\mu \in \Lambda^1(\mathcal{H})$ telles que $\mathcal{F}_{(i)}^{(4)} = \star_{(4)} \mathcal{K}_{(i)}$ pour $i=1,2$. Ainsi, l'équation \eqref{sum1} devient
\BE \sum_{i=1}^2 \kappa_i \left[J_{(i)\mu} J_{(i)\nu} + K_{(i)\mu} K_{(i)\nu} \right] = \frac{1}{4} h_{\mu\nu} \sum_{i=1}^2 \kappa_i\left[ \left(J_{(i)}^{(4)}\right)^2 + \left(K_{(i)}^{(4)}\right)^2 \right] \ . \label{sum2} \EE
Il est désormais clair que l'espace d'Einstein le plus simple $\mathcal{H}=T^4$ est une solution satisfaisant \eqref{sum2}. En effet, si $(x,y,z,w)$ désigne les coordonnées intrinsèques de $T^4$, nous pouvons choisir
\BE \mathcal{J}_{1}^{(4)} = \frac{Q}{\sqrt{\kappa_1}}\mathrm{d}x \qquad \mathcal{K}_{1}^{(4)} = \frac{Q}{\sqrt{\kappa_1}}\mathrm{d}y \qquad \mathcal{J}_{2}^{(4)} = \frac{Q}{\sqrt{\kappa_2}}\mathrm{d}z \qquad \mathcal{K}_{2}^{(4)} = \frac{Q}{\sqrt{\kappa_2}}\mathrm{d}w \EE
ou de manière équivalente 
\BE\mathcal{J}_{1}^{(4)} = \frac{Q}{\sqrt{\kappa_1}}\mathrm{d}x \quad \mathcal{F}_{1}^{(4)} = \frac{Q}{\sqrt{\kappa_1}}\mathrm{d}x\wedge\mathrm{d}z\wedge\mathrm{d}w \quad \mathcal{J}_{2}^{(4)} = \frac{Q}{\sqrt{\kappa_2}}\mathrm{d}z \quad \mathcal{F}_{2}^{(4)} = \frac{Q}{\sqrt{\kappa_2}}\mathrm{d}x\wedge\mathrm{d}y\wedge\mathrm{d}z \EE
où $Q$ est une constante. Finalement, cette configuration fournit un trou noir avec deux champs axioniques où la métrique est donnée par
\BE \mathrm{d}s^2 = - V(r)\mathrm{d}t^2 + \frac{\mathrm{d}r^2}{V(r)} + r^2\left(\mathrm{d}x^2+\mathrm{d}y^2+\mathrm{d}z^2+\mathrm{d}w^2\right)\EE
avec le potentiel
\BE V(r)=\frac{r^2}{12\alpha}\left[1\pm\sqrt{1+\frac{12\alpha\Lambda}{5}+\frac{3\alpha m}{r^5}-\frac{24\alpha Q^2}{r^6}}\right]  \ . \EE
Ce trou noir est asymptotiquement localement AdS et possède en particulier une limite en gravité d'Einstein. En effet, lorsque $\alpha\rightarrow 0$, la branche d'Einstein fournit le potentiel $V(r)=-\frac{\Lambda}{10}r^2-\frac{m}{8}\frac{1}{r^3}+\frac{Q^2}{r^4}$. Nous restaurons ainsi les trous noirs façonnés par plusieurs champs rencontrés au chapitre précédent.

			\subsubsection{Trou noir avec quatre champs scalaires avec $\mc H=T^4$}
Par analogie avec l'exemple précédent, nous pouvons construire une solution statique de type trou noir avec quatre champs scalaires en considérant la théorie suivante 
\BE S^{(6)} = \int_\mathcal{M} \mathrm{d}^6 x \sqrt{-g^{(6)}} \left[ R - 2\Lambda + \alpha \hat{G} - \sum_{i=1}^4\kappa_i \partial_A\phi_{(i)}\partial^A\phi_{(i)} \right]  \ . \EE
En choisissant de nouveau un horizon plat, nous construisons la solution
\BE \mathrm{d}s^2 = - V(r)\mathrm{d}t^2 + \frac{\mathrm{d}r^2}{V(r)} + r^2\left(\mathrm{d}x^2+\mathrm{d}y^2+\mathrm{d}z^2+\mathrm{d}w^2\right) \EE
avec le potentiel
\BE V(r)=\frac{r^2}{12\alpha}\left[1\pm\sqrt{1+\frac{12\alpha\Lambda}{5} + \frac{8\alpha\lambda^2}{r^2} + \frac{3\alpha m}{r^5}}\right]  \EE
et les champs scalaires
\BE \phi_{(1)}=\frac{\lambda x}{\sqrt{\kappa_1}} \qquad \phi_{(2)}=\frac{\lambda y}{\sqrt{\kappa_2}} \qquad \phi_{(3)}=\frac{\lambda z}{\sqrt{\kappa_3}} \qquad \phi_{(4)}=\frac{\lambda w}{\sqrt{\kappa_4}}  \EE
où $\la$ est une constante. Il est intéressant de considérer la limite de cette solution en gravité d'Einstein, nous trouvons pour la branche d'Einstein le potentiel suivant
\BE V(r)=-\frac{\Lambda}{10}r^2-\frac{\lambda^2}{3}-\frac{m}{8}\frac{1}{r^3} \ .\label{hair1} \EE
Comme nous avons déjà pu l'observer dans le chapitre précédent, la géométrie est celle d'un trou noir hyperbolique bien que l'horizon soit plat. Finalement, nous aurions aussi pu considérer à la fois des champs axioniques et scalaires pour trouver un potentiel de la forme : 
\BE V(r)=\frac{r^2}{12\alpha}\left[1\pm\sqrt{1+\frac{12\alpha\Lambda}{5} + \frac{8\alpha\lambda^2}{r^2} + \frac{3\alpha m}{r^5}-\frac{24 \alpha Q^2}{r^6}}\right] \ . \EE
			
			\subsubsection{Solution générale pour la classe II avec toutes les $p$-formes}
Comme dernier exemple de la classe II, nous considérons le cas général mettant en jeu un champ scalaire, l'interaction électromagnétique et une 3-forme via l'action : 
\BE S^{(6)} = \int_\mathcal{M} \mathrm{d}^6 x \sqrt{-g^{(6)}} \left[ R - 2\Lambda + \alpha \hat{G} - \kappa_1 F_{A}F^{A} - \frac{\kappa_2}{2} F_{AB}F^{AB} - \frac{\kappa_3}{6} F_{ABC}F^{ABC} \right] \ . \EE
Le point que nous aimerions montrer est que la combinaison de ces champs de matière conduit à un potentiel $V$ assez général. Tout d'abord, nous exprimons le potentiel $V$ à l'aide de l'équation $(uv)$ \eqref{uvII}. En définissant $p=\alpha \hat{G}^{(4)}-\frac{\kappa_2}{2}F^2$, $q=\kappa_1\left(\partial\phi\right)^2$ et $t=\kappa_3\left(J^2+\frac{1}{6}H^2 \right)$, nous avons 
\BE V(r) = \frac{R^{(4)}}{12} + \frac{r^2}{12\alpha}\left[
1\mp\sqrt{1 + \frac{12\alpha\Lambda}{5} + \frac{2\alpha q}{r^2} + \frac{\alpha^2\left(R^{(4)}\right)^2-6\alpha p}{r^4} + \frac{3\alpha m}{r^5} - \frac{6\alpha t}{r^6} - \frac{2\alpha\kappa_2 Q^2}{r^8}}
 \right] \label{complex1} \EE
où $m$ est une constante d'intégration. C'est juste une condition nécessaire pour $V$. Par ailleurs, les équation $(\mu\nu)$ \eqref{munuII} fournissent l'équation
\BE S^{(4)}_{\mu\nu}\left[1+\frac{4\alpha B^{1/4}}{B'}\left(B^{1/2}\frac{U'}{U}\right)'\right]= \kappa_1\bar{T}_{\phi\mu\nu} + \kappa_2 B^{-1/2} \bar{T}_{F\mu\nu} + \kappa_3 B^{-1} \left(\bar{T}_{J\mu\nu}+\bar{T}_{H\mu\nu}\right) \EE
où les tenseurs $T_{\phi\mu\nu}$, $T_{F\mu\nu}$ et $T_{H\mu\nu}$ sont donnés par \eqref{TM} avec $p=1,2$ et 3 respectivement ; $T_{J\mu\nu}$ par \eqref{TE} avec $p=3$ et $S^{(4)}_{\mu\nu}=R^{(4)}_{\mu\nu}-\frac{1}{4}R^{(4)}h_{\mu\nu}$. Nous avons aussi noté par $\bar{T}_{\phi\mu\nu}$ la partie sans trace du tenseur $T_{\phi\mu\nu}$ et de même pour les autres tenseurs. Puis, après deux intégrations successives par rapport à $r$, il existe deux tenseurs symétriques $X_{\mu\nu}$ et $Y_{\mu\nu}$ de trace nulle tels que
\BE \label{complex2} S^{(4)}_{\mu\nu}V(r) = \frac{S^{(4)}_{\mu\nu}-\kappa_1\bar{T}_{\phi\mu\nu}}{12\alpha}r^2 - \frac{\kappa_2}{2\alpha}\bar{T}_{F\mu\nu}\ln r + \frac{X_{\mu\nu}}{8\alpha}\frac{1}{r} - \frac{\kappa_3}{4\alpha}\left(\bar{T}_{J\mu\nu}+\bar{T}_{H\mu\nu}\right)\frac{1}{r^2}-\frac{1}{32\alpha}Y_{\mu\nu} \ . \EE
Nous devons ensuite comparer \eqref{complex1} et \eqref{complex2}. En supposant $\al<0$ et $ce > 0$, nous pouvons résumer les différentes conditions pour avoir une solution générale avec $S_{\mu\nu}^{(4)}\neq 0$ :
\BE V(r)=\frac{1-c}{12\alpha}r^2 + \frac{a}{4\alpha} + \frac{e}{4\alpha r^2} \qquad\text{avec}\qquad c=\pm\sqrt{1+\frac{12\alpha\Lambda}{5}} \ ,\qquad  e=\pm\sqrt{\frac{-2\alpha\kappa_2}{9}}|Q| \ ,\EE
\BE h^{\rho\sigma}F_{\mu\rho}F_{\nu\sigma} = \frac{1}{4}h_{\mu\nu}\left(F^{(4)}\right)^2 \ ,\EE
\BE G^{(4)}_{\mu\nu} + \frac{3a}{4\alpha} h_{\mu\nu} = \frac{\kappa_1}{c}T_{\phi\mu\nu} = -\frac{\kappa_3}{e}\left(T_{H\mu\nu}+T_{J\mu\nu}\right) \ ,\label{einsteinX}\EE
\BE \left(C^{(4)}\right)^2 = \frac{4\kappa_1^2}{3c^2}\left[(\partial\phi)^2\right]^2 + \frac{\kappa_2}{2\alpha}\left(F^{(4)}\right)^2 + \frac{ec}{\alpha^2} \ ,\EE
\BE \alpha \hat{G}^{(4)}-\frac{\kappa_2}{2}\left(F^{(4)}\right)^2 = a R^{(4)} + \frac{ec}{\alpha} - \frac{3a^2}{2\alpha} \ . \EE
De plus, chaque $p$-forme doit résoudre sa propre équation du mouvement. Quant à la valeur de $a$, elle est déterminée par la trace de l'équation d'Einstein \eqref{einsteinX} :
\BE a=\frac{\alpha}{3}\left(R^{(4)} - \frac{\kappa_1}{c}\left(\partial\phi\right)^2\right) \ . \EE
			
			\subsubsection{Corde noire en classe III}
Considérons une charge électrique $J^{(4)}$ émanant d'un champ électromagnétique $(p=2)$ avec un champ magnétique nul, $\mathcal{F}^{(4)}=0$, dans la classe III. Considérons le cas simple pour lequel $\la=0$ dans \eqref{liouville} et $\be=1$. Puis, si nous effectuons une double rotation de Wick sur la métrique $g$, nous obtenons alors un espace-temps dit de \ita{Kaluza-Klein}. La métrique $h$ sur $\mc H$ est désormais lorentzienne et il y a deux directions "courbes" pour compléter la métrique de l'espace-temps \eqref{ds1}. Considérons simplement un ansatz statique et à symétrie sphérique, planaire ou hyperbolique pour $h$ :
\BE h=-f(r) \dd t^2+\frac{\dd r^2}{f(r)}+r^2\left( \frac{\mathrm{d}\chi^2}{1-\kappa\chi^2} + \chi^2\mathrm{d}\theta^2 \right) \ . \EE
L'équation \eqref{trace2} fournit également une relation $\al$, $\La$ et $J^{(4)}$. Puis, la seule équation qui reste à résoudre est \eqref{teos} qui fournit
\BE f(r)=\kappa+\frac{r^2}{4\alpha}\left(1\pm \sqrt{\frac{4}{3}(1+2\alpha\Lambda)+\frac{\alpha^{3/2} \mu}{r^3}-\frac{\alpha^2 q}{r^4} + \frac{16\alpha^2\kappa^2}{r^4} }\right) \ . \EE
Cette solution se réduit à celle de \cite{Maeda:2006iw,Maeda:2006hj} récemment étudiée quand nous posons $4\alpha \Lambda=-1$ ; nous évitons donc un réglage fin à l'aide de l'inclusion de la charge électrique $J^{(4)}$. Quant au cas $\la\neq 0$, nous devons résoudre l'équation d'Einstein \eqref{ein3} mais elle est incompatible avec la contrainte scalaire \eqref{teos} pour notre ansatz sur $h$.
			
			\subsubsection{Solution cosmologique en classe III}	
Terminons par un dernier exemple en considérant un champ scalaire $(p=1)$ avec un potentiel $V(\phi)$ en classe III après avoir effectué une double rotation de Wick pour $\be=1$. Nous souhaitons ici étudier le cas où $h$ correspond à la métrique de FLRW en présence d'un fluide parfait généré par le champ scalaire. Nous commençons donc avec l'action suivante
\BE S^{(6)} =  \int_\mathcal{M} \mathrm{d}^6 x \sqrt{-g^{(6)}} \left[ R - 2\Lambda + \alpha \hat{G} - \kappa\left(\partial_A\phi \partial^A \phi + 2V(\phi) \right) \right] \ .\EE
La variation par rapport à la métrique de cette action fournit l'équation d'Einstein
\BE G_{AB} + \Lambda g_{AB} - \alpha H_{AB} = \kappa T_{AB} \EE
avec le tenseur énergie-impulsion
\BE T_{AB} = \partial_A \phi \partial_B \phi - g_{AB}\left[\frac{1}{2}\partial^C \phi \partial_C \phi + V(\phi) \right] \EE
et sa variation par rapport au champ scalaire donne l'équation de Klein-Gordon $\Box\phi=V'(\phi)$. Nous allons considérer le cas $\la=1$ pour lequel l'espace perpendiculaire à $\mc H$ est plat et nous ainsi une cosmologie dite de \ita{Kaluza-Klein} avec deux dimensions supplémentaires \eqref{ds2}. La métrique $h$ doit vérifier l'équation d'Einstein suivante
\BE G_{\mu\nu} + \Lambda g_{\mu\nu} = \kappa\left[\partial_\mu \phi \partial_\nu \phi - h_{\mu\nu}\left(\frac{1}{2}\partial^\rho \phi \partial_\rho \phi + V(\phi) \right)\right]\EE
avec la contrainte additionnelle $\alpha\hat{G}^{(4)}+2\left[\Lambda+\kappa V(\phi)\right]=0$ qui provient de \eqref{constraint2}. Nous aurions pu inclure la constante cosmologique dans le potentiel mais pour garder les mêmes notations nous laissons cela ainsi. En utilisant la contrainte précédente, il est loisible d'écrire l'équation d'Einstein sous la forme suivante pour laquelle la dépendance en $\La$ n'est plus explicite : 
\BE G_{\mu\nu} - \frac{\alpha}{2}\hat{G}^{(4)} h_{\mu\nu}= \kappa\left[\partial_\mu \phi \partial_\nu \phi - \frac{1}{2} h_{\mu\nu}\partial^\rho \phi \partial_\rho \phi \right] \ . \label{cosmo} \EE
Puis, nous imposons la métrique de FLRW
\BE  h=-\mathrm{d}t^2 + a^2(t)\left(\mathrm{d}x^2+\mathrm{d}y^2+\mathrm{d}z^2\right) \EE 
où $a(t)$ représente le facteur d'échelle. Ainsi l'équation \eqref{cosmo} fournit les deux équations
\begin{align} \label{rory1} 3 H^2 =& \frac{\kappa}{2}\dot{\phi}^2 - 12\alpha H^2 \frac{\ddot{a}}{a} \\
-H^2-2\frac{\ddot{a}}{a} =& \frac{\kappa}{2}\dot{\phi}^2 + 12\alpha H^2 \frac{\ddot{a}}{a}
\label{rory2} \end{align}
où nous avons introduit le paramètre d'Hubble $H=\dot{a}/a$ et, d'après les hypothèses habituelles en cosmologie, le champ scalaire ne dépend que de $t$. Par conséquent, en définissant les quantités $\rho=\frac{1}{2}\dot{\phi}^2 - 12\frac{\alpha}{\kappa} H^2 \frac{\ddot{a}}{a}$ et $P=\frac{1}{2}\dot{\phi}^2 + 12\frac{\alpha}{\kappa} H^2 \frac{\ddot{a}}{a}$, nous reconnaissons les équations de Friedmann 
\BE H^2 = \frac{\kappa}{3}\rho \qquad\text{et}\qquad \frac{\ddot{a}}{a} = -\frac{\kappa}{6}\left(\rho+3P\right) \ . \EE
De plus, l'équation d'état en $\rho$ et $P$ est fixé ici par 
\BE \left(1+4\alpha\kappa\rho\right)P = \left(1-\frac{4}{3}\alpha\kappa\rho\right)\rho  \ . \EE
Enfin, en traitant la dérivée du champ scalaire comme un terme de source, nous pouvons combiner les équations \eqref{rory1} et \eqref{rory2} pour obtenir les deux branches suivantes
\BE H^2=\frac{1}{8\alpha}\left(2\alpha\kappa \dot\phi^2 -1\pm \sqrt{1-\frac{4\alpha}{3}\kappa\dot\phi^2+4\alpha^2\kappa^2 \dot\phi^4}\right) \ . \EE

		\subsection{Conclusion}
Dans cette section, nous avons considéré des solutions de la théorie de Lovelock en dimension 6 en présence de $p$-formes \eqref{action}. Nous avons étudié une large classe de métrique \eqref{metric} en laissant arbitraire la géométrie intrinsèque d'un horizon possible $\mc H$ et nous avons  observé l'influence d'une double rotation de Wick sur cette classe de solutions \eqref{bstrings}. Notre principale hypothèse se porte sur le secteur de la matière dans le but de préserver les conditions $\mc E_{uu}=\mc E_{vv}=0$ qui ont permis dans une étude précédente \cite{Bogdanos:2009pc} d'intégrer complètement le problème dans le vide. Pour ces hypothèses, nous avons montré l'intégrabilité de ce problème en classifiant les solutions en trois classes avec notamment des conditions restrictives sur la géométrie de $\mc H$. En fait, la situation est plus contraignante que celle de la gravité d'Einstein où $\mc H$ est généralement un espace d'Einstein.
\bs

Nous avons vu que les solutions de la classe I sont généralement dégénérées et existent à la limite de Born-Infeld qui relie les coefficients de la théorie de Lovelock :  $5+12\alpha\Lambda=0$. Cependant, l'inclusion de matière ou la réduction de la symétrie sur $\mc H$ permet de lever cette dégénérescence. 

Même en présence de matière, les solutions de la classe II sont localement statiques et mettent en jeu des solutions de trous noirs sous certaines conditions. Ces géométries incluent des solutions statiques connues pour cette théorie comme celle de  Boulware et Deser \cite{Boulware:1985wk}, de R. Cai \cite{Cai:2001dz} et d'autres récemment découvertes dans \cite{Dotti:2007az,Bogdanos:2009pc,Maeda:2010qz}. Le résultat important est que si nous restreignons les solutions à celles qui sont asymptotiquement plates alors la seule solution possible est le trou noir de Boulware et Deser en présence d'une charge électrique. Une exception à ce résultat a lieu par l'inclusion d'un champ magnétique tel que $W=0$ \eqref{miles}. Dans ce dernier cas, le potentiel du trou noir se comporte comme une solution asymptotiquement plate mais la géométrie de l'horizon n'est pas celle de la sphère, $\mathcal{H}\neq S^4$, à cause de la présence de la charge magnétique. Ainsi, la théorie de Lovelock permet de lever la dégénérescence de la géométrie de l'horizon contrairement à la théorie d'Einstein en dimensions supplémentaires. En présence d'une charge magnétique, nous avons aussi vu en particulier que l'horizon $\mc H$ peut être un espace d'Einstein-Kähler ; alors qu'en l'absence de cette charge magnétique, $\mc H$ est un espace d'Einstein avec le scalaire $\left(C^{(4)}\right)^2$ constant.

Les solutions de la classe III sont également statiques et recouvrent notamment des limites déjà connues \cite{Maeda:2006iw,Maeda:2006hj} en théorie d'Einstein-Gauss-Bonnet. Précisons que l'inclusion de matière supprime le réglage fin entre $\al$ et $\La$ en particulier.
\bs

Finalement, nous avons présenté des exemples précis pour chaque classe de solutions pour comprendre leurs propriétés. Nous avons vu que de nouvelles géométries peuvent être construites dans la classe I sans être dégénérées. Quant à la classe II, la restriction sur la géométrie de $\mc H$ nous a conduit à construire des solutions de trous noirs avec plusieurs champs scalaires ou champs axioniques. Ces nouvelles solutions généralisent celles du chapitre précédent dans le cadre de la théorie d'Einstein-Gauss-Bonnet par une construction similaire et constituent ainsi les premiers exemples de trous noirs mettant en jeu des champs axioniques. Ils sont asymptotiquement localement AdS et la géométrie intrinsèque de leur horizon est plate. Nous avons aussi exhibé dans la classe III une solution de type corde noire et une solution cosmologique avec un champ scalaire qui dépend du temps. 

Pour conclure, nous avons remarqué que ce qui est perçu comme une $p$-forme en dimensions supplémentaires possède une interprétation complètement différente en dimension quatre. Par exemple, nous avons vu que la charge électrique en dimension six se comporte comme une constante cosmologique effective dans l'exemple de la corde noire, tandis que l'inclusion de plusieurs champs scalaires modifie le terme de courbure dans le potentiel d'un trou noir.

\bs
Dans le chapitre suivant, nous allons revenir à la dimension quatre en incluant un champ scalaire couplé conformément à la gravitation d'Einstein et nous étudierons notamment pour cette théorie l'influence des champs axioniques, qui nous ont permis jusqu'à présent de découvrir de nouvelles solutions intéressantes.

\chapter{Trous noirs en présence d'un champ scalaire conforme\label{chapter-conforme}}	

Nous allons dans ce chapitre nous tourner vers une dernière généralisation possible de la Relativité Générale en dimension quatre. Nous porterons notre intérêt à une théorie dans laquelle un champ scalaire, responsable de la modification de la gravitation, est couplé conformément à la gravitation d'Einstein. Nous ferons des rappels sur les trous noirs connus dans cette théorie et nous étudierons l'influence de champs axioniques qui nous permettrons d'exhiber une nouvelle solution \cite{Bardoux}.  

\minitoc
	\section{Présentation de la théorie\label{conforme}}
Avant de décrire la théorie de la gravitation modifiée en présence d'un champ scalaire conforme, nous allons présenter une extension plus générale de la théorie d'Einstein via les \ita{théories tenseur-scalaire} dans lesquelles la gravitation est induite par la métrique et par un degré de liberté supplémentaire porté par un champ scalaire. Les précurseurs de ces théories sont P. Jordan, M. Fierz, C. Brans et R. Dicke.
	
			\subsection{Les théories tenseur-scalaire de la gravitation\label{tenseur-scalaire}}
Nous allons suivre les présentations pédagogiques \cite{peter2005cosmologie} et \cite{EspositoFarese:2000ij} pour décrire ces théories. Le lecteur intéressé par ce sujet pourra consulter également les articles cités dans ces références et le livre de C. Will \cite{will1993theory}. Nous allons voir qu'il existe deux types de représentations pour ces théories.
				\subsubsection{La représentation de Jordan}		
Dans la représentation de Jordan, l'action des théories tenseur-scalaire prend usuellement la forme suivante
\BE S[g_{\mu\nu},\varphi,\psi] = \frac1{16\pi G_*} \int_\mc M \sqrt{-g} \lb F(\varphi)R - Z(\varphi) g^{\mu\nu}\p_\mu\varphi\p_\nu\varphi - 2U(\varphi) \rb \dd^4 x + S_m[g_{\mu\nu},\psi]  \label{jordan-frame}\EE
où $G_*$ est la constante de gravitation "nue", $F$ et $Z$ sont des fonctions sans dimension de $\phi$ avec $F>0$ et $U$ est le potentiel du champ scalaire $\varphi$. De plus, $S_m[g_{\mu\nu},\psi]$ représente l'action des champs de matière, que nous notons collectivement $\psi$, qui sont couplés minimalement à la métrique $g_{\mu\nu}$; le champ scalaire $\varphi$ n'est pas en jeu dans cette action en particulier. Par conséquent, ces théories préservent l'universalité de la chute libre. Remarquons que par une redéfinition du champ scalaire $\varphi$, il est toujours possible de remplacer les fonctions $F$ et $Z$ par une seule fonction inconnue. Une paramétrisation souvent utilisée est celle de \ita{Brans-Dicke} dans laquelle
\BE F(\varphi)=\varphi \qquad\text{et}\qquad Z(\varphi)=\frac{\om(\varphi)}{\varphi} \ .\EE

				\subsubsection{La représentation d'Einstein}
Il existe également une autre représentation dite \ita{d'Einstein} pour décrire les théories tenseur-scalaire. Cette représentation est déduite à partir de la précédente par la transformation conforme suivante
\BE g^*_{\mu\nu} = F(\varphi) g_{\mu\nu}\EE
et en utilisant les redéfinitions suivantes
\begin{align}
\lp \frac{\dd \varphi_*}{\dd \varphi} \rp^2 &= \frac{3}{4}\lp \frac{\dd \ln F(\varphi)}{\dd \varphi} \rp^2 + \frac{Z(\varphi)}{2F(\varphi)}  \ ,\\
A(\varphi_*) &= F^{-1/2}(\varphi) \ ,\\
2V(\varphi_*) &= U(\varphi) F^{-2}(\varphi) \ .
\end{align}
De manière générale, sous une transformation conforme $\tilde{g}_{\mu\nu} = \Om^2 g_{\mu\nu}$ où $\Om$ est une fonction $\mc C^\infty$ et strictement positive, le scalaire de Ricci se transforme selon l'égalité \cite{wald1984general} : 
\BE \tilde{R} = \Om^{-2} \lb R - 6 \Box \ln \Om - 6 g^{\mu\nu} (\p_\mu \ln \Om)( \p_\nu \ln \Om )\rb \ .  \EE
Ce résultat permet alors de montrer que l'action \eqref{jordan-frame} écrite en représentation de Jordan peut se réécrire, avec ici $\Om^2=F$, sous la forme 
\BE S[g^*_{\mu\nu},\varphi_*,\psi] = \frac1{16\pi G_*} \int_\mc M \sqrt{-g_*} \lb R_* - 2 g^{\mu\nu}_*\p_\mu\varphi_*\p_\nu\varphi_* - 4V(\varphi_*) \rb \dd^4 x + S_m[A^2(\varphi_*) g^*_{\mu\nu},\psi] \ .  \EE
Toutes les quantités étoilées sont définies dans cette représentation. L'intérêt de cette représentation est d'obtenir le terme d'Einstein-Hilbert et le terme cinétique du champ scalaire afin de mettre en évidence que les degrés de liberté de spin 0 et 2 correspondent aux perturbations de $\varphi_*$ et $g_{\mu\nu}^*$ respectivement. Nous notons que le champ scalaire $\varphi_*$ couple aux champs de matière $\psi$ dans cette représentation. Introduisons pour la suite les quantités suivantes : 
\BE \al(\varphi_*)  = \frac{\dd \ln A}{\dd \varphi_*} \qquad\text{et}\qquad \be(\varphi_*) = \frac{\dd \al}{\dd \varphi_*} \ ;\EE
la Relativité Générale est ainsi le point $(\al,\be)=(0,0)$ dans l'espace des théories tenseur-scalaire.
		
				\subsubsection{La constante de gravitation}
L'action \eqref{jordan-frame} permet de définir une constante de gravitation effective :
\BE G_\text{eff} = \frac{G_*}{F} = G_* A^2 \ . \EE
Par ailleurs, il est possible de montrer que la force mesurée dans une expérience de type Cavendish entre deux masses $m_1$ et $m_2$ proches séparées d'une distance $r$ est de norme $G_\text{cav} \frac{m_1 m_2}{r^2}$ avec 
\BE G_\text{cav} = G_* A^2 \lp 1+\al^2 \rp = \frac{G_*}{F} \lp \frac{2ZF+4 (\dd F/\dd\varphi)^2}{2ZF+3 (\dd F/\dd\varphi)^2} \rp \ . \label{cavendish}\EE
Le premier terme $G_* A^2$ correspond à l'échange d'un graviton alors que le second $\al^2 G_* A^2$ correspond à l'échange d'un scalaire. Précisons que pour des distances bien supérieures à l'inverse de la masse du champ scalaire, qui est déterminée par $\left. \frac{\dd^2 V}{\dd\varphi_*}\right|_{\phi_0^*}$ au minimum $\phi_0^*$ du potentiel, la contribution du champ scalaire au niveau du potentiel newtonien est négligeable. Notons aussi que dans la paramétrisation de Brans-Dicke, l'expression de cette constante prend une forme simple
\BE G_\text{cav} = \frac{G_*}{\varphi} \frac{2\om+4}{2\om+3} \ . \label{cavendish2} \EE
N'oublions pas que les théories tenseur-scalaire sont assez contraintes par les tests de la gravitation dans le système solaire. Les paramètres dits post-newtoniens, que nous ne définissons pas volontairement ici, permettent d'obtenir aujourd'hui la contrainte suivante $\al_0^2 < 2.10^{-4}$, ce qui implique que $G_\text{cav}$ et $G_\text{eff}$ ne diffèrent que de $0.02\%$. De plus, l'observation des pulsars binaires fournit l'inégalité $\be_0 > -4,5$. L'indice $_0$ correspond ici à la valeur actuelle des quantités $\al$ et $\be$. Dans la suite, nous allons présenter une théorie spécifique parmi les théories tenseur-scalaire.

			\subsection{Théorie en présence d'un champ scalaire invariant conforme\label{conformal}}
Il est bien connu que les équations de Maxwell dans le vide sont invariantes sous une transformation conforme alors que ce n'est pas le cas pour un champ scalaire libre. En revanche, il est possible de construire une théorie tenseur-scalaire particulière pour laquelle les équations du mouvement du champ scalaire restent invariantes sous une transformation conforme avec l'action suivante 
\BE S[g_{\mu\nu},\phi,\psi] = \int_\mc M \sqrt{-g} \lp \frac{R}{16\pi G} - \frac12 \p_\al \phi \p^\al \phi - \frac1{12}R\phi^2 \rp \dd^4 x + S_m[g_{\mu\nu},\psi]  \label{conformal-action} \EE
écrite dans la représentation de Jordan. Concernant les équations du mouvement de cette théorie, la variation de l'action $S[g_{\mu\nu},\phi,\psi]$ par rapport à la métrique fournit l'équation d'Einstein
\BE G_{\al\be} = 8\pi G \lp T_{\al\be}^{(s)} + T_{\al\be}^{(m)} \rp \label{EOMgab1}\EE
avec le tenseur énergie-impulsion suivant pour la partie de l'action mettant en jeu le champ scalaire
\BE T_{\al\be}^{(s)} = \p_\al \phi \p_\be \phi - \frac12 g_{\al\be} \p_\ga \phi \p^\ga \phi + \frac16 \lp g_{\al\be}\Box - \n_\al \n_\be + G_{\al\be} \rp \phi^2 \label{EOMgab2}\EE
alors que $T_{\al\be}^{(m)} = \frac{-2}{\sqrt{-g}} \frac{\de S_m}{\de g^{\al\be}} $ désigne le tenseur énergie-impulsion de la matière. Puis, la variation de l'action $S[g_{\mu\nu},\phi,\psi]$ par rapport au champ scalaire fournit l'équation de Klein-Gordon généralisée
\BE \Box\phi = \frac16 R \phi \ . \label{Klein-Gordon}\EE
La propriété importante de \eqref{conformal-action} est que l'action $S^{(s)}=\int_\mc M \sqrt{-g} \lp - \frac12 \p_\al \phi \p^\al \phi - \frac1{12}R\phi^2 \rp \dd^4 x$ est invariante sous la transformation conforme
\begin{align}
 g_{\al\be} &\mapsto \tilde{g}_{\al\be} = \Om^2 g_{\al\be} \\
\phi &\mapsto \tilde{\phi} = \Om^{-1} \phi 
\end{align}
avec $\Om$ une fonction $\mc C^\infty$ et strictement positive comme précédemment. Puisque l'action $S^{(s)}$ est invariante conforme alors l'équation de Klein-Gordon \eqref{Klein-Gordon} est invariante aussi sous la transformation conforme précédente et nous avons la propriété intéressante $T_{\al}^{(s)\al } = 0$. Par conséquent, en l'absence de matière, le scalaire de Ricci est nul, $R=0$, ce qui conduit simplement à l'équation $\Box\phi=0$ pour le champ scalaire. Précisons également qu'un terme de la forme $\int_\mc M \sqrt{-g} \phi^4 \dd^4 x$ est invariant conforme aussi et peut être ajouté à l'action \eqref{conformal-action}.
\bs

Afin d'être complet, il est loisible de réécrire l'action \eqref{conformal-action} en utilisant la paramétrisation de Brans-Dicke. Pour cela, nous introduisons la redéfinition
\BE \varphi = 1 - \frac{4\pi G}{3}\phi^2 \ , \EE
ce qui permet de réécrire l'action sous la forme
\BE S[g_{\mu\nu},\varphi,\psi] = \frac{1}{16\pi G}\int_\mc M \sqrt{-g} \lp \varphi R - \frac{\om(\varphi)}{\varphi} \p_\al \varphi \p^\al \varphi \rp \dd^4 x + S_m[g_{\mu\nu},\psi] \EE
avec la fonction 
\BE \om(\varphi)=\frac32 \frac{\varphi}{1-\varphi} \ . \EE
La constante de gravitation mesurée dans une expérience de Cavendish est alors donnée dans ce cas par 
\BE  G_\text{cav} =\frac{4-\varphi}{3\varphi} G\EE
d'après l'équation \eqref{cavendish2}. Enfin, en effectuant la transformation conforme de la métrique $\tilde{g}_{\mu\nu}=\lp 1 - \frac{4\pi G}{3} \phi^2 \rp g_{\mu\nu}$ lorsque $\phi^2 < \frac{3}{4\pi G} $ et la redéfinition $\Ga=\sqrt{\frac{3}{4\pi G}} \mathrm{Arctanh}\lp \sqrt{\frac{4\pi G}{3}} \phi \rp$, nous obtenons la représentation d'Einstein de l'action \eqref{conformal-action} : 
\BE S[\tilde{g}_{\mu\nu},\Ga,\psi] = \frac{1}{16\pi G}\int_\mc M \sqrt{-\tilde{g}} \lp  \tilde{R} - \tilde{g}^{\al\be} \p_\al \Ga \p_\be \Ga \rp \dd^4 x + S_m[g_{\mu\nu},\psi] \ , \EE
qui se réduit à l'action d'Einstein-Hilbert en présence d'un champ scalaire libre. Dans la suite, nous allons présenter des solutions du vide, puis en présence d'une constante cosmologique et de champs axioniques.

	\section{Solutions du vide}
		\subsection{La solution de BBMB}
Une solution de la théorie \eqref{conformal-action} dans le vide est bien connue. Elle est décrite par la métrique statique et à symétrie sphérique suivante
\BE \dd s^2 = - \lp 1-\frac{m}{r} \rp^2 \dd t^2 + \frac{\dd r^2}{\lp 1-\frac{m}{r} \rp^2} + r^2 \lp \dd\theta^2 + \sin^2\theta \dd\varphi^2 \rp \label{RN-ext} \EE
et par le champ scalaire
\BE \phi = \sqrt{\frac{3}{4\pi G}} \frac{m}{r-m} \ .  \label{BBMB-scalar}\EE
Cette solution, dont la métrique est celle de Reissner-Nordstrom extrémale, a été découverte par N. Bocharova et al. \cite{Bocharova:1970} et redécouverte par J. Bekenstein \cite{Bekenstein:1974sf} au début des années 70. Elle peut être facilement généralisée en présence de l'interaction électromagnétique en ajoutant à l'action \eqref{conformal-action} le terme de Maxwell \BE - \frac1{8\pi} \int_\mc M \mc F \w \star \mc F \EE
où $\mc F$ est la 2-forme de Faraday. Ainsi, la métrique est toujours décrite par \eqref{RN-ext} et par les champs suivants
\BE \phi = \sqrt{\frac{3}{4\pi G}} \frac{\sqrt{m- G q^2}}{r-m} \qquad\text{et}\qquad \mc F = q \dd r \w \dd t \ . \EE
La généralisation en présence d'une charge magnétique est triviale et se trouve dans \cite{Virbhadra:1993st}. Au premier abord, cette solution peut sembler pathologique à cause de la divergence du champ scalaire sur l'horizon, localisé à la coordonnée radiale $r=m$. Néanmoins, en suivant une suggestion de B. DeWitt, J. Bekenstein étudia dans \cite{Bekenstein:1975ts} les géodésiques d'une particule-test de masse $\mu$ et couplée au champ scalaire avec une action de la forme
\BE S = - \int_P^Q (\mu + f \phi) \sqrt{- g_{\mu\nu} \frac{\dd x^\mu}{\dd \la} \frac{\dd x^\nu}{\dd \la} } \dd \la \EE
où $f$ est une constante de couplage et $x^\mu(\la)$ désigne la trajectoire de la géodésique paramétrisée par $\la$ entre deux points $P$ et $Q$ de l'espace-temps. Il montra ainsi que le temps propre d'une telle particule est proportionnel à $\ln(r-m)$, c'est-à-dire qu'une particule-test ne peut s'approcher de l'horizon que de manière asymptotique. Seules les particules-tests neutres vis-à-vis du champ scalaire $\phi$ peuvent donc traverser l'horizon. Il étudia également les forces de marées et montra que ces dernières restent bornées en s'approchant de l'horizon. En ce sens, le caractère singulier de cet horizon est absent malgré la divergence du champ scalaire sur l'horizon. Concernant la stabilité de cette solution K. Bronnikov et al. affirment que la solution de BBMB est instable \cite{Bronnikov:1978mx} alors que P. McFadden et al. soutiennent le contraire \cite{McFadden:2004ni}.
\bs

Pour clore cette sous-section, notons que J. Bekenstein dans \cite{Bekenstein:1974sf} a également présenté des solutions cosmologiques, c'est-à-dire pour des métriques de type FLRW, en présence de radiation décrit par le tenseur énergie-impulsion d'un fluide parfait \eqref{fluide-parfait} pour lequel $P=\rho/3$. Pour certaines solutions, la singularité de type Big-Bang est remplacée par un "rebond". Le lecteur intéressé pourra également consulter \cite{Bekenstein:1975ww}.
		
		\subsection{Le problème de Weyl}
Avant de présenter de nouvelles solutions en présence d'une constante cosmologique et de champs axioniques, nous allons explorer le problème dit de \ita{Weyl} pour la théorie \eqref{conformal-action} dans le vide, c'est-à-dire que nous allons considérer les métriques qui sont \ita{statiques et à symétrie axiale.} Nous imposerons également ces symétries au champ scalaire $\phi$.
\bs

Tout d'abord, rappelons qu'un espace-temps est dit \ita{stationnaire} s'il existe un groupe d'isométries à un paramètre $\si_t : \mathbb{R}\times\mc M \rightarrow \mc M$ dont les orbites sont des courbes de genre-temps. Notons $\xi$ le vecteur de Killing de genre-temps qui est le générateur infinitésimal de cette transformation. De manière similaire, un espace-temps est dit \ita{à symétrie axiale} s'il existe un groupe d'isométries à un paramètre $\chi_\varphi$ dont les orbites sont des courbes fermées de genre-espace. De même, nous notons $\psi$ le vecteur de Killing de genre-espace qui génère cette transformation.

Puis, un espace-temps est dit \ita{stationnaire et à symétrie axiale} s'il possède les deux groupes d'isométries précédents et si 
\BE \si_t \circ \chi_\varphi =\chi_\varphi \circ \si_t \ . \EE
Cette dernière condition se traduit par la commutativité des vecteurs $\xi$ et $\psi$, c'est-à-dire $ [\xi,\psi]=0 $. Ce qui permet d'introduire le système de cooordonnées $(x^0=t,x^1=\varphi,x^2,x^3)$ tel que $\xi=\p_t$ et $\psi=\p_\varphi$. Par conséquent, nous pouvons écrire la métrique sous la forme suivante
\BE \dd s^2 = g_{\mu\nu}(x^2,x^3) \dd x^\mu \otimes \dd x^\nu \ . \EE
De plus, nous imposons les symétries de l'espace-temps au champ scalaire $\phi$, c'est-à-dire que $\pounds_\xi \phi =\pounds_\psi \phi =0$, par conséquent $\phi=\phi(x^2,x^3)$.
Avant de simplifier la métrique, donnons le résultat suivant démontré dans \cite{wald1984general} : 
\begin{proposition}[Frobenius] Soit $\xi$ et $\psi$ deux vecteurs de Killing qui commutent tels que\\
(i) $\xi_{[\mu} \psi_\nu \n_\rho \xi_{\si]}$ et $\xi_{[\mu} \psi_\nu \n_\rho \psi_{\si]}$ s'annulent chacun en au moins un point de l'espace-temps (ce qui est en particulier vrai si $\xi$ ou $\psi$ s'annule en un point) et  \\
(ii) $\xi^\mu R_\mu^{\ [\nu} \xi^\rho \psi^{\si]} = \psi^\mu R_\mu^{\ [\nu} \xi^\rho \psi^{\si]}=0$ \\
alors les sous-espaces bidimensionnels, des espaces tangents de chaque point de la variété, qui sont engendrés par les vecteurs orthogonals à $\xi$ et $\psi$ sont intégrables, c'est-à-dire qu'ils sont tangents à une variété bidimensionnelle.
\end{proposition}
Si l'espace-temps est asymptotiquement plat, alors il existe un axe de rotation sur lequel $\psi$ s'annule et ainsi l'hypothèse (i) est vérifiée. En revanche, la seconde condition n'est pas toujours vérifiée pour la théorie \eqref{conformal-action}. En effet, en utilisant les équations du mouvement, nous avons dans le système de coordonnées précédent
\BE \lp \phi^2 - \frac{3}{4\pi G}\rp R_i^\mu = \lp \p^\nu \phi \p_\nu \phi \rp \delta_i^\mu \qquad\text{avec}\qquad i=t,\varphi \ .\EE
Les conditions du résultat précédent sont donc vérifiées si $\phi^2 \neq \frac{3}{4\pi G}$. Nous utiliserons alors différents systèmes de coordonnées pour décrire une solution de \eqref{conformal-action}. Puis, en utilisant le fait que toute variété différentielle bidimensionelle munie d'une métrique est conformément plate, nous pouvons écrire la métrique sous la forme suivante dans les coordonnées $(t,\varphi,x^2,x^3)$ à condition de vérifier les hypothèses (i) et (ii) : 
\BE
g_{\mu\nu}=
\left(\begin{array}{cccc}
-V & W & 0 & 0 \\ W & X & 0 & 0 \\ 0 & 0 & Z^2 & 0 \\ 0 & 0 & 0 & Z^2
\end{array}\right) \label{lewis-papapetrou}
\EE
avec $V = - \xi^\mu\xi_\mu$ , $W = \psi^\mu\xi_\mu$ , $X= \psi^\mu\psi_\mu$ et $Z$ est une fonction $\mc C^\infty$  des variables $x^2$ et $x^3$. Désormais, nous restreignons notre attention au problème de Weyl, c'est-à-dire que nous imposons en plus la condition de Frobenius \eqref{frobenius}, $\xi^\flat \w \dd \xi ^\flat = 0$, pour avoir un espace-temps statique, ce qui correspond à prendre $W=0$. Nous en déduisons alors le résultat suivant
\begin{proposition}[Weyl-I] Si $\phi^2\neq\frac{3}{4\pi G}$ alors toute solution du problème de Weyl de l'action \eqref{conformal-action} s'écrit sous la forme 
\BE \dd s^2 = - e^{2\la} \dd t^2 + e^{2(\nu-\la)} \lp \dd r^2 + \dd z^2 \rp + \al^2 e^{-2\la} \dd\varphi^2 \EE
où $\la,\nu,\al$ et le champ scalaire $\phi$ sont des fonctions des coordonnées $r$ et $z$.
\end{proposition} 
Dans la suite, nous utiliserons les coordonnées complexes $u=\frac{r-iz}{2}$ et $v=\frac{r+iz}{2}$. Lorsque $\phi^2 < \frac{3}{4\pi G}$, il est judicieux d'introduire les fonctions
\begin{align}
\be &= \lp 1-\frac{4\pi G}{3}\phi^2 \rp \al \\
e^{2\om} &= \lp 1-\frac{4\pi G}{3}\phi^2\right)e^{2\la} \\
e^\chi &= \left(1-\frac{4\pi G}{3}\phi^2\right)e^\nu 
\end{align}
et la redéfinition suivante pour le champ scalaire 
\BE \ga = \sqrt{3} \mathrm{Arctanh} \lp \sqrt{\frac{4\pi G}{3}} \phi \rp \EE 
afin d'obtenir le résultat suivant
\begin{proposition}[Weyl-II]\label{Weyl-II}
Si $\phi^2 < \frac{3}{4\pi G}$ alors toute solution statique et à symétrie axiale de l'action \eqref{conformal-action} dans le vide admet la métrique suivante
\BE \mathrm{d}s^2 = \cosh^2\left(\frac{\gamma}{\sqrt{3}}\right)\left[-e^{2\omega}\mathrm{d}t^2 + 4e^{2(\chi-\omega)}\mathrm{d}u\mathrm{d}v + \beta^2 e^{-2\omega}\mathrm{d}\varphi^2\right]\label{weyl-scalar}\EE
avec les équations du mouvement
\begin{align}
\beta_{,uv} &= 0 \label{beta-harmonique}\\
\gamma_{,uv} + \frac{1}{2}\left(\gamma_{,u}\frac{\beta_{,v}}{\beta}+\gamma_{,v}\frac{\beta_{,u}}{\beta}\right) &= 0 \\
\omega_{,uv} + \frac{1}{2}\left(\omega_{,u}\frac{\beta_{,v}}{\beta}+\omega_{,v}\frac{\beta_{,u}}{\beta}\right) &= 0 \\
\chi_{,uv} + \omega_{,u}\omega_{,v} + \gamma_{,u}\gamma_{,v} &=0\\
2\frac{\beta_{,u}}{\beta}\chi_{,u} -\frac{\beta_{,uu}}{\beta} &= 2\omega_{,u}^2 + 2\gamma_{,u}^2 \qquad (u \leftrightarrow v) \ .
\end{align}
\end{proposition}
Notons que l'avant-dernière équation se déduit des autres facilement. Nous pouvons ainsi reconnaître les équations du mouvement du problème de Weyl dans le cadre de la théorie d'Einstein dans le vide en dimension 5. En effet, dans ce contexte, la métrique peut être écrite sous la forme \cite{robin,Charmousis:2006fx}
\BE \mathrm{d}s^2_5 = e^{2\chi}\beta^{-2/3}\left(\mathrm{d}r^2+\mathrm{d}z^2\right) + \beta^{2/3}\sum_{\mu=0}^2 \eta_{\mu\mu}e^{2B_\mu}\left(\mathrm{d}x^\mu\right)^2\EE
où chaque fonction dépend seulement des coordonnées $r$ et $z$, $B_0+B_1+B_2=0$ et 
\BE
\eta_{\mu\nu}=
\left(\begin{array}{ccc}
-1 & 0 & 0 \\ 0 & 1 & 0 \\ 0 & 0 & 1
\end{array}\right) \ .
\EE
Puis, l'équation d'Einstein, $R_{\mu\nu}=0$, se réduit alors au système d'équations suivant
\begin{align}
\beta_{,uv} &= 0 \\
\frac{1}{\beta}\vec{\nabla}.\left(\beta\vec{\n} B_\mu\right) &= 0\\
\chi_{,uv} + \frac{1}{2}\vec{B}_{,u}.\vec{B}_{,v} &=0\\
2\frac{\beta_{,u}}{\beta}\chi_{,u} -\frac{\beta_{,uu}}{\beta} &= \vec{B}_{,u}.\vec{B}_{,u} \qquad (u \leftrightarrow v) \ .
\end{align}
où $\vec{B}$ désigne le vecteur de composantes $B_\mu$ dans une base euclidienne orthonormée, le produit scalaire associé est noté par un point et nous avons introduit le gradient $\vec{\n}=\vec{u}_r \p_r + \vec{u}_z \p_z$ et le laplacien $\Delta=\vec{\n}.\vec{\n}$ du plan euclidien $\mathbb{R}^2$. Enfin, la correspondance entre le problème de Weyl pour la théorie \eqref{conformal-action} dans le vide et le problème de Weyl en théorie d'Einstein dans le vide en dimension 5 se fait par les identifications suivantes
\BE B_0 = -\frac{2\ga}{\sqrt{3}} \qquad B_1 = \frac{\ga}{\sqrt{3}} + \om \qquad B_2 = \frac{\ga}{\sqrt{3}} - \om  \ .\EE
Par ailleurs, lorsque $\phi^2 > \frac{3}{4\pi G}$, nous introduisons, avec les \ita{mêmes notations}, les fonctions suivantes
\BE \beta = \lp\frac{4\pi G}{3}\phi^2 -1 \rp\alpha\EE
\BE e^{2\omega} = \left(\frac{4\pi G}{3}\phi^2 -1\right)e^{2\lambda} \EE
\BE e^\chi = \left(\frac{4\pi G}{3}\phi^2 -1\right)e^\nu \EE
et nous redéfinissons le champ scalaire par
\BE \gamma = \sqrt{3}\mathrm{Arctanh}\lp \sqrt{\frac{3}{4\pi G}}\frac{1}{\phi} \rp\EE
pour obtenir le résultat similaire suivant
\begin{proposition}[Weyl-III]
Si $\phi^2 > \frac{3}{4\pi G}$ alors toute solution statique et à symétrie axiale de l'action \eqref{conformal-action} dans le vide admet la métrique suivante
\BE \mathrm{d}s^2 = \sinh^2\left(\frac{\gamma}{\sqrt{3}}\right)\left[-e^{2\omega}\mathrm{d}t^2 + 4e^{2(\chi-\omega)}\mathrm{d}u\mathrm{d}v + \beta^2 e^{-2\omega}\mathrm{d}\varphi^2\right]\EE
avec les équations du mouvement
\begin{align}
\beta_{,uv} &= 0 \\
\gamma_{,uv} + \frac{1}{2}\left(\gamma_{,u}\frac{\beta_{,v}}{\beta}+\gamma_{,v}\frac{\beta_{,u}}{\beta}\right) &= 0 \\
\omega_{,uv} + \frac{1}{2}\left(\omega_{,u}\frac{\beta_{,v}}{\beta}+\omega_{,v}\frac{\beta_{,u}}{\beta}\right) &= 0 \\
\chi_{,uv} + \omega_{,u}\omega_{,v} + \gamma_{,u}\gamma_{,v} &=0\\
2\frac{\beta_{,u}}{\beta}\chi_{,u} -\frac{\beta_{,uu}}{\beta} &= 2\omega_{,u}^2 + 2\gamma_{,u}^2 \qquad (u \leftrightarrow v) \ .
\end{align}
\end{proposition}
 
		\subsection{La structure de barres}
			
			\subsubsection{La structure de barre en Relativité Générale}
C'est H. Weyl qui a trouvé la solution générale statique et à symétrie axiale de l'équation d'Einstein dans le vide \cite{Weyl:1917gp}. Dans ce paragraphe, nous allons rappeler la structure de barre utilisée en Relativité Générale en dimension 4 pour construire des solutions statiques et à symétrie axiale dans le vide. Avec de telles symétries, nous pouvons toujours réécrire la métrique sous la forme diagonale suivante
\BE \mathrm{d}s^2 = -e^{2\lambda}\mathrm{d}t^2 + e^{2(\nu-\lambda)}\left(\mathrm{d}r^2+\mathrm{d}z^2\right) + \alpha^2 e^{-2\lambda}\mathrm{d}\varphi^2 \label{weyl-RG}\EE
où les fonctions $\la$, $\nu$ et $\al$ dépendent des coordonnées $r$ et $z$. Puis, il est utile d'introduire les coordonnées complexes  $u=\frac{r-iz}{2}$ et $v=\frac{r+iz}{2}$ pour écrire l'équation d'Einstein sous la forme
\begin{align}
\alpha_{,uv} &= 0 \label{harmonique}\\
\lambda_{,uv} + \frac{1}{2}\left(\lambda_{,u}\frac{\alpha_{,v}}{\alpha}+\lambda_{,v}\frac{\alpha_{,u}}{\alpha}\right) &= 0 \label{laplace} \\
\nu_{,uv} + \lambda_{,u}\lambda_{,v} &=0\\
2\frac{\alpha_{,u}}{\alpha}\nu_{,u} -\frac{\alpha_{,uu}}{\alpha} &= 2\lambda_{,u}^2 \qquad (u \leftrightarrow v) \ . \label{quadrature}
\end{align}
Comme précédemment l'avant dernière équation se déduit des autres équations. D'après l'équation \eqref{harmonique}, nous pouvons choisir $\al=r$. Ainsi, l'équation \eqref{laplace} devient une équation de Laplace, pour le potentiel dit de Weyl $\la$, écrite en coordonnées cylindriques
\BE \la_{,rr} + \frac{1}{r}\la_{,r} + \la_{,zz} =0 \ . \EE
C'est une équation différentielle linéaire ; nous pouvons par conséquent superposer différentes solutions. Notons que dans la limite newtonienne, la fonction $\la$ correspond au potentiel newtonien. Après cela, la dernière équation \eqref{quadrature}, qui encode le caractère non linéaire de l'équation d'Einstein, se résout explicitement pour donner la fonction $\nu$.
\bs

Dans la suite, nous allons interpréter la fonction $\la$ comme une solution de l'équation de Laplace, $\Delta \la = - 4\pi \rho$ avec un terme de source $\rho$, qui représente une distribution de "matière" située sur l'axe $r=0$. Ainsi, par l'utilisation des fonctions de Green, nous pouvons montrer que le potentiel de Weyl prend la forme suivante : 
\BE \lambda(\vec{r}) = - \int\frac{\rho(\vec{x})}{||\vec{r}-\vec{x}||}\mathrm{d}^3\vec{x} \ .\EE
\bs

Avant de continuer, proposons un intermède en gravitation newtonienne.  Si $\la$ désigne le potentiel gravitationnel, il vérifie alors l'équation de Poisson : $\Delta\lambda = -4\pi\rho$ $(G=1)$ où $\rho$ désigne la densité de masse. En particulier, si nous considérons une distribution de matière localisée, dans des coordonnées cylindriques $(r,z)$, sur l'axe $r=0$ entre $z=z_1$ et $z=z_2(>z_1)$ avec une densité uniforme $\si_\la$ par unité de longueur, nous trouvons alors le potentiel
\BE \lambda(r,z) = -\sigma_\lambda\int_{z_1}^{z_2}\frac{\mathrm{d}\tilde{z}}{\sqrt{r^2+(z-\tilde{z})^2}} + \text{cst}= \sigma_\lambda\ln\left(\frac{\sqrt{r^2+(z-z_1)^2}-(z-z_1)}{\sqrt{r^2+(z-z_2)^2}-(z-z_2)}\right) +\text{cst} \EE
et si $z_2 = +\infty$ :
\BE \lambda(r,z) = \sigma_\lambda\ln\left(B\sqrt{r^2+(z-z_1)^2}-B(z-z_1)\right) +\text{cst} \EE
où $B^{-1}$ est une constante qui a la dimension d'une longueur.

			\subsubsection{La structure de barre de la solution de Schwarzschild et les multi-trous noirs}
Nous allons exposer dans ce paragraphe la structure de barre de la solution de Schwarzschild. Tout d'abord, nous redonnons la métrique de cette solution
\BE \mathrm{d}s^2 = -\left(1-\frac{2m}{\rho}\right)\mathrm{d}t^2 + \frac{\mathrm{d}\rho^2}{1-\frac{2m}{\rho}} + \rho^2\left(\mathrm{d}\theta^2+\sin^2\theta\mathrm{d}\varphi^2\right) \ . \EE
Dans le but d'écrire cette métrique sous la forme \eqref{weyl-RG}, il est naturel de l'exprimer dans les coordonnées $\left(t,u,\theta,\varphi\right)$ où $u$ est donnée par $\cosh^2(u/2) = \frac{\rho}{2m}$. Puis, le changement de coordonnées suivant
\BE z = m \cosh u \cos\theta\EE
\BE r = m \sinh u \sin\theta\EE
permet de réécrire la solution de Schwarzschild sous la forme \eqref{weyl-RG} dans les coordonnées $\left(t,r,z,\varphi\right)$ avec
\BE e^{2\lambda} = \frac{\sqrt{r^2+(z+m)^2}-(z+m)}{\sqrt{r^2+(z-m)^2}-(z-m)} \EE
et
\BE e^{2\nu} = \frac{\sqrt{r^2+(z-m)^2}\sqrt{r^2+(z+m)^2}+(z-m)(z+m)+r^2}{2\sqrt{r^2+(z-m)^2}\sqrt{r^2+(z+m)^2}} \ .\EE
Le potentiel de Weyl $\la$ est alors solution de l'équation de Laplace avec une source placée sur l'axe $r=0$ entre $z=-m$ et $z=m$ avec une densité newtonienne $\sigma_\lambda = 1/2$. Signalons que dans les coordonnées de Schwarzschild $\left(t,\rho,\theta,\varphi\right)$, cette distribution est localisée sur l'horizon des événements. \bs

Maintenant, nous allons montrer comment nous pouvons construire une solution de $n$ trous noirs \cite{multiBH} avec des masses arbitraires $(m_i)_{1\leq i \leq n}$ où $n\in\mathbb{N}^*$. A l'aide de la structure de barre de la solution de Schwarzschild, nous construisons $\la$ comme la somme de potentiels de Weyl générés par $n$ barres, qui ne se chevauchent pas, placées sur l'axe $r=0$. Puis, la fonction $\nu$ sera alors simplement déterminée par
\BE \nu_{,r} = r\left(\lambda_{,r}^2-\lambda_{,z}^2\right) \qquad\text{et}\qquad\nu_{,z} = 2r\lambda_{,r}\lambda_{,z} \ . \label{nu-schw}\EE
Plus explicitement, nous considérons $n$ barres centrées en $z=a_i$ et de longueur $2m_i$ pour $i\in\llbracket 1,n\rrbracket$. Il est ensuite judicieux d'introduire les distances suivantes 
\BE z_i^\pm = z-a_i \pm m_i \qquad\text{et}\qquad R_i^\pm  = \sqrt{r^2+\left(z_i^\pm\right)^2} \ .\EE
Ainsi, nous contruisons le potentiel de Weyl par
\BE \la = \sum_{i=1}^n \la_i \EE
avec
\BE \lambda_i = \frac{1}{2}\ln\left(\frac{R_i^+ - z_i^+}{R_i^- - z_i^-}\right) \ .\EE
Enfin, nous trouvons
\BE \nu = \sum_{1\leq i,j \leq n} \nu_{ij} + \nu_0^{(n)} \EE
avec
\BE \nu_{ij} = \frac{1}{4}\ln\left(\frac{(R_i^+ R_j^- + z_i^+ z_j^-  + r^2)(R_i^- R_j^+ + z_i^- z_j^+ + r^2)}{(R_i^+ R_j^+ + z_i^+ z_j^+ + r^2)(R_i^- R_j^- + z_i^- z_j^- + r^2)}\right) \ . \EE
Nous devons être prudent avec la fonction $\nu$ près de l'axe $r=0$. En effet, nous exigeons que la surface d'un disque infinitésimal divisée par le carré de son périmètre tend vers $\frac1{4\pi}$ sur l'axe $r=0$, ce qui se traduit par $\lim\limits_{r \to 0} \nu(r,z) = 0$. En général, il y a des singularités coniques lorsque des solutions sont superposées.

Pour $n=1$, cette condition est vérifée avec $\nu_0^{(1)}=0$ sauf sur la barre. En revanche, lorsque nous considérons une seconde barre pour modéliser un second trou noir, nous ne pouvons plus avoir un espace-temps statique avec un axe $r=0$ régulier à cause de l'attraction gravitationnelle entre les trous noirs. Considérons une première barre placée entre  $z=c_1$ et $z=c_2$ et une seconde entre $z=c_3$ et $z=c_4$ avec $c_1<c_2<c_3<c_4$. Le calcul de la fonction $\nu$ sur l'axe donne
\BE
\nu(0,z)=
\left\{\begin{aligned}
&\nu_0^{(2)} \qquad\forall z<c_1 \quad\text{ou}\quad z>c_4\\
&\nu_0^{(2)} + \ln\left(\frac{(c_4-c_1)(c_3-c_2)}{(c_3-c_1)(c_4-c_2)}\right) <\nu_0^{(2)} \qquad\forall z\in]c_2,c_3[
\end{aligned}\right. 
\EE
Ainsi, si $\nu_0^{(2)}=0$, cela signifie que l'axe $r=0$ est régulier à l'extérieur des barres alors que $\nu(0,z)<0$ entre les deux ; ce qui peut être interprété par la présence d'un support qui maintient les deux barres dans un équilibre statique. De manière similaire, si nous choisissons $\nu_0^{(2)}$ de telle façon que l'axe $r=0$ soit régulier entre les deux, il y a alors une singularité conique à l'extérieur des barres, qui peut être interprétée par la présence de deux cordes qui maintiennent les trous noirs dans une configuration statique \cite{Charmousis:2003wm}.  
\bs

Cette interprétation en termes de barre peut également être menée en présence de l'interaction électromagnétique. Ainsi, la solution de Reissner-Nordstrom \eqref{RN} est interprétée par une barre de longueur $2\sqrt{M^2-Q^2}$ \cite{Emparan:2001bb}. Par conséquent, la présence de la charge électrique diminue la taille de la barre pour une masse $M$ donnée. En particulier, la barre se réduit à un point pour le cas extrémal. A ce sujet, R. Emparan étudia notamment un "dipole" constitué de deux trous noirs extrémaux mais de charges opposées dans la théorie d'Einstein-Maxwell avec et sans la présence d'un dilaton \cite{Emparan:1999au}. 

La structure de barre permet aussi de décrire une autre solution connue en Relativité Générale : la C-métrique \cite{Kinnersley:1970zw}. Cette solution permet de décrire deux trous noirs uniformément accélérés dans des directions opposées. La force responsable de l'accélération se manifeste par la présence d'une singularité conique. La structure de barres de la C-métrique a été déterminée dans \cite{godfrey1972horizons,bonnor1983sources}. Pour l'obtenir, il suffit d'ajouter, à la barre de la solution de Schwarzschild, une barre semi-infinie de densité newtonienne $1/2$ qui ne se superpose pas à la précédente. Le potentiel de Weyl $\la$ se met alors sous la forme
\BE \la = \frac12 \ln \lp \frac{\sqrt{r^2+(z+m)^2}-(z+m)}{\sqrt{r^2+(z-m)^2}-(z-m)} \rp +  \frac12 \ln \lp A \sqrt{r^2+\lp z-\frac1{2A} \rp^2} - A\lp z-\frac1{2A} \rp \rp   \EE
où $A$ est le paramètre d'accélération. En particulier, si nous considérons exclusivement la barre semi-infinie, alors nous décrivons l'espace-temps de Rindler. Pour une dérivation plus récente de la structure de barre, nous renvoyons le lecteur vers \cite{cornish1995interpretion} et \cite{cornish1995interpretation} pour le cas neutre et le cas chargé respectivement et vers \cite{Hong:2003gx} aussi.  Par ailleurs, il est aussi possible d'ajouter une barre semi-infinie aux solutions de multi-trous noirs pour décrire un ensemble de trous noirs en accélération \cite{Dowker:2001dg}.

Le problème de Weyl peut également être posé en théorie d'Einstein en dimension $D \geq 4$, ce qui a permis de déterminer de nouvelles solutions et notamment un anneau noir en dimension 5 dont la topologie de l'horizon des événements est celle de $S^2 \times S^1$ \cite{Emparan:2001wk}.

			\subsubsection{La structure de barres de la solution de BBMB}
Dans ce paragraphe, nous allons revenir au problème de Weyl de la théorie \eqref{conformal-action} dans le vide. Nous proposons, dans ce qui suit, de déterminer la structure de barres de la solution de BBMB, puis nous exhiberons des solutions de multi-trous noirs analogues à ceux du paragraphe précédent. Pour autant que nous le sachions, ces résultats non publiés sont nouveaux. La solution de BBMB est donnée par la métrique
\BE \mathrm{d}s^2 = -\left(1-\frac{m}{\rho}\right)^2\mathrm{d}t^2 + \frac{\mathrm{d}\rho^2}{\left(1-\frac{m}{\rho}\right)^2} + \rho^2\left(\mathrm{d}\theta^2+\sin^2\theta\mathrm{d}\varphi^2\right) \EE
avec le champ scalaire
\BE \phi = \sqrt{\frac{3}{4\pi G}}\frac{m}{\rho-m} \ .\EE
Dans le cas $\rho>2m$ pour lequel $\phi\in \left] 0,\sqrt{\frac{3}{4\pi G}} \right[$, le résultat \ref{Weyl-II} est valable et il est loisible d'effectuer les changements de coordonnées $\left(t,\rho,\theta,\varphi\right) \rightarrow \left(t,u,\theta,\varphi\right)$
avec 
\BE u = \ln\left(\frac{\rho-m}{m}\right) \EE
puis $\left(t,u,\theta,\varphi\right) \rightarrow \left(t,r,z,\varphi\right)$ avec 
\begin{align}
z &= 2m \cosh u \cos\theta \\
r &= 2m \sinh u \sin\theta
\end{align}
dans le but de réécrire la métrique sous la forme \eqref{weyl-scalar}.  Par conséquent, avec le choix possible $\be=r$ d'après \eqref{beta-harmonique}, nous avons la métrique
\BE \mathrm{d}s^2 = \cosh^2\left(\frac{\gamma}{\sqrt{3}}\right)\left[-e^{2\omega}\mathrm{d}t^2 + e^{2(\chi-\omega)}\left(\mathrm{d}r^2+\mathrm{d}z^2\right) + r^2 e^{-2\omega}\mathrm{d}\varphi^2\right]\EE
avec
\BE e^{4\omega} = e^{-\frac{4}{\sqrt{3}}\ga} = \frac{\sqrt{r^2+(z+2m)^2}-(z+2m)}{\sqrt{r^2+(z-2m)^2}-(z-2m)} \EE
et
\BE e^{2\chi} = \frac{\sqrt{r^2+(z-2m)^2}\sqrt{r^2+(z+2m)^2}+(z-2m)(z+2m)+r^2}{2\sqrt{r^2+(z-2m)^2}\sqrt{r^2+(z+2m)^2}} \ .\label{chi}\EE

Puisque les fonctions $\om$ et $\ga$ satisfont des équations de Laplace, nous pouvons de nouveau développer le formalisme précédent des barres, une associée au potentiel $\om$ et l'autre à $\ga$. Ces potentiels sont ici solutions de l'équation de Laplace avec une source placée sur l'axe $r=0$ entre $z=-2m$ et $z=2m$ avec une densité newtonienne $\si_\om = 1/4$ et $\si_\ga = -\sqrt{3}/4$ pour $\om$ et $\ga$ respectivement. Signalons que dans les coordonnées $\left(t,\rho,\theta,\varphi\right)$, ces distributions sont localisées en $\rho=2m$.

Pour les autres valeurs de $\rho$, nous résumons nos résultats dans le tableau ci-dessous \ref{rod-structure-BBMB} où nous utilisons le même changement de coordonnées entre $\left(t,u,\theta,\varphi\right)$ et $\left(t,r,z,\varphi\right)$. Nous avons ainsi trois systèmes de coordonnées pour décrire la solution. Concernant la fonction $\chi$, nous trouvons la même expression que l'équation \eqref{chi} pour chaque système de coordonnées et, comme pour le cas de la solution de Schwarzschild, $\lim\limits_{r \to 0} \chi(r,z) = 0$.
\begin{table}[H]
$$
\begin{array}{|c|c|c|c|c|c|} \hline
           & \rho \rightarrow u & \text{facteur conforme} & \si_\om  & \si_\ga    & \text{lieux des barres} \\
\hline
\rho>2m    & u = \ln\left(\frac{\rho-m}{m}\right)  & \cosh^2\left(\frac{\ga}{\sqrt{3}}\right)       & \frac{1}{4}    & -\frac{\sqrt{3}}{4} & \rho=2m \\
\hline
m<\rho<2m  & u = -\ln\left(\frac{\rho-m}{m}\right) & \sinh^2\left(\frac{\ga}{\sqrt{3}}\right)       & \frac{1}{4}    & -\frac{\sqrt{3}}{4} & \rho=2m \\
\hline
0<\rho<m   & u = -\ln\left(\frac{m-\rho}{m}\right) & \sinh^2\left(\frac{\ga}{\sqrt{3}}\right)       & -\frac{1}{4}   & \frac{\sqrt{3}}{4}  & \rho=0  \\
\hline
\end{array}
$$
\caption{Structure de barres de la solution de BBMB}\label{rod-structure-BBMB}
\end{table}
\bs

De la même manière que précédemment, nous allons montrer comment construire une solution représentant $n$ trous noirs de BBMB avec des masses arbitraires $\left(m_i\right)_{1 \leq i \leq n}$ où $n\in\mathbb{N}^*$. A l'aide de la structure de barres de la solution de BBMB, nous construisons chaque potentiel $\om$ et $\ga$ comme la somme de potentiels de Weyl générés par $n$ barres, qui ne se chevauchent pas, placées sur l'axe $r=0$. Puis, la fonction $\chi$ sera alors simplement déterminée par
\BE \chi_{,r} = r\left(\omega_{,r}^2-\omega_{,z}^2 + \gamma_{,r}^2-\gamma_{,z}^2\right) \qquad\text{et}\qquad\chi_{,z} = 2r\left(\omega_{,r}\omega_{,z}+\gamma_{,r}\gamma_{,z}\right) \label{chi2} \EE
d'après le résultat \ref{Weyl-II}. Comme précédemment, nous considérons $n$ barres centrées en $z=a_i$ et de longueur $4m_i$ pour $i\in\llbracket 1,n\rrbracket$. Il est ensuite judicieux d'introduire les distances suivantes avec les \ita{mêmes notations} que le paragraphe précédent en changeant $m_i$ en $2m_i$ : 
\BE z_i^\pm = z-a_i \pm 2m_i \qquad\text{et}\qquad R_i^\pm  = \sqrt{r^2+\left(z_i^\pm\right)^2} \ . \EE
Ainsi, nous construisons les potentiels de Weyl sous la forme
\BE \omega = \sum_{i=1}^n \omega_i  \qquad\text{et}\qquad \gamma = \sum_{i=1}^n \gamma_i \EE
avec
\BE\omega = \frac{1}{4}\ln\left(\frac{R_i^+ - z_i^+}{R_i^- - z_i^-}\right)\EE
et avec la condition
\BE \forall i \in\llbracket 1,n\rrbracket \ , \frac{\gamma_i}{\sqrt{3}} = -\omega_i \ .\EE
Ainsi, les équations \eqref{chi2} deviennent
\BE
\chi_{,r} = r\left[\left(2\omega_{,r}\right)^2-\left(2\omega_{,z}\right)^2\right] \qquad\text{et}\qquad
\chi_{,z} = 2r\left(2\omega_{,r}\right)\left(2\omega_{,z}\right) \ .
\EE
Par conséquent, $2\om$ se comporte comme le potentiel de Weyl $\la$ de la solution de Schwarzschild \eqref{nu-schw}, d'où
\BE \chi = \sum_{1\leq i,j \leq n} \chi_{ij} + \chi_0^{(n)} \EE
avec
\BE \chi_{ij} = \frac{1}{4}\ln\left(\frac{(R_i^+ R_j^- + z_i^+ z_j^- + r^2)(R_i^- R_j^+ + z_i^- z_j^+ + r^2)}{(R_i^+ R_j^+ + z_i^+ z_j^+ + r^2)(R_i^- R_j^- + z_i^- z_j^- + r^2)}\right) \EE
Enfin, nous ne répétons pas la discussion sur les singularités coniques qui est exactement la même que celle du paragraphe précédent. \bs
	
Pour clore ce paragraphe, signalons qu'il existe une généralisation de la C-métrique chargée en présence d'un champ scalaire conforme et d'une constante cosmologique \cite{Charmousis:2009cm,Anabalon:2009qt}. Il est alors loisible de déterminer la structure de barres de la solution de BBMB avec un paramètre d'accélération $A$ à partir des résultats de \cite{Charmousis:2009cm}. Nous ne donnons pas les détails de cette dérivation mais simplement le résultat. Les potentiels de Weyl de cette solution de BBMB avec un paramètre d'accélération sont
\begin{align}
\om &= \frac{1}{4}\ln\left(\frac{\sqrt{r^2+(z+2m)^2}-(z+2m)}{\sqrt{r^2+(z-2m)^2}-(z-2m)}\right) + \frac{1}{2}\ln\left(A\sqrt{r^2+\left(z-\frac{1}{2A}\right)^2}-A\left(z-\frac{1}{2A}\right)\right) \\
\ga &= -\frac{\sqrt{3}}{4}\ln\left(\frac{\sqrt{r^2+(z+2m)^2}-(z+2m)}{\sqrt{r^2+(z-2m)^2}-(z-2m)}\right)  \ .
\end{align}
La structure est donc la même que celle de la solution de BBMB avec en plus une barre semi-infinie de densité newtonienne $1/2$ associée au potentiel de Weyl $\om$. En particulier, nous restaurons la structure de barres de la solution de BBMB lorsque l'accélération tend vers 0. Ce résultat n'est pas surprenant au regard de la structure de barre associée à la C-métrique en Relativité Générale que nous avons décrit dans le paragraphe précédent.
	
		\subsection{Solutions statiques et à symétrie sphérique}

			\subsubsection{Influence de la densité newtonienne dans la structure de barres}
Dans ce paragraphe, nous allons étudier l'influence de la densité newtonienne à partir de la structure de barres de la solution de BBMB. C'est-à-dire que nous allons considérer une densité newtonienne quelconque pour chaque barre sans modifier leur longueur. Dans le cas $\phi^2 < \frac{3}{4\pi G}$, nous pouvons écrire la métrique sous la forme
\BE \mathrm{d}s^2 = \cosh^2\left(\frac{\gamma}{\sqrt{3}}\right)\left[ -e^{2\omega}\mathrm{d}t^2 + e^{2(\chi-\omega)}\left(\mathrm{d}r^2+\mathrm{d}z^2\right) + r^2 e^{-2\omega}\mathrm{d}\varphi^2\right]\EE
et nous choisissons les potentiels de Weyl suivants : 
\BE \omega = \frac{\alpha}{4}\ln\left(\frac{\sqrt{r^2+(z+2m)^2}-(z+2m)}{\sqrt{r^2+(z-2m)^2}-(z-2m)}\right) \EE
\BE \gamma = -\frac{\sqrt{3}\beta}{4}\ln\left(\frac{\sqrt{r^2+(z+2m)^2}-(z+2m)}{\sqrt{r^2+(z-2m)^2}-(z-2m)}\right) \EE
de densité newtonienne $\al/4$ et $-\sqrt{3}\be/4$ respectivement. Nous restaurons alors la solution de BBMB avec le couple $(\al,\be)=(1,1)$. Dans ce cas, nous rappelons que le champ scalaire est donné par $\phi=\sqrt{\frac{3}{4\pi G}}\tanh\lp \frac{\ga}{\sqrt{3}} \rp$. Puis, il reste simplement à déterminer la fonction $\chi$ et nous trouvons
\BE \chi = \frac{\alpha^2+3\beta^2}{8}\ln\left(\frac{\sqrt{r^2+(z-2m)^2}\sqrt{r^2+(z+2m)^2}+(z-2m)(z+2m)+r^2}{2\sqrt{r^2+(z-2m)^2}\sqrt{r^2+(z+2m)^2}}\right) \ . \EE
Pour cette fonction $\chi$, nous avons toujours $\lim\limits_{r \to 0} \chi(r,z) = 0$ où $z\in\mathbb{R}\setminus[-2m,2m]$. Pour les coordonnées de Weyl $\left(t,r,z,\varphi\right)$, nous avons $t\in\mathbb{R}$, $\varphi\in[0,2\pi[$ et$(r,z)\in\mathbb{R}^2\setminus\mathscr{D}$ où $\mathscr{D}$ représente la barre $\left\{(0,z)\in\mathbb{R}^2/-2m\leq z \leq 2m\right\}$. En s'inspirant des changements de coordonnées du paragraphe précédent, nous effectuons le changement suivant
\BE z = 2m\cosh u \cos \theta \qquad\text{et}\qquad r = 2m\sinh u \sin \theta \EE
où $\theta\in[0,\pi]$ et $u>0$. En fait, la valeur $u=0$ correspond à la barre $\mathscr{D}$. Après cela, nous effectuons un dernier changement de coordonnée
\BE u = \ln\left(\frac{\rho-m}{m}\right)  \qquad\text{avec}\qquad \rho>2m \ .\EE
Ainsi, nous trouvons la métrique suivante
\begin{multline}
\mathrm{d}s^2 = -\frac{1}{4}\left[\left(1-\frac{2m}{\rho}\right)^{\frac{\alpha+\beta}{2}}+\left(1-\frac{2m}{\rho}\right)^{\frac{\alpha-\beta}{2}}\right]^2 \mathrm{d}t^2
+ \frac{1}{4}\left[\left(1-\frac{2m}{\rho}\right)^{\frac{\beta-\al}{2}}+\left(1-\frac{2m}{\rho}\right)^{\frac{-\alpha-\beta}{2}}\right]^2 \\ \times
\frac{\rho^2\left(\rho-2m\right)^2}{\left(\rho-m\right)^2}
\left[
\left[\cos^2\theta+\left[\frac{\left(\rho-m\right)^2+m^2}{\rho\left(\rho-2m\right)}\right]^2\sin^2\theta\right]^{1-\frac{\alpha^2+3\beta^2}{4}}
\left[\frac{\mathrm{d}\rho^2}{\left(\rho-m\right)^2}+\mathrm{d}\theta^2\right]
+\sin^2\theta\mathrm{d}\varphi^2
\right] \notag
\end{multline}
et le champ scalaire
\BE \phi = \sqrt{\frac{3}{4\pi G}} \frac{1-\left(1-\frac{2m}{\rho}\right)^\beta}{1+\left(1-\frac{2m}{\rho}\right)^\beta} \ .\EE
En particulier, nous avons une famille de solutions statiques et à symétrie sphérique pour la condition $\al^2+3\be^2=4$. Nous pouvons paramétriser cette famille avec le paramètre $\ep$ en le définissant par $\al=2\cos\epsilon $ et $ \beta = \frac{2}{\sqrt{3}}\sin\epsilon $. Nous verrons dans le prochain paragraphe que la condition $\alpha^2+3\beta^2=4$ n'est pas suffisante pour avoir toutes les solutions statiques à symétrie sphérique de la théorie \eqref{conformal-action} dans le vide. De cette façon, nous obtenons la forme isotropique de la métrique pour cette famille
\begin{multline}
\mathrm{d}s^2 = -\frac{1}{4}\left[\left(1-\frac{2m}{\rho}\right)^{\frac{\alpha+\beta}{2}}+\left(1-\frac{2m}{\rho}\right)^{\frac{\alpha-\beta}{2}}\right]^2 \mathrm{d}t^2 \\
+ \frac{1}{4}\left[\left(1-\frac{2m}{\rho}\right)^{\frac{-\alpha+\beta}{2}}+\left(1-\frac{2m}{\rho}\right)^{\frac{-\alpha-\beta}{2}}\right]^2
\frac{\rho^2\left(\rho-2m\right)^2}{\left(\rho-m\right)^4}
\left[
\mathrm{d}\rho^2+
\left(\rho-m\right)^2
\left(\mathrm{d}\theta^2+\sin^2\theta\mathrm{d}\varphi^2\right)
\right] . \notag
\end{multline}
La solution de BBMB correspond à $\ep=\pi/3$ alors que celle de Schwarzschild est obtenue pour $\ep=0$ avec $2m$ comme paramètre de masse.
\bs

Dans le cas $\phi^2 > \frac{3}{4\pi G}$, nous pouvons effectuer exactement la même construction en changeant le facteur conforme de la métrique par $\sinh^2\left(\frac{\gamma}{\sqrt{3}}\right)$ avec $\phi=\sqrt{\frac{3}{4\pi G}}\coth\lp \frac{\ga}{\sqrt{3}} \rp$ et en effectuant le changement de coordonnée  $u = -\ln\left(\frac{\rho-m}{m}\right)$ où $\rho\in\left]m,2m\right[$. Nous déterminons ainsi la métrique suivante dans ce cas
\begin{multline}
\mathrm{d}s^2 = -\frac{1}{4}\left[\left(\frac{2m}{\rho}-1\right)^{\frac{\alpha+\beta}{2}}-\left(\frac{2m}{\rho}-1\right)^{\frac{\alpha-\beta}{2}}\right]^2 \mathrm{d}t^2
+ \frac{1}{4}\left[\left(\frac{2m}{\rho}-1\right)^{\frac{\beta-\al}{2}}-\left(\frac{2m}{\rho}-1\right)^{\frac{-\alpha-\beta}{2}}\right]^2 \\ \times
\frac{\rho^2\left(\rho-2m\right)^2}{\left(\rho-m\right)^2}
\left[
\left[\cos^2\theta+\left[\frac{\left(\rho-m\right)^2+m^2}{\rho\left(\rho-2m\right)}\right]^2\sin^2\theta\right]^{1-\frac{\alpha^2+3\beta^2}{4}}
\left[\frac{\mathrm{d}\rho^2}{\left(\rho-m\right)^2}+\mathrm{d}\theta^2\right]
+\sin^2\theta\mathrm{d}\varphi^2
\right] \notag
\end{multline}
et le champ scalaire
\BE \phi = \sqrt{\frac{3}{4\pi G}} \frac{1+\left(\frac{2m}{\rho}-1\right)^\beta}{1-\left(\frac{2m}{\rho}-1\right)^\beta} \ .\EE
Comme précédemment, la condition $\al^2+3\be^2=4$ engendre une famille de solutions statiques à symétrie sphérique dont la métrique est sous la forme isotropique suivante
\begin{multline}
\mathrm{d}s^2 = -\frac{1}{4}\left[\left(\frac{2m}{\rho}-1\right)^{\frac{\alpha+\beta}{2}}-\left(\frac{2m}{\rho}-1\right)^{\frac{\alpha-\beta}{2}}\right]^2 \mathrm{d}t^2 \\
+ \frac{1}{4}\left[\left(\frac{2m}{\rho}-1\right)^{\frac{-\alpha+\beta}{2}}-\left(\frac{2m}{\rho}-1\right)^{\frac{-\alpha-\beta}{2}}\right]^2
\frac{\rho^2\left(\rho-2m\right)^2}{\left(\rho-m\right)^4}
\left[
\mathrm{d}\rho^2+
(\rho-m)^2\left(\mathrm{d}\theta^2+\sin^2\theta\mathrm{d}\varphi^2\right)
\right] . \notag
\end{multline}

Un autre paramètre à étudier est l'influence de la taille des barres ; cependant, nous n'avons pas abouti à des résultats intéressants.

				\subsubsection{Solutions statiques et à symétrie sphérique issues du problème de Weyl}
En partant du résultat \ref{Weyl-II} valable pour $\phi^2 < \frac{3}{4\pi G}$ avec $\be=r$, nous recherchons toutes les solutions statiques et à  symétrie sphérique. En particulier, nous souhaitons écrire la métrique dans des coordonnées isotropiques $\left(t,\rho,\theta,\varphi\right)$. Il n'est alors pas difficile de montrer que le changement de coordonnées suivant 
\BE r = \left(a\rho-\frac{b}{\rho}\right)\sin\theta \EE
\BE z = \left(a\rho-\frac{b}{\rho}\right)\cos\theta \EE
permet de déterminer toutes les solutions statiques et à symétrie sphérique de l'action \eqref{conformal-action} avec la métrique
\BE \mathrm{d}s^2 = \cosh^2\left(\frac{\gamma}{\sqrt{3}}\right)\left[-e^{2\omega}\mathrm{d}t^2 + \left(a-\frac{b}{\rho^2}\right)^2 e^{-2\omega} \left(\mathrm{d}\rho^2+\rho^2\mathrm{d}\theta^2+\rho^2\sin^2\theta\mathrm{d}\varphi^2\right) \right] \EE
où $\om$ et $\ga$ dépendent seulement de la coordonnée radiale $\rho$ et $(a,b)\neq(0,0)$. Quant aux équations du mouvement, elles deviennent simplement
\begin{align}
\left[(a\rho^2-b)\gamma_{,\rho}\right]_{,\rho} &= 0 \\
\left[(a\rho^2-b)\omega_{,\rho}\right]_{,\rho} &= 0 \\
(a \rho^2-b)^2\left(\omega_{,\rho}^2+\gamma_{,\rho}^2\right) &= 4ab \ .
\end{align}
Il est donc aisé d'intégrer ce système et de montrer que la métrique s'écrit sous la forme suivante après un changement de coordonnées
\begin{multline}
\mathrm{d}s^2 = \frac{1}{1-\psi^2}\left[-\left(\frac{\bar{\rho}-m}{\bar{\rho}+m}\right)^{2\cos\epsilon}\mathrm{d}\bar{t}^2 \right. \\ \left. +
\left(\frac{\bar{\rho}-m}{\bar{\rho}+m}\right)^{-2\cos\epsilon}\left(1-\frac{m^2}{\bar{\rho}^2}\right)^2
\left(\mathrm{d}\bar{\rho}^2+\bar{\rho}^2\mathrm{d}\theta^2+\bar{\rho}^2\sin^2\theta\mathrm{d}\varphi\right)
\right] 
\end{multline}
et le champ scalaire est
\BE \psi = \sqrt{\frac{3}{4\pi G}} \frac{1-\delta\left(\frac{\bar{\rho}-m}{\bar{\rho}+m}\right)^{\frac{2}{\sqrt{3}}\sin\epsilon}}{1+\delta\left(\frac{\bar{\rho}-m}{\bar{\rho}+m}\right)^{\frac{2}{\sqrt{3}}\sin\epsilon}}  \EE
où $m$, $\epsilon$ et $\delta$ paramètrisent cette famille de solutions. Pour le cas $\phi^2 > \frac{3}{4\pi G}$, nous avons un résultat similaire. En fait, ces solutions de la théorie \eqref{conformal-action} ont déjà été découvertes par C. Barcelo et M. Visser \cite{Barcelo:1999hq} à partir des solutions statiques et à symétrie sphérique de la théorie d'Einstein en présence d'un champ scalaire libre \cite{PhysRevLett.20.878,Wyman:1981bd,Virbhadra:1997ie} et en utilisant la transformation conforme entre ces deux théories. Néanmoins, nous pouvons décrire désormais la structure de barres de ces solutions. Cette famille est engendrée par deux barres de même longueur $4m$ avec une densité newtonienne $\frac{\cos\epsilon}{2}$ et $\frac{-\sin\epsilon}{2}$ pour $\om$ et $\ga$ respectivement. Quant au troisième paramètre $\delta$, il est donné par $\delta = e^{-\frac{2}{\sqrt{3}}\gamma_\infty}$ où $\gamma_\infty$ est une constante que nous pouvons toujours ajouter au potentiel de Weyl $\ga$ en plus du terme logarithmique associé à la barre ; alors que pour le potentiel de Weyl $\om$, il est toujours possible de prendre $\omega_\infty=0$ par un changement de coordonnées. Nous retrouvons en particulier les solutions du paragraphe précédent pour $\delta=0$.

		\subsection{Solutions stationnaires et à symétrie axiale}
			\subsubsection{La méthode d'Ernst}
		
Dans les paragraphes précédents, nous avons restreint notre attention au problème de Weyl. Désormais, nous allons considérer les solutions stationnaires à symétrie axiale de la théorie \eqref{conformal-action} dans le vide en supprimant l'hypothèse $W=0$ dans la métrique \eqref{lewis-papapetrou}. Nous trouvons ainsi un résultat similaire à celui de A. Papapetrou \cite{Papapetrou:1953zz} en Relativité Générale : 

\begin{proposition}[Papapetrou]
Si $\phi^2 < \frac{3}{4\pi G}$ alors toute solution stationnaire et à symétrie axiale de l'action \eqref{conformal-action} dans le vide admet la métrique suivante
\BE \mathrm{d}s^2 = \cosh^2\left(\frac{\gamma}{\sqrt{3}}\right)\left[-e^{2\omega}\left(\mathrm{d}t+A\mathrm{d}\varphi\right)^2 + 4e^{2(\chi-\omega)}\mathrm{d}u\mathrm{d}v + \beta^2 e^{-2\omega}\mathrm{d}\varphi^2\right]\EE
avec les équations du mouvement
\begin{align}
\beta_{,uv} &= 0 \label{papapetrou1} \\
A_{,uv} - \frac{1}{2}\left(A_{,u}\frac{\beta_{,v}}{\beta}+A_{,v}\frac{\beta_{,u}}{\beta}\right) + 2 A_{,u} \omega_{,v} + 2 A_{,v} \omega_{,u} &= 0 \label{imaginary}\\
\omega_{,uv} + \frac{1}{2}\left(\omega_{,u}\frac{\beta_{,v}}{\beta}+\omega_{,v}\frac{\beta_{,u}}{\beta}\right) + \frac{e^{4\omega}}{2\beta^2}A_{,u} A_{,v} &= 0 \label{real}\\
\chi_{,uv} + \omega_{,u}\omega_{,v} + \gamma_{,u}\gamma_{,v} + \frac{e^{4\omega}}{4\beta^2}A_{,u} A_{,v} &=0 \label{papapetrou4} \\
\gamma_{,uv} + \frac{1}{2}\left(\gamma_{,u}\frac{\beta_{,v}}{\beta}+\gamma_{,v}\frac{\beta_{,u}}{\beta}\right) &= 0 \\
2\frac{\beta_{,u}}{\beta}\chi_{,u} -\frac{\beta_{,uu}}{\beta} &= 2\omega_{,u}^2 + 2\gamma_{,u}^2 - \frac{e^{4\omega}}{2\beta^2} A_{,u}^2 \qquad (u \leftrightarrow v) \ .
\end{align}
\end{proposition}
Toutes les fonctions qui interviennent dans ce résultat ne dépendent que de $u$ et $v$ avec toujours $u=\frac{r-iz}{2}$ et $v=\frac{r+iz}{2}$. Nous pouvons également établir un résultat similaire pour $\phi^2 > \frac{3}{4\pi G}$. Comme précédemment, l'équation \eqref{papapetrou4} se déduit des autres. Dans la suite, nous allons présenter la méthode de F. Ernst \cite{Ernst:1967wx} qui permet notamment de dériver avec élégance la solution de Kerr à partir de la solution de Schwarzschild en Relativité Générale. Cette méthode se généralise aussi pour dériver la solution Kerr-Newman \cite{Ernst:1967by}. Nous avons exploré cette direction dans le but de déterminer une possible version en rotation de la solution de BBMB. Nous n'avons pas abouti à un tel résultat, néanmoins nous allons dériver une version NUT de la solution de BBMB. En suivant la méthode de F. Ernst, nous introduisons un champ $\Om(r,z)$ vérifiant
\BE A_{,r} = \beta e^{-4\omega} \Omega_{,z} \qquad\text{et}\qquad A_{,z} = -\beta e^{-4\omega} \Omega_{,r} \ .\EE
Puis, la fonction complexe $ \mathcal{E} = e^{2\omega} + i \Omega $ vérifie alors l'équation différentielle suivante
\BE \frac{1}{\beta}\vec{\nabla}.\left(\beta\vec{\nabla}\mathcal{E}\right) = \frac{\left(\vec{\nabla}\mathcal{E}\right)^2}{\mathrm{Re}\left(\mathcal{E}\right)} \label{ernst1}\EE
connue sous le nom d'\ita{équation d'Ernst} où nous rappelons que $\vec{\n}=\vec{u}_r \p_r + \vec{u}_z \p_z$ et $\Delta=\vec{\n}.\vec{\n}$ sont respectivement le gradient et le laplacien du plan euclidien $\mathbb{R}^2$. Les parties réelle et imaginaire de cette équation correspondent à l'équation \eqref{real} et à l'équation \eqref{imaginary} respectivement. Après cela, nous introduisons la fonction $\xi$ par
\BE \mc E = \frac{\xi-1}{\xi+1} \ . \EE
L'équation d'Ernst \eqref{ernst1} s'écrit alors sous la forme
\BE \left(\xi\xi^*-1 \right)\frac{1}{\beta}\vec{\nabla}.\left(\beta\vec{\nabla}\xi\right) = 2\xi^*\left(\vec{\nabla}\xi\right)^2 \label{ernst2}\EE
qui contient manifestement une invariance sous $\mathrm{U}(1)$, c'est-à-dire que si $\xi$ est une solution alors $\forall \alpha\in\mathbb{R}$,  $e^{i\alpha}\xi$  en est une aussi. Désormais, nous adoptons le choix possible $\beta=r$ d'après \eqref{papapetrou1} et nous introduisons les coordonnées $(x,y)$ définies par
\begin{align}
r &= \sigma\sqrt{x^2-1}\sqrt{1-y^2} \\
z &= \sigma x y
\end{align}
où $x$ et $y$ sont sans dimension et $\si$ est une constante qui a la dimension d'une longueur. Nous avons en particulier
\BE \mathrm{d}r^2 + \mathrm{d}z^2 = \sigma^2 (x^2-y^2) \left(\frac{\mathrm{d}x^2}{x^2-1} + \frac{\mathrm{d}y^2}{1-y^2} \right) \ .\EE
Pour la solution de BBMB, nous trouvons $\si=2m$ et 
\BE e^{2\omega} = \left(\frac{x-1}{x+1}\right)^\delta = \frac{x\pm\sqrt{x^2-1}-1}{x\pm\sqrt{x^2-1}+1}  \qquad\text{avec}\qquad \delta=1/2 \ .\EE
En Relativité Générale, nous avons un résultat similaire pour le potentiel de Weyl $\la$ avec $\delta=1$ pour la solution de Schwarzschild, ce qui correspond à $\xi=x$. Or, si $\xi(x,y)$ est une solution de \eqref{ernst2} alors $\xi(y,x)$ en est une aussi. Il est donc loisible de regarder une combinaison linéaire de ces solutions et nous pouvons montrer que $\xi = x\cos\lambda+ i y \sin\lambda$ est une solution de \eqref{ernst2} dans le cadre de la Relativité Générale. C'est ainsi que la solution de Kerr peut être construite. Pour la solution de BBMB, $\xi=x\pm\sqrt{x^2-1}$, la question est alors la suivante : existe-t-il un choix judicieux de $\xi$ pour générer une version en rotation de la solution de BBMB ? Nous n'avons pas réussi à répondre à cette question. Connaissant $\xi$, nous en déduisons $\mc E$, $e^{2\om}$ et $\Om$, puis nous déterminons la fonction $A$ par
\BE A_{,x} = \sigma(1-y^2)e^{-4\omega}\Omega_{,y} \qquad\text{et}\qquad A_{,y} = \sigma(1-x^2)e^{-4\omega}\Omega_{,x} \ . \label{rotation}\EE
Finalement, la fonction $\chi$ est déterminée par
\begin{align}
\chi_{,x} &= \frac{1-y^2}{4(x^2-y^2)e^{4\omega}}
\left[
x(x^2-1)B_{xx} -x(1-y^2)B_{yy} -2y(x^2-1)B_{xy}
\right] \label{chix}
\\
\chi_{,y} &= \frac{x^2-1}{4(x^2-y^2)e^{4\omega}}
\left[
y(x^2-1)B_{xx} -y(1-y^2)B_{yy} +2x(1-y^2)B_{xy}
\right] \label{chiy}
\end{align}
où
\BE B_{ij} = \left(e^{2\omega}\right)_{,i} \left(e^{2\omega}\right)_{,j} + \Omega_{,i} \Omega_{,j} + e^{4(\omega-\gamma)}\left(e^{2\gamma}\right)_{,i} \left(e^{2\gamma}\right)_{,j}  \ .\EE
Concernant la fonction $\ga$, il est par exemple possible de garder celle correspondant à la solution de BBMB, ce que nous ferons avec la solution du paragraphe suivant. Le lecteur intéressé par ces méthodes pourra consulter \cite{Harmark:2005vn} pour un traitement en dimensions supplémentaires par exemple.

					\subsubsection{Une version NUT de la solution de BBMB}
En utilisant la méthode d'Ernst présentée au paragraphe précédent, nous allons générer une version NUT de la solution de BBMB. Pour la solution de BBMB, nous avons vu que $\xi=x\pm\sqrt{x^2-1}$. Or, d'après l'invariance sous $\mathrm{U}(1)$ des solutions de \eqref{ernst2}, $\forall \la \in \mathbb{R}$, $\xi=e^{i\la}\lp x\pm\sqrt{x^2-1} \rp$ est une solution dont la fonction complexe associée est 
\BE \mc E = \frac{\sqrt{x^2-1}}{x+\cos\la} + i \frac{\sin\la}{x+\cos\la} \ . \EE
Les équations \eqref{rotation} permettent alors de déterminer la fonction $A$ sachant que $A=0$ lorsque $\la=0$ : 
\BE A = 2m y \sin\la \ .  \EE
Puis, nous choisissons de garder le même champ scalaire que celui de la solution de BBMB, nous avons ainsi 
\BE e^{2\ga} = \lp \frac{x+1}{x-1} \rp^{\sqrt{3}/2} \ . \EE
Par conséquent, les équations \eqref{chix}-\eqref{chiy} deviennent $\chi_{,x}=\frac{x(1-y^2)}{(x^2-1)(x^2-y^2)}$, $\chi_{,y}=\frac{y}{x^2-y^2}$ et fournissent la fonction $\chi$
\BE e^{2\chi} = \frac{x^2-1}{x^2-y^2} \ , \EE
qui est la même que celle de la solution de BBMB. Finalement, avec les changements de coordonnées $x=\cosh u $ et $y=\cos\theta$, puis avec $e^u=\frac{\rho}{m}-1$, nous obtenons la métrique suivante
\BE \mathrm{d}s^2 = - U(\rho)^{-1} \left(\mathrm{d}t + 2m\sin\lambda\cos\theta\mathrm{d}\varphi \right)^2+ U(\rho)\mathrm{d}\rho^2 + R(\rho)^2\left(\mathrm{d}\theta^2+\sin^2\theta\mathrm{d}\varphi^2\right) \EE
avec les fonctions
\BE U(\rho) = 1+2\frac{\cos\lambda}{w(\rho)}  + \frac1{w(\rho)^2} \qquad\text{et}\qquad R(\rho)^2 = m^2 \lb 1+2w(\rho)\cos\lambda  + w(\rho)^2 \rb\EE
où $ w(\rho) = \rho/m -1 $. Enfin, le changement de coordonnées $\rho\rightarrow r = \rho + m \lp \cos\la-1 \rp $ et l'introduction du paramètre NUT, $n=m\sin\la$, permet d'écrire la métrique sous une forme plus transparente
\begin{multline}
 \dd s^2 = - \frac{\lp r - \sqrt{m^2-n^2}  \rp^2}{r^2+n^2} \lp \dd t + 2n\cos\theta\dd\varphi \rp^2 \\ + \frac{r^2+n^2}{\lp r - \sqrt{m^2-n^2}  \rp^2} \dd r^2 + \lp r^2+n^2 \rp \left(\mathrm{d}\theta^2+\sin^2\theta\mathrm{d}\varphi^2\right) \label{NUT1} 
\end{multline} 
avec le champ scalaire suivant
\BE \phi = \sqrt{\frac{3}{4\pi G}} \frac{m}{r-\sqrt{m^2-n^2}} \label{NUT2} \ .\EE 
Il reste bien évidemment à analyser cette solution.
	
			\subsubsection{Une version NUT de la solution de BBMB avec constante cosmologique\label{NUT-BBMB-4}}
Nous exposons brièvement dans ce paragraphe une nouvelle solution que nous avons déterminée récemment et qui demande à être analysée dans les détails. Elle constitue la généralisation de la solution du paragraphe précédent en présence de la constante cosmologique $\La$. Nous considérons donc la théorie suivante
\BE S = \int_\mc M \sqrt{-g} \lp \frac{R-2\La}{16\pi G} - \frac1{12}R\phi^2 - \frac12 \p_a \phi \p^a \phi - \al \phi^4 \rp \dd^4 x \ . \label{Lambda-conformal-action} \EE
La variation de cette action par rapport à la métrique donne l'équation d'Einstein
\BE G_{ab} + \La g_{ab} = 8\pi G \lb \p_a \phi  \p_b \phi - \frac12 g_{ab} \p_c \phi \p^c \phi + \frac16 \lp g_{ab}\Box - \n_a \n_b + G_{ab} \rp \phi^2- \al g_{ab} \phi^4 \rb \EE
et sa variation par rapport au champ scalaire fournit l'équation du mouvement
\BE \Box\phi = \frac16 R \phi + 4\al \phi^3 \EE
qui est invariante sous une transformation conforme. En fixant la constante de couplage $\al$ à la valeur $\al = - \frac29 \pi \La G $,
la métrique
\BE \dd s^2 = - \frac{f(r)}{r^2+n^2} \lp \dd t + 2n\cos\theta\dd\varphi \rp^2 + \frac{r^2+n^2}{f(r)} \dd r^2 + \lp r^2+n^2 \rp \lp \dd\theta^2 + \sin^2\theta\dd\varphi^2 \rp \EE
où
\BE f(r) = - \frac{\La}{3} r^4 + \lp 1- 2n^2\La \rp r^2 + 2\sqrt{m^2-n^2} \lp \frac43 n^2 \La -1 \rp r - \frac{\La}{3}n^4 + (m^2-n^2)\lp 1- \frac43 n^2 \La \rp \EE
avec le champ scalaire 
\BE \phi = \sqrt{\frac{3}{4\pi G}} \frac{m}{r-\sqrt{m^2-n^2}} \EE		
est une solution de \eqref{Lambda-conformal-action}. Pour $\La=0$, nous restaurons la solution du paragraphe précédent \eqref{NUT1}-\eqref{NUT2} ; et pour le paramètre NUT nul ($n=0$), nous retrouvons la solution de C. Mart\'\i{}nez, R. Troncoso and J. Zanelli \cite{Martinez:2002ru} qui est à la généralisation de la solution de BBMB en présence d'une constante cosmologique positive.
		
		\section{Conditions de jonction pour une hypersurface de gen\-re-temps\label{jonctions}}
Nous allons dans cette section aborder un thème totalement différent de la théorie \eqref{conformal-action}. Nous proposons de dériver les conditions de jonction pour une hypersurface de genre-temps de la théorie \eqref{conformal-action}. Ce thème est présenté dans le but d'étudier l'effondrement gravitationnel d'une coquille de matière avec la géométrie et le champ scalaire de la solution de BBMB à l'extérieur de cette coquille. Nous n'avons pas abouti à un tel résultat, nous nous contenterons simplement de donner les conditions de jonction de ce problème. \bs

En électrostatique, les conditions de jonction, ou plus couramment appelées les \ita{relations de passage}, sont bien connues. Soit $\Si$ une surface conductrice de densité surfacique de charge $\si$ séparant deux régions, notées $(1)$ et $(2)$. Les relations de passage expriment alors le fait que la composante tangentielle à $\Si$ du champ électrique $\vec{E}$ est continue
\BE \lp \vec{E}_{(2)} - \vec{E}_{(1)} \rp \w  \vec{n} = \vec{0}\EE
alors que pour la composante normale à $\Si$, il y a un saut du champ électrique à travers cette surface donné par
\BE \lp \vec{E}_{(2)} - \vec{E}_{(1)} \rp . \  \vec{n} = \frac{\si}{{\epsilon_0}} \ .\EE
Cette section a donc pour but de donner l'analogue de ces relations pour la théorie \eqref{conformal-action}. Nous restaurerons les conditions de jonction bien connues de la Relativité Générale en éteignant le champ scalaire $\phi$. Avant d'aborder la dérivation qui va suivre, le lecteur pourra revenir à la sous-section \ref{hypersurface} dans laquelle nous avons présenté les quantités intrinsèques et extrinsèques qui décrivent une hypersurface $\Si$ plongée dans une variété $\mc M$. \bs

Nous avons vu dans la sous-section \ref{conformal} que la théorie \eqref{conformal-action} peut être écrite dans la représentation d'Einstein. Les conditions de jonction dans cette représentation ont déjà été données dans \cite{Barrabes:1997kk} pour des théories tenseur-scalaire. Nous allons ici suivre la présentation pédagogique d'E. Poisson \cite{poisson2004relativist} basée sur \cite{Barrabes:1991ng} pour déduire les conditions de jonction de la théorie \eqref{conformal-action} dans la représentation de Jordan. Nous considérons une hypersurface de genre-temps $\Si$ qui sépare l'espace-temps en deux régions $\mc V^+$ et $\mc V^-$. Nous notons $g_{\al\be}^\pm$ la métrique dans $\mc V^\pm$ exprimée dans un système de coordonnées $x^\al_\pm$ et $\phi^\pm$ le champ scalaire conforme de chaque région. En autorisant la présence d'une distribution de matière, une fine couche en $\Si$, nous recherchons les conditions de jonction qui relient les "sauts" du champ gravitationnel et du champ scalaire dans la direction orthogonale à $\Si$ au tenseur énergie-impulsion de la couche.

Pour cela, nous installons sur chaque côté de l'hypersurface un système de coordonnées $y^a$ et nous notons $n^\al$ le vecteur unitaire normale à $\Si$ de genre-espace pointant de $\mc V^-$ vers $\mc V^+$. Puis, nous supposons qu'un système de coordonnées continues $x^\al$ peut être introduit des deux côtés de $\Si$ dans un ouvert contenant $\Si$. Nous devons aussi introduire les vecteurs $e_a^\al = \frac{\p x^\al}{\p y^a}$ tangents aux courbes $x^\al = x^\al(y^a)$ contenues dans $\Sigma$. Ces vecteurs sont en particulier normaux à $n^\al$, c'est-à-dire $e_a^\al n_\al = 0$. De cette manière, la métrique induite sur $\Si$ est donnée par $h_{ab}=g_{\al\beta} e_a^\al e_b^\beta$ qui est un scalaire sous un changement de coordonnées d'espace-temps $x^\al \rightarrow x'^\al$. De plus, nous imaginons l'hypersurface $\Si$ percée par une congruence de géodésiques qui intersectent $\Si$ orthogonalement. Nous notons par $l$ la distance propre le long de ces géosésiques entre un point de l'espace-temps et l'hypersurface $\Si$ avec $l<0$ pour un point de $\mc V^-$ et $l>0$ pour un point de $\mc V^+$ par convention et nous supposons la distance $l$ continue en $\Si$. Un déplacement infinitésimal hors de l'hypersurface le long d'une de ces géodésiques est alors décrit par $\dd x^\al = n^\al \dd l$, ainsi $n_\al = \p_\al l $ puisque $n_\al n^\al =1$. Nous notons aussi par $\delta(l)$ la distribution de Dirac et par $\Theta(l)$ celle de Heaviside définie par $+1$ si $l>0$, 0 si $l<0$ et elle est indéterminée si $l=0$. Nous ajoutons enfin la notation $\left[ A \right] = A(\mc V^+)_{|\Sigma} - A(\mc V^-)_{|\Sigma}$ pour toute quantité tensorielle $A$. En particulier, nous avons $\left[ n^\al \right] = 0 = \left[ e^\al_a \right]$. \bs

La stratégie consiste à exprimer la métrique et le champ scalaire sous la forme
\BE g_{\al\beta} = \Theta(l)g^+_{\al\beta} + \Theta(-l)g^-_{\al\beta} \EE
\BE \phi = \Theta(l)\phi^+ + \Theta(-l)\phi^-  \EE
où tous les tenseurs sont exprimés dans les coordonnées $x^\al$. Ainsi, la dernière égalité conduit à 
\BE \p_\al\phi = \Theta(l)\p_\al\phi^+ + \Theta(-l)\p_\al\phi^- + \delta(l) n_\al \left[ \phi \right] \ . \label{dphi}\EE
En fait, si nous calculons $\p_\al\phi \p_\beta\phi$, alors le dernier terme du membre de droite \eqref{dphi} va générer un terme proportionnel à $\Theta(l) \delta(l)$ qui n'est pas une distribution. Pour cette raison, nous imposons la \textit{première condition de jonction} suivante 
\BE \left[ \phi \right] = 0 \ , \EE
qui exprime simplement la continuité du champ scalaire en $\Si$. De manière similaire, si nous calculons les symboles de Christoffel $\Ga_{\al\be}^\ga$, alors le dernier terme du membre de droite de 
\BE \p_\ga g_{\al\be} = \Theta(l)\p_\ga g_{\al\be}^+ + \Theta(-l)\p_\ga g_{\al\be}^- + \delta(l) n_\ga \left[ g_{\al\be} \right] \EE
va générer aussi un terme proportionnel à $\Theta(l) \delta(l)$ et nous devons alors imposer $\lb g_{\al\be} \rb = 0$, d'où $\lb g_{\al\beta} e_a^\al e_b^\beta \rb =0$ puisque $ \left[ e^\al_a \right]=0$, ce qui conduit à la \textit{seconde condition de jonction}
\BE \left[ h_{ab} \right] = 0 \EE
comme il est courant en Relativité Générale. Nous soulignons que ces deux conditions sont exprimées indépendamment du système de coordonnées $x^\al$ ou $x^\al_\pm$. \bs

Après cela, nous allons déterminer la contribution distributionnelle des équations du mouvement \eqref{EOMgab1}-\eqref{Klein-Gordon} en autorisant une distribution de matière $S_{\al\beta}$ en $\Sigma$, c'est-à-dire en décomposant le tenseur énergie-impulsion de la matière sous la forme
\BE T^{(m)}_{\al\beta} = \Theta(l)T^{(m)+}_{\al\beta} + \Theta(-l)T^{(m)-}_{\al\beta}  + \delta(l)S_{\al\beta} \ . \EE
Tout d'abord, la seconde condition de jonction implique que le tenseur de Riemann soit donné par
\BE R^\al_{\ \be\ga\de} = \Theta(l) R^\al_{+\be\ga\de} + \Theta(-l) R^\al_{-\be\ga\de} + \delta(l) \lp \lb \Ga^\al_{\be\delta} \rb n_\ga - \lb \Ga^\al_{\be\ga} \rb n_\delta \rp \ . \EE
Or, puisque la discontinuité de $\p_\gamma g_{\al\beta}$ est nécessairement orthogonale à l'hypersurface $\Si$, il existe un champ tensoriel $\kappa_{\al\beta}$ tel que $\left[ \p_\gamma g_{\al\beta} \right] = \kappa_{\al\beta} n_\gamma $. Après un peu d'algèbre, nous trouvons que la partie distributionnelle du tenseur d'Einstein est $\frac{1}{2}\delta(l)E_{\al\beta}$ avec
\BE E_{\al\beta} = \kappa_{\mu\beta} n_\al n^\mu + \kappa_{\al\mu} n_\beta n^\mu - \kappa n_\al n_\beta - \kappa_{\al\beta} - g_{\al\beta} \left( \kappa_{\mu\nu} n^\mu n^\nu -\kappa \right) \EE
où $\kappa=\kappa_\al^\al$. Nous précisons que $E_{\al\beta}$ est tangent à $\Si$ puisque $E_{\al\beta} n^\al=0$. Par conséquent, nous avons la décomposition suivante $E^{\al\beta}=E^{ab}e_a^\al e_b^\beta$ où $E_{ab}=E_{\al\beta} e^\al_a e^\beta_b $ est un 3-tenseur symétrique. Puis, en introduisant la courbure extrinsèque, que nous avons définie dans la sous-section \ref{hypersurface}, 
\BE K_{ab} = \nabla_{(\al} n_{\beta)} e^\al_a e^\beta_b  = \frac{1}{2}\left( \pounds_n g_{\al\beta} \right) e^\al_a e^\beta_b  \ , \EE
nous trouvons 
\BE E_{ab} =  2\left[ K \right] h_{ab} - 2\left[ K_{ab} \right] \EE
où $K=K_a^a$. Par ailleurs, concernant la contribution distributionnelle provenant de l'équation d'Einstein \eqref{EOMgab1}-\eqref{EOMgab2}, il y a seulement le terme $\left(g_{\al\beta}\Box - \n_\al \n_\beta \right)\phi^2$ qui fournit un terme distributionnel sous la forme $\delta(l) F_{\al\beta}$ avec $F_{\al\beta}=\left( g_{\al\beta}-n_\al n_\beta \right) \left[ \pounds_n \phi^2 \right]$. Comme précédemment, le tenseur $F_{\al\beta}$ est tangent à $\Si$, ainsi nous avons la décomposition suivante $F^{\al\beta}=F^{ab}e_a^\al e_b^\beta$ où $F_{ab}=F_{\al\beta} e^\al_a e^\beta_b $ est un 3-tenseur symétrique qui est égal à 
\BE F_{ab} = h_{ab} \left[ \pounds_n \phi^2 \right] \ . \EE
En collectant toutes les parties distributionnelles de l'équation \eqref{EOMgab1}-\eqref{EOMgab2}, nous obtenons la \textit{troisième condition de jonction}
\BE \left( 1-\frac{4\pi G}{3}\phi^2 \right) \left( \left[ K \right] h_{ab} - \left[ K_{ab} \right] \right)
= \frac{4\pi G}{3} \left[\pounds_n \phi^2 \right] h_{ab} + 8\pi G S_{ab}  \EE
où  $S_{ab}$ représente le tenseur énergie-impulsion de la fine couche de matière donné par $S_{ab}=S_{\al\beta} e^\al_a e^\beta_b$ puisque $S_{\al\beta}$ est tangent à $\Si$. \bs

Finalement, nous déterminons de la même façon la contribution distributionnelle de la seconde équation du mouvement \eqref{Klein-Gordon}, ce qui fournit la \ita{quatrième condition de jonction}
\BE \left[\pounds_n \phi \right]  = - \frac{\phi}{3} \left[ K \right]   \ .  \EE
Au passage, nous retrouvons bien les conditions de jonction bien connues en Relativité Générale 
\BE  \lb h_{ab} \rb = 0 \qquad\text{et}\qquad \left[ K \right] h_{ab} - \left[ K_{ab} \right] = 8\pi G S_{ab} \ .\EE

	\section{Trous noirs en présence de champs axioniques\label{sec4:axionique}}
Nous allons dans cette section présenter des trous noirs statiques et chargés en présence d'une constante cosmologique négative en dimension quatre, décrits dans \cite{Bardoux}. Ces solutions possèdent un cheveu scalaire secondaire et constituent la version plate des solutions de trous noirs de Mart\'\i nez-Troncoso-Zanelli, qui sont, jusqu'à présent, connues avec un horizon dont la géométrie intrinsèque est courbe. Notre version plate est rendue possible grâce à la présence de deux charges axioniques égales et distribuées de façon homogène pour vêtir l'horizon plat. Ces solutions sont données dans la représentation de Jordan et d'Einstein et leurs principales propriétés sont décrites. L'aspect thermodynamique de ces solutions sera reporté à la section \ref{transition de phases}. Nous exposerons essentiellement deux branches de solutions : le trou noir axionique de Reissner-Nordstrom-AdS chauve, c'est-à-dire sans champ scalaire, et le nouveau trou noir portant un cheveu scalaire secondaire. Motivé par les récentes applications dans le domaine des supraconducteurs via l'holographie, nous montrerons à la sous-section \ref{transition de phases} qu'il existe une température critique pour laquelle le trou noir chauve subit une transition de phase du second ordre pour acquérir un cheveu scalaire décrit par notre nouvelle solution lorsque la température diminue.

		\subsection{Introduction}
Nous avons expliqué à la sous-section \ref{no-hair} que les théorèmes d'unicité contraignent sévèrement la géométrie intrinsèque de l'horizon d'un trou noir en dimension quatre. En revanche, nous avons vu, à la sous-section \ref{violer}, qu'il existe, en dimension supplémentaire, des trous noirs asymptotiquement plats en rotation avec une topologie non sphérique pour l'horizon, comme les solutions d'anneaux noirs par exemple \cite{Emparan:2001wn}. La  présence d'une constante cosmologique négative pour un espace-temps de dimension quatre permet de contourner les théorèmes d'unicité en introduisant une échelle supplémentaire, autorisant ainsi la construction de trous noirs statiques dont la courbure intrinsèque de l'horizon est nulle ou négative. Ces solutions sont qualifiées de trous noirs topologiques \cite{Lemos:1994xp,Mann:1996gj,Vanzo:1997gw} et sont facilement généralisées en dimension supplémentaire \cite{Birmingham:1998nr}.

Vêtir un trou noir avec des champs de matière est difficile d'après la conjecture \textit{no-hair} initialement énoncée par J. Wheeler. L'idée générale derrière cette conjecture est que l'effondrement gravitationnel en un trou noir d'une distribution de matière conduit nécessairement à une solution stationnaire caractérisée par ses charges mesurées à l'infini, comme la masse et la charge électrique de la solution de Reissner-Nordstrom. Nous renvoyons le lecteur à la sous-section \ref{no-hair} pour plus de détails sur ce sujet. Récemment, des trous noirs axioniques, que nous avons présentés au chapitre \ref{BH in higher D}, asymptotiquement localement AdS ont été découverts \cite{Bardoux:2012aw}. Ces trous noirs possèdent un horizon des événements plat et étendu ou toroïdal et compact. Ces solutions sont bien chauves puisqu'il existe une charge définie à l'infini associée au champ axionique décrit par une 3-forme.

Construire des trous noirs chevelus est un sujet intéressant, non seulement pour tester ou contourner la conjecture \textit{no-hair}, mais aussi pour fournir de possibles applications holographiques. Une famille de trous noirs chevelus en dimension quatre, découverte initialement par C. Mart\'\i{}nez, R. Troncoso and J. Zanelli (MTZ), avec une constante cosmologique peut être obtenue avec un champ scalaire couplé minimalement \cite{Martinez:2004nb,Martinez:2006an} ou avec un champ scalaire conforme \cite{Martinez:2002ru,Martinez:2005di}. Les trous noirs MTZ n'ont pas une constante d'intégration indépendante associée au champ scalaire, qui constitue ainsi un cheveu secondaire et qui modifie indirectement la géométrie par son couplage non trivial à la gravitation. En fait, comme nous avons pu le voir au début de ce chapitre, cette solution a d'abord été découverte en l'absence de la constante cosmologique \eqref{RN-ext}-\eqref{BBMB-scalar} dans \cite{Bocharova:1970,Bekenstein:1974sf} et sa particularité principale est que le champ scalaire diverge sur l'horizon. Précisons que la solution de BBMB écrite dans la représentation d'Einstein est singulière puisque la transformation conforme pour passer d'une représentation à une autre est singulière avant d'atteindre l'horizon. La constante cosmologique fournit une échelle supplémentaire, qui permet de déplacer la singularité du champ scalaire à l'intérieur du trou noir. Une constante cosmologique positive est possible en présence d'un horizon des événements sphérique \cite{Martinez:2002ru} alors qu'une constante cosmologique négative exige un horizon hyperbolique \cite{Martinez:2005di}. Pour une discussion concernant la stabilité de ces solutions, nous renvoyons le lecteur vers \cite{Winstanley:2004ay,Harper:2003wt}. Contrairement au cas des trous noirs topologiques chauves, les solutions connues en présence d'un champ scalaire conforme avec un horizon plat sont singulières. Ceci est un problème puisque les récentes applications en AdS/CFT exigent d'utiliser des trous noirs avec un horizon plat, afin d'étudier des matériaux supraconducteurs. Les supraconducteurs à hautes températures ne peuvent être décrits par la traditionnelle théorie BCS (Bardeen-Cooper-Schrieffer) pour calculer leurs propriétés de transport alors que leurs versions duales gravitationnelles fournissent une approche prometteuse pour étudier ce problème (voir par exemple les revues \cite{Horowitz:2010gk,Hartnoll:2009sz,Herzog:2009xv}). Dans ce chapitre, nous allons contourner ce problème en exposant des trous noirs avec des horizons plats en ajoutant des champs axioniques à la théorie \eqref{conformal-action}. Nous montrerons qu'en chargeant ces trous noirs de façon homogène, des solutions régulières peuvent être construites. En effet, l'addition de deux champs axioniques distribués de manière homogène dans les directions de l'horizon engendre un terme de courbure effectif dans le potentiel du trou noir rendant l'espace-temps régulier \cite{Bardoux:2012aw}. Les solutions que nous présentons ici possèdent un horizon plat, un cheveu scalaire secondaire et partagent des propriétés quasiment identiques à celles des trous noirs avec horizons hyperboliques étudiés dans \cite{Martinez:2005di,Koutsoumbas:2009pa,Martinez:2010ti}.

Ces solutions axioniques de trous noirs chargés en présence d'un champ scalaire constituent une phase supplémentaire par rapport aux trous noirs axioniques de Reissner-Nordstrom-AdS avec horizons plats. Une question naturelle, qui a une application dans le domaine des supraconducteurs holographiques, émerge : étant donné un ensemble de conditions de bord, lequel de ces deux trous noirs domine thermodynamiquement ? Comme nous allons le voir, la situation est en fait similaire à celle des trous noirs hyperboliques \cite{Martinez:2005di,Koutsoumbas:2009pa,Martinez:2010ti}, mais il y a ici un intérêt supplémentaire pour des applications holographiques \cite{Horowitz:2010gk,Hartnoll:2009sz,Hartnoll:2008vx,Hartnoll:2008kx,Herzog:2009xv} puisque la géométrie intrinsèque de l'horizon est plate.

Dans la prochaine sous-section, nous donnerons la théorie que nous considérons et les équations du mouvement qui en résultent. Puis, à la sous-section \ref{chevelus-BH}, nous donnerons les nouveaux trous noirs axioniques avec un horizon plat en présence d'un cheveu scalaire secondaire. Leurs charges et leurs propriétés thermodynamiques seront déduites en utilisant le formalisme hamiltonien dans le chapitre \ref{thermo}, consacré à l'aspect thermodynamique des théories gravitationnelles, à la sous-section \ref{sec::ham} et l'analyse des transitions de phase aura lieu à la sous-section \ref{sec::phase}.

		\subsection{Présentation de la théorie}
Nous considérons l'action suivante
\begin{multline} 
 \mc S =\frac{1}{16\pi G} \int_{\mc M}  \sqrt{-g} \left[\left( 1-\frac{4\pi G}{3}\phi^2\right) R-2\La \right] \dd^4x \\
-  \int_{\mc M}  \sqrt{-g}\left[\frac{1}{2} \left( 1-\frac{4\pi
 G}{3}\phi^2\right)^{-1}\sum_{i=1}^2  \frac{1}{3!} H_{abc}^{(i)} H^{(i)abc}  + \frac{F_{ab}F^{ab}}{16\pi} + \frac{1}{2}\p_a \phi \p^a \phi +
 \al\phi^4  \right] \dd^4x \ . \label{action-4}
\end{multline}
Comme nous avons pu le voir au début de ce chapitre, en l'absence des champs axioniques, le couplage non trivial entre la gravitation et le champ scalaire via le terme $\frac{1}{12}\phi^2R$ permet d'avoir une équation pour le champ scalaire invariante conforme et a donné la solution de BBMB \cite{Bekenstein:1975ts,Bocharova:1970}. Par ailleurs, le facteur devant le scalaire de Ricci permet de définir une constante de gravitation effective
\BE \label{newton-4} G_\text{eff}=\frac{G}{ 1-\frac{4\pi G}{3}\phi^2} \ . \EE
L'ajout de la constante cosmologique et du terme en $\phi^4$, qui préserve toujours l'invariance conforme du champ scalaire, a notamment donné la solution de MTZ \cite{Martinez:2002ru} (en l'absence des 3-formes). Ce qui est nouveau ici est la présence des champs axioniques représentés par deux champs de 3-formes $\mc H^{(i)} = \frac{1}{3!} H^{(i)}_{abc} \dd x^a \w \dd x^b \w \dd x^c $ ($i=1,2$) issues de deux potentiels de Kalb-Ramond $\mc B^{(i)}$ tel que $\mc H^{(i)} = \dd \mc B^{(i)}$. Notons qu'ils sont également couplés non minimalement en mettant en jeu le champ scalaire $\phi$. Quant à la 2-forme $\mc F = \frac{1}{2} F_{ab} \dd x^a \w \dd x^b = \dd \mc A$, elle correspond à la 2-forme usuelle de Faraday avec son potentiel $\mc A$. Rappelons que le terme mettant en jeu $\mc F$ est invariant conforme en dimension 4.
\bs

Il est peut-être plus judicieux d'écrire l'action avec un couplage minimal pour le champ scalaire. En effectuant la transformation conforme suivante $\tilde{g}_{ab}=\left( 1 - \frac{4\pi G}{3}\phi^2 \right) g_{ab}$ et la redéfinition du champ scalaire $\Psi = \sqrt{\frac{3}{4\pi G}}\,\text{Arctanh}\left(\sqrt{\frac{4\pi G}{3}}\phi\right)$, l'action \eqref{action-4} devient
\BE \mc S = \int_{\mc M}  \sqrt{-\tilde{g}}\left[\frac{\tilde{R}}{16\pi G}   - \frac{1}{2} \sum_{i=1}^2  \frac{1}{3!} H_{abc}^{(i)} H^{(i)abc}  - \frac{F_{ab}F^{ab}}{16\pi} - \frac{1}{2}\p_a \Psi \p^a \Psi - V(\Psi)  \right] \dd^4x
\label{minimal-4}\EE
où le potentiel est donné par
\BE V(\Psi) = \frac{\La}{8\pi G}\left[ \cosh^4\left(\sqrt{\frac{4\pi G}{3}}\Psi\right) + \frac{9\al}{2\pi\La G} \sinh^4\left(\sqrt{\frac{4\pi G}{3}}\Psi\right) \right] \ .  \EE
Notons que si le facteur conforme, qui apparaît dans la constante de gravitation effective \eqref{newton-4}, n'est pas strictement positif alors les deux représentations ne sont pas équivalentes. Nous reviendrons sur cette remarque en étudiant les solutions de l'action \eqref{action-4}.
\bs

Donnons désormais les équations du mouvement dans la première représentation. La variation de \eqref{action-4} par rapport à la métrique fournit l'équation d'Einstein
\begin{multline}
\label{einstein-4}
 \left(1-\frac{4\pi G }{3}\phi^2\right)  G_{ab}  + \La g_{ab}  = 8\pi G \left(1-\frac{4\pi G }{3}\phi^2\right)^{-1} \sum_{i=1}^2 \left(\frac{1}{2}H_{acd}^{(i)} H_b^{(i)cd} - \frac{1}{12}g_{ab} H_{cde}^{(i)} H^{(i)cde} \right) \\
 + 2 G \left( F_{ac}F_b^{\ c} - \frac{1}{4}g_{ab}F_{cd} F^{cd} \right)  + 8\pi G \left(  \p_a \phi \p_b \phi - \frac{1}{2}g_{ab} \p_c \phi \p^c \phi \right) \\
 + \frac{4\pi G}{3}\left( g_{ab} \Box - \n_a \n_b \right)\phi^2 - 8\pi G \al g_{ab} \phi^4.
\end{multline}
Puis, les variations de \eqref{action-4} par rapport à $\phi$, $\mc A$ et $\mc B^{(i)}$ donnent
\BE \Box\phi = \frac{R}{6}\phi + \frac{4\pi G}{3} \phi \left( 1 - \frac{4\pi G}{3} \phi^2 \right)^{-2} \sum_{i=1}^2  \frac{1}{3!} H_{abc}^{(i)} H^{(i)abc} + 4\al \phi^3\ , \label{eqphi-4}\EE
\BE \n_a F^{ab} = 0 \qquad\text{et}\qquad  \n_a \left[ \left( 1 - \frac{4\pi G}{3} \phi^2 \right)^{-1} H^{(i)abc} \right] =0
\EE
respectivement. Une importante propriété qui découle du couplage conforme est la suivante : en déterminant le scalaire de Ricci par la trace de l'équation d'Einstein et en le remplaçant dans l'équation du mouvement du champ scalaire \eqref{eqphi-4}, nous trouvons
\BE \Box\phi = \frac{2}{3}\La\phi + 4\al \phi^3 \ . \label{boxphi-4}\EE
C'est aussi l'équation du mouvement qui émane de la théorie \eqref{action-4} en l'absence des champs axioniques. Par conséquent, nous pouvons essayer de trouver une solution en relation avec une solution chevelue de la théorie non-axionique.

			\subsection{Trous noirs axioniques chevelus\label{chevelus-BH}}
Pour une constante cosmologique négative $\La = -3/l^2$, la théorie \eqref{action-4} admet la solution
\BE \dd s^2=-V(r)\dd t^2+\frac{\dd r^2}{V(r)}+r^2\left(\dd x^2+\dd y^2\right)\qquad\text{avec}\qquad
V(r)=\frac{r^2}{l^2}-p^2\left(1+\frac{G\mu}{r}\right)^2.
\label{hairy-4}\EE
Notons que le potentiel $V(r)$ du trou noir est celui associé au trou noir hyperbolique de \cite{Martinez:2005di} même si les sections définies par $t$ et $r$ constants sont plates. Le point crucial ici est que les 3-formes génèrent un terme de courbure effectif dans le potentiel $V(r)$ \cite{Bardoux:2012aw} permettant ainsi d'obtenir une solution régulière comme nous allons le voir. Nous avons en fait besoin de deux champs axioniques pour préserver l'homogénéité de l'horizon bidimensionnel \cite{Bardoux:2012aw}. Si l'horizon est compact alors la charge axionique, reliée au paramètre $p$ comme nous le verrons à la sous-section \ref{sec::ham}, ne peut être normalisée à 1 et nous gardons ainsi cette constante tout au long de notre analyse. Concernant le secteur de la matière, le cocktail de champs est donné par\footnote{L'action \eqref{action-4} est invariante sous la symétrie $\phi\mapsto-\phi$. Nous choisirons donc un seul signe pour le champ $\phi$.}
\BE
\phi=\frac{1}{\sqrt{2\al l^2}}\,\frac{G \mu}{r+G \mu}\ , \qquad
\mc F =-\frac{q}{r^2}\,\dd t \w \dd r \ ,  \qquad
\mc H^{(i)}=-\frac{p}{\sqrt{8 \pi G}}\left( 1 - \frac{4\pi G}{3} \phi^2 \right)\dd t \w \dd r \w \dd x^i \ .
\label{hairyfields-4}\EE
En particulier, le champ scalaire est réel à condition que la constante de couplage $\al$ soit positive. De plus, les diverses constantes doivent satisfaire la relation\footnote{A l'aide de la dualité électromagnétique, nous pouvons généraliser la solution pour inclure un paramètre représentant une charge magnétique $g$. Nous obtenons alors la contrainte $q^2+g^2= p^2\mu^2G\left(\frac{2\pi G}{3\al\ell^2}-1\right)$ à la place de \eqref{q-4} avec la 2-forme de Faraday donnée par $\mc F =-\frac{q}{r^2}\,\dd t \w \dd r + g\,\dd x \w \dd y$.}
\BE q^2= p^2\mu^2G\left(\frac{2\pi G}{3\al l^2}-1\right) \label{q-4}\EE
qui réduit le nombre de constantes d'intégration indépendantes à deux. Par conséquent, cette contrainte exige l'inégalité
\BE 0<\al\leq\frac{2\pi G}{3 l^2}  \label{alpha-4}\EE
où la borne supérieure est atteinte pour le cas neutre $q=0$. Par ailleurs, puisque qu'il n'y a pas de constante d'intégration indépendante associée au champ scalaire $\phi$, la solution possède alors un cheveu scalaire secondaire. La solution décrit ainsi un trou noir asymptotiquement localement AdS avec un horizon plat et un cheveu secondaire. Il est intéressant de noter que, pour $p=0$, nous restaurons une des solutions de \cite{Martinez:2005di} avec un horizon plat. Par conséquent, le trou noir que nous présentons ici est la généralisation de cette solution.
\bs

L'analyse des horizons et de la structure causale est identique à celle donnée dans \cite{Martinez:2005di}. Nous répétons donc simplement les principaux points et nous renvoyons le lecteur à cette référence pour plus de détails et notamment pour les diagrammes de Penrose. La métrique a une singularité de courbure en $r=0$ puisque la courbure scalaire y diverge en particulier. Par ailleurs, le champ scalaire diverge en $r=-G\mu$ si $\mu$ est négatif. Si le paramètre $\mu$ est positif alors le trou noir possède un unique horizon des événements en $r_+ = \frac{|p| l}{2} \left( 1 + \sqrt{1+ \frac{4 G \mu}{|p| l} } \right)$ ; alors que si $ 0 >G \mu > -\frac{|p| l}{4} $, il y a dans ce cas trois horizons en $   r_\pm = \frac{|p| l}{2} \left( 1 \pm \sqrt{1+ \frac{4 G \mu}{|p| l} } \right) $ et en $ r_{--} = \frac{|p| l}{2} \left( -1 + \sqrt{1- \frac{4 G \mu}{|p| l} } \right) $. Dans ce dernier cas, $r_{--}$ correspond aussi à un horizon des événements et nous avons alors "un trou noir à l'intérieur d'un trou noir" \cite{Martinez:2005di}. Enfin, la divergence du champ scalaire se trouve dans la région comprise entre $r_{--}$ et $r_-$ et il existe un trou noir extrémal pour $r_\text{ext}=\frac{|p| l}{2} $. Nous avons donc la situation suivante
\BE 0<r_{--}<-G \mu<r_-<r_\text{ext}<r_+\label{bound1-4} \ . \EE
Quand $G\mu < -\frac{|p| l}{4}$, nous avons un seul horizon des événements localisé en $r=r_{--}$ mais le champ scalaire explose avant d'atteindre l'horizon ; cependant des particules-tests qui ne sont pas couplées à ce champ ne sont pas influencées par cette divergence.

Il existe aussi un point pour lequel la constante de gravitation effective diverge. Il est donné par
$r_N=G \mu \left( -1+\sqrt{\frac{2 \pi G}{3\alpha l^2}} \right)$ pour $\mu>0$ et $r_N=-G \mu \left( 1+\sqrt{\frac{2 \pi G}{3\alpha l^2}} \right)$ pour $\mu<0$. Ce point correspond à une singularité de courbure pour la solution écrite dans la représentation d'Einstein puisque le facteur conforme de la métrique est nul en ce point. En fait, nous avons $r_N<r_+$ si et seulement si
\BE - \frac{\sqrt{\frac{3\al l^2}{2\pi G}}}{\left( 1+ \sqrt{\frac{3\al l^2}{2\pi G}} \right)^2 } < \frac{G \mu }{|p| l} < \frac{\sqrt{\frac{3\al l^2}{2\pi G}}}{\left( 1- \sqrt{\frac{3\al l^2}{2\pi G}} \right)^2 }\ . \label{ghost-4} \EE
Par conséquent, seules les solutions qui vérifient ces deux inégalités sont des trous noirs dans la représentation d'Einstein \eqref{minimal-4}. Dans ce cas, la métrique de la solution est
\BE \dd s^2=\left(1-\frac{2\pi G}{3\alpha l^2}\frac{G^2\mu^2}{(r+G\mu)^2}\right)\left[-V(r)\dd t^2+\frac{\dd r^2}{V(r)}+r^2\left(\dd x^2 +\dd y^2\right)\right] \label{bhminimal-4} \EE
avec la même fonction $V(r)$ que l'équation \eqref{hairy-4}. Nous verrons à la sous-section \ref{sec::ham} que ce point $r=r_N$ correspond à un changement de signe de l'entropie des trous noirs \cite{Martinez:2005di} pour les solutions écrites dans la représentation \eqref{action-4}. \bs

Ces trous noirs chevelus possèdent deux limites intéressantes. Considérons, tout d'abord, les limites  $\al\rightarrow0$ et $\mu\rightarrow0$ en gardant le rapport $\mu^2/\al$ constant. La contrainte \eqref{q-4} fixe alors cette constante à la valeur $3 l^2q^2/(2\pi G^2p^2)$ et la solution atteint la configuration suivante
\BE \dd s^2=-\lp\frac{r^2}{ l^2}-p^2\rp\dd t^2+\lp\frac{r^2}{ l^2}-p^2\rp^{-1}\dd r^2+r^2\lp\dd x^2+\dd y^2\rp \label{alpha=0}\EE
\BE
\mc F=-\frac{q}{r^2}\,\dd t\w\dd r\ ,\qquad
\phi=\sqrt{\frac3{4\pi}}\,\frac{q}{pr}\ ,\qquad
\mc H^{(i)}=-\frac{p}{\sqrt{8\pi G}}\lp1-\frac{4\pi G}{3}\phi^2\rp \dd t\w\dd r\w\dd x^i \ .
\EE
Cette solution avec $\al=0$ décrit un trou noir asymptotiquement localement AdS avec un horizon des événements plat en $r_+=|p| l$. Le champ scalaire est singulier à l'origine ($r=0$) et la constante de gravitation effective diverge en $r_N=\sqrt G\,|q/p|$. Pour une charge électrique faible, $\sqrt G\,|q|<p^2 l$, ce point est à l'intérieur de l'horizon. En prenant une autre limite pour laquelle les champs axioniques et de Maxwell s'annulent, $p\rightarrow0$ et $q\rightarrow 0$ en gardant $q/p$ constant, la métrique \eqref{alpha=0} se réduit à la métrique de l'espace AdS dans les coordonnées de Poincaré avec un champ scalaire non trivial singulier en $r=0$ dont le tenseur énergie-impulsion est nul. Cette solution a déjà été obtenue et discutée dans \cite{Martinez:2005di}.

La seconde limite que nous pouvons prendre est $p\rightarrow0$ en gardant $\tilde\mu=p\mu$ constant. La valeur de $\tilde\mu$ est donnée par la contrainte \eqref{q-4}. La configuration des champs qui en résulte est donc
\BE \dd s^2=-\lp\frac{r^2}{l^2}+\frac{G_{\rm eff}q^2}{r^2}\rp\dd t^2+\lp\frac{r^2}{l^2}+\frac{G_{\rm eff}q^2}{r^2}\rp^{-1}\dd r^2+r^2\lp\dd x^2+\dd y^2\rp \label{p0limit}\EE
\BE
\mc F=-\frac{q}{r^2}\,\dd t\w\dd r\ ,\qquad
\phi=\frac1{\sqrt{2\al l^2}}\ ,\qquad
\mc H^{(i)}=0 \ .
\EE
Le champ scalaire constant $\phi$ entraîne une constante de gravitation effective $G_{\rm eff}$ négative puisque $\al\leq2\pi G/3l^2$. Ainsi, la métrique décrit un trou noir avec un horizon localisé à la coordonnée $r_+=(-G_{\rm eff}q^2 l^2)^{1/4}>0$ avec une singularité de courbure en $r=0$. C'est un trou noir dont le potentiel est similaire à la solution de Reissner-Nordstrom-AdS sans masse avec un cheveu secondaire sous forme d'un champ scalaire constant $\phi$. La singularité en $r=0$ est entourée par un horizon des événements au lieu d'être nue car la constante de gravitation effective est négative ici. En revanche, à cause de cela, ce trou noir n'engendre pas une nouvelle solution dans la représentation d'Einstein\footnote{C'est cependant possible d'aller dans la représentation d'Einstein en redéfinissant le champ scalaire par $\Psi = \sqrt{\frac{3}{4\pi G}}\,\mathrm{Arctanh}\left(\sqrt{\frac{4\pi G}{3}}\frac1\phi\right)$. Cela changera le potentiel $V(\Psi)$ mais engendrera surtout le mauvais signe devant le terme cinétique associé à la 2-forme de Faraday} \eqref{minimal-4}. \bs

Finalement, il y a d'autres solutions de la théorie \eqref{action-4} avec les mêmes comportements asymptotiques que nous devons prendre en compte pour l'analyse thermodynamique. En effet, il existe un phénomène de transition de phase entre le trou noir hyperbolique chevelu de \cite{Martinez:2005di} et le trou noir de  Reissner-Nordstrom-AdS hyperbolique, comme cela est détaillé dans \cite{Martinez:2010ti}. Par conséquent, il est légitime de comparer notre nouvelle solution avec le trou noir axionique de Reissner-Nordstrom-AdS trouvé dans \cite{Bardoux:2012aw} et discuté à la sous-section \ref{axionique} . Cette dernière solution de \eqref{action-4} est donnée par
\BE \dd s^2= - V(r)\,\dd t^2 + \frac{\dd r^2}{V(r)} + r^2 \lp\dd x^2+\dd y^2\rp\ ,\qquad
V(r)=\frac{r^2}{l^2}-p^2-\frac{2G\mu}{r} + \frac{G q^2}{r^2}\, \label{RN1}\EE
\BE
\phi=0\ ,\qquad
\mc F=-\frac{q}{r^2}\,\dd t\w\dd r\ ,\qquad
\mc H^{(i)}=-\frac{p}{\sqrt{8\pi G}}\,\dd t \w\dd r\w\dd x^i \ .
\label{RN2}\EE
Plus généralement, toutes solutions de la théorie d'Einstein-Maxwell avec une constante de gravitation $G$ et deux champs de 3-formes $H^{(i)}$, qui ont été étudiées dans \cite{Bardoux:2012aw},
\BE \mc S=\frac{1}{16\pi G} \int_{\mc M}  \sqrt{-g} \lb R-2\La- \frac{4\pi G}3 \lp H_{(1)}^2+H_{(2)}^2\rp-GF^2\rb \dd^4x\label{noscalar}\EE
sont solutions de la théorie \eqref{action-4}. 

Une classe de solutions plus générale peut être déterminée en exigeant un champ scalaire $\phi$ constant. L'équation \eqref{boxphi-4} fixe cette constante à être
\BE \phi=\frac{1}{\sqrt{2\al l^2}}\ .\label{constscalar}\EE
Puis, les équations du mouvement pour les champs restants se réduisent à celles de la théorie suivante
\BE \mc S =\frac{1}{16\pi G_{\rm eff}} \int_{\mc M}  \sqrt{-g} \lb R-2\La
-\frac{4\pi G_{\rm eff}}3 \left( 1-\frac{4\pi G}{3}\phi^2\right)^{-1}\lp H_{(1)}^2+H_{(2)}^2\rp-G_{\rm eff}F^2\rb \dd^4x \EE
et ainsi, toute solution de \eqref{noscalar} conduit à une nouvelle solution avec un champ scalaire constant \eqref{constscalar}, avec le remplacement $G\mapsto G_{\rm eff}$ et en redéfinissant les 3-formes. A l'aide de cette méthode, le trou noir axionique de Reissner-Nordstrom-AdS conduit à la solution
\BE
\dd s^2= - V(r)\,\dd t^2 + \frac{\dd r^2}{V(r)} + r^2 \lp\dd x^2+\dd y^2\rp \ ,\qquad
V(r)=\frac{r^2}{l^2}-p^2-\frac{2G_{\rm eff}\mu}{r} + \frac{G_{\rm eff} q^2}{r^2} \, \label{phiRN1}
\EE
\BE
\phi=\frac{1}{\sqrt{2\al l^2}}\ ,\qquad
\mc F=-\frac{q}{r^2}\,\dd t\w\dd r\ ,\qquad
\mc H^{(i)}=-\frac{p}{\sqrt{8\pi G}}\,\dd t \w\dd r\w\dd x^i \ .
\label{phiRN2}
\EE
Quand $p=\mu=0$, cette solution se réduit à la solution \eqref{p0limit}, connectant ainsi cette famille de solutions à celle constituée de trous noirs chevelus \eqref{hairy-4}. Enfin, précisons que cette solution \eqref{phiRN1}-\eqref{phiRN2} possède une constante de gravitation effective négative comme la solution \eqref{p0limit}. \bs

Dans cette section, nous avons présenté des trous noirs en présence d'un cheveu scalaire secondaire qui sont asymptotiquement localement AdS et possèdent des charges axioniques et électrique. La caractéristique principale de ces solutions est de représenter la version plate des solutions hyperboliques de MTZ \cite{Martinez:2002ru}. La géométrie intrinsèque plate de l'horizon est réalisée ici grâce aux deux champs axioniques. Au chapitre suivant, nous déterminerons les charges et les caractéristiques thermodynamiques de ces nouvelles solutions à la sous-section \ref{sec::ham}. Puis, nous étudierons la transition de phase entre ces solutions et la version axionique de la solution de Reissner-Nordstrom-AdS à la sous-section \ref{sec::phase}.

\chapter{Thermodynamique des théories gravitationnelles\label{thermo}}
Ce chapitre a pour objectif de décrire l'aspect thermodynamique des trous noirs vêtus de $p$-formes présentées au chapitre \ref{BH in higher D} et d'exposer l'existence de transition de phases pour les nouvelles solutions du chapitre précédent en présence d'un champ scalaire conforme et de deux champs axioniques. Ces deux analyses se retrouvent dans les références \cite{Bardoux:2012aw} et \cite{Bardoux} respectivement.
\minitoc
	\section{Trous noirs et thermodynamique}
		\subsection{Les lois de la mécanique des trous noirs}
Nous avons présenté au premier chapitre, à la sous-section \ref{no-hair}, des théorèmes d'unicité : partant d'un effondrement gravitationnel correspondant à un grand nombre de degrés de liberté, un état final de type trou noir est seulement décrit par un faible nombre de paramètres correspondant à des charges mesurées à l'infini en accord avec ces théorèmes d'unicité. Cette perte d'information signale un problème majeur dans la physique des trous noirs. Une résolution a été d'associer une entropie à ces objets. C'est cet aspect thermodynamique que nous allons décrire dans la suite à travers les \ita{lois de la mécanique des trous noirs}. Pour une présentation plus complète et précise, nous renvoyons le lecteur vers \cite{Wald:1999xu} par exemple. \bs

Au début des années 70, J. Bekenstein proposa \cite{Bekenstein:1973ur} que la non décroissance de l'aire d'un trou noir, démontrée par S. Hawking \cite{Hawking:1971tu}, est analogue au second principe de la thermodynamique, qui dit que, pour tout processus physique, l'entropie d'un système ne peut pas diminuer. Il proposa alors qu'il existe un concept d'entropie pour un trou noir encodé par l'aire de l'horizon de ce dernier. J. Bekenstein obtint notamment une formule analogue au premier principe de la thermodynamique pour le trou noir de Kerr \cite{Bekenstein:1973ur}. Puis, ces résultats furent érigés sous forme de lois par J. Bardeen, B. Carter et S. Hawking pour toute solution stationnaire et à symétrie axiale des équations d'Einstein contenant un trou noir en présence de matière \cite{Bardeen:1973gs}. Nous allons donner ces lois dans la suite.

		\subsubsection{La loi zéro}
Considérons un horizon de Killing $\mc H$ avec un champ de vecteur de Killing $\xi^a$ normal à $\mc H$. Puisque que $\n^a (\xi^b \xi_b)$ est aussi normal à l'horizon, ce vecteur doit être proportionnel à $\xi^a$ en chaque point de $\mc H$. Par conséquent, il existe une fonction $\ka$ définie sur $\mc H$ et supposée positive qui vérifie
\BE \n^a (\xi^b \xi_b) = - 2 \ka \xi^a \ . \label{surface-gravity-1}\EE
Cette quantité $\ka$ est appelée la \textit{gravité de surface}. La loi zéro affirme que la gravité de surface est constante. L'analogue thermodynamique de cette loi est que la température d'un corps en équilibre thermique est constante. J. Bardeen, B. Carter et S. Hawking ont démontré cette loi pour toute solution des équations d'Einstein dont la matière vérifie la condition d'énergie dominante \cite{Bardeen:1973gs}. A partir de \eqref{surface-gravity-1}, il n'est pas difficile d'obtenir une formule donnant directement la gravité de surface 
\BE \ka^2 = - \frac12 \lp \n^a \xi^b \rp \lp \n_a \xi_b \rp \ . \label{surface-gravity-2}\EE

Par exemple, pour une métrique statique de la forme
\BE \dd s^2 = - V(r)\dd t^2 + \frac{\dd r^2}{V(r)} + r^2 \si_{ij}(y^k) \dd y^i \dd y^j \ ,\EE
nous trouvons que la gravité de surface est simplement donnée par $\ka = \frac{|V'(r_h)|}2$ où $r_h$ désigne la coordonnée radiale de l'horizon de Killing. \bs

Il est relativement simple de démontrer que $\ka$ est une constante dans le cas d'un trou noir présentant une \textit{sphère de bifurcation}, qui est une surface bidimensionnelle de genre-espace sur laquelle le vecteur de Killing $\xi$ s'annule. Par exemple, pour la solution de Schwarzschild, le vecteur de Killing de genre-temps est $\xi=-\frac{U}{4M}\p_U + \frac{V}{4M}\p_V$ dans les coordonnées de Kruskal-Szekeres $(U,V)$ \eqref{Kruscal-metric} ; il est ainsi nul sur la sphère de bifurcation donnée par $(U,V)=(0,0)$. De manière générale, pour tout horizon de Killing présentant une sphère de bifurcation, les équations \eqref{surface-gravity-2} et \eqref{killing-riemann} donnent la relation
\BE 2 \ka \p_ a \ka = \lp \n^b \xi^c \rp R_{bcad} \xi^d \ .  \EE
Le caractère constant de $\ka$ se déduit alors en projetant cette relation sur les vecteurs tangents à l'horizon et en utilisant la sphère de bifurcation \cite{poisson2004relativist}.

		\subsubsection{La première loi}
La première loi de la mécanique des trous noirs fournit la variation infinitésimale $\de M$ de la masse entre deux solutions stationnaires et à symétrie axiale dont les vecteurs de Killing qui génèrent ces symétries sont notés $\p_t$ et $\p_\phi$. Nous définirons la masse $M$ d'une solution au moment d'aborder le formalisme hamiltonien à la section \ref{hamiltonian-formalism}. Dans le cadre de la Relativité Générale en présence de l'interaction électromagnétique, la première loi est donnée par
\BE \de M = \frac{\ka}{8\pi} \de A + \Om \de J + \Phi \de Q \label{first-law}\EE
où $A$ représente l'aire de l'horizon, $\Om$ est appelée la \textit{vitesse angulaire du trou noir}, $J$ représente le moment angulaire du trou noir, $Q$ est la charge électrique du trou noir et $\Phi$ son potentiel électrique associé. La constante $\Om$ apparaît dans le vecteur de Killing normal à l'horizon $\xi=\p_t + \Om \p_\phi$. Pour dériver cette loi, J. Bardeen, B. Carter et S. Hawking ont d'abord établi des formules générales pour la masse et le moment angulaire d'une solution stationnaire à symétrie axiale en présence de matière. Puis, ils ont différencié leurs résultats en présence d'un fluide parfait pour obtenir cette première loi. 

Pour le cas particulier de la famille de solutions de Reissner-Nordstrom \eqref{RN} paramétrisée par $(M,Q)$, il est rapide de dériver cette loi. L'horizon du trou noir est localisé à la coordonnée de type Schwarzschild $r=r_+=M+\sqrt{M^2-Q^2}$. La gravité de surface et l'aire de l'horizon valent $\ka=\frac{r_+ - M}{r_+^2}$ et $A=4\pi r_+^2$ respectivement. En considérant, l'aire comme une fonction des variables $M$ et $Q$, sa différentiation redonne alors \eqref{first-law} où le potentiel électrique vaut $\Phi=Q/r_+$.

Quand nous aborderons le formalisme hamiltonien à la section \ref{hamiltonian-formalism}, nous présenterons une dérivation plus générale de la première loi en suivant \cite{Wald:1993ki,Sudarsky:1992ty}. Cette loi suggère de nouveau une analogie avec le premier principe de la thermodynamique $\de E = T \delta S - P \de V + \cdots$ avec des notations évidentes.

	\subsubsection{La seconde loi}
Cette seconde loi correspond au \textit{théorème de l'aire} \cite{Hawking:1971tu} qui est le pendant du second principe de la thermodynamique. Ce théorème dit que l'aire $A$ de l'horizon des événements d'un trou noir ne peut jamais décroître dans le temps pour une solution des équations d'Einstein en présence de matière vérifiant la condition d'énergie de genre-lumière. 

Une conséquence de ce résultat est que si deux trous noirs coalescent alors l'aire de l'horizon final est supérieure à la somme des aires des horizons initiaux. 

		\subsection{Est-ce une simple analogie ?}
Nous venons de présenter une analogie entre les lois de la mécanique des trous noirs et les principes de la thermodynamique. Si nous remplaçons formellement dans le premier principe de la thermodynamique $E$ par $M$, $T$ par $\al \ka$ et $S$ par $\frac{A}{8\pi\al}$, où $\al$ est une constante, alors nous retrouvons la première loi de la mécanique des trous noirs. Il est remarquable que $E$ et $M$ représentent la même quantité physique, ce qui sera encore plus lumineux en utilisant le formalisme hamiltonien à la section \ref{hamiltonian-formalism}. 
Néanmoins, alors qu'un trou noir a une température nulle dans le cadre de la Relativité Générale puisque qu'il se comporte comme un absorbeur parfait, S. Hawking expliqua en 1975 que, par des effets quantiques de créations de particules\cite{Hawking:1974sw}, un trou noir rayonne comme un corps noir à une température 
\BE T = \frac{\ka}{2\pi} \ , \label{hawking-temperature}\EE
ce phénomène est appelé la \textit{radiation d'Hawking}. Ce qui permet ainsi de fixer la constante $\al$ introduite précédemment et d'associer une entropie $S$ au trou noir :
\BE S=\frac{A}{4} \ . \EE
Pour un trou noir de Schwarzschild de $10$ masses solaires, la température est de l'ordre de $T \simeq 6.10^{-9}K$!

La dernière étape pour montrer que la première loi de la mécanique des trous noirs n'est pas seulement une simple analogie, mais une identité, est de montrer qu'il existe une théorie dans laquelle il est possible d'interpréter l'entropie d'un trou noir et d'y associer la valeur $A/4$ en comptant le nombre de micro-états associés au trou noir. Il existe des progrès récents à ce problème à la fois en théorie des cordes pour des trous noirs extrémaux ou presque extrémaux \cite{Strominger:1996sh} et en gravité quantique à boucles dans laquelle des géométries quantiques jouent le rôle de micro-états \cite{Rovelli:1996dv,Kaul:2012pf}. Nous ne donnerons bien évidemment pas plus de détails sur ces tentatives, qui vont bien au-delà de cette thèse.
\bs

Pour clore cette section, indiquons que la seconde loi présente un défaut. Considérons de la matière qui tombe dans le trou noir. Durant ce phénomène, l'entropie initiale de la matière est perdue et il n'y a aucune compensation d'un point de vue classique. L'entropie de l'univers contenant un trou noir décroît donc dans le temps, ce qui est contraire au seconde principe de la thermodynamique. 

Cependant, si nous tenons compte des phénomènes quantiques mis en jeu durant l'évaporation d'un trou noir, il y a alors de l'entropie qui est créée hors du trou noir bien que l'aire de l'horizon diminue (ce qui ne contredit pas le théorème de l'aire étant donné que le tenseur énergie-impulsion associé à la radiation d'Hawking ne vérifie pas la condition d'énergie de genre-lumière). Par ailleurs, en jetant de la matière dans le trou noir, l'aire de l'horizon augmente d'après la première loi, même si l'entropie de la région extérieure au trou noir diminue. Ces considérations nous amènent par conséquent à reformuler la seconde loi de la façon suivante \cite{Bekenstein:1973ur,Bekenstein:1974ax}. L'\textit{entropie} dite \textit{généralisée} $S'=S+A/4$ ne décroît jamais dans le temps, pour tout processus physique,
\BE \de S' \geq 0  \ . \EE 

Dans la suite, nous allons déterminer la masse et les charges d'une solution de trou noir par une méthode langrangienne. Puis, nous dériverons de nouveau ces résultats par une méthode hamiltonienne. Nous appliquerons notamment ces méthodes aux solutions que nous avons rencontrées dans les chapitres précédents.
		\section{L'approche thermodynamique par l'intégrale de chemin}
		\subsection{En Relativité Générale}
Nous allons présenter dans cette sous-section une approche de la thermodynamique des trous noirs en utilisant l'intégrale de chemin euclidienne appliquée à la Relativité Générale. Dans cette approche, la \textit{fonction de partition canonique du champ gravitationnel} est définie comme l'intégrale de chemin euclidienne suivante \cite{Gibbons:1976ue}
\BE Z(\be) = \int \mc D[g]  \mc D[\psi]  e^{-I[g,\psi]} \ , \EE
où $I[g,\psi]$ désigne l'action euclidienne de la théorie étudiée et $\psi$ correspond à d'éventuels champs de matière. Cette somme a lieu sur l'ensemble des géométries $g$ riemaniennes vérifiant certaines conditions aux bords qui définissent précisément l'ensemble thermodynamique comme nous verrons dans la suite. Par conséquent, l'action $I[g,\psi]$ choisie doit être celle qui génère les équations du mouvement en tenant compte de ces conditions aux bords dans le principe variationnel. De plus, le temps euclidien utilisé possède une période identifiée à l'inverse de la température $\be$. 
\bs

Il est remarquable que cette identification possède une interprétation géométrique qui consiste à supprimer la  singularité conique, d'une métrique euclidienne de trou noir, située à l'horizon de Killing. En effet, considérons par exemple une classe de métriques statiques à symétrie sphérique possédant un horizon de Killing donnée par la métrique
\BE g = V(r) \dd \tau^2 + \frac{\dd r^2}{V(r)} + r^2 \lp \dd\theta^2 + \sin^2\theta\dd\phi^2 \rp  \ ,\EE
après avoir effectué une rotation de Wick $\tau=-i t$. Le vecteur de Killing de genre-temps $\p_t$ devient ainsi un vecteur de Killing de genre-espace $\p_\tau$. L'horizon de Killing est localisé à la coordonnée radiale $r_h$ vérifiant $V(r_h)=0$ et nous considérons $r \geq r_h$ pour assurer le caractère euclidien de cette solution. L'idée consiste à observer la métrique au voisinage de l'horizon ; nous effectuons pour cela le changement de coordonnée 
\BE r \rightarrow \ep(r)=\int_{r_h}^r \frac{\dd\tilde{r}}{\sqrt{V(\tilde{r})}} \ . \EE
Ainsi, un développement limité donne
\BE V(r) = \frac{\ep^2}4 \lp V'(r_h) \rp ^2 + \mc O(\ep^3) \EE
et la métrique se met alors sous la forme suivante au voisinage de l'horizon
\BE g = \dd\ep^2 + \frac{\ep^2}4 \lp V'(r_h) \rp ^2  \dd\tau^2 + r_h^2 \dd \Om^2 \ . \EE
Afin d'avoir une géométrie plate sans singularité conique, la période $\be$ de la coordonnée $\tau$ est donc fixée à la valeur
\BE \be = \frac{4\pi}{|V'(r_h)|} \ .\EE
Par ailleurs, le calcul de l'inverse de la température $\be=2\pi/\ka$ \eqref{hawking-temperature} redonne bien le même résultat. Précisons également que le développement limité précédent est valable à condition que $V'(r_h)\neq 0$ puisque la distance radiale propre $\ep$ diverge lorsque $r$ tend vers $r_h$ si $V'(r_h)=0$, ce qui est le cas pour un trou noir extrémal. L'image du cigare infini pour représenter un trou noir euclidien n'est donc pas valable pour un trou noir extrémal.\bs

Revenons à l'évaluation de la fonction de partition canonique. La contribution dominante dans l'intégrale de chemin provient des solutions des équations du mouvement. Nous allons donc approximer sa valeur par la somme des contributions classiques
\BE Z(\be)=\sum_i e^{-I\lb g^{(i)},\psi^{(i)} \rb} \EE
où le couple $\lp g^{(i)},\psi^{(i)} \rp$ désigne une solution des équations du mouvement. Cette procédure est qualifiée d'\textit{approximation semi-classique}. L'énergie libre $F$ associée à une solution, définie par $Z=e^{-\be F}$, est donc donnée par
\BE F=I/\be \EE
où $I$ désigne l'action euclidienne évaluée sur la solution étudiée. Puis, en considérant la fonction de partition comme $Z=\sum_\text{états} e^{-\be E_n}$ avec $p_n=\frac{e^{-\be E_n}}{Z}$ la probabilité pour le système d'être dans l'état $n$, il est loisible de déterminer l'énergie d'une solution de la manière suivante
\BE E = \sum_n p_n E_n = - \frac{\p \ln Z}{\p \be} = \frac{\p I}{\p\be} \ . \label{energy5}\EE
Cette énergie définit ainsi \textit{la masse} de la solution. Puis, l'entropie est obtenue par
\BE S = - \sum_n p_n \ln p_n = \be \lp  E- F \rp = \be \frac{\p I}{\p\be} - I \ . \label{entropy5}\EE
Ces résultats sont équivalents à supposer la validité de la première loi de la thermodynamique $\dd E = T \dd S$ avec $F=E-TS$ où nous avons identifié l'énergie $E$ à la masse de la solution. Ce point de vue semble être plus judicieux puisque nous ne sommes pas en mesure de définir l'état $n$ du système en l'absence d'une théorie quantique de la gravitation. \bs

Présentons explicitement cette méthode dans le vide dans le cadre de la Relativité Générale pour la solution de Schwarzschild en dimension 4 \cite{Gibbons:1976ue}. Tout d'abord, l'action euclidienne de la Relativité Générale est donnée par
\BE I = -  \frac1{16\pi} \lp \int_\mc M R \sqrt{g} \dd^4 x + 2 \int_{\p\mc M} K \sqrt{h} \dd^3 x \rp \ . \label{euclidean-action}\EE
Nous avons inclus ici le terme de Gibbons-Hawking-York \eqref{GHY} puisque nous faisons le choix de sommer, dans l'intégrale de chemin, sur les solutions  dont la métrique induite est fixe sur le bord $\p M$. Le signe moins dans \eqref{euclidean-action} provient de la rotation de Wick : un facteur $i$ est issu de la mesure d'intégration et un autre est dû au déterminant de la métrique qui est désormais positif. La frontière $\p M$ correspond ici à une section du cigare semi-infini, paramétrisée par $r=R$, qui sera par la suite envoyée à l'infini $(R\rightarrow\infty)$. Cette frontière correspond à $\varep=1$ dans les notations de l'équation \eqref{GHY}. Nous reviendrons dans la suite sur le facteur $16\pi$ dans \eqref{euclidean-action}. La métrique euclidienne de la solution de Schwarzschild, obtenue après la rotation de Wick $\tau=-i t$, est donnée par
\BE g = V(r)\dd\tau^2 + \frac{\dd r^2}{V(r)} + r^2 \lp \dd\theta^2 + \sin^2\theta\dd\phi^2 \rp \label{metric-5}\EE
où le potentiel du trou noir est $V(r)=1-2M/r$. Le temps euclidien $\tau$ est de période $\be$, qui est donnée par l'inverse de la température, ainsi $\be=8\pi M$ pour cette solution. Puisque nous travaillons dans le vide, le scalaire de Ricci est nul et l'action euclidienne $I$ est simplement déterminée par le terme de bord. Pour évaluer ce terme, nous introduisons le vecteur unitaire normal à $\p M$ donné par $u=\sqrt{V(R)} \p_r$. La trace de la courbure extrinsèque du bord $\p M$ est alors
\BE K = \n_\al u^\al = \frac{V'(R)}{2\sqrt{V(R)}} + 2\frac{\sqrt{V(R)}}{R} \ .\EE
Par conséquent, l'action euclidienne associée à la solution de Schwarzschild est 
\BE I = \frac{\be}{2} \lp 3M - 2R \rp \ . \EE

Bien évidemment le problème majeur de ce résultat est que $I$ diverge lorsque $R\rightarrow\infty$. En effet, l'action euclidienne \eqref{euclidean-action} est généralement bien définie pour des géométries spatialement compactes mais diverge pour des géométries spatialement non compactes. G. Gibbons et S. Hawking donnent alors la prescription suivante. Pour définir l'action euclidienne d'une géométrie non compacte, nous allons la comparer par rapport à celle évaluée sur une solution de référence $\lp g_0, \psi_0\rp$. Ainsi, l'action "physique" $I_P$ que nous allons considérer est 
\BE I_P \lb g, \psi \rb = I \lb g, \psi \rb  - I \lb g_0, \psi_0 \rb \ ;\EE
de cette façon, l'action physique est nulle pour la solution de référence. De plus, nous imposons que les champs $\lp g, \psi\rp$ de la solution considérée et ceux de la solution de référence $\lp g_0, \psi_0\rp$ coïncident sur la frontière $\p M$. Par conséquent, dans le cadre de la Relativité Générale, la métrique induite sur le bord $\p\mc M$ doit être la même pour la solution considérée et la solution de fond ; l'action physique est donc donnée par 
\BE I_P \lb g, \psi \rb = - \frac1{8\pi}\int_{\p\mc M}\sqrt{h}\lp K - K_0 \rp \dd^3 x \EE 
pour une solution du vide. Il existe néanmoins une ambiguïté sur le choix d'une solution de référence, puisque différentes solutions de fond sont admissibles\footnote{Un autre problème de la méthode de l'intégrale de chemin apparaît au-delà de l'approximation semi-classique. G. Gibbons, S. Hawking et M. Perry \cite{Gibbons:1978ac,Gibbons:1978ji} ont trouvé que des perturbations conformes de la métrique diminuent toujours la valeur de l'action euclidienne et rend alors l'intégrale de chemin divergente. Le lecteur trouvera plus de détails dans  la thèse de R. Monteiro \cite{Monteiro:2010cq} qui explique notamment le lien entre ce problème et l'instabilité thermodynamique locale.}. En ce qui nous concerne ici, nous allons choisir la solution asymptotique de la solution de Schwarzschild comme solution de référence, c'est-à-dire l'espace euclidien à quatre dimensions. Puisque cette solution ne présente pas de singularité conique, nous pouvons assigner au temps euclidien n'importe quelle période $\be_0$. Or, la métrique induite sur le bord $\p M$ de la solution considérée $h=V(R)\dd\tau^2+R^2\dd^2\Om$ et celle de la solution de référence $h_0=V_0(R_0)\dd\tau_0^2+R_0^2\dd^2\Om$ sont les mêmes. Ainsi, nous avons $R=R_0$ et 
\BE V(R)\be = V_0(R)\be_0  \ ,\EE
ce qui correspond à l'égalité des températures, dites de Tolman, mesurées par un détecteur statique à la coordonnée $r=R$. Ici, nous avons simplement $V_0(R)=1$, d'où $\be_0 = \be \lp 1- \frac{M}{R} + \mc O (M/R) \rp$. L'action euclidienne physique pour la solution de Schwarzschild vaut alors finalement
\BE I_P = \frac{\be M}{2} \ ,  \EE
après avoir envoyé le bord à l'infini et l'énergie libre associée est $F=M/2$. Nous en déduisons donc l'entropie de Bekenstein $S=A/4$ et l'énergie de la solution $E=M$. C'est cette dernière égalité qui justifie le choix du facteur $16\pi$ dans l'action d'Einstein-Hilbert. Nous avons déterminé ces résultats en appliquant les formules \eqref{energy5}-\eqref{entropy5} sur la famille de solutions de Schwarzschild paramétrisée par $M$, c'est-à-dire que nous avons appliqué la première loi de la thermodynamique le long de cette famille. Il existe cependant une dérivation plus générale de l'entropie en considérant des variations de $\be$ hors de la solution considérée \cite{Banados:1993qp,Carlip:1993sa}.
\bs

Pour clore cette sous-section, nous allons brièvement présenter l'influence d'une constante cosmologique négative $\La=-3/l^2$ qui induit notamment une transition de phase entre l'espace AdS et le trou noir de Schwarzschild-AdS. Ce phénomène a été étudié pour la première fois par S. Hawking et D. Page dans \cite{Hawking:1982dh}. Le trou noir de Schwarzschild-AdS est donné par la métrique \eqref{metric-5} avec le potentiel $V(r)=1 - 2M/r + r^2/l^2$. La température en fonction de la coordonnée radiale $r_h$ de l'horizon est
\BE T = \frac{3 r_h}{4\pi l^2} + \frac1{4\pi r_h} \ .\EE
Par conséquent, cette classe de trous noirs existe pour des températures supérieures à $T_\text{min}=\frac{\sqrt{3}}{2\pi l}$. Nous pouvons distinguer deux branches de solutions : de "petits" trous noirs pour $r_h<l/\sqrt{3}$ qui ressemblent aux trous noirs de Schwarzschild et de "larges" trous noirs pour $r_h>l/\sqrt{3}$, spécifiques à l'espace AdS. Par ailleurs, en effectuant le même calcul que précédemment en tenant compte de la constante cosmologique et en choisissant l'espace AdS comme espace de référence, nous déterminons l'énergie libre associée à cette classe de solutions
\BE F = \frac{r_h}{4}\lp 1 - \frac{r_h^2}{l^2} \rp \ . \EE
Il est intéressant de remarquer que la fonction $F$ est négative pour $r_h>l$. Ce qui signifie que les trous noirs "larges" sont thermodynamiquement favorisés par rapport à l'espace AdS pour une température supérieure à la \textit{température dite de Hawking-Page} $T_\text{HP}=\frac{1}{\pi l}$. Il existe ainsi une transition de phase du premier ordre à la température $T=T_\text{HP}>T_\text{min}$. Notons que ce phénomène n'est plus présent si la courbure de l'horizon n'est pas strictement positive. En effet, si nous remplaçons la 2-sphère de l'horizon par un espace à courbure constante dont la courbure scalaire vaut $2\ka$, l'énergie libre devient $ F = \frac{r_h}{4}\lp \ka - \frac{r_h^2}{l^2} \rp $.
\bs

Dans la suite, nous allons appliquer cette méthode aux solutions que nous avons rencontrées au chapitre \ref{BH in higher D} avec la prise en compte de dimensions supplémentaires et la présence de champs de matière.

		\subsection{En présence de $p$-formes\label{p-charges-1}}
Nous avons construit au chapitre \ref{BH in higher D} une large classe de trous noirs en présence d'une ou de plusieurs $p$-formes en dimension quelconque $D=n+3$. Nous adopterons donc les mêmes notations que ce chapitre \ref{BH in higher D} dans la suite. Ces trous noirs sont solutions de l'action suivante
\BE S = S_0 + \sum_{p=1}^D \sum_{k=1}^{N_p} S_M^{(p,k)} \label{action1-5}\EE
où $S_0$ est l'action d'Einstein-Hilbert avec constante cosmologique donnée par \eqref{action0}, $N_p$ est le nombre de $p$-formes $H_{[p]}^{(k)}$ et le secteur de la matière consiste en une somme de lagrangiens libres de la forme \eqref{SM} : 
\BE  S_M^{(p,k)} = -\frac{1}{16\pi G} \int_{\mc M}   \sqrt{-g} \frac{1}{2p!} \lp H_{[p]}^{(k)}\rp^2 d^Dx \ . \EE
Nous avons montré qu'il existe des solutions statiques de cette théorie avec une métrique de la forme \eqref{ansatz} dont le potentiel \eqref{V} est déterminé par la fonction
\BE  \ep(r)=\sum_{p,k} \ep_{(p,k)}(r) \qquad\text{où}\qquad\ep_{(p,k)}(r) =  \frac{\mc E^{(k)2}_{[p-2]}}{2(p-2)! r^{n-2p+5}} +  \frac{\mc B^{(k)2}_{[p]}}{2p! r^{2p-n-1}} \EE
d'après le principe de superposition exposé à la sous-section \ref{multiple} où nous rappelons que $\mc E^{(k)}_{[p-2]}$ et $\mc B^{(k)}_{[p]}$ sont les polarisations électriques et magnétiques associées à $H_{[p]}^{(k)}$ respectivement (voir \eqref{H}). Ainsi, l'expression du potentiel du trou noir est
\BE V(r) = \kappa - \frac{r_0^n}{r^n} + \frac{r^2}{ l^2} + \frac{1}{2(n+1)} \sum_{p,k} \left[ \frac{\mc E^{(k)2}_{[p-2]}}{(p-2)! (n-2p+4) r^{2(n-p+2)}} +  \frac{\mc B^{(k)2}_{[p]}}{p! (2p-n-2) r^{2(p-1)}}  \right] \ . \label{genV-5}\EE
Pour simplifier la discussion, nous avons omis les termes logarithmiques qui apparaissent quand $2p=n+4$ et $2p=n+2$ pour un espace-temps de dimension impaire. La partie électrique de la série en puissance de $r$ commence avec un terme en $1/r^{2n}$ quand $p=2$ et se termine avec un terme en $r^2$ quand $p=n+3$, c'est-à-dire comme un terme de constante cosmologique. Quant à la partie magnétique, elle commence avec un terme en $1/r^0$ quand $p=1$, comme un terme de courbure, et se termine avec un terme en $1/r^{2n}$ quand $p=n+1$. 

Si $r_0$ est suffisamment grand, la fonction $\eqref{genV-5}$ a toujours au moins une racine positive. Notons $r_h$ la plus grande racine du potentiel $V(r)$. La solution présente alors un horizon des événements en $r=r_h$ et l'espace-temps contient un trou noir. Sa température, qui est proportionnelle à la gravité de surface de l'horizon le plus extérieur, est donnée par 
\BE T=\frac{\left|V'(r_h)\right|}{4\pi} =  \frac{n\kappa}{4 \pi r_h} + \frac{(n+2)r_h}{4\pi l^2} - \frac{1}{8\pi (n+1)} \sum_{p,k} \left( \frac{\mc E^{(k)2}_{[p-2]}}{(p-2)! r_h^{2n-2p+5}} +  \frac{\mc B^{(k)2}_{[p]}}{p! r_h^{2p-1}} \right) \ . \EE
Pour des raisons de clarté, nous restreignons désormais notre attention au cas électrique et nous autorisons toutes les $p$-formes dont le rang satisfait $2p\leq n+3$. Lorsque cette inégalité est vérifiée, les composantes du champ de force correspondant, $H^{(k)}_{tr i_1 \ldots i_{p-2}}$, s'annulent à l'infini spatial, ce qui permet d'obtenir des potentiels thermodynamiques finis par la méthode que nous allons présenter. Nous associons un potentiel électrique $A_{[p-1]}^{(k)}$ de rang $p-1$ à chaque $p$-forme vérifiant $H_{[p]}^{(k)}=\dd \mc A_{[p-1]}^{(k)}$. Un choix possible est le suivant
\BE \mc A^{(k)}_{t i_1\ldots i_{p-2}} = \Phi^{(k)}_{i_1 \ldots i_{p-2}} + \frac{1}{(n-2p+4)r^{n-2p+4}} \mc E^{(k)}_{i_1 \ldots i_{p-2}} \ .\EE
$ \Phi^{(k)}_{i_1 \ldots i_{p-2}}$ fournit ici la valeur du potentiel électrique à l'infini spatial. Puis, nous fixons le choix de jauge en imposant que le potentiel électrique s'annule sur l'horizon localisé en $r=r_h$ en prenant
\BE \Phi^{(k)}_{i_1 \ldots i_{p-2}} = - \frac{1}{(n-2p+4)r_h^{n-2p+4}} \mc E^{(k)}_{i_1 \ldots i_{p-2}} \ .\EE
De manière équivalente, nous pouvons imposer que la quantité $\lp\mc A_{[p-1]}^{(k)}\rp^2$, qui est un scalaire, soit finie sur l'horizon, comme cela a été discuté pour la première fois dans \cite{Gibbons:1976ue} pour le cas électromagnétique ($p=2$). Pour notre cas restreint, nous avons 
\BE T = \frac{n\kappa}{4 \pi r_h} + \frac{(n+2)r_h}{4\pi l^2} - \frac{1}{8\pi (n+1)} \sum_{p,k}  (n-2p+4)^2 r_h^{3-2p} \frac{\lp\Phi^{(k)}_{[p-2]}\rp^2}{(p-2)!} \ .\label{genT-5}\EE
Dans l'ensemble grand-canonique dans lequel les potentiels $\Phi^{(k)}_{[p-2]}$ sont gardés fixes, $r_h=r_h \lp M,\Phi^{(k)}_{[p-2]} \rp $ est une fonction de la masse $M$ et des potentiels $\Phi^{(k)}_{[p-2]}$ de la solution. La température est donnée par une série de puissances impaires de $r_h$ qui commence en $1/r_h$ quand $p=2$, contribuant ainsi au terme de courbure, et qui se termine en $1/r_h^{2n+3}$ quand $p=n+3$. Une importante conséquence de cela est que la seule façon d'avoir deux branches de trous noirs, des "petits" et des "larges" comme pour les trous noirs de Reissner-Nordstrom-AdS, est de considérer un horizon avec une courbure scalaire positive ($\ka>0$). Dans ce cas, il est possible d'observer des transitions de phase entre ces deux familles. Par exemple, le cas électromagnétique ($p=2$) admet un riche diagramme de phase, pour des trous noirs possédant des horizons sphériques, qui est similaire au système liquide-gaz de Van der Waals-Maxwell \cite{Chamblin:1999tk,Chamblin:1999hg}. Par ailleurs, les trous noirs dont l'horizon est plat ne subissent pas de transition de phase.
\bs

L'énergie libre de ces trous noirs peut être évaluée en utilisant l'approche de l'intégrale de chemin \cite{Gibbons:1976ue}, présentée à la sous-section précédente, dans laquelle la fonction de partition dans un ensemble thermodynamique est identifiée à l'intégrale de chemin euclidienne évaluée dans l'approximation du point selle avec des conditions au bord qui déterminent l'ensemble thermodynamique. Plus précisément, nous analyserons les trous noirs dans l'ensemble grand-canonique dans lequel les potentiels électriques associés aux $p$-formes sont gardés fixes à l'infini spatial\footnote{L'ensemble canonique consisterait à garder fixe les $p$-charges associées aux $p$-formes où ces $p$-charges, qui généralisent la notion de charge électrique pour $p>2$, seront définies par la suite.}. Les solutions euclidiennes dont la métrique est donnée par \eqref{ansatz} avec le potentiel \eqref{genV-5}, obtenues par une rotation de Wick, présentent généralement une singularité conique. Néanmoins, en identifiant la période $\be$ du temps imaginaire $\tau=-it$ à l'inverse de la température $\be=1/T$, où $T$ est la température donnée par \eqref{genT-5}, nous éliminons cette singularité conique.

Les conditions au bord, ou de manière équivalente le choix de l'ensemble thermodynamique, imposent les termes de bord qui doivent être ajoutés à l'action dans le but d'obtenir un principe variationnel bien défini. Comme d'habitude, l'action d'Einstein-Hilbert doit être complétée par le terme de bord de Gibbons-Hawking-York \eqref{GHY} ; ce qui donne l'action euclidienne $I = - S - S_{GH}$ avec
\BE S_{GH}  = \frac{1}{8\pi G} \int_{\p \mc M} K \ .\EE
$K$ désigne ici la trace de la courbure extrinsèque de la frontière $\p \mc M$ de l'espace-temps. Nous rappelons que ce terme de bord est nécessaire si nous autorisons des variations de la métrique pour lesquelles la métrique induite sur le bord est fixe pour établir l'équation d'Einstein. Par ailleurs, il n'y a pas de terme de bord pour le secteur de la matière puisque cette action est bien définie en gardant les potentiels électriques fixes sur le bord. En effet, si nous considérons une $p$-forme exacte $H$, c'est-à-dire qu'il existe une $(p-1)$-forme $A$ telle que $H = \dd A$, alors la variation de l'action $S_H = \int_\mc M H \w \star H$ par rapport au potentiel $A$ donne
\BE \delta  S_H = \int_{\p \mc M} \delta A \w \star H + 2(-1)^p \int_\mc M \delta A \w \dd \star H \ . \EE
Le second terme du membre de droite de cette égalité fournit l'équation du mouvement. Nous allons écrire plus explicitement le terme de bord. Pour cela, nous introduisons une base orthonormale de 1-formes $\{ e^A \}$, la métrique se décompose alors sous la forme $g=\eta_{AB} e^A \otimes e^B $, la $p$-forme $H$ s'écrit $H = \frac1{p!} H_{A_1 \ldots A_p} e^{A_1} \w \ldots \w e^{A_p} $ et de même pour le potentiel $A$. Ainsi, en utilisant la relation \eqref{tetrad}, nous avons
\BE \delta S_H = \frac{2}{(p-1)!} \int_{\p \mc M} \delta A_{A_1 \ldots A_{p-1}} H^{B A_1 \ldots A_{p-1}} e^*_B  + 2(-1)^p \int_\mc M \delta A \w \dd \star H \EE
ou  dans le langage des composantes
\begin{multline}
\delta S_H =
-\frac{2}{(p-1)!} \int_\mc M \sqrt{-g}\,\nabla_B H^{B A_1\ldots A_{p-1}}\delta A_{A_1\ldots A_{p-1}} \dd^{D}x \\
+\frac{2}{(p-1)!}\int_{\p \mc M}  H^{B A_1\ldots A_{p-1}}
\delta A_{A_1\ldots A_{p-1}} \dd \Si_B \ .
\end{multline}
Par conséquent, cette action fournit un principe variationnel bien défini dans l'ensemble grand-canonique. Cependant, l'action euclidienne $I$ évaluée sur une solution est généralement divergente. Pour extraire la partie finie, nous utilisons la méthode qui consiste à calculer l'action euclidienne d'une solution par rapport à l'action euclidienne d'une solution de référence $\mc M_0$. Nous régularisons l'action $I$ en intégrant sur une région finie de l'espace-temps avec une frontière $\p \mc M$ et en soustrayant l'action euclidienne évaluée sur une région finie de l'espace $\mc M_0$ de telle façon que la métrique induite et les potentiels électriques sur le bord de $\mc M_0$ coïncident avec ceux de $\p M$. Puis, la frontière $\p \mc M$ est envoyée à l'infini spatial.

Puisque nous travaillons dans l'ensemble grand-canonique, la solution de fond est choisie de telle façon que sa température (de Tolman) coïncide avec celle de la solution étudiée et de même pour les potentiels électriques à l'infini spatial. Un bon candidat comme solution de fond ici est l'espace AdS euclidien avec les potentiels électriques constants $\Phi^{(k)}_{[p-2]}$. En effet, cette solution ne contient pas de singularité conique, c'est-à-dire que n'importe quelle périodicité $\be$ peut être utilisée pour le temps euclidien. Par conséquent, nous trouvons le potentiel de Gibbs $G$ suivant
\BE \beta G = I - I_0  = \frac{\beta V_{\mc H}}{16\pi G} \left[  \kappa r_h^n - \frac{r_h^{n+2}}{l^2}+ \frac{1}{2(n+1)} \sum_{p,k}  (3-2p ) (n-2p+4) r_h^{n-2p+4} \frac{ \Phi^{(k)2}_{[p-2]} }{(p-2)! }   \right] \EE
où l'indice $_0$ se réfère à la solution de référence et $V_{\mc H}=\int \sqrt{\sigma} \dd^{D-2} y$ est l'aire des sections $\mc H$. En supposant la première loi de la thermodynamique
\BE \delta M= T \delta S - \sum_{p,k} \frac{1}{(p-2)!} \Phi^{(k)}_{i_1 \ldots i_{p-2}}\delta Q_{(k)}^{i_1 \ldots i_{p-2}} \ , \EE
nous obtenons l'entropie, la $p$-charge et la masse du trou noir 
\begin{align}
S &=  - \left. \frac{\p G}{\p T} \right|_{ \{ \Phi^{(k)}_{[p-2]} \}_{p,k}} = \frac{V_{\mc H}}{4G}\,r_h^{n+1}\\  
Q_{(k)}^{i_1 \ldots i_{p-2}} &=  (p-2)! \left. \frac{\p G}{\p \Phi^{(k)}_{i_1 \ldots i_{p-2}} } \right|_{ \left\{ T,  \Phi^{(k')}_{i'_1 \ldots i'_{p'-2}}  \neq  \Phi^{(k)}_{i_1 \ldots i_{p-2}} \right\} }   = \frac{V_{\mc H}}{16\pi G} \mc E_{(k)}^{i_1 \ldots i_{p-2}} \label{p-charges-5} \\
M&=G + TS -  \sum_{p,k} \frac{1}{(p-2)!} Q_{(k)}^{i_1 \ldots i_{p-2}}  \Phi^{(k)}_{i_1 \ldots i_{p-2}} = \frac{V_{\mc H}}{16\pi G}(n+1) r_0^n \label{masse-5}
\end{align}
respectivement. Il est facile de vérifier que l'entropie est celle de Bekenstein-Hawking et que la masse est en accord avec une analyse asymptotique de la métrique. Quant aux $p$-charges, elles sont simplement proportionnelles à la polarisation électrique $\mc E$. Dans la prochaine section, nous allons dériver de nouveau ces résultats par une approche hamiltonienne qui donne notamment une interprétation des $p$-charges.
	\section{Le formalisme hamiltonien\label{hamiltonian-formalism}}
		\subsection{En Relativité Générale}
Nous allons dans cette sous-section donner une définition de l'énergie \textit{totale} du champ gravitationnel à travers le formalisme hamiltonien. En effet, une définition \textit{locale} de l'énergie semble être impossible puisque, d'après le principe d'équivalence, il est toujours possible d'adopter un système de coordonnées dans un voisinage suffisamment petit afin d'effacer l'effet de la gravitation. Par ailleurs, ce formalisme hamiltonien est également utilisé comme point de départ dans la procédure de quantification canonique. Nous suivrons, à l'aide de \cite{wald1984general}, la démarche de S. Hawking et G. Horowitz \cite{Hawking:1995fd} dans la suite, en prenant soin de dériver les divers termes de bords qui permettent de définir l'énergie totale d'une solution. 

			\subsubsection{Dérivation de l'hamiltonien}
La première étape de ce formalisme consiste à décomposer l'espace-temps $\mc M$ en "espace et temps". Nous supposons qu'il existe un difféomorphisme 
$\phi : \mc M \rightarrow \Sigma \times I$, avec $I \subset \mathbb{R}$ et $\Si$ une variété riemanienne de dimension 3, tel que les hypersurfaces $\Sigma_t = \phi^{-1}\left( \Sigma \times \{t\} \right)$ soient de genre-espace et les courbes $ \phi^{-1}\left(  \{x\} \times I \right)$ soient de genre-temps. Un vecteur tangent $t^\mu$ à ces courbes peut alors être décomposé en une composante normale et une autre parallèle à $\Si_t$ de la façon suivante 
\BE t^\mu = N n^\mu + N^\mu \label{flow-time}\EE
où $n^\mu$ est le vecteur unitaire normal à la surface $\Sigma_t$. La fonction $N$ est appelée le \ita{laps} (le \ita{lapse} en anglais) et $N^\mu$ est appelé le vecteur \ita{déplacement} (le \ita{shift} en anglais). Le vecteur $t^\mu$ est donc utilisé pour identifier chaque $\Si_t$ à $\Si_0$ ; ainsi la métrique induite sur $\Si_t$, donnée par $h_{\mu\nu}=g_{\mu\nu}+n_\mu n _\nu$, est vue comme la variable dynamique en Relativité Générale. Le laps et le vecteur déplacement ne sont pas dynamiques, ils indiquent simplement comment le système évolue dans le temps. 

Indiquons qu'un système de coordonnées $x^i$ sur $\Si$ induit naturellement des coordonnées sur $\mc M$. En effet, si un point $P$ de $\Si$ a les coordonnées $x^i$ alors nous assignons les coordonnées $(t,x^i)$ au point $\phi^{-1}(P,t)$. Ainsi, le vecteur tangent aux courbes $ \phi^{-1}\left(  \{x\} \times I \right)$ paramétrisées par $t$ est donnée par $t=t^\mu \p_\mu=\p_t$. Par conséquent, la métrique s'écrit
\BE g = - \lp N^2 - N^i N_i\rp \dd t^2 + 2 N_i \dd x^i \dd t + g_{ij} \dd x^i \dd x^j  \EE
dans ce système de coordonnées ou sous la forme
\BE g = - N^2 \dd t^2 + g_{ij}\lp \dd x^i + N^i \dd  t\rp \lp \dd x^j + N^j \dd  t\rp \label{ADM-metric-5}\EE
où la somme sur les indices latins va de 1 à 3.

\bs
Après avoir expliqué la foliation de l'espace-temps, nous décomposons le bord de $\mc M$ en deux hypersurfaces initiale et finale de genre-espace, notées $\Sigma_{t_i}$ et $\Sigma_{t_f}$ respectivement, et en un bord de genre-temps $^3B$ avec un vecteur normal unitaire $u^\mu$. Ce dernier bord sera éventuellement envoyé à l'infini spatial. De plus, nous notons par $S_t$ le bord de $\Sigma_t$ et nous imposons la condition d'orthogonalité\footnote{Le lecteur pourra trouver dans \cite{Hawking:1996ww} le cas où cette condition d'orthogonalité est supprimée.} $u^\mu n_\mu = 0$ sur $^3B$. La métrique induite sur $S_t$ est $\si_{\mu\nu}=h_{\mu\nu}-u_\mu u _\nu$.

\BF 
\BC \hspace{3cm}\includegraphics[scale=0.5]{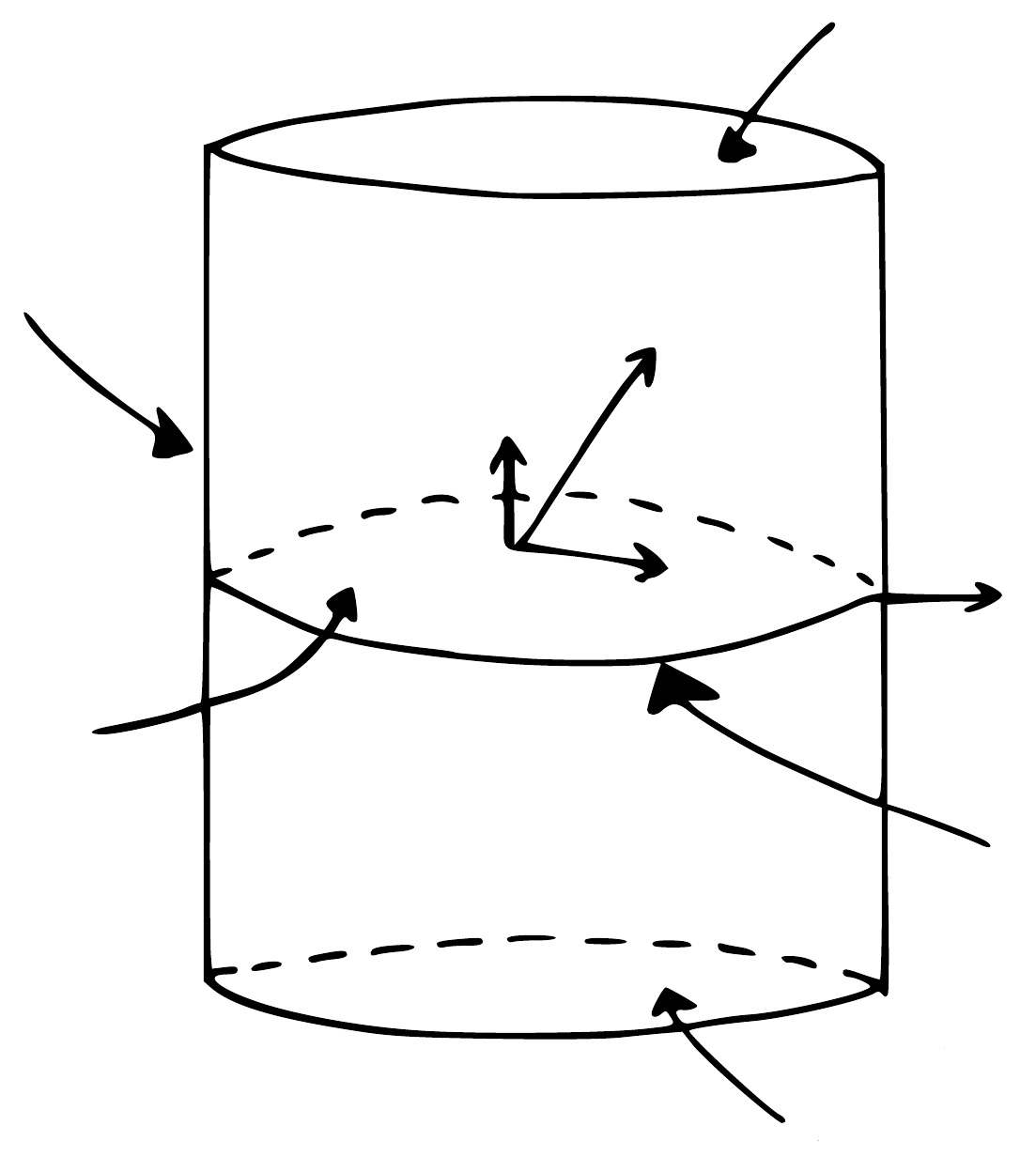}
\begin{picture}(100,100)
\put(-40,3){$\Si_{t_i}$}
\put(-33,173){$\Si_{t_f}$}
\put(-164,64){$\Si_{t}$}
\put(-8,46){$S_t$}
\put(-174,134){$^3B$}
\put(-6,87){$u^\mu$}
\put(-98,112){$n^\mu$}
\put(-59,87){$N^\mu$}
\put(-61,129){$t^\mu$}
\end{picture} 
\caption{Foliation de l'espace-temps} \EC
\EF

Rappelons que l'action de la Relativité Générale dans le vide est donnée par
\BE S = \al \lp  \int_{\mc M}R + 2\varep\int_{\p \mc M} K \rp \label{action-RG-5} \EE
où $\varep=1$ si $\p \mc M$ est de genre-temps et $\varep=-1$ si $\p \mc M$ est de genre-espace, $\al$ est une constante que nous déterminerons dans la suite et $K$ est la trace de la courbure extrinsèque du bord $\p\mc M$. Le terme de bord, présent dans \eqref{action-RG-5}, permet d'avoir un principe variationnel bien défini dans lequel la métrique induite du bord est fixe. 
\bs

Désormais, le but est de décomposer la géométrie intrinsèque de $\mc M$ à l'aide de la géométrie intrinsèque de $\Si_t$ et de sa géométrie extrinsèque. Précisons qu'il sera utile dans la suite de voir $h^\nu{}_{\mu}$ comme la projection de l'espace tangent de $\mc M$ en un point $P$ sur l'espace tangent de $\Sigma_t$ en ce même point ; ainsi la projection est donnée par $T_{\parallel}^{\mu_1 \ldots \mu_k}{}_{\nu_1 \ldots \nu_l} = h_{\ \rho_1}^{\mu_1} \cdots \ h_{\ \rho_k}^{\mu_k} h_{\nu_1}^{\ \sigma_1}\cdots h_{\nu_l}^{\ \sigma_l}  T^{\rho_1\ldots\rho_k}{}_{\sigma_1\ldots\sigma_l}$ pour tout tenseur $T^{\mu_1\ldots\mu_k}{}_{\nu_1\ldots\nu_l}$ de $\mc M$ de type $(k,l)$. Tout d'abord, c'est un bon exercice de montrer que le tenseur de Riemann de $\Si_t$, $R^{(3)\ \sigma}_{\mu\nu\rho} $, admet la décomposition suivante, appelée \textit{première relation de Gauss-Codacci},
\BE R^{(3)\ \sigma}_{\mu\nu\rho}  = R^{\ \ \ \ \ \sigma}_{\parallel\mu\nu\rho} - K_{\mu\rho} K_\nu^{\ \sigma} + K_{\nu\rho} K_\mu^{\ \sigma} \EE
où $K_{\mu\nu}=h_\mu^{\ \rho}\n_\rho n_\nu$ est la courbure extrinsèque de $\Si_t$. Il s'ensuit alors la relation suivante
\BE   R^{(3)} = 2 G_{\mu\nu} n^\mu n^\nu - K^2 + K_{\mu\nu} K^{\mu\nu} \label{Gauss-Codacci-5}\EE
où $ R^{(3)}$ est le scalaire de Ricci de $\Si_t$ et $K=K_\mu^{\ \mu}$. Par ailleurs, la définition \eqref{Riemann0} du tenseur de Riemann de $\mc M$ permet de déterminer la relation suivante (voir \cite{wald1984general})
\BE R_{\mu\nu} n^\mu n^\nu = K^2 - K_{\mu\nu} K^{\mu\nu} - \n_\mu \lp n^\mu \n_\nu n^\nu \rp + \n_\nu \lp n^\mu \n_\mu n^\nu \rp \ .\EE
Enfin, en utilisant les deux résultats précédents et en remarquant que le scalaire de Ricci se décompose sous la forme $R=2\lp G_{\mu\nu} - R_{\mu\nu} \rp n^\mu n^\nu $, nous trouvons 
\BE S = \al\int_\mc M \lb  R^{(3)} - K^2 + K_{\mu\nu} K^{\mu\nu} + 2 \n_\mu \lp n^\mu \n_\nu n^\nu \rp - 2 \n_\nu \lp n^\mu \n_\mu n^\nu  \rp \rb + 2\al\varep\int_{\mc \p M} K \ . \label{bords-5}\EE
Or, d'après le théorème de Stokes \ref{th-Stokes} et en notant $r^\mu$ le vecteur unitaire normal à $\mc \p M$ (qui vaut donc soit $n^\mu$ ou $u^\mu$), nous avons
\BE \int_\mc M \n_\mu \lp n^\mu \n_\nu n^\nu \rp  = \varep \oint_{\mc \p M}  r_\mu n^\mu \n_\nu n^\nu = - \varep\int_{\Si_{t_i}\cup\Si_{t_f}} \n_\nu n^\nu = - \varep\int_{\Si_{t_i}\cup\Si_{t_f}} K \EE
puisque $u^\mu n_\mu=0$ sur $^3B$ et
\BE \int_\mc M \n_\nu \lp n^\mu \n_\mu n^\nu \rp  = \varep \oint_{\mc \p M}  r_\nu  n^\mu \n_\mu n^\nu = \varep \int_{^3B} u_\nu n^\mu \n_\mu n^\nu \ . \EE 
Par conséquent, les termes de bords de $\Si_{t_i}$ et $\Si_{t_f}$ se simplifient dans \eqref{bords-5} et l'action se réduit à
\BE S = \al\int_\mc M \lp  R^{(3)} - K^2 + K_{\mu\nu} K^{\mu\nu} \rp + 2\al\int_{^3B}  \underbrace{ \n_\mu u^\mu - u_\nu n^\mu \n_\mu n^\nu}_{=h^{\mu\nu}\n_\mu u_\nu} \ .
\EE
Puis, en introduisant $k_{\mu\nu}=\si_\mu^{\ \rho} D_\rho u_\nu $ la courbure extrinsèque de $S_t$ où $D_\mu$ désigne la dérivée covariante de $\Si_t$ associée à $h_{\mu\nu}$ qui est simplement déterminée à partir de la dérivée covariante de $\mc M$ associée à $g_{\mu\nu}$ de la façon suivante
\BE D_\rho T^{\mu_1 \ldots \mu_k}{}_{\nu_1 \ldots \nu_l} = \lp \n_\rho T^{\mu_1 \ldots \mu_k}{}_{\nu_1 \ldots \nu_l}\rp  _\parallel\EE
pour tout tenseur $T^{\mu_1 \ldots \mu_k}{}_{\nu_1 \ldots \nu_l}$ de $\mc M$ de type $(k,l)$. Ainsi, il est de loisible de remarquer que $h^{\mu\nu}\n_\mu u_\nu=k_\mu^{\ \mu}=k$. Nous aboutissons alors à l'action suivante
\BE S = \int_{t_i}^{t_f} \lb \int_{\Si_t} \underbrace{\al N \sqrt{h} \lp  R^{(3)} - K^2 + K_{\mu\nu} K^{\mu\nu} \rp }_{=\mc L} \dd^3 x + 2\al\int_{S_t} \sqrt{\si} N k \dd^2x \rb \dd t \ .
\EE
\bs

Après cela, nous définissons le \textit{moment canonique conjugué à $h_{\mu\nu}$} par $\pi^{\mu\nu}=\frac{\p\mc L}{\p \dot{h}_{\mu\nu}}$ où la dérivée temporelle $\dot{h}_{\mu\nu}$ de la métrique induite sur $\Si_t$ est définie par $\dot{h}_{\mu\nu}=\lp \pounds_t h_{\mu\nu}\rp_\parallel$, $t$ étant le vecteur de composante $t^\mu$ \eqref{flow-time} qui décrit la foliation de l'espace-temps. Un autre bon exercice est de montrer l'égalité suivante
\BE 2 N K_{\mu\nu} = \dot{h}_{\mu\nu} - D_\mu N_\nu - D_\nu N_\mu  \EE
qui permet alors de déterminer le moment conjugué de $h_{\mu\nu}$
\BE \pi^{\mu\nu} = \al \sqrt{h}\lp K^{\mu\nu} - K h^{\mu\nu} \rp \ . \label{moment-5}\EE
Nous pouvons aussi remarquer que $\mc L$ ne contient pas de dérivée temporelle de $N$ et $N^\mu$, ce qui confirme que le laps et le vecteur déplacement ne sont pas des variables dynamiques. En notant $\pi=\pi_\mu^{\ \mu}$, il est ensuite facile de réécrire l'action sous la forme
\begin{align}
 S &= \int_{t_i}^{t_f} \lb \int_{\Si_t} \lb \pi^{\mu\nu}\dot{h}_{\mu\nu} - \frac{N}{\al\sqrt{h}} \lp \pi^{\mu\nu}\pi_{\mu\nu} - \frac{\pi^2}{2} - \al^2 h R^{(3)}  \rp - 2\pi^{\mu\nu}D_\mu N_\nu \rb \dd^3x + 2\al\int_{S_t} N k \rb\dd t \label{hamilton1}\\ 
&=\int_{t_i}^{t_f} \lb \int_{\Si_t} \lp \pi^{\mu\nu}\dot{h}_{\mu\nu} - \mc H \rp \rb \dd t 
\end{align}
où nous avons défini la \textit{densité hamiltonienne} $\mc H=\mc H[h_{\mu\nu},\pi^{\mu\nu}]$ dans la seconde égalité. Cette définition permet de restaurer les équations du mouvement à l'aide des équations d'Hamilton $\dot{h}_{\mu\nu} = \frac{\de \mc H}{\de \pi^{\mu\nu}}$ et $\dot{\pi}^{\mu\nu} = -\frac{\de \mc H}{\de h_{\mu\nu}}$ puisque la variation de l'action $S$ est donnée par
\BE \de S = \int_{t_i}^{t_f} \lb \int_{\Si_t} \lp \dot{h}_{\mu\nu} - \frac{\de\mc H}{\de \pi^{\mu\nu}} \rp \de \pi^{\mu\nu} - \lp \dot{\pi}^{\mu\nu} + \frac{\de\mc H}{\de h_{\mu\nu}} \rp \de h_{\mu\nu} \rb \dd t + \int_{\Si_{t_f}} \pi^{\mu\nu} \underbrace{\de h_{\mu\nu}}_{=0}- \int_{\Si_{t_i}} \pi^{\mu\nu} \underbrace{\de h_{\mu\nu}}_{=0} \label{variational-principal-5} \ .
\EE
Les deux termes de bords  sont nuls puisque nous avons choisi de fixer la métrique induite du bord de $\mc M$. Les équations du mouvement, données par $\de S=0$, sont alors équivalentes aux équations d'Hamilton. Enfin, en réécrivant un des termes de \eqref{hamilton1} sous la forme suivante
\BE \int_{\Si_t} \pi^{\mu\nu}D_\mu N_\nu \dd^3x = \int_{S_t} \frac{\pi^{\mu\nu}}{\sqrt{h}} u_\mu N_\nu - \int_{\Si_t} \sqrt{h} N_\mu D_\nu \lp \frac{\pi^{\mu\nu}}{\sqrt{h}} \rp  \dd^3x \ ,\EE
l'\textit{hamiltonien} de la Relativité Générale, $H=\int_{\Si_t} \mc H$, est
\BE  H = \int_{\Si_t}  \sqrt{h} \lp N \mc C + N^\mu \mc C_\mu \rp\dd^3 x + 2 \int_{S_t} \sqrt{\si} \lp -\al N k + N^\mu u^\nu \frac{\pi_{\mu\nu}}{\sqrt{h}} \rp \dd^2x  \label{hamiltonien-RG} \EE
avec
\begin{align}
\mc C &= \frac1{\al h} \lp \pi^{\mu\nu}\pi_{\mu\nu} - \frac{\pi^2}{2} \rp - \al R^{(3)} =  \frac1{\al\sqrt{h}} G_{\mu\nu\rho\si}\pi^{\mu\nu}\pi^{\rho\si} - \al R^{(3)} \\
\mc C_\mu &= - 2 D_\nu \lp \frac{\pi_\mu^{\ \nu}}{\sqrt{h}} \rp 
\end{align}
où nous avons introduit le tenseur\footnote{Ce tenseur connu sous le nom de \textit{métrique du super-espace} \cite{DeWitt:1967yk} intervient dans l'équation de Wheeler-DeWitt.} $G_{\mu\nu\rho\si}$ qui vaut
\BE G_{\mu\nu\rho\si} = \frac1{2\sqrt{h}}\lp h_{\mu\rho}h_{\nu\si} + h_{\mu\si}h_{\nu\rho} - h_{\mu\nu}h_{\rho\si} \rp  \ . \EE
Le laps $N$ et le vecteur déplacement $N^\mu$ jouent ici le rôle de multiplicateur de Lagrange et génèrent les contraintes $\mc C= 0$ et $\mc C_\mu=0$ pour une solution des équations du mouvement. En effet, en utilisant les équations \eqref{Gauss-Codacci-5} et \eqref{moment-5}, il est facile de réécrire la contrainte $\mc C$ sous la forme
\BE \mc C = -2\al G_{\mu\nu}n^\mu n^\nu \EE
qui est nulle sur une solution. De même, en dérivant la \textit{seconde relation de Gauss-Codacci}
\BE D_\nu K^\nu_{\ \mu} - D_\mu K = R_{\nu\rho} h_\mu^{\ \nu} n^\rho \EE
et en utilisant \eqref{moment-5}, nous trouvons
\BE \mc C_\mu = - 2 \al G_{\nu\rho} h_\mu^{\ \nu} n^\rho \ .\EE
qui est également nulle sur une solution. Notons qu'avec ce formalisme, il est simple de déterminer le nombre de degrés de liberté de la théorie. Il y a 6 degrés de liberté qui proviennent de la métrique spatiale et 6 autres des moments conjugués associés. Or, il y a 4 équations de contraintes et 4 autres degrés de liberté sont éliminés par l'invariance des équations du mouvement sous les changements de coordonnées. Il y a donc 4 degrés de liberté en chaque point de l'espace-temps.

			\subsubsection{Energie totale}
Nous sommes désormais en mesure de définir l'\textit{énergie totale d'une solution} comme la valeur de l'hamiltonien $H$ \eqref{hamiltonien-RG} évaluée sur une solution des équations du mouvement. Puisque les contraintes évaluées sur une solution sont nulles, l'énergie totale est alors simplement donnée par l'intégrale sur $S_t$ \cite{Hawking:1995fd}. Cependant, cette expression est généralement divergente. Pour contourner ce problème, nous allons la comparer à une solution de référence $g_0$ dont la métrique induite sur le bord $^3B$ coïncide avec celle de la solution considérée. Nous choisissons aussi une solution de référence dont les moments conjugués $\pi_0^{\mu\nu}$ sont nuls pour simplifier et la foliation est choisie telle que $N=N_0$ sur $^3B$. De cette façon, l'énergie totale d'une solution est donnée par
\BE  E(t^\mu) = - 2 \int_{S_t} \lb \al N \lp k - k_0 \rp - N^\mu u^\nu \frac{\pi_{\mu\nu}}{\sqrt{h}}  \rb  \EE
où $k_0$ est la courbure extrinsèque du bord de la solution de référence. Soulignons que cette expression dépend du choix de la foliation à travers le vecteur $t^\mu=N n^\mu+N^\mu$ et de la surface $S_t$. Cependant, étant donné un vecteur de Killing $\xi^\mu$, l'énergie totale $E(\xi^\mu)$ est une \textit{charge}, c'est-à-dire que cette expression est indépendante de la surface $S_t$ choisie. Cette conservation a été montrée par J. Brown et J. York dans \cite{Brown:1992br} en adoptant une approche à la Hamilton-Jacobi pour définir cette charge\footnote{Suite à cet article, ces deux auteurs ont aussi formulé dans \cite{Brown:1992bq} le formalisme hamiltonien de la gravitation dans un ensemble microcanonique afin de définir la densité d'état pour déterminer en particulier l'entropie d'un trou noir.}. \bs

Appliquons ce résultat à la solution de Schwarzschild en calculant la charge $E$ associée au vecteur de Killing de genre-temps $\p_t$ dans les coordonnées de Schwarzschild dans le but de fixer la valeur de la constante $\al$ en facteur de l'action $S$. La métrique est donnée par
\BE g = - V(r)\dd t^2 + \frac{\dd r^2}{V(r)} + r^2\lp\dd\theta^2 + \sin^2\theta\dd\phi^2 \rp\EE
avec le potentiel $V(r)=1-2M/r$. Puisque nous choisissons $t=t^\mu \p_\mu = \p_t$, nous pouvons déterminer le laps $N=\sqrt{V}$ et le vecteur déplacement est nul, $N^\mu=0$, d'après \eqref{ADM-metric-5}. Le vecteur unitaire normal à $\Si_t$ vaut donc $n=n^\mu \p_\mu = \frac{1}{\sqrt{V}}\p_t$ et, en choisissant le bord $^3B$ à la coordonnée radiale $r=R$, nous avons $u=u^\mu \p_u = \sqrt{V(R)}\p_r$. Nous enverrons le bord $^3B$ à l'infini spatial dans la suite. Ainsi, en définissant la \textit{masse} $m$ d'une solution comme étant la valeur de l'énergie $E$ associée au vecteur de Killing de genre-temps et unitaire à l'infini spatial, nous avons
\BE m = - 2 \al \sqrt{V(R)} 4 \pi R^2 \lp k - k_0 \rp\EE
où nous choisissons l'espace-temps de Minkowski comme solution de référence. Or, la trace de la courbure extrinsèque de $S_t$ est
\begin{align}
k =& \si_\mu^{\ \rho} D_\rho u^\mu =  \si_\mu^{\ \rho} h_\rho^{\ \al} h^\mu_{\ \be} \n_\al u^\be  = \si_\be^{\ \al} \n_\al u^\be = \si_\be^{\ \al} \Ga_{\al\rho}^\be u^\rho \\
=& \frac12 \si_\be^{\ \al} g^{\be\ga} \lp g_{\rho\ga,\al} + g_{\al\ga,\rho} - g_{\al\rho,\ga} \rp u^\rho \\
=& \frac12 \si^{\al\ga} u^\rho g_{\al\ga,\rho} = \frac12 \sqrt{V} \si^{\al\be} \si_{\al\be,r} \\
=& \frac{2}{R}\sqrt{1-2M/R}
\end{align}
et $k_0=\frac{2}{R}$. Ainsi, dans la limite où $R\rightarrow\infty$, la masse vaut $m=16\al\pi M$. Par conséquent, pour être en accord avec la limite newtonienne, la constante $\al$ est fixée à la valeur suivante
\BE \al = \frac1{16\pi} \ . \EE

Pour conclure, signalons que la définition de l'énergie totale que nous venons de donner est en accord avec celle de R. Arnowitt, S. Deser, et C. Misner \cite{Arnowitt:1962hi} dans le cas d'espace-temps asymptotiquement plats et avec celle de L. Abbott et S. Deser \cite{Abbott:1981ff} pour des espace-temps asymptotiquement AdS. Cet accord entre les diverses définitions de l'énergie est exposé dans \cite{Hawking:1995fd}.

			\subsubsection{Dérivation de la première loi selon R. Wald} 
Dans la dérivation de l'énergie totale, nous avons simplement tenu compte du bord $S_t$. Cependant, pour des géométries de trous noirs, l'espace $\Si_t$ possède également un bord intérieur, il y aura donc une contribution supplémentaire provenant de ce bord dans l'expression de l'énergie dans ce cas. Dans cette brève sous-section, nous proposons de dériver la première loi de la mécanique des trous noirs en utilisant le formalisme hamiltonien que nous venons de décrire en suivant la présentation de R. Wald \cite{Wald:1993ki}. Considérons un trou noir stationnaire et à symétrie axiale solution des équations d'Einstein dans le vide. La surface spatiale $\Si_t$ possède désormais une surface $S_t$ à l'infini spatial et un autre bord interne donné par la surface de bifurcation $S$. Notons par $\p_t$ et $\p_\phi$ les vecteurs de Killing de genre-temps et de genre-espace respectivement qui génèrent les symétries de la solution et introduisons le vecteur $\xi=\p_t+\Om\p_\phi$ qui s'annule sur $S$, définissant ainsi la vitesse angulaire $\Om$ de l'horizon. De plus, le vecteur $\p_t$ est choisi unitaire à l'infini spatial et le vecteur $\p_\phi$ est une rotation à l'infini spatial ; fixant ainsi la normalisation des vecteurs de Killing. Comme précédemment, lorsque nous avons défini la masse comme une charge, nous imposons que le vecteur $t=t^\mu \p_\mu$ \eqref{flow-time} vérifie $t=\xi$. \bs

La procédure de R. Wald pour générer la première loi consiste à perturber l'hamiltonien \eqref{hamiltonien-RG} entre une solution initiale donnée par $\lp h_{\mu\nu} , \pi^{\mu\nu} \rp$ et une solution finale infiniment proche donnée par $\lp h_{\mu\nu} + \de h_{\mu\nu} , \pi^{\mu\nu}+\de\pi^{\mu\nu} \rp$. Par conséquent, nous obtenons
\BE \de H = \underbrace{\de \lp  \int_{\Si_t}  \sqrt{h} \lp N \mc C + N^\mu \mc C_\mu \rp\dd^3 x \rp}_{=0} + \de m - \Om \de J - \frac{\ka}{8\pi} \de A \ . \EE
où la masse $m$ est donnée par $E(\p_t)$, le moment angulaire $J$ par $E(\p_\phi)$ et le dernier terme $\frac{\ka}{8\pi} \de A$, avec $\ka$ la gravité de surface et $A$ l'aire de la surface de bifurcation $S$, provient de la prise en compte du bord interne $S$. La dérivation de ce dernier terme est donnée explicitement dans la référence \cite{Sudarsky:1992ty} dans laquelle la première loi est généralisée au cas de la théorie d'Einstein-Yang-Mills. La variation du terme de contraintes est effectivement nulle puisque les contraintes évaluées sur les solutions initiale et finale sont nulles.  Par ailleurs, nous avons
\BE \de H = \int_{\Si_t} \lp \frac{\de\mc H}{\de h_{\mu\nu}} \de h_{\mu\nu} + \frac{\de\mc H}{\de \pi^{\mu\nu}} \de \pi^{\mu\nu}  \rp \ . \EE
Or, comme nous avons pu le voir, à travers l'équation \eqref{variational-principal-5}, les équations d'Hamilton sont vérifiées sur les solutions des équations du mouvement : $\dot{h}_{\mu\nu} = \frac{\de \mc H}{\de \pi^{\mu\nu}}$ et $\dot{\pi}^{\mu\nu} = -\frac{\de \mc H}{\de h_{\mu\nu}}$. Ainsi, puisque les solutions considérées sont stationnaires, nous avons $\de H=0$. Finalement, nous obtenons la première loi de la mécanique des trous noirs 
\BE \de m  = \frac{\ka}{8\pi} \de A + \Om \de J \ . \EE
Pour plus de précisions sur des détails que nous avons esquivés, nous renvoyons vers \cite{Wald:1993ki,Sudarsky:1992ty}. Dans la prochaine section, nous dériverons de nouveau de tels résultats en présence de champ de $p$-formes et en tenant compte des dimensions supplémentaires pour des solutions statiques. 
%
%

		\subsection{En présence de $p$-formes\label{p-charges-2}}
Dans cette sous-section, nous proposons de restaurer les résultats de la sous-section \ref{p-charges-1} par une approche hamiltonienne \cite{Hawking:1995fd} pour souligner que les $p$-charges sont des quantités qui proviennent du bord de $\mc M$ à l'infini spatial. Nous considérons alors l'action
\BE S'=\alpha \int_{\mc M}  \sqrt{-g} \left[ R - 2\La -\frac{\kappa}{p!}H^2_{[p]} \right] \dd^D x + 2\alpha\varep \int_{\p \mc M} K \EE
qui fournit un problème variationnel dans lequel la métrique induite et le potentiel associé à $H_{[p]}=\dd A_{[p-1]}$ sont fixes sur le bord $\p \mc M$ (la frontière $^3B$ de la sous-section précédente est remplacée par une frontière $^{D-1}B$ de dimension $D-1$). Après cela, nous suivons la procédure standard \cite{Hawking:1995fd,wald1984general}, présentée à la sous-section précédente, pour dériver la formulation hamiltonienne de cette théorie. Les moments canoniques conjugués à la métrique spatiale $h_{\mu\nu}$ et au potentiel $A_{ \mu_2 \ldots \mu_p}$ sont
\BE \pi_G^{\mu\nu} = \alpha \sqrt{h} \left( K^{\mu\nu} - K h^{\mu\nu} \right) \qquad\text{et}\qquad
\pi_{H}^{ \mu_2 \ldots \mu_p} = \frac{2\kappa\alpha\sqrt{h}}{(p-1)!} \left( n_{\mu_1} H^{\mu_1 \mu_2 \ldots \mu_p} \right)_{\parallel} \EE
où $K_{\mu\nu}$ est la courbure extrinsèque de $\Sigma_t$ et le symbole $\parallel$ désigne l'opérateur de projection sur $\Sigma_t$ défini précédemment. Il en résulte l'hamiltonien suivant
\begin{multline}
 H = \int_{\Sigma_t} \sqrt{h} \left (N \mc C +  N^\mu \mc C_\mu + t^{\mu_1} A_{\mu_1 \mu_2\ldots\mu_{p-1}} \mc C^{\mu_2\ldots\mu_{p-1}}\right) \dd^{D-1}x\\
 + \int_{S_t} \sqrt{\si} \left[ -2 \alpha N k + 2 N^\mu u^\nu \frac{\pi_{G\mu\nu}}{\sqrt{h}} + (p-1) \frac{\pi_{H}^{ \mu_2\ldots\mu_p}}{\sqrt{h}} t^{\mu_1} A_{\mu_1  \mu_3\ldots\mu_p} u_{\mu_2} \right]  \dd^{D-2}x \label{hamiltonien-5}
\end{multline}
où les contraintes, qui évaluées sur une solution sont nulles, sont données par
\begin{align}
\mc C &= \frac{1}{\alpha h}\left(\pi_{G\mu\nu} \pi_G^{\mu\nu}  + \frac{\pi_G^2}{2-D} \right)  - \alpha R^{(D-1)} + 2\alpha\Lambda + \frac{(p-1)!}{4\kappa\alpha h} \pi_{H \mu_2\ldots\mu_p} \pi_{H}^{\mu_2\ldots\mu_p} \nonumber\\
&\qquad + \frac{\kappa \alpha}{p!} H_{\parallel\mu_1\ldots\mu_p} H_{\parallel}^{ \mu_1\ldots\mu_p}
 	= - 2 \alpha \left( G_{\mu\nu} + \Lambda g_{\mu\nu} - \kappa T_{\mu\nu} \right) n^\mu n^\nu \\
\mc C_\mu &= - 2 D_\nu \left( \frac{\pi_{G\mu}^\nu}{\sqrt{h}} \right)   + \frac{\pi_{H \mu_2\ldots\mu_p}}{\sqrt{h}} H_{\parallel \mu  \mu_2\ldots\mu_p}
= - 2 \alpha \left( G_{\nu\rho} + \Lambda g_{\nu\rho}  - \kappa T_{\nu\rho} \right) h_\mu^{\ \nu} n^\rho\\
\mc C^{\mu_2\ldots\mu_{p-1}} & = -(p-1)D_\mu \left( \frac{\pi_H^{\mu \mu_2\ldots\mu_{p-1}}}{\sqrt{h}} \right) = - \frac{2\alpha \kappa}{(p-2)!} \left( n_{\mu_p} \nabla_{\mu_1} H^{ \mu_2\ldots\mu_p \mu_1 } \right)_{\parallel} \ .
\end{align}
avec $R^{(D-1)}$ le scalaire de Ricci de $\Sigma_t$. Dans le terme de bord de \eqref{hamiltonien-5}, $k$ désigne la trace de la courbure extrinsèque de $S_t$ donnée par $k =  \si_\mu{}^{\nu}D_\nu u^\mu$ avec $D_\mu$ la dérivée covariante sur $\Sigma_t$. Nous soulignons que la présence des termes de bords est crucial pour établir la première loi de la thermodynamique. Nous avons simplement considérer ici la frontière ${}^{D-1}B$ ; pour un trou noir noir, il existe aussi un bord intérieur et donc un terme de bord supplémentaire dans \eqref{hamiltonien-5}. Le résultat présenté ici est en accord avec celui de \cite{Copsey:2005se} dans lequel les termes de bords sont dérivés en variant l'hamiltonien.

Par conséquent, nous pouvons définir l'énergie totale $E$ de la solution comme l'hamiltonien évalué pour cette solution par rapport à une solution de référence. Pour l'action \eqref{action1-5}, nous trouvons
\BE E = \int_{S_t}  \sqrt{\si} \left[ \frac{-N}{8\pi G}  \left( k - k_0 \right)  + \sum_{p,k} \frac{p-1}{\sqrt{h}} u_{\mu_2} \pi_{H_{[p]}^{(k)}}^{ \mu_2\ldots\mu_p}  t^{\mu_1} A^{(k)}_{\mu_1\mu_3\ldots\mu_p}    \right] d^{n+1}x
\EE
en considérant une foliation telle que $N=N_0$ sur $^{D-1}B$ et telle que le vecteur déplacement soit nul pour simplifier ; l'indice $_0$ se réfère toujours à la solution de fond. Après cela, nous imposons
\BE E = M + \sum_{p,k} \frac{1}{(p-2)!} Q_{(k)}^{i_1\ldots i_{p-2}}  \Phi^{(k)}_{i_1\ldots i_{p-2}} \EE
où $M$ est la masse de la solution et $ Q_{(k)}^{i_1\ldots i_{p-2}}$ est la $p$-charge associée au potentiel électrique $\Phi^{(k)}_{i_1\ldots i_{p-2}}$. De cette façon, nous déduisons l'expression suivante pour la masse
\BE M = -\frac{1}{8\pi G}  \int_{S_t} \sqrt{s} N \left( k - k^0 \right)\dd^{n+1}x \EE
et pour les charges, nous avons
\BE Q_{(k)}^{i_1\ldots i_{p-2}}  \Phi^{(k)}_{i_1\ldots i_{p-2}} = \frac{1}{16\pi G}  \int_{S_t} \sqrt{s}     \left( n_{\mu_1} H_{(k)}^{\mu_1 \mu_2\ldots \mu_p} \right)_{\parallel} u_{\mu_2}  t^\nu A^{(k)}_{\nu  \mu_3\ldots \mu_p} \dd^{n+1}x \ .\EE
En particulier, pour les trous noirs présentés au chapitre \ref{BH in higher D} solutions de \eqref{action1-5} possédant le vecteur de Killing de genre-temps $\p_t$, nous trouvons en choisissant $t=t^\mu \p_\mu = \p_t$ : 
\BE M = \frac{V_{\mc H}}{16\pi G}(n+1) r_0^n \qquad\text{et}\qquad Q_{(k)}^{i_1\ldots i_{p-2}}=\frac{V_{\mc H}}{16\pi G}\mc E_{(k)}^{i_1\ldots i_{p-2}}\ . \EE
Nous obtenons ainsi une masse et des $p$-charges en accord avec celles que nous avons déterminées par la méthode de l'intégrale de chemin \eqref{p-charges-5}-\eqref{masse-5}. Cependant, dans cette approche hamiltonienne, nous n'avons pas besoin de supposer la validité de la première loi de la thermodynamique ; au contraire, en introduisant un terme supplémentaire dû à la présence de l'horizon, il est possible de montrer en suivant \cite{Wald:1993ki,Sudarsky:1992ty} que la première loi de la mécanique des trous noirs
\BE \delta M = \frac{\kappa}{8\pi}\delta A  - \sum_{p,k} \frac{1}{(p-2)!} \Phi_{(k)}^{i_1\ldots i_{p-2}}  \delta Q^{(k)}_{i_1\ldots i_{p-2}} \EE
est vérifiée, où $\kappa$ et $A$ sont la gravité de surface et l'aire de l'horizon respectivement.
	\section{Thermodynamique en présence d'un champ scalaire conforme\label{transition de phases}}
Nous analysons dans cette section l'aspect thermodynamique de la nouvelle solution de la théorie \eqref{action-4}, présentée au chapitre précédent, mettant notamment en jeu deux champs axioniques et un champ scalaire. Nous déterminerons d'abord les charges de cette solution par une approche hamiltonienne. Puis, nous étudierons un phénomène de transition de phase entre cette solution et la version axionique de la solution de Reissner-Nordstrom-AdS présentée à la sous-section \ref{axionique}.

			\subsection{Analyse hamiltonienne des charges\label{sec::ham}}
Dans l'approche euclidienne, la fonction de partition pour un ensemble thermodynamique est identifiée à l'intégrale de chemin euclidienne évaluée sur une section euclidienne de la solution classique  \cite{Gibbons:1976ue}. La période $\be$ du temps imaginaire $\tau=-it$ est identifiée à l'inverse de la température et l'énergie libre (ou le potentiel thermodynamique) $W$ est reliée à l'action euclidienne par $I_E=\be W$.

Par simplicité et sans perte de généralité, nous restreignons l'espace de configuration aux métriques statiques présentant la symétrie planaire pour les sections $(t,r)$ constantes. Nous considérons par conséquent la famille de métriques euclidiennes qui s'écrivent sous la forme
\BE \dd s_E^2 = N^2(r) f(r) \dd \tau^2 + \frac{\dd r^2}{f(r)} + r^2 \left( \dd x^2 + \dd y^2 \right) \label{emetric-5}\EE
et les champs satisfont l'ansatz
\BE
\phi = \phi(r)\ , \qquad
\mc A = A(r)\,\dd t\ , \qquad
\mc B^{(i)} = B_i(r)\,\dd t \w \dd x^i \ .
\EE
Nous adoptons bien évidemment les notations de la section \ref{sec4:axionique}. Asymptotiquement, nous imposons que les champs se comportent sous la forme suivante
\BE
\begin{array}{l@{\qquad}l}
\displaystyle f(r)=\frac{r^2}{l^2}+f_0+\frac{f_1}r+\mc O\lp\frac1{r^2}\rp\ ,
&\displaystyle N(r)=N_0+\mc O\lp\frac1{r^2}\rp \ ,
\vphantom{\lp\frac1{r^2_{p_p}}\rp}\\
\displaystyle A(r)=A_0+\frac{4\pi N_0\pi_A}r+\mc O\lp\frac1{r^2}\rp\ ,
&\displaystyle B_i(r)=-2rN_0\pi_{B_i}+B^0_i+\mc O\lp\frac1r\rp \ ,
\vphantom{\lp\frac1{r^2_{p_p}}\rp}\\
\displaystyle \phi(r)=\frac{\phi_1}{r}+\frac{\phi_2}{r^2}+\mc O\lp\frac1{r^3}\rp \ .
\end{array}
\label{falloff-5}\EE
Ici $\{f_0, f_1,N_0,\phi_1,\phi_2,A_0,\pi_A,B^0_i,\pi_{B_i}\}$ est un ensemble de constantes qui détermine le comportement asymptotique des champs avec les contraintes supplémentaires suivantes
\BE
f_0=-16\pi G\,\lp\pi_{B_1}^2+\pi_{B_2}^2\rp \qquad\text{et}\qquad
\phi_2=-\sqrt{2\al l^2}\,\phi_1^2
\label{constraints-5}\EE
que nous avons obtenues en résolvant asymptotiquement les équations du mouvement. Les développements de $f$ et $N$ assurent que l'espace-temps soit asymptotiquement localement AdS et nous pouvons aussi poser $N_0=1$ par un changement de coordonnée du temps $\tau$.

Pour garantir la présence d'un trou noir parmi la famille de métriques \eqref{emetric-5}, nous imposons à la fonction $f$ de s'annuler en $r=r_h$ où $r_h$ est la plus grande racine de $f$. Dans le formalisme euclidien, l'horizon agit comme une seconde frontière sur laquelle nous devons restreindre le comportement des champs. Nous avons en particulier
\BE f(r)=f'(r_h)(r-r_h)+\mc O\lp(r-r_h)^2\rp \ .\EE
En exigeant l'absence de singularité conique en $r=r_h$, nous relions la valeur des champs sur l'horizon à la température inverse selon l'équation
\BE N(r_h)f'(r_h)=\frac{4\pi}\be \ . \label{conical-5}\EE
Finalement, nous imposons que les autres champs de matière $\phi(r)$, $A(r)$ et $B^{(i)}(r)$ admettent une limite finie sur l'horizon que nous nommons $\phi_h$, $A_h$ et $B^{(i)}_h$ respectivement.

Ces conditions aux bords définissent l'espace de configuration pour notre analyse hamiltonienne et nous soulignons que les deux solutions qui nous intéressent, le trou noir chevelu \eqref{hairy-4}-\eqref{hairyfields-4} et le trou noir axionique de Reissner-Nordstrom-AdS \eqref{RN1}-\eqref{RN2}, appartiennent à cet espace. 
\bs

Pour cette classe de métriques et ces conditions aux bords que nous venons de définir, l'action euclidienne prend la forme
\BE \mc S_E = \si\be \int_{r_h}^{\infty}\!dr\lp N\mc H-A\pi_A'-2B_1\pi_{B_1}'-2B_2\pi_{B_2}'\rp + \mc Q  \label{action-euclidienne-5} \EE
où $\mc Q$ est un terme de bord et $\si$ est l'aire des sections transverses définies par $t$ et $r$ constants, provenant de l'intégration le long des directions $x$ et $y$. Bien que l'aire $\si$ soit formellement infinie, nous pouvons toujours compactifier l'horizon en un 2-tore et la rendre finie. Concernant la contrainte hamiltonienne, elle vaut
\begin{multline}
\mc H=2\lp1-\frac{4\pi G}3\phi^2\rp\lp\pi_{B_1}^2+\pi_{B_2}^2\rp+\frac{2\pi}{r^2}\pi_A^2
+\frac{r^2}{8\pi G}\lb\lp1-\frac{4\pi G}3\phi^2\rp\lp\frac{f'}r+\frac f{r^2}\rp+\La\rb \\
+\frac{r^2}6\lb
f\phi'^2-\lp f'+\frac{4f}r\rp\phi\phi'-2f\phi\phi''+6\al\phi^4
\rb \ .\qquad\qquad\qquad\qquad\qquad\quad
\end{multline}
Les quantités $\pi_A$ et $\pi_{B_i}$ désignent ici la $r$-composante du moment conjugué de $A_\mu$ et la $(ri)$-composante du moment conjugué de $B^{(i)}_{\mu\nu}$ respectivement\footnote{Comme nous le verrons, ces moments conjugués sont identiques aux constantes qui apparaissent dans le développement de $A(r)$ et $B_i(r)$. C'est pourquoi nous utilisons la même notation.}. Avec notre ansatz, elles sont égales à 
\BE \pi_A=-\frac{r^2A'}{4\pi N} \qquad\text{et}\qquad \pi_{B_i}=-\frac{B_i'}{2N\lp1-\frac{4\pi G}3\phi^2\rp}\ .\EE

Pour avoir un problème variationnel bien défini, l'action doit être une fonction différentiable par rapport aux variables canoniques $\{A,\pi_A,B_i,\pi_{B_i},f,\phi\}$ sur l'espace de configuration défini par les conditions aux bords \eqref{falloff-5}-\eqref{conical-5} (voir \cite{Regge:1974zd}).   Ainsi, en calculant la variation totale $\delta S_E$ de l'action, le terme $\delta\mc Q$ qui représente la variation du terme de bord doit compenser les termes de bords qui proviennent de la variation du premier terme du membre de droite de \eqref{action-euclidienne-5} conduisant aux équations du mouvement. Il s'ensuit alors que la variation du terme de bord est donnée par
\begin{multline}
\delta\mc Q=\si\be\lb
A\,\delta\pi_A+2B_1\,\delta\pi_{B_1}+2B_2\,\delta\pi_{B_2}
-\frac{rN}{8\pi G}\lp1-\frac{4\pi G}3\phi^2-\frac{4\pi G}3r\phi\phi'\rp\delta f
\right.\qquad \\
\left.
-\frac{r^2N}6\lp4f\phi'+f'\phi+\frac{2N'}Nf\phi\rp\delta\phi
+\frac{r^2N}3f\phi\,\delta\phi'
\rb_{r_h}^\infty \ .
\end{multline}
Sur l'horizon, la variation des champs est donnée par
\BE \delta f|_{r_h}=-f'(r_h)\delta r_h \qquad\text{et}\qquad \delta \phi|_{r_h}=\delta\phi(r_h)-\phi'(r_h)\delta r_h\ .\EE
Nous régularisons le terme de bord en prenant une frontière asymptotique en $r=R$. Puis, en définissant une constante de Newton effective évaluée sur l'horizon $G_h$ par
\BE G_h=\frac{G}{1-\frac{4\pi G}3\phi^2(r_h)} \EE
et en considérant les conditions asymptotiques \eqref{falloff-5} pour les champs, nous obtenons
\begin{multline}
\delta\mc Q=\beta\si\,\Phi\,\delta\pi_A+2\be\si\,\Psi_1\,\delta\pi_{B_1}+2\be\si\,\Psi_2\,\delta\pi_{B_2}
-\delta\lp\frac{\si r_h^2}{4G_h}\rp
-\frac{\be\si}{8\pi G}\,\delta f_1\qquad\qquad\qquad \\
-\be\si\,R\lp4\pi_{B_1}\delta\pi_{B_1}+4\pi_{B_2}\delta\pi_{B_2}+\frac{\delta f_0}{8\pi G}\rp
+\frac{\be\si}{3\ell^2}\lp2\phi_2\delta\phi_1-\phi_1\delta\phi_2\rp
+\mc O\lp\frac1R\rp\ .
\end{multline}
Nous avons défini ici les différences de potentiels $\Phi$ et $\Psi_i$ entre l'infini spatial et l'horizon des événements pour les champs de Maxwell et axioniques par
\BE \Phi=A_0-A(r_h) \qquad\text{et}\qquad \Psi_i=B^0_i-B_i(r_h) \ .\EE
Les variations qui apparaissent ici ne sont pas toutes indépendantes et, en vertu des contraintes \eqref{constraints-5}, elles sont reliées par
\BE \delta f_0=-16\pi G\,\delta\lp\pi_{B_1}^2+\pi_{B_2}^2\rp \qquad\text{et}\qquad\phi_1\delta\phi_2=2\phi_2\delta\phi_1 \ .\EE
Par conséquent, le terme linéairement divergent dans $\delta\mc Q$ est nul et nous pouvons alors envoyer la frontière $r=R$ à l'infini spatial pour obtenir 
\BE \delta\mc Q=\beta\si\,\Phi\,\delta\pi_A+2\be\si\,\Psi_1\,\delta\pi_{B_1}+2\be\si\,\Psi_2\,\delta\pi_{B_2}-\delta\lp\frac{\si r_h^2}{4G_h}\rp
-\frac{\be\si}{8\pi G}\delta f_1 \ .
\label{deltaQ-5}\EE
Avec les conditions aux bords que nous utilisons, nous gardons $\beta$, $\Phi$ et $\Psi_{i}$ fixes et ainsi $\delta\mc Q$ est une variation totale à partir de laquelle nous pouvons déduire le terme de bord $\mc Q$. Cela prouve que l'action euclidienne est différentiable sur l'espace de configuration avec nos conditions aux bords et que les solutions euclidiennes de la théorie \eqref{action-4} sont des extrema de $S_E$. L'annulation de la variation par rapport à $N$ impose la contrainte hamiltonienne $\mc H=0$. Les potentiels $A$ et $B_i$ jouent aussi le rôle de multiplicateurs de Lagrange, ce qui établit que les moments conjuguées $\pi_A$ et $\pi_{B_i}$ sont des constantes du mouvement, en accord avec la loi de Gauss. Enfin, les variations par rapport à $f$ et $\phi$ fournissent les équations d'Einstein restantes et l'équation du mouvement du champ scalaire.
\bs

Puisque le premier terme du membre de droite de \eqref{action-euclidienne-5} s'écrit sous forme de contrainte, il s'annule alors pour une solution de la théorie \eqref{action-4} et l'action euclidienne évaluée sur une telle solution est complètement déterminée par la valeur de $\mc Q$
\BE \mc S_E = -\frac{\mc A_h}{4G_h}-\frac{\be\si}{8\pi G}f_1+\be\si\lp
\Phi\pi_A+2\Psi_1\pi_{B_1}+2\Psi_2\pi_{B_2}\rp \EE
où nous avons introduit l'aire de l'horizon $\mc A_h =\sigma r_h^2$. Avec nos conditions aux bords, nous travaillons dans l'ensemble grand-canonique et le potentiel thermodynamique associé $W$ est relié à l'action euclidienne par $\mc S_E=\be W$. Nous pouvons donc déterminer les charges et les quantités thermodynamiques suivantes par la procédure habituelle
\BE
S=\be^2\left.\frac{\p W}{\p\be}\right|_{\Phi,\Psi^{(i)} } =\frac{\mc A_h}{4G_h}\ , \quad
Q=-\left.\frac{\p W}{\p \Phi} \right|_{\be,\Psi^{(i)}  } =-\sigma\pi_A\ ,\quad
Q_i=-\left.\frac{\p W}{\p\Psi_i}\right|_{\be,\Phi,\Psi^{(j)}_{j\neq i}}=-2\sigma\pi_{B_i} \ .
\label{entropy-5}\EE
Nous observons alors que les quantités $\pi_A$ et $\pi_{B_i}$ sont les densités de charges portées par le champ de Maxwell et les champs de 3-formes. La masse du trou noir est obtenue par la transformation de Legendre suivante qui amène dans l'ensemble micro-canonique
\BE M = W+TS+\Phi Q+\Psi_1Q_1+\Psi_2Q_2=-\frac{\sigma}{8\pi G}f_1 \ .\label{mass-5}\EE
Pour des variations le long d'une famille de solutions, nous avons $\delta\mc Q=0$ et \eqref{deltaQ-5} fournit alors la première loi de la thermodynamique \cite{Wald:1993ki} 
\BE \delta M = T\,\delta S +\Phi\,\delta Q+\Psi_1\,\delta Q_1+\Psi_2\,\delta Q_2 \ .\EE
Une importante remarque est que, à cause du couplage non-minimal entre la gravitation et le champ scalaire $\phi$, l'entropie du trou noir diffère de la formule de Bekenstein-Hawking par la présence de la constante de Newton effective $G_h$ dans \eqref{entropy-5} à la place de la constante de gravitation. La raison est que, dans la représentation d'Einstein, la métrique présente un facteur conforme \eqref{bhminimal-4} et l'aire de l'horizon est alors donnée par $\mc A^{\text{min}}_h=(G/G_h)\si r_h^2$, conduisant ainsi à la formule habituelle $\mc A_h^\text{min}/4G$ pour l'entropie dans cette représentation (voir aussi la discussion dans \cite{Martinez:2005di}).
\bs

Nous pouvons désormais appliquer nos résultats aux trous noirs de la section \ref{sec4:axionique} en commençant par le trou noir chevelu \eqref{hairy-4}-\eqref{hairyfields-4}. Sa métrique donne  $f_1=-2p^2G\mu$ et la masse \eqref{mass-5} est alors
\BE M=\frac\si{4\pi}\,p^2\mu \ .\EE
Sa température est obtenue en exigeant l'absence de singularité conique pour la métrique euclidienne en $r=r_h$, conformément à l'équation \eqref{conical-5},  
\BE T=\frac{V'(r_h)}{4\pi}=\frac{1}{\pi\ell^2}\left(r_h-\frac{|p| l}{2}\right) \EE
et son entropie est donnée par
\BE S=\frac{\si r_h^2}{4G_{h}} \ . \EE
Les charges de Maxwell et axioniques peuvent être déterminées par leurs développements asymptotiques et en utilisant \eqref{entropy-5} pour trouver
\BE Q=\frac{\si q}{4\pi} \qquad\text{et}\qquad Q_i=\frac{\si p}{\sqrt{8\pi G}} \ .\EE
Les potentiels correspondants sont donnés par
\BE \Phi=\frac q{r_h} \qquad\text{et}\qquad \Psi_i=-\frac p{\sqrt{8\pi G}}\lp r_h+\frac{2\pi G}{3\al l^2}\frac{G^2\mu^2}{r_h+G\mu}\rp \ .\EE
\bs

L'autre solution en compétition dans cet ensemble est le trou noir axionique de Reissner-Nordstrom-AdS \eqref{RN1}-\eqref{RN2} pour lequel le champ scalaire $\phi$ est nul. Cette fois-ci $f_1=-2G\mu$ et sa masse est donnée par 
\BE M=\frac\si{4\pi}\mu \ .\EE
Sa température et son entropie sont
\BE T=\frac1{2\pi l^2}\lp r_h+\frac{G\mu l^2}{r_h^2}-\frac{Gq^2 l^2}{r_h^3}\rp \qquad\text{et}\qquad S=\frac{\si r_h^2}{4G} \ .\EE
L'entropie est celle de Bekenstein-Hawking puisque $G_h=G$ pour cette solution. Enfin, les charges et les potentiels sont  
\BE
Q=\frac{\si q}{4\pi}\ ,\qquad
Q_i=\frac{\si p}{\sqrt{8\pi G}}\ ,\qquad
\Phi=\frac q{r_h}\ ,\qquad
\Psi_i=-\frac{pr_h}{\sqrt{8\pi G}}\ .
\EE
Nous concluons cette section avec une dernière remarque au sujet des solutions avec un champ scalaire $\phi$ constant données par \eqref{phiRN1}-\eqref{phiRN2}. Il semble qu'il ne soit pas possible de trouver un ensemble thermodynamique commun pour comparer ces solutions avec les précédentes, en effet les conditions aux bords du champ scalaire doivent être modifiées.

			\subsection{Transitions de phase\label{sec::phase}}
Avec les résultats de la sous-section précédente, nous pouvons attaquer l'étude des diagrammes de phase des trous noirs précédents. Par simplicité, nous travaillerons dans un ensemble thermodynamique dans lequel les quantités $(T,\Phi,Q_i)$ sont fixes. Ainsi, le potentiel thermodynamique $\mc G$ pertinent, aussi appelé énergie libre de Gibbs, est obtenu par la transformation de Legendre de $W$ suivante
\BE \mc G(T,\Phi,Q_1,Q_2)=W+\Psi_1 Q_1+\Psi_2 Q_2=M-TS-\Phi Q \ . \EE
En s'inspirant de \cite{Martinez:2010ti}, il est judicieux d'introduire la constante sans dimension 
\BE a=\frac{2\pi G}{3\al l^2} \qquad\text{avec}\qquad a\geq1  \EE
et de travailler avec les autres quantités adimensionnées définies par\footnote{Les charges $Q_1$ et $Q_2$ ne sont pas indépendantes dans cet ensemble ; pour les solutions que nous avons, elles doivent être égales, $Q_1=Q_2$, et par conséquent nous avons simplement $\varpi=|p|$. Nous les noterons généralement $Q_i$.}
\BE \varpi=\frac{\sqrt{4\pi G}}{\si}\sqrt{Q_1^2+Q_2^2} \qquad\text{et}\qquad \xi=\frac{2\pi l T}{\varpi} \ . \EE
Dans ces variables, la condition \eqref{ghost-4} traduisant la positivité de la constante de Newton effective, et donc de l'entropie aussi, pour la solution chevelue \eqref{hairy-4}-\eqref{hairyfields-4} impose les inégalités suivantes sur $\xi$
\BE \frac{\sqrt a-1}{\sqrt a+1}<\xi<\frac{\sqrt a+1}{\sqrt a-1} \ . \label{xireg-5}\EE
De plus, cette solution de trou noir existe seulement pour des paramètres qui satisfont \eqref{q-4}. Cette contrainte, écrite avec les variables thermodynamiques de l'ensemble que nous considérons, est
\BE \Phi^2=\frac{\varpi^2}{4G}(a-1)(\xi-1)^2  \label{consphi-5}\EE
et peut alors être utilisée pour éliminer le potentiel de Maxwell $\Phi$. Après un peu d'algèbre, le potentiel de Gibbs pour la solution chevelue de trou noir \eqref{hairy-4}-\eqref{hairyfields-4} est donné par
\BE \mc G_{\rm chevelu}=-\frac{\si l\varpi^3}{32\pi G} \lb (\xi+1)^2+a(\xi-1)^2\rb \ .\EE
Un calcul analogue pour la solution axionique du trou noir de Reissner-Nordstrom-AdS, en se restreignant aux solutions vérifiant la contrainte \eqref{consphi-5}, conduit au potentiel de Gibbs suivant
\BE \mc G_{\rm RN}=-\frac{\si l\varpi^3}{108\pi G}\lb
\xi^3+\frac{9\xi}{2}\lp1+\frac14(a-1)(\xi-1)^2\rp+\lp\xi^2+3\lp1+\frac14(a-1)(\xi-1)^2\rp\rp^{3/2}\rb.
\EE

Notons que la charge axionique adimensionnée $\varpi$ apparaît seulement par un facteur $\varpi^3$ dans $\mc G_{\rm chevelu}$ et $\mc G_{\rm RN}$. Ainsi, dans cet ensemble, le diagramme de phase ne dépend que de la variable adimensionnée $\xi$ et est déterminé par le signe de $\Delta\mc G=\mc G_{\rm RN} - \mc G_{\rm chevelu}$. C'est la solution qui possède l'énergie libre de Gibbs la plus faible qui est favorisée. Une observation intéressante est que, pour le cas $\varpi=1$, la différence d'énergie libre $\Delta\mc G$ se réduit à la différence d'énergie libre entre  la solution de Reissner-Nordstrom-AdS avec un horizon hyperbolique et le trou noir hyperbolique chevelu, comme cela est reporté dans \cite{Martinez:2010ti}. La raison est que les champs axioniques modifient de façon effective la courbure de l'horizon (voir \cite{Bardoux:2012aw}) et les  potentiels des trous noirs $f(r)$ coïncident pour les solutions axioniques avec des horizons plats où $\varpi=1$ et pour les solutions non-axioniques avec des horizons hyperboliques. Les propriétés thermodynamiques, et en particulier le diagramme de phase, pour les trous noirs vérifiant $\varpi=1$ sont donc les mêmes que celles des trous noirs hyperboliques obtenues dans \cite{Martinez:2010ti}. Par conséquent, puisque la dépendance en $\varpi$ a lieu seulement par un facteur commun $\varpi^3$ dans les énergies libres, les propriétés thermodynamiques peuvent être déduites par un changement d'échelle de $\xi$, ou de manière équivalente de la température, pour toutes les autres valeurs $\varpi\neq1$.

\begin{figure}[tb]
\centerline{\includegraphics[scale=1.3]{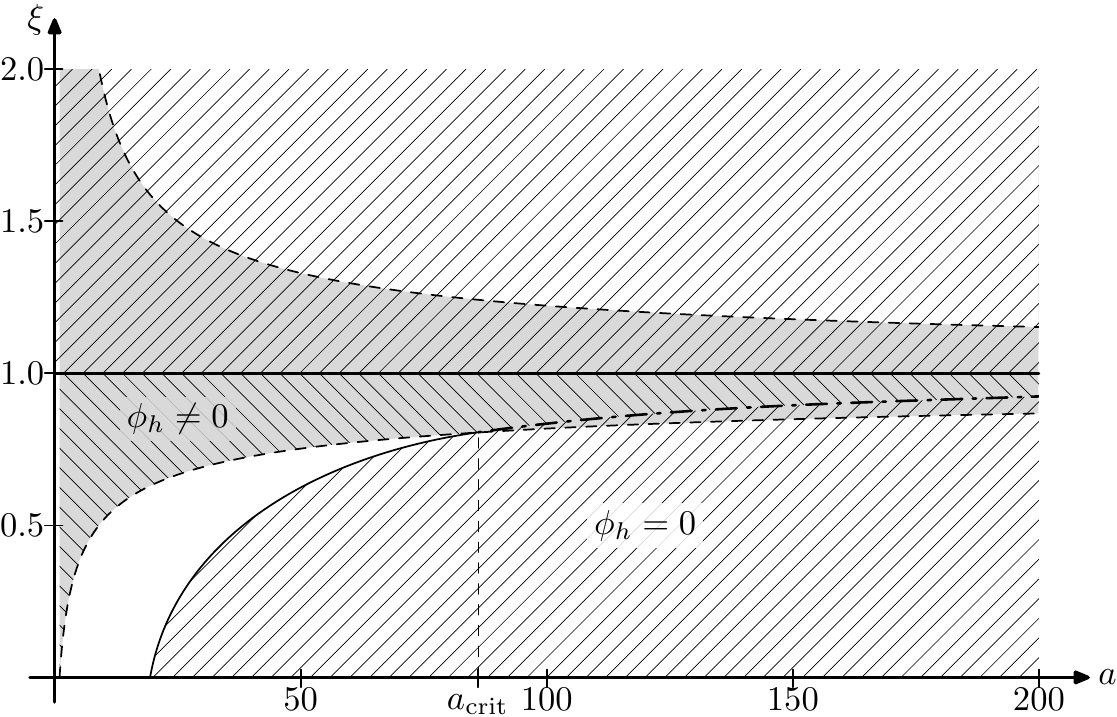}}
\caption[]{{\sc Phases en fonction de la constante de couplage $a$.} Dans la région grisée, les trous noirs chevelus dont l'entropie est positive sont en compétition avec les trous noirs axioniques de Reissner-Nordstrom-AdS. Ils dominent dans la région hachurée par
\protect\raisebox{-2pt}{\includegraphics[width=1.8em]{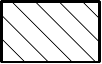}} ; tandis que les trous noirs axioniques de Reissner-Nordstrom-AdS dominent dans l'autre région hachurée par \protect\raisebox{-2pt}{\includegraphics[width=1.8em]{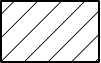}}. Ils sont séparés par une transition de phase du second ordre, représentée par la ligne continue en $\xi=1$ ; et à des températures inférieures par une transition de phase du premier ordre, représentée par une ligne en pointillés. Dans la région blanche, les trous noirs chevelus ont une entropie négative et une énergie libre inférieure à celle des trous noirs axioniques de Reissner-Nordstrom-AdS.}
\label{fig1}\end{figure}

Tout d'abord, remarquons que $\Delta\mc G=0$ pour $\xi=1$, ce qui correspond à une température critique
\BE T_{\rm crit}=\sqrt{\frac G\pi}\frac{\sqrt{Q_1^2+Q_2^2}}{\si l} \EE
pour laquelle une transition de phase du second ordre a lieu. Pour $T>T_{\rm crit}$, le trou noir axionique de Reissner-Nordstrom-AdS, pour lequel le champ scalaire est nul, domine. En abaissant la température en dessous de la température critique, c'est la solution chevelue qui domine pour $T<T_{\rm crit}$ ; le champ scalaire est ainsi à une valeur finie $\phi_h$ sur l'horizon, qui peut être notamment utilisé comme paramètre d'ordre de la transition (voir figure \ref{fig1}).

\begin{figure}[tb]
\centerline{\quad\includegraphics[scale=1.1]{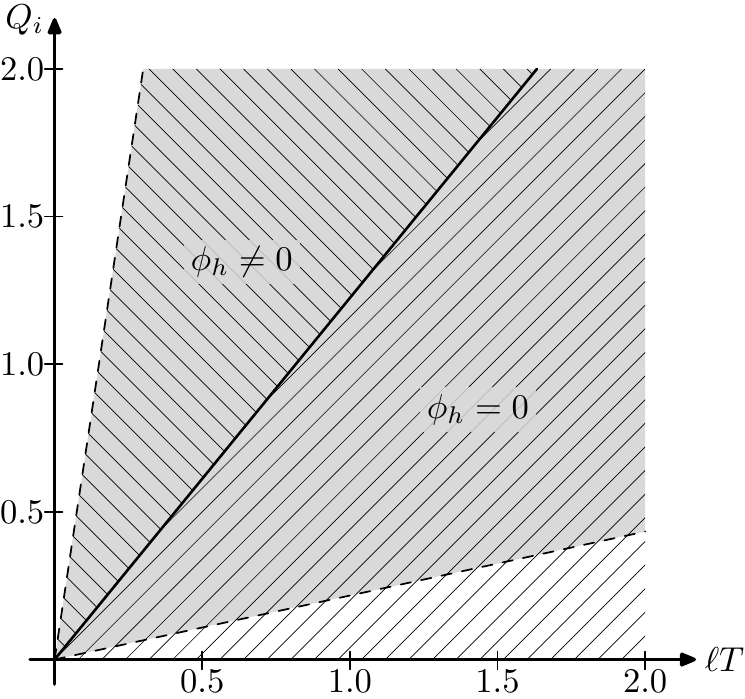}\hfill\includegraphics[scale=1.1]{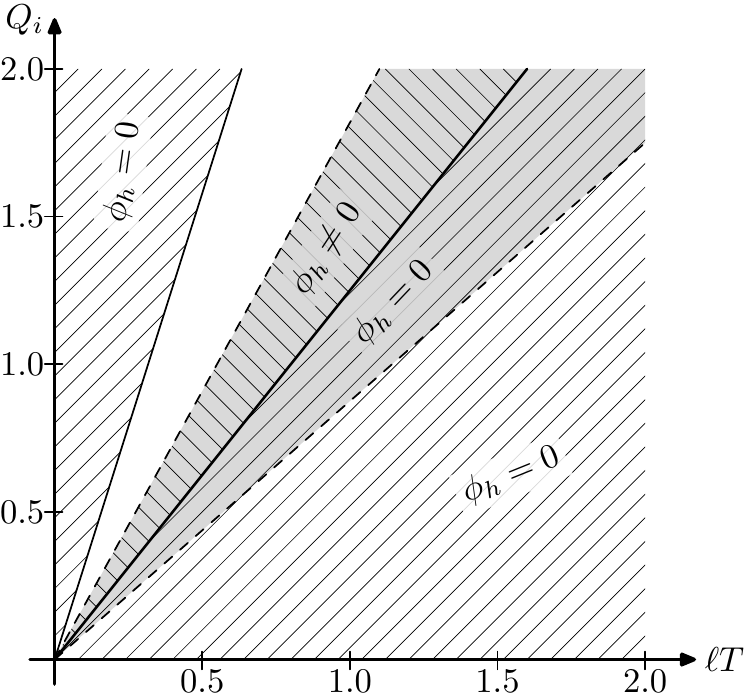}\quad}
\caption[]{{\sc Diagramme de phase dans le plan $(T,Q_i)$.} 
Pour $a$ faible et lorsque la température est diminuée, nous passons d'une phase dominée par un trou noir axionique de Reissner-Nordstrom-AdS à une phase dominée par le trou noir chevelu, à travers une transition de phase du second ordre (représentée par une ligne continue épaisse dans les diagrammes) et le paramètre d'ordre $\phi_h$ prend une valeur non nulle durant ce processus. En continuant à diminuer la température, le trou noir chevelu possède une entropie négative et une énergie libre inférieure à celle du trou noir axionique de Reissner-Nordstrom-AdS (la région blanche). 
Ceci est illustré pour $a=2$ par le diagramme de gauche et reste valide pour $1<a<3 (3 + 2\sqrt{3})$. Pour $3 (3 + 2\sqrt{3})<a<a_{\rm crit}$ avec $a_{\rm crit}\simeq 86$ et à très basses températures, nous trouvons une autre phase dominée par les trous noirs chauves pour lesquels $\phi_h=0$. Afin d'illustrer cette situation, le cas $a=30$ est représenté par la figure de droite. Dans ces diagrammes, la région grisée indique la région dans laquelle le trou noir chevelu et celui chauve coexistent, la région hachurée par \protect\raisebox{-2pt}{\includegraphics[width=1.8em]{hairy.pdf}} correspond à la phase chevelue pour laquelle $\phi_h\neq0$ et l'autre région hachurée par \protect\raisebox{-2pt}{\includegraphics[width=1.8em]{rn.pdf}} correspond au trou noir axionique de Reissner-Nordstrom-AdS où $\phi_h=0$.}
\label{fig2}\end{figure}

\begin{figure}[htb]
\centerline{\includegraphics[scale=1.3]{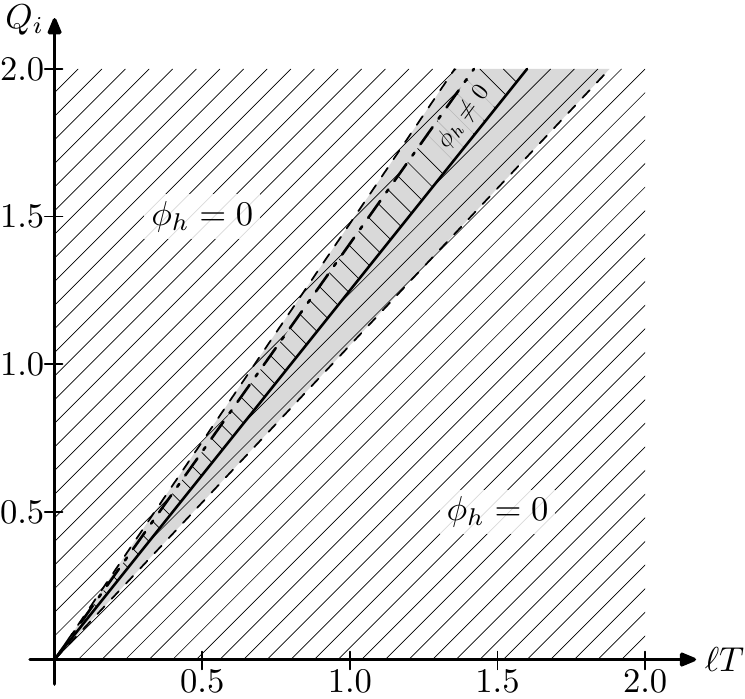}}
\caption[]{{\sc Diagramme de phase dans le plan $(T,Q_i)$ pour une faible constante de couplage $\al$ ($a>a_{\rm crit}$).} 
Dans ce cas, tout point du diagramme de phase est dominé par un trou noir dont l'entropie est positive : le trou noir axionique de Reissner-Nordstrom-AdS dans la région hachurée par \protect\raisebox{-2pt}{\includegraphics[width=1.8em]{rn.pdf}} et le trou noir chevelu dans la région hachurée par \protect\raisebox{-2pt}{\includegraphics[width=1.8em]{hairy.pdf}}. La transition de phase à haute température est du second ordre (ligne continue) et celle à basse température est du premier ordre (ligne en pointillés). Dans la région grisée, les deux phases coexistent. Le diagramme représente ici le cas $a=150$.}
\label{fig3}\end{figure}

Lorsque la température continue à diminuer, une seconde transition de phase ($\Delta\mc G=0$) apparaît quand $a>3(3+2\sqrt{3})\simeq 19.4$, ce qui signale que c'est de nouveau le trou noir axionique de Reissner-Nordstrom-AdS qui est thermodynamiquement favorisé. Tant que $a>a_{\rm crit}$ avec $a_{\rm crit} \simeq 86$ \cite{Martinez:2010ti}, elle correspond à une véritable transition de phase de premier ordre. Cependant, pour des valeurs inférieures du paramètre $a$, la phase chevelue possède une entropie négative, puisque les inégalités \eqref{xireg-5} ne sont pas vérifiées, et il n'est alors pas clair que l'état d'équilibre est correct dans cette région.

Enfin, les diagrammes de phase sont résumés dans les figures \ref{fig2} et \ref{fig3} pour différentes valeurs des constantes de couplage dans le plan $(T,Q_i)$ où $Q_i$ désigne la charge axionique.
\bs

En somme, nous avons présenté deux familles de solutions de la théorie \eqref{action-4} à la section \ref{sec4:axionique} : une famille de trous noirs présentant un cheveu scalaire secondaire \eqref{hairy-4}-\eqref{q-4} et une autre qui est la version axionique des trous noirs de Reissner-Nordstrom-AdS \eqref{RN1}-\eqref{RN2}. Ces deux familles de solutions possèdent la particularité d'avoir un horizon plat grâce à l'introduction de deux champs axioniques dans la théorie. Cette caractéristique les rend ainsi intéressant en vue de possibles applications dans le domaine des supraconducteurs via l'holographie \cite{Koutsoumbas:2009pa,Myung:2010rb}. Cependant, nous n'avons pas exploré cette direction. A travers la dernière section de ce chapitre, nous avons présenté les propriétés thermodynamiques de ces solutions qui sont en compétition. En fonction de la différence d'énergie libre de ces solutions, nous avons donné les différents diagrammes de phase dans le plan $(T,Q_i)$ pour diverses valeurs de la constante de couplage $\al$. Retenons que dans le cas électriquement neutre ($a=1$), la solution de Reissner-Nordstrom-AdS qui domine à haute température subit une transition de phase du second ordre de telle façon que ce soit le trou noir chevelu qui domine à basse température ; le champ scalaire se comporte ainsi comme l'aimantation d'un matériau durant une transition paramagnétique/ferromagnétique.

\chapter*{Perspectives \& Conclusions\markboth{\textit{Perspectives \& Conclusions}}{Perspectives \& Conclusions}} 
\phantomsection\addcontentsline{toc}{chapter}{\protect\numberline{}Perspectives \& Conclusions} 

En somme, nous avons exposé dans cette thèse de nouvelles solutions de trous noirs pour des théories modifiées de la gravitation. Après avoir présenté la théorie de la Relativité Générale au premier chapitre, nous avons considéré au chapitre \ref{BH in higher D} la théorie d'Einstein en dimension quelconque en présence d'une constante cosmologique négative et en incluant des champs de $p$-formes libres. Ces champs nous ont notamment permis de modeler à notre guise la géométrie intrinsèque de l'horizon des événements d'un trou noir en fonction du contenu en champs choisis. En particulier, la présence de champs axioniques permet de construire des trous noirs dont la géométrie intrinsèque de l'horizon est plate alors que la fonction potentiel associée est celle d'un trou noir dont la géométrie de l'horizon est hyperbolique en l'absence de ces champs axioniques. C'est cette dernière remarque que nous avons largement exploitée dans les travaux que nous avons exposés à la fois dans le cadre de la théorie d'Einstein-Gauss-Bonnet au chapitre \ref{EGB} et en tenant compte aussi de la présence d'un champ scalaire couplé conformément à la gravitation au chapitre \ref{chapter-conforme}.

Il serait intéressant de voir si les trous noirs topologiques rencontrés au chapitre \ref{BH in higher D} admettent une version en rotation. Dans le cadre de la Relativité Générale en présence d'une constante cosmologique négative, des trous noirs topologiques en rotation sont bien connus pour être des solutions du vide \cite{Klemm:1997ea}. Est-il possible de vêtir ces derniers trous noirs par deux champs axioniques afin de mettre en rotation nos nouvelles solutions ? Ceci serait une première étape avant de rechercher toutes les géométries de type D décrites par la métrique de Pleba\'nski-Demia\'nski en présence de champs axioniques. Un autre sujet intéressant à étudier serait l'influence des champs de $p$-formes sur la stabilité des nouvelles solutions du chapitre \ref{BH in higher D}. Dans le vide en théorie d'Einstein en dimension quelconque, ces trous noirs sont connus pour être instables \cite{Gibbons:2002th}, est-ce toujours le cas en présence de matière ?

Concernant les solutions rencontrées au chapitre \ref{EGB}, un projet possible serait d'étendre la classification des solutions en dimension quelconque dans le cadre de la théorie de Lovelock en l'absence de matière dans un premier temps. Nous nous attendons à des contraintes supplémentaires sur la géométrie intrinsèque de l'horizon dans ce cas.

Cette thèse nous a également permis de présenter quelques résultats nouveaux qui demandent à être analysés plus en détails avec notamment les structures de barres dans le cadre de la gravitation d'Einstein en présence d'un champ scalaire conforme au chapitre \ref{chapter-conforme}. En plus, de donner la structure de barres de solutions déjà connues dans ce contexte, nous avons généré une version NUT de la solution de BBMB à l'aide de ce formalisme à la sous-section \ref{NUT-BBMB-4}. Il serait intéressant de reprendre cette dérivation en préservant la relation $\ga=-\sqrt{3}\om$ entre les potentiels des barres et d'étendre à la solution de BBMB la dérivation de F. Ernst qui permet de dériver la solution de Kerr à partir de la solution de Schwarzschild en Relativité Générale.
\bs

Je laisse ici ces quelques pistes d'investigations à tous lecteurs intéressés par la gravitation. 

\newpage
\addcontentsline{toc}{chapter}{Bibliographie}
\bibliography{biblio}
\bibliographystyle{utphys}
\end{document}